\documentclass{uiucthesis2021}

\usepackage{enumerate}

\usepackage{subfiles}

\usepackage{comment}
\usepackage[normalem]{ulem}

\usepackage{graphicx, color}
\usepackage{bmpsize}
\usepackage{ascmac}

\usepackage[hang,small,bf]{caption}
\usepackage[subrefformat=parens]{subcaption}
\usepackage{physics,slashed}
\usepackage{amsmath,amssymb,mathrsfs,amsthm}
\usepackage{mathtools}
\allowdisplaybreaks

\numberwithin{equation}{section}
\numberwithin{figure}{section}

\def\rmM{{\rm M}}

\def\rmO{{\rm O}}

\def\calD{{\cal D}}
\def\calE{{\cal E}}

\def\calG{{\cal G}}
\def\calH{{\cal H}}

\def\calL{{\cal L}}

\def\calO{{\cal O}}
\def\calP{{\cal P}}

\def\calS{{\cal S}}

\def\calU{{\cal U}}
\def\calV{{\cal V}}

\def\MeV{{\rm ~ MeV}}
\def\GeV{{\rm ~ GeV}}
\def\fm{{\rm ~ fm}}
\def\km{{\rm ~ km}}

\def\rmi{{\rm i}}

\def\msbar{{\overline{\rm MS}}}
\def\diag{{\rm diag \,}}

\def\braket#1{\left\langle #1 \right\rangle}

\def\norm<#1|#2>{\left\langle #1 \middle| #2 \right\rangle}
\def\expv<#1|#2|#3>{\left\langle #1 \middle|\, #2 \,\middle| #3 \right\rangle}

\usepackage{bm}
\def\vx{{\bm{x}}}
\def\vy{{\bm{y}}}
\def\vk{{\bm{k}}}
\def\vp{{\bm{p}}}

\def\psibar{\overline{\psi}\,}
\def\Phibar{\overline{\Phi}\,}
\def\ubar{\overline{u}\,}
\def\dbar{\overline{d}\,}
\def\qbar{\overline{q}\,}

\def \alert#1 {{\color{magenta} {#1}}}

\def \redflag#1{\textcolor{red}{#1}}
\def \blueflag#1{\textcolor{blue}{#1}}
\def \redout#1{\redflag{\sout{#1}}}
\def \blueout#1{\blueflag{\sout{#1}}}

\def \todo#1{\textcolor{red}{\bf{TODO: #1}}}

\def\figref#1{Fig.\,\ref{#1}}
\def\eqref#1{Eq.(\ref{#1})}
\def\eqsref#1#2{Eqs.(\ref{#1})-(\ref{#2})}

\def\Det{{\rm Det}}

\renewcommand{\redflag}[1]{#1}
\renewcommand{\blueflag}[1]{#1}
\renewcommand{\blueout}[1]{}
\renewcommand{\redout}[1]{}

\usepackage{afterpage}
\newcommand\blankpage{%
    \null
    \thispagestyle{empty}%
    \addtocounter{page}{-1}%
    \newpage}

\usepackage[utf8]{inputenc}
\usepackage{CJKutf8}
\usepackage{csquotes}
\usepackage{microtype}

\usepackage[
    backend=bibtex,
    sorting=none,
    style=phys,
    biblabel=brackets,
    pageranges=false,
    url=false,
    hyperref=auto,
    giveninits=false,
    sortcites=true,
    isbn=false,
    doi=false,
    eprint=true,
    giveninits=true
]{biblatex}
\usepackage[
    bookmarksdepth=3,
    linktoc=all,
    colorlinks=true,
    urlcolor=magenta,
    linkcolor=magenta,
    citecolor=blue,
    bookmarks=true,
    bookmarksnumbered=true,
    breaklinks,
]{hyperref}

\newcommand\fH[1]{\sbox0{#1}\dimen0=\ht0 \advance\dimen0 -1ex
  \sbox2{\'{}}\sbox2{\raise\dimen0\box2}%
  {\ooalign{\hidewidth\kern.1em\copy2\kern-.5\wd2\box2\hidewidth\cr\box0\crcr}}}

\newcounter{counterforappendices}

\begin{document}

\titleEN{Analysis of QCD at finite isospin density: on the relationship between quark degrees of freedom in hadrons and equation of state}
\titleJP{\begin{CJK}{UTF8}{ipxm}有限アイソスピン密度系のQCDの解析：ハドロン中のクォーク自由度と状態方程式の関係について\end{CJK}}
\author{Ryuji Chiba}
\department{Physics}
\degreeyear{2023}
\maketitle

\afterpage{\blankpage}
\frontmatter
\begin{abstract}

\begin{comment}
The equation of state (EOS) of cold and dense QCD matter with both hadron and quark degrees of freedom is closely related to the Neutron Star (NS) physics,
the sign problem of the lattice QCD prevent the understanding of the transition region from hadrons to quarks.
\end{comment}
%
In this thesis
we investigate the quark contribution to the equation of state (EOS) of the isospin QCD matter 
using the two-flavor quark meson model at finite isospin density.
This model includes the quark degrees of freedom through the lowest order of the loop correction.
The renormalizability of the model ensures that it is free from high density artifacts present in the cutoff model, such as the NJL model.
This model describes the crossover of the pion condensate from the Bose-Einstein condensation (BEC) phase at low density to the Bardeen-Cooper-Schrieffer (BCS) phase at high density.
In the absence of the quark degrees of freedom
the pion condensate behaves as the bosonic object,
but the quark substructure becomes important and suppress the pion condensate by Pauli blocking as density increase.

As the isospin density increases,
the EOS rapidly becomes stiff and approaches to the quark matter
even before pion starts to overlap.
Consequently,
the sound velocity exceeds the conformal value $c_s^2 = 1/3$,
forming a peak structure
and then relax to the conformal value from above at high density.
This is in good agreement with the recent lattice QCD result.
In contrast, the perturbative QCD result suggests that the sound velocity approaches to the conformal value from below.
This discrepancy comes from the non-perturbative effects arising from the pion condensate or the quark-antiquark correlation near the Fermi surface in the quark meson model.
We also investigated the trace anomaly as another measure of conformality
and found that the non-perturbative correction can make the trace anomaly negative,
in contrast to the recent conjectures based on the perturbative QCD.
This discrepancy highlights the importance of non-perturbative corrections even at high density,
where perturbative treatments are believed to work well. 

The effects of finite temperature on sound velocity were also examined.
On the isentropic trajectory, where $s/n_I$ is fixed,
which is good approximation of the supernova explosion or heavy ion collision,
the thermal quarks were found to suppress the sound velocity and smear out the peak structure.
The thermal mesons are considered to be important at zero density from lattice QCD,
which is confirmed in this model calculation.
The analysis is further extended to the condensed phase at high density, 
where it is found that thermal mesons do not significantly contribute to the EOS.

\end{abstract}

%\begin{dedication}
%To My Family.
%\end{dedication}
\begin{acknowledgments}
First of all, I am deeply grateful to my supervisor, Prof. Toru Kojo, 
for his invaluable guidance and expertise throughout the entire research process. 
His mentorship has been instrumental in shaping the direction and quality of this thesis.

I extend my sincere appreciation to Prof. Daiki Suenaga, 
for his instruction and discussions on the finite temperature effects and the hadron mass spectrum.

I would like to thank Prof. B. B. Brandt and Prof. G. Endr\fH{o}di 
for kindly providing us with their lattice data in Ref.\cite{Brandt:2022aa},
and Prof. R. Abbott and his collaborators 
for their kindness in providing the lattice data in Ref.\cite{Abbott:2023aa}.

I would also like to express my gratitude to all the member of the nuclear theory group in Tohoku University
for their collaborative spirit and insightful contributions. 
Special thanks to Yugo Kurebayashi 
for the invaluable discussions and for reviewing the manuscript of this thesis.

Heartfelt thanks go to my family members for their unending love, understanding, and support. 
I am especially grateful to my father for providing me with the opportunity to pursue higher education and for his long-time support.
\end{acknowledgments}

\afterpage{\blankpage}
{
    \hypersetup{linkcolor=black}  % disable link coloring locally
    \tableofcontents
}

\mainmatter

\chapter{Introduction}
\label{ch:intro}
Quantum chromodynamics (QCD) is a part of the standard model of the particle physics
which describes the strong interaction between quarks and gluons.
The confinement does not allow the colored objects
and the quarks are confined in the hadrons at low energies.
As the energy increases,
in the low temperature and high density region,
the LQCD calculation is not available because of the sign problem.
The experimental approach is also difficult.

Such cold and dense region is related to the Neutron Star (NS) physics.
The macroscopic properties such as masses and radii of NSs are related to the equation of states (EOS) of the dense QCD matter
which reflects the microscopic matter properties.
There are several studies of EOS at low density based on the nuclear theory and at high density based on the perturbative QCD.
But in the middle of the two different degrees of freedom,
the theoretically unified treatment is difficult and the large uncertainty still remains.

In this thesis 
we discuss the cold and dense QCD matter EOS with both the hadron and the quark degrees of freedom
based on the isospin QCD.
Isospin QCD is the QCD with zero baryon chemical potential $\mu_B$ and the finite isospin chemical potential $\mu_I$.
It gains the attention as the laboratory of the finite density QCD
because the LQCD calculation at the finite density is available for the isospin QCD,
and the phenomenological concepts can be examined by the effective model approach in the comparison to the LQCD results.

The outline of this chapter is followings.
In Sec.$\,$\ref{sec:intro-qcd-phase} provides a brief overview of the QCD phase diagram.
Sec.$\,$\ref{sec:intro-ns} reviews dense QCD matter in Neutron Star and its connection to observations.
The importance of the sound velocity in constructing the EOS is discussed in Sec.$\,$\ref{sec:intro-sound-velocity}.
Sec.$\,$\ref{sec:intro-theoretical-finite-density} reviews the theoretical approaches to the finite density QCD from low density region to the high density limit.
The main themes of this thesis are summarized in the final section.

\section{QCD phase diagram}
\label{sec:intro-qcd-phase}
\begin{figure}[thpb]
	\centering
	\includegraphics[width=0.8\textwidth]{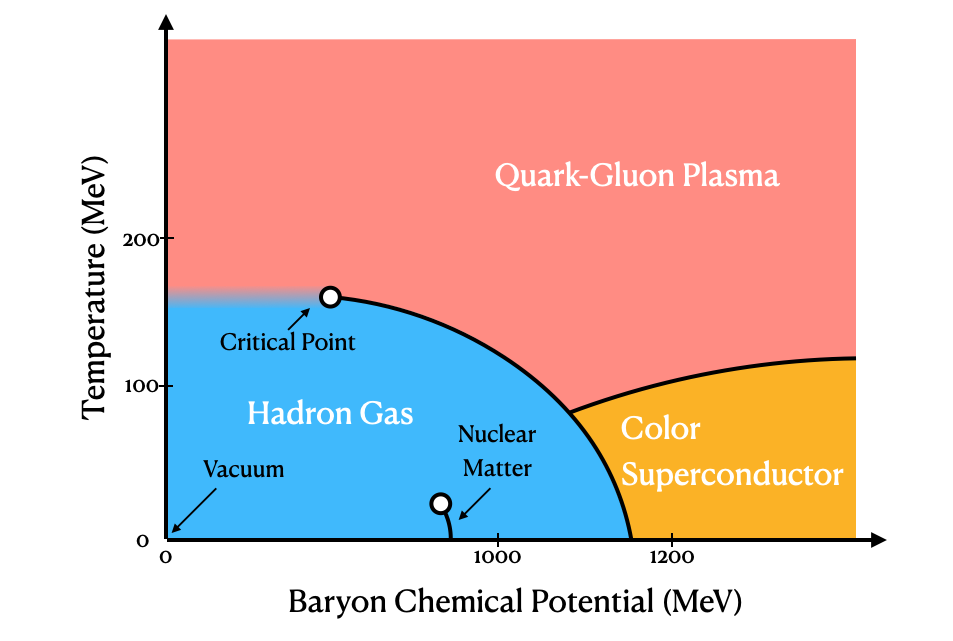}
	\caption{
		An example of QCD phase diagram.
	}
	\label{fig:intro-phase-diagram}
\end{figure}

The QCD phase is usually characterized by the temperature and baryon chemical potential
(\figref{fig:intro-phase-diagram}).
In low temperature and low density region the color confinement is realized
and hadrons are the effective degrees of freedom.
This region is called the hadronic phase.
As increase the temperature with keeping small baryon density,
the hadronic phase turns into the quark-gluon plasma (QGP).
This phenomenon is a consequence of the asymptotic freedom \cite{Gross:1973id},
wherein the coupling constant of QCD $\alpha_s$ becomes small at high density and the quarks are deconfined.
The transition from a hadronic phase to a QGP is considered to be crossover in the low-density region \cite{Aoki_2006,Bhattacharya:2014ara}
and it temperature is around $T_c = 155- 160 \MeV$ \cite{AOKI200646,Aoki_2009,Bazavov:2011nk}.
Increasing the density the transition changes from crossover to first-order phase transition.

\begin{figure}[thpb]
	\centering
	\includegraphics[width=0.8\textwidth]{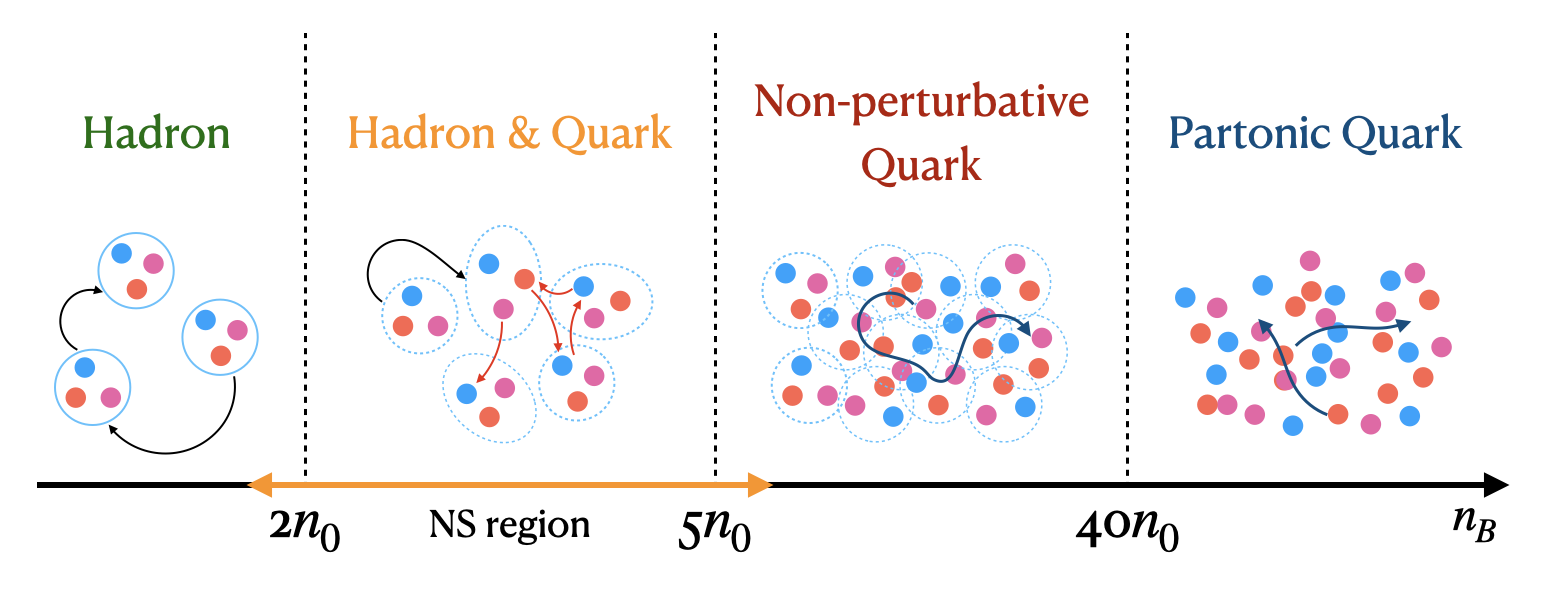}
	\caption{
		The effective degrees of freedom at the various density.
		The black lines illustrates the mesonic interactions and the red lines are the quark interactions.
		In the non-perturbative and partonic quark region
		quarks can propagate in the medium which is illustrated by the blue lines.
	}
	\label{fig:intro-effective-dof}
\end{figure}

Increasing the density with keeping low temperature also changes the effective degrees of freedom 
(\figref{fig:intro-effective-dof}).
In low density region the quarks are confined and does not have direct contribution to the dynamics.
Here the hadrons are the effective degrees of freedom and baryons interacts each other through the meson exchange.

At the density range $n_B \simeq 2 - 5 n_0$,
with the nuclear saturation density $n_0 \simeq 0.16 \fm^{-3}$,
the quark exchange among hadrons appears while the baryon still does not overlap.
Here both hadrons and quarks have the comparable role
and it is quite difficult to consider the effective degrees of freedom.
The transition from confined hadronic phase to deconfined quark phase has been considered to be the first-order phase transition.
But recently the possibility of the crossover transition or {\it quark hadron continuity} \cite{Alford:aa, Schaefer:aa}
is also considered as a plausible scenario.

In contrast,
in the higher density region the relevant degrees of freedom is clear.
In the dense region $5n_0 < n_B < 40 n_0$
the quark degrees of freedom are expected to be dominant.
The hadron overlap and the percolation of hadrons should take place \cite{CELIK1980128,BAYM1979131,Satz:aa},
then the quarks are able to propagate in the hadronic medium.
But the QCD coupling constant $\alpha_s$ is not much small to use the perturbation theory.
Here the constituent quark picture in the hadron physics is still effective
and some analysis based on the quark model and the renormalization group (RG) approach are conducted.
In denser region $n_B > 40 n_0$
the quarks interact weakly because of the asymptotic freedom
and the perturbative QCD (pQCD) approach is applicable.

\section{Neutron star}
\label{sec:intro-ns}
Low density and high temperature QCD is related to the early universe and heavy ion collision experiments.
After the birth of the universe and the big bang,
 the universe was filled with the QGP.
As the universe expanded and cooled down,
the QGP transitioned to the hadronic phase.
Heavy ion collision experiments attempt to recreate the QGP in the laboratory
by colliding heavy ions at high energy.

In contrast to the high temperature and low density,
cold and dense QCD matter has not yet been realized in a laboratory setting. 
However,
 neutron stars provide a natural laboratory for studying dense QCD matter.
NS is a compact object with the radius $R \simeq 10 \km$ and the mass $M \simeq 2 M_\odot$.
It forms through the gravitational collapse of a massive star and subsequent supernova explosion.
During this process, most of the electrons are captured by protons.
As a result, the NS is primarily composed of neutrons.

\begin{figure}
	\centering
	\includegraphics[width=0.8\textwidth]{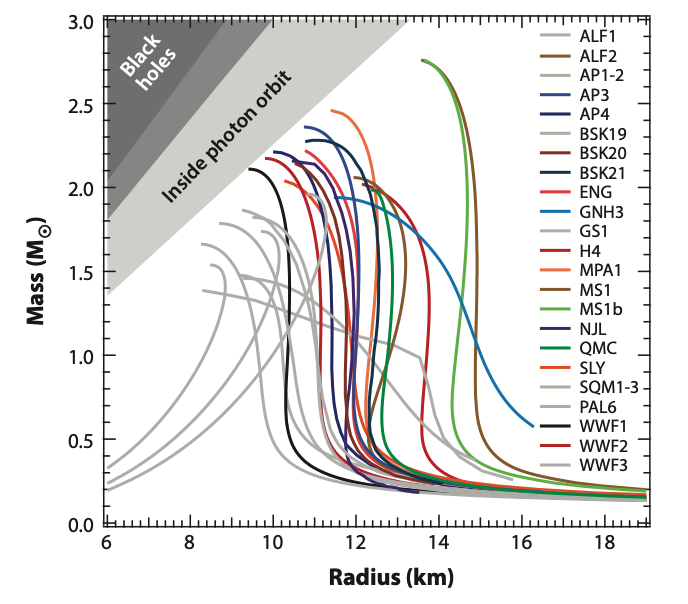}
	\caption{
		The neutron star mass-radius (MR) relation.
		Each lines corresponds to the single mass-radius relation for given EOS.
		This figure is originally from \cite{Ozel:2016oaf}.
	}
	\label{fig:intro-mr}
\end{figure}

The macroscopic properties of NSs are characterized by the mass and radius.
The relations between these properties is known as the mass-radius relation (M-R relation) as shown in \figref{fig:intro-mr}.
By observing individual NSs,
we can gather information about this relations.
This information is connected to the equation of states (EOS) of the NS matter.
The EOS is the relation between the pressure and energy density which corresponds to the stiffness of the matter.
The Einstein equation or its special case, Tolman-Oppenheimer-Volkoff (TOV) equation
\cite{Tolman:1939jz,Oppenheimer:1939ne} (see also Appendix \ref{app:TOV}),
provides the one-by-one correspondence between the EOS and M-R relation.

Recent observations discovered the NSs with the mass around $2 M_\odot$;
the binary millisecond pulsar J1614-2230 with $M = 1.928 \pm 0.017 M_\odot$ \cite{Fonseca_2016}
(the original value is $1.97 \pm 0.04 M_\odot$ \cite{Demorest_2010}),
and the pulsar J0348+0432 with $M = 2.01 \pm 0.04 M_\odot$ \cite{Antoniadis:2013aa}.
The radius of the J0740+6620 is observed by the Neutron Star Interior Composition Explorer (NICER),
which concluded as $R = 12.33^{+0.76}_{-0.81} \km$ or $R = 12.18^{+0.56}_{-0.79} \km$ \cite{Raaijmakers_2021}.
The huge mass and the small radius suggests the presence of the dense matter $n = \calO(1) n_0$.

Current observations of NSs suggest that the EOS of the dense matter is of the so-called {\it soft-to-stiff} type \cite{Kojo:2020krb};
the EOS is soft (small pressure at fixed energy density) at low density and becomes stiff (large pressure at fixed energy density) at high density.
In addition to the information of M-R relation,
recent observation of the gravitational wave from the binary neutron star merger GW170817 \cite{LIGOScientific:2017vwq}
provides the information of the stiffness of the EOS at $1-3 n_0$ \cite{Huang:2022aa,Radice:2017aa,Annala:2018aa,Annala:2019aa}.
The analysis of EOS using the bayesisan framework with astronomical data both of the M-R relation and the gravitational wave
are conducted, for instance \cite{Steiner:2010fz,Xie:2020rwg,Ayriyan:2021prr,Biswas:2021pvm}.

\begin{figure}[tpb]
	\centering
	\includegraphics[width=0.6\textwidth]{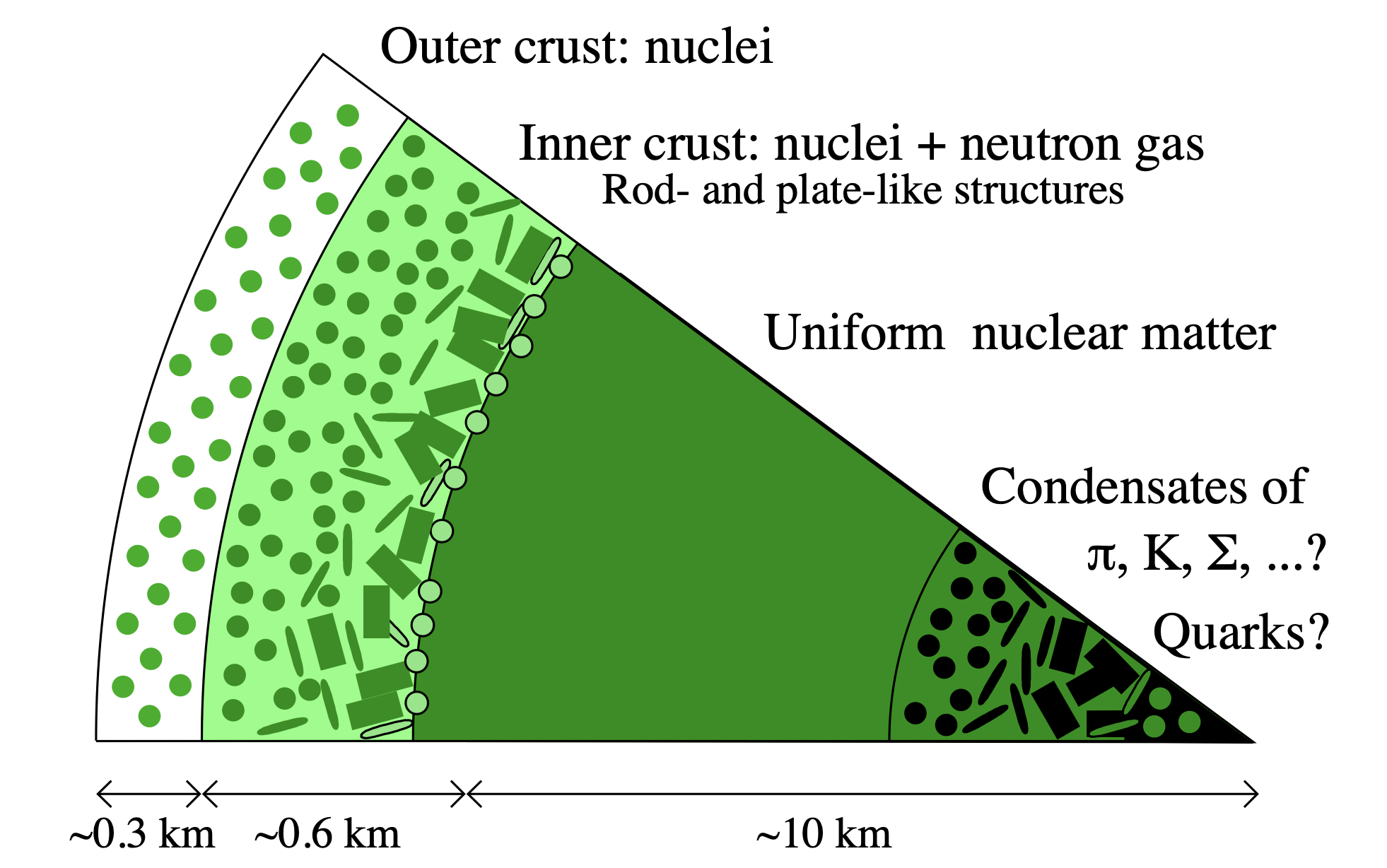}
	\caption{
		The inner structure of the NS.
		This figure is originally from \cite{Heiselberg:2002dk}.
	}
	\label{fig:intro-ns-structure}
\end{figure}

The matter in the NSs varies with depth and density (see \figref{fig:intro-ns-structure}).
The surface with $\sim 1 \km$ or density $n_B \lesssim 0.5 n_0$ is called crust,
wherein the nuclei, neutrons and electrons are coexist.
This region has the critical role on the character as the pulsar, quasi-periodic oscillation and so on,
but does not affect the mass of the NSs.
In the deeper and denser region, central core, the matter is considered as the uniform nucleon matter.
Its density reaches $n_B \sim 2 n_0$
and this region has the important role on the NS mass.
For larger NSs,
exotic phases with hyperon and mesons, or and quarks are considered to appear at the center of NS.

From the microscopic point of view,
it is important to establish the EOS of the dense matter from hadronic or quark degrees of freedom.
The low density EOS is firmly established around the nuclear saturation density $n_0$.
The first method contributing to the low density EOS is the potential model \cite{Akmal:1998cf,Togashi:2017mjp}
and chiral perturbation theory (Ch.PT) \cite{Drischler:2017wtt}.
The calculation is mainly conducted for pure neutron matter and symmetric nuclear matter wherein the proton and neutron are equally populated.
The second method is the relativistic mean field (RMF) theory \cite{Serot:1991st,Shen:1998gq,Oertel:2016bki}
where the baryons are treated as the fundamental degrees of freedom
and the parameters of the model are determined from the properties of the nuclear matter
such as binding energy, saturation density, symmetry energy and so on.

But these hadronic approach lack the quark degrees of freedom
and can not extend to the high density $n_B > 2n_0$.
Also they could not totally explain the NS observations.
Until 2010 the most of the NSs are observed as the mass around {\it canonical} value $1.4-1.5 M_\odot$ \cite{Hulse:1974eb,Taylor:1992,Thorsett_1999}
and the theoretical predictions supported these results.
Thus discovery of the $2.0 M_\odot$ NS \cite{Demorest_2010} has the huge impact on the dense QCD EOS.
The existence of the large mass NS indicates the stiff EOS at high density
and forces us to find the new mechanism of the stiffening.

As increasing the density,
the heavy hadrons such as hyperons can appear
but this gives rise to the serious problems.
The hyperons are the baryons with the strangeness and the mass is larger than the nucleon.
In the low density region they are not populated because of the large mass,
but as growing the density the lower energy states of nucleons or $u$- and $d$-quarks are filled
and then hyperons or $s$-quarks can appear.
But the two-body interactions among hyperons and nucleons are considered to be attractive.
For example
$\Lambda$-nuclei and nucleon ($\Lambda N$) interaction \cite{HASHIMOTO2006564},
$\Xi N$ interaction \cite{Nakazawa:2015joa}
and $\Lambda\Lambda$ interaction \cite{Takahashi:2001nm}
are all attractive.
This means inclusion of the hyperons leads the softening of EOS
and does not explain the existence of the $2.0 M_\odot$ NS.
This is so-called {\it hyperon puzzle} \cite{Chatterjee:2015aa,Weise:2023cua}.
One possible prescription to this puzzle is the repulsive three-body or many-body interaction among nucleons and hyperons
and some research has been conducted that incorporate the many body effects \cite{Zuo:2002aa,Gandolfi:2009aa}.

The necessity of the many-body interaction is considered as the signature of the emergence of the quark degrees of freedom.
As mentioned in the previous section,
the quark degrees of freedom gains the importance in the high density region $n_B > 2 n_0$
but the behavior of the quark matter varies with the density,
which makes the comprehensive understanding of the matter difficult.
One approach is to use the low and high density regions as boundary conditions and interpolate the intermediate region smoothly based on the quark-hadron crossover scenario.
In this case, the densities $n_B = 2n_0$ and $5n_0$ are chosen as the boundaries.

Despite the well-established nuclear theory at $n_B < 2n_0$ and pQCD at $n_B > 40n_0$,
there is still uncertainty in the intermediate density region $n_B \simeq 2 - 40 n_0$,
particularly in the NS region $n_B \simeq 2-5 n_0$.
Therefore, it is important to understand the dynamics of the QCD matter at intermediate density
especially the stiffening of the EOS.
As an increasing amount of data on NSs, both quality and quantity, is being accumulated, 
it is crucial to construct the EOS of dense matter from a microscopic perspective.

\section{Sound velocity}
\label{sec:intro-sound-velocity}
When constructing the EOS,
we have to consider the another constraint in addition to the $2.0 M_\odot$ constraint.
That is the causality constraint though the sound velocity $c_s$.
Sound velocity is defined by the slope of the pressure $P$ as a function of energy density $\varepsilon$
as $c_s^2 = \pdv*{P}{\varepsilon}$,
which contains the information of the stiffening of the EOS.
As the sound velocity of the matter must be smaller than the light velocity,
the causality constraint is given by $c_s \leq 1$ by setting the light velocity $c = 1$.
The massive NS favors the stiff EOS at high density,
but taking the causality constraint into account,
the too rapid stiffening of the EOS is prohibited.

As we can calculate the sound velocity for given pressure as a function of energy,
it is useful to examine what degrees of freedom contribute to the stiffening of the EOS.
First we consider a simple parametrization of the expression of the energy density $\varepsilon$ 
as a function of baryon number density $n_B$ as
\begin{align}
	\varepsilon = a n_B^{4/3} + b n_B^\alpha.
\end{align}
The first term is the contribution of the relativistic free quark gas
and the next term is the interaction term.
As the chemical potential and pressure are given by $\mu_B = \pdv*{\varepsilon}{n_B}$ and $P = \mu_B n_B - \varepsilon$,
we obtain
\begin{align}
	P = \frac{1}{3} \varepsilon + b \qty(\alpha - \frac{4}{3}) n_B^\alpha.
\end{align}

In the non-relativistic limit $a \ll b$ the second term becomes dominant
and the repulsive interaction ($b>0$) leads the stiffening of the EOS in the case of $\alpha > 4/3$ .
The value of $\alpha$ corresponds to the $\alpha$-body interactions in dilute limit,
thus we can conclude that $N\geq 2$-body repulsive interactions contribute to the stiffening.

On the contrary,
the attractive interactions for $\alpha < 4/3$ also can lead the stiff EOS.
One example is the pairing effects of two quark and/or antiquark near the Fermi surface
which energy is $\sim n_B^{2/3} \sim p_F^2 \sim \mu^2$ for the Fermi momentum $p_F$ and the chemical potential $\mu$.
Such correlation is considered in the context of the isospin QCD and two-color QCD (${\rm QC_2D}$, \cite{Cotter:2012mb,Iida:2019rah,Kojo:2021hqh})
and can be extended to the three quarks \cite{McLerran:2007qj}.
This density dependence shows that the attractive interaction $b<0$ can lead the stiffening of the EOS against the naive expectation.
This suggests that the density dependence of the interaction is important for the stiffening of EOS
as well as whether the interaction is attractive or repulsive.

Finally in the relativistic limit $a \gg b$
the sound velocity approaches to the {\it conformal value} $c_s^2 = 1/3$.
The pQCD predict that the sound velocity approaches to the conformal value from below \cite{Leonhardt:2019fua}.

Recent LQCD results and the NS observation suggest the sound velocity exceed $c_s^2 =1/3$ around $n_B \sim 2-5 n_0$
and approaches to the conformal value at high density \cite{Altiparmak:2022bke, Abbott:2023aa}.
This means that the sound velocity makes the peak structure
relating with the stiffening of the matter and the {\it soft-to-stiff} type EOS.

In the end,
we give a comment on the $\mu^2$ term comes from the quark and antiquark pairing.
This term should be included in the pressure as $\Lambda^2 \mu^2$ with energy scale $\Lambda$.
In the high density limit $\mu \to \infty$
the remaining energy scale is $\Lambda_{\rm QCD}$ which has the non-analytic form for the coupling constant $\alpha_s$ as
\begin{align}
	\Lambda_{\rm QCD} = \exp\qty(- \frac{2\pi}{\beta_0 \alpha_s(\mu^2)}).
\end{align}
Thus the mere perturbation theory can not include such $\mu^2$ term.
It is useful to check how the non-perturbative $\mu^2$ term contributes to the sound velocity.
Starting with the parametric expression of the pressure
\begin{align}
	P = a_0 \mu^4 + a_2 \mu^2,
\end{align}
we can find
\begin{align}
	c_s^2 = \frac{4a_0\mu^3 + 2a_2\mu}{12a_0\mu^3 + 2a_2\mu} = \frac{1}{3}\qty[1 + \frac{4a_0}{12a_0\mu^2 + 2a_2}].
\end{align}
The extracted factor $1/3$ is the conformal value and the second term is the correction from the $\mu^2$ term.
This shows that the $\mu^2$ term with positive coefficient $a_2$ can make the sound velocity larger than the conformal value.
Thus the $c_s^2$ can approaches to the conformal value from above because of the non-perturbative corrections.

The sound velocity is not just the constraint of the EOS,
but also reflects the microscopic physics of the dense matter.
In this sense
the examination of the behavior of the sound velocity based on the microscopic physics
provides the important information of the physics of the dense QCD matter.

\begin{comment}
\begin{figure}[tpb]
	\centering
	\includegraphics[width=0.8\textwidth]{figures/intro/sound_velocity_map.png}
	\caption{
		The sound velocity of the dense QCD matter as a function of baryon number density $n_B$.
		\todo{this figure must be replaced as soon as possible.}
	}
	\label{fig:intro-sound-velocity-map}
\end{figure}
\end{comment}

\section{Isospin QCD}
\label{sec:intro-isospin-qcd}
To explore the dense QCD matter,
we would increase the density of the conserved charge such as the baryon density
by increasing the corresponding chemical potential.
In the case of the two-flavor QCD matter in the NS focusing only on the strong interactions,
we have to consider the baryon chemical potential $\mu_B$ and isospin chemical potential $\mu_I$.
Non-zero isospin chemical potential corresponds to the isospin asymmetric matter.
The realistic setting is $\mu_B \gg \mu_I$,
however,
the LQCD calculation is not available with finite $\mu_B$ because of the sign problem.
In contrast, 
in the idealized setting $\mu_B = 0$ and $\mu_I \neq 0$, isospin QCD \cite{Splittorff:2000mm,Son:2000xc,Son:2000by},
the LQCD is sign-problem free and the calculation is available.
Thus the isospin QCD is the useful laboratory to test the concepts of the finite density QCD
and would provide the important insight of the dense matter.

\begin{figure}[htpb]
	\centering
	\includegraphics[width=0.7\linewidth]{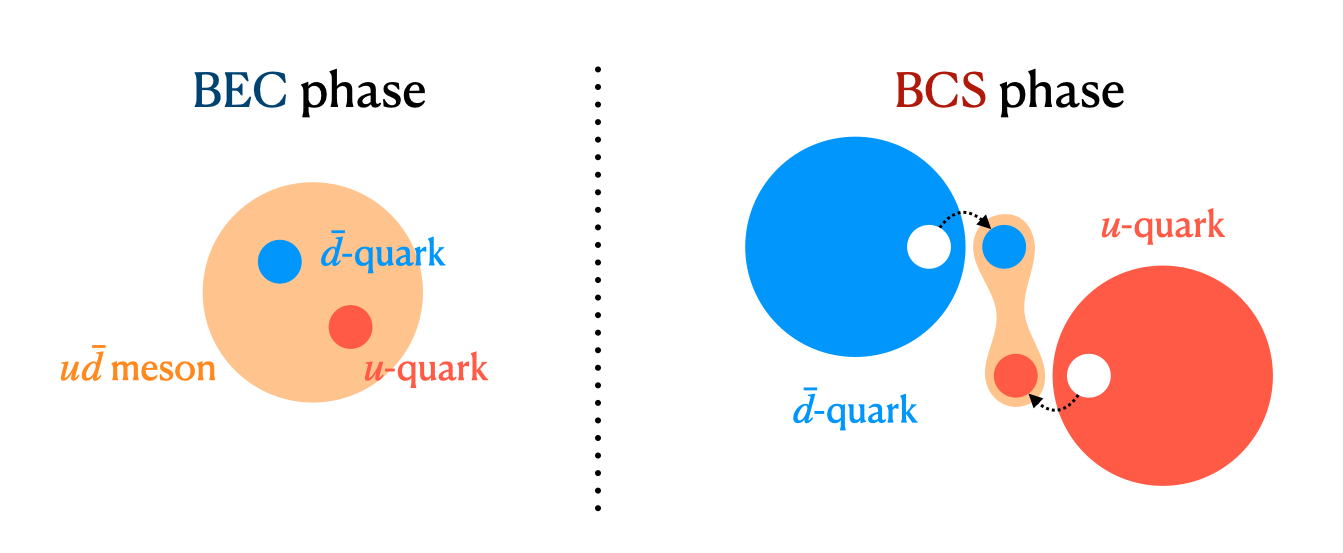}
	\caption{
		The $u$-quark and $\dbar$-quark pairing near the Fermi surface
		which corresponds to the $u\dbar$ meson or $\pi^+$.
		The $u$-quark hole and $\dbar$-quark particle pairing corresponds to the $\pi^-$.
	}
	\label{fig:intro2-massgap-pairing}
\end{figure}

The transition of the effective degrees of freedom is similar to the case of finite baryon density 
illustrated in \figref{fig:intro-effective-dof}.
As increasing the isospin chemical potential and reached to $\mu_I = m_\pi/2$,
the matter start with the hadronic phase
where the mesons with positive isospin charge start to be populated and increase the isospin density $n_I$.
Growing the isospin density,
the Fermi surface of the $u-$ and $\dbar-$quarks are formed
and the pion condensate is interpreted as the pairing of the quarks near the Fermi surface \figref{fig:intro2-massgap-pairing}.
In this region the quark substructure becomes important.
The transition from the hadronic to quark phase is discussed in the analogy of the crossover from Bose-Einstein-condensation (BEC) to Bardeen-Cooper-Schrieffer (BCS).
This crossover has the important role in explaining the stiffening of the dense QCD matter in the NSs.
In the denser region the matter approaches to the non-perturbative quark matter
and finally the pQCD approach is applicable.

\section{Theoretical approach to finite density}
\label{sec:intro-theoretical-finite-density}
In this section we will briefly review the theoretical approaches to the finite density QCD
from low density region to the high density limit.

The starting point is the QCD Lagrangian,
\begin{align}
	\calL_{\rm QCD} = \psibar (i\gamma_\mu D^\mu-M) \psi - \frac{1}{4}\tr F_{\mu\nu}F^{\mu\nu},
\end{align}
wherein $\psi$ is the quark field, $A_\mu$ is the gauge field (gluon field),
$D_\mu = \partial_\mu -igA_\mu$ is the covariant derivative 
and $F_{\mu\nu} = \partial_\mu A_\nu - \partial_\nu A_\mu + ig [A_\mu, A_\nu]$ is the field strength tensor.
The gauge field $A_\mu$ is the adjoint representation of $\rm SU(3)$ gauge group
and also written as $A_\mu = A_\mu^a T^a$ with the generator $T^a$ of $\rm SU(3)$ or Gell-Mann matrices.
We can define the left- and right-handed quarks via projection operator $P_{\rm L/R}$ as
\begin{align}
	\psi_L = P_L \psi = \frac{1 + \gamma_5}{2}\psi, ~~~ \psi_R = P_R \psi = \frac{1 - \gamma_5}{2}\psi.
\end{align}
Then the chiral symmetry is the invariance under the transformation
\begin{align}
	\psi_L \mapsto \psi_L' = e^{-i\theta_L} \psi_L, \notag\\
	\psi_R \mapsto \psi_R' = e^{-i\theta_R} \psi_R
\end{align}
for independent parameter $\theta_{L/R}$.
For massless quarks $M=0$ the chiral symmetry is exact,
but the chiral condensate breaks the chiral symmetry spontaneously.
Thus the original symmetry $\rm SU_L(3)\times SU_R(3) \times U(1)_V$ is broken down to $\rm SU_V(3) \times U(1)_V$
and the massless pions are emerged as the Nambu-Goldstone bosons.
The low energy effective models including the chiral symmetry are constructed
to respect the chiral symmetry and its breaking.

\subsection{Chiral perturbation theory}
Chiral perturbation theory is the low energy effective field theory of QCD on the basis of the chiral symmetry.
In low energy region the relevant degrees of freedom are the pions and nucleons.
The effective Lagrangian is expanded by the power of $Q/\Lambda_{\chi}$ for the pion mass $Q$ and the chiral symmetry breaking scale $\Lambda_\chi \sim 4\pi f_\pi \sim 1.2 \GeV$ \cite{Manohar:1983md}.
It has the form of
\begin{align}
	\calL_{\rm eff} = \calL_{\pi\pi} + \calL_{\pi N} + \dots
\end{align}
wherein $\calL_{\pi\pi}$ and $\calL_{\pi N}$ are the pion-pion and pion-nucleon interactions.
Each interaction terms are expanded by the number of the derivatives or pion mass as
\begin{align}
	\calL_{\pi\pi} = \calL_{\pi\pi}^{(2)} + \calL_{\pi\pi}^{(4)} + \dots, ~~~ \calL_{\pi N } = \calL_{\pi N }^{(1)} + \calL_{\pi N }^{(2)} + \calL_{\pi N }^{(3)} + \dots.
\end{align}

The extension of the finite density is based on the Hamiltonian formalism.
In the grand canonical ensemble the finite density is included by the term of $\mu_i Q_i$ as $\calH_{\rm dense} = \calH_{\rm vacuum} - \mu_i Q_i$
where $Q_i$ is the conserved Noether charge for the symmetry of Lagrangian
\begin{align}
	Q_i = \int d^3x J_i^0(x)
\end{align}
with the conserved current $J_i^\mu(x)$.
The detailed derivation of the finite density Lagrangian is out of our scope
so see Ref.\cite{Adhikari:2019aa} for instance.
Since chiral perturbation theory does not include the quark degrees of freedom,
and since the expansion is valid only for the low energy region $Q/\Lambda_\chi < 1$,
this approach is limited to the low density region $n_B < 2 n_0$.

\subsection{Lattice QCD}
Another approach to the low density QCD is the Lattice QCD.
In the formulation of the quantum field theory
the observable are calculated as the weighted average of the operator within the path integral formalism,
which is given by
\begin{align}
	\braket{O} \propto \int\calD A_\mu \calD \psibar \calD \psi  ~ O(A_\mu,\overline{q},q) e^{-iS_{\rm QCD}}
\end{align}
with the QCD action $S_{\rm QCD}$.
The path integral corresponds to the integral over infinite number of variables
and can not be calculated analytically.
To avoid this problem the LQCD calculate this path integral using the Monte Carlo method.
In the LQCD the imaginary-time is introduced and the space-time metric is changed to the Euclidean from Minkovski.
Then the observances are calculated as
\begin{align}
	\braket{O} \propto \int\prod_{n,\mu}d U_{n,\mu}\prod_n d\overline{q}_n dq_n ~ O(U,\overline{q},q)e^{-S_{\rm QCD}}.
\end{align}
Here $U_{n,\mu}$ is the link variable which corresponds to the gauge field $A_\mu$,
and $q_n$ corresponds to the quark field.
In the Monte Carlo calculation the average over the ensemble $(U,q)$ generated by the probability $e^{-S_{\rm QCD}}$
instead of the weighted average.
This statement is valid for the zero-chemical potential,
but in the presence of the quark chemical potential $S_{\rm QCD}$ has the complex phase
and $e^{-S_{\rm QCD}}$ can no longer be interpreted as the probability to generate the ensemble.
This is the {\it sign problem} of LQCD.

However in the case of the isospin QCD and the two-color QCD, 
these complex phases for each Fermion degrees of freedom cancelled out
and the sign problem is absent.
In such system the LQCD calculation for finite density is possible
and the EOS has been calculated,
which provide the first principle guide to the effective model approach.

\subsection{Nambu-Jona-Lasino model}
In the region $5 n_0 < n_B < 40n_B$ where the non-perturbative quarks are dominant,
the effective model approach is used to examine the microscopic physics and construct the EOS.
One of the most successful model is the Nambu-Jona-Lasino (NJL) model
which Lagrangian has the form of
\begin{align}
	\calL_{\rm NJL} = \psibar\qty(i\slashed{\partial} - m)\psi + \frac{G}{N_f}\qty[\qty(\psibar T_a \psi)^2 + \qty(\psibar i\gamma_5 T_a\psi)^2]
\end{align}
wherein $m$ is the current quark mass and $G \sim 1\GeV$ is the coupling constant.
The NJL model is used as the boundary condition of the EOS \cite{Baym:2019iky} in the high density region $n_B \simeq 5 n_0$.
But the NJL model lacks the meson degrees of freedom
and the application to the hadronic phase at the low density is limited.
In addition
the NJL model is the cutoff theory and the sound velocity converges to $c_s^2 \to 1$ instead of the conformal value $c_s^2 = 1/3$,
which means that the NJL model can not be used for the high density limit.

\subsection{Quark-meson model}
\label{ssec:intro-qm-model}
In order to explore the wide density range from low to high density region
the two-flavor quark-meson (QM) \cite{Ueda:2013sia,Kamikado:2012bt,Stiele:2013pma} model has gained the attention.
This is the effective model of isospin QCD 
which contains the $u$- and $d$-quarks and spin-0 mesons $\vec\phi = (\sigma,\vec\pi)$.
The change of the degrees of freedom is considered to be similar to the real QCD;
mesons are the effective degrees of freedom at low density region,
and the quark degrees of freedom gently appear as the density increases,
and the partonic picture of quarks are valid at the high density limit,
as shown in \figref{fig:intro-effective-dof}.
This similarity allows us to test the phenomenological concepts of QCD in the isospin QCD
and discuss the real QCD with the help of the isospin QCD.
As previously mentioned it is another advantage of isospin QCD 
that the LQCD calculation is possible 
and that we can discuss the model calculations in comparison using the LQCD results as a laboratory.
The similar analysis of isospin QCD had been conducted by NJL model,
but the QM model is cutoff-free and the extension to the high density region is possible
on contrast to the NJL model.

\subsection{Perturbative QCD}
At the high density limit
the perturbative description is applicable and pQCD is frequently used.
The observable is calculated by the perturbative expansion of the coupling constant $\alpha_s$.
For the calculation of EOS,
the next-to-next-to-leading order (${\rm N^2LO}$) calculation has performed \cite{Freedman:1976ub,Freedman:1976dm} in the late 1970's
and now the ${\rm N^3LO}$ calculation \cite{Gorda:2023mkk} is conducted.
The validity of perturbative calculation can be checked by the renormalization scale dependence
and the ${\rm N^2LO}$ examination concluded that the perturbation approach is valid for $\mu_q > 1 \GeV$ or $n_B > 40 n_0$ \cite{Kurkela:2009gj}.

\section{Subject of this research}
\label{sec:intro-subject}
Phenomenological analysis of the isospin QCD at finite density provides the important information of the dense QCD matter.
In particular, 
a unified description of the matter from low to high density in terms of quark degrees of freedom
is essential for understanding the matter properties based on the quark-hadron continuity.

In this thesis 
we study the EOS of the isospin QCD matter from low to high density using two-flavor quark meson model.
We focus on the transition of the effective degrees of freedom from hadrons to quarks
and the impact of the quark substructure on the EOS and the sound velocity.

This thesis is organized as follows.
In Chap.$\,$\ref{chap:thermo} we review the thermodynamics in the quantum field theory.
We introduce the effective potential and the thermodynamic potential via the effective action and the field condensates.
Then we give the relation between the thermodynamic potential and other thermodynamic quantities
including the sound velocity.
In Chap.$\,$\ref{chap:model} we introduce the QM model and its extension to the finite density.
In the first section we focus on the extension to the finite density and introduce the mean field approximation.
In the next section we extend the model to the finite temperature.
Also we introduce the Polyakov loop to incorporate the quark confinement.
In Chap.$\,$\ref{chap:zeroT} we will show the numerical results of the EOS with vanishing temperature.
The effects of thermal quarks and mesons on the EOS are examined in Chap.$\,$\ref{chap:finiteT}.
Finally, we close with a summary in Chap.$\,$\ref{chap:summary}.

\chapter{Thermodynamics in quantum field theory}
\label{chap:thermo}

\section{Effective potential}
\label{sec:thermo-effective-potential}
In quantum field theory, the partition function is defined by the path integral as
\begin{align}
Z[J] = \int \calD\phi~ \exp\qty[ i \int d^4 x \qty(\calL(\phi) + J(x) \phi(x))]
\end{align}
where $\phi$ is a field and $J$ is an external source.
In the presence of the external source $J$,
we can define the generating function of the connected Green's function 
\begin{align}
	Z[J] \propto e^{iW[J]}, ~~~ W[J] = \sum_{n=0}^\infty \frac{1}{n!}\int dx_1...dx_n ~ G^{(n)}(x_1,...,x_n)J(x_1)...J(x_n).
\end{align}
The effective action is defined by the Legendre transformation of the generating function as
\begin{align}
	\Gamma[\varphi_v] = W[J] - \int d^4 x J(x) \varphi_v(x),
\end{align}
where $\varphi_v(x)$ is defined by
\begin{align}
	\varphi_v(x) := \fdv{W[J]}{J(x)}.
\label{eq:def_vev}
\end{align}

Before mentioning the physical meaning of $\varphi_v(x)$,
we define the effective action
\footnote{
	The detailed discussion of the effective action is found in the following references;
	\begin{itemize}
		\item Peskin, M.,\& Schroeder, D. (1995). {\it An Introduction To Quantum Field Theory}. Westview Press. $\S 11.3$
		\item Weinberg, S. (1995). {\it The Quantum Theory of Fields (Volume 2)}. Cambridge University Press. $\S 16.1$
		\item Srednicki, M. (2007). {\it Quantum Field Theory}. Cambridge University Press. $\S 21$
		\item \begin{CJK}{UTF8}{ipxm}坂本, 眞. (2020). {\it 場の量子論(II) : ファインマン・グラフとくりこみを中心にして}. 裳華房. $\S 13$\end{CJK}
	\end{itemize}
}.
as Legendre transformation of $W[J]$.
\begin{align}
\Gamma[\varphi_v] := W[J] - \int dx J(x)\varphi_v(x)
\end{align}
and this $\Gamma[\varphi_v]$ should be source independent.
\begin{align}
\fdv{\Gamma[\varphi_v]}{J(x)} = 0.
\end{align}
We should notice that $\varphi(x)$ is given by the functional form of $J(x)$ through \eqref{eq:def_vev},
and vice versa;
\begin{align}
J(x) = - \fdv{\Gamma[\varphi_v]}{\varphi_v(x)}.
\label{eq:J_from_Gamma}
\end{align}

Now we consider the physical meaning of $\varphi_v(x)$.
To begin with, let us see the path integral formulation of the generating function $W[J]$.
\begin{align}
e^{iW[J]} = \frac{Z[J]}{Z[0]} = \frac{\int\calD\phi~ e^{i\int dx \qty(\calL[\phi]+J\phi)}}{\int\calD\phi~ e^{i\int dx \calL[\phi]}}.
\end{align}
This means $W[J]$ is the logarithm of the ratio of the partition function with and without the external source.
\begin{align}
W[J] = \frac{1}{i}\qty(\ln Z[J] - \ln Z[0]).
\end{align}
By taking the functional derivative for $J(x)$ we get $\varphi_v$ in terms of partition function $Z[J]$ as 
\begin{align}
\varphi_v(x) = \fdv{W[J]}{J(x)} = \frac{1}{i}\frac{1}{Z[J]}\fdv{Z[J]}{J(x)} = \frac{\int\calD\phi~ \phi(x) ~ e^{i\int dx \qty(\calL[\phi]+J\phi)}}{\int\calD\phi~ e^{i\int dx \calL[\phi]}}.
\end{align}
This is nothing but the vacuum expectation value of the field $\phi(x)$
with the Lagrangian density $\calL[\phi] + J(x)\phi(x)$.
Going a little further,
the vacuum expectation value of the field $\phi(x)$ with the Lagrangian density $\calL[\phi]$ is $\varphi_v(x)$ in the limit of $J(x) \to 0$.
\begin{align}
\lim_{J(x)\to 0}\varphi_v(x) = \braket{\phi(x)}.
\end{align}

Next,
we consider the expansion of the effective action.
In the analogy of $J(x)$ expansion of the generating function $W[J]$ we find
\begin{align}
\Gamma[\varphi_v] = \sum_{n=0}^\infty \frac{1}{n!}\int dx_1...dx_n ~ \Gamma^{(n)}(x_1,...,x_n)\varphi_v(x_1)...\varphi_v(x_n).
\end{align}
These $\Gamma^{(n)}(x_1,...,x_n)$ are n-point vertex function.
$\Gamma^{(n)}$ itself has the important role on the renormalization,
however, we do not focus on this expansion.
Instead, we consider the expansion of $\Gamma[\varphi_v]$ in terms of the derivative of $\varphi_v$ as
\begin{align}
\Gamma[\varphi_v] = \int dx ~ \qty[-\calE(\varphi_v) + X(\varphi_v)(\partial_\mu\varphi_v)^2 + Y(\varphi_v)(\partial_\mu\varphi_v)^4 + ...]
\end{align}
where $\calE(\varphi_v), X(\varphi_v), Y(\varphi_v)$ are the function (not functional) of $\varphi_v$.
Here only $\calE(\varphi_v)$ contains the non-functional contribution of the effective action
which is called {\it effective potential} $V_{\rm eff}$.

Now let us assume the vacuum expectation value $\varphi_v$ is space-time independent.
When taking $J(x)\to 0$ in addition to this condition,
\eqref{eq:J_from_Gamma} yields
\begin{align}
0 = \int dx \pdv{V_{\rm eff}(\varphi_v)}{\varphi_v} = \calV_4 \pdv{V_{\rm eff}(\varphi_v)}{\varphi_v}
\end{align}
with the space-time volume $\calV_4$.
This condition for the vacuum expectation value $\varphi_v$ is called the {\it gap equation}.
This equation is nothing but the condition for the minimum of the effective potential $V_{\rm eff}(\varphi_v)$
and showing how the vacuum expectation value $\varphi_v$ should be determined.

We defined the effective potential through the effective action $\Gamma[\varphi_v]$,
however,
we can also get the effective potential using the path integral formalism by shifting the filed $\phi$ as $\phi \to \tilde\phi + \varphi_v$.
We consider the partition function $Z[J]$.
\begin{align}
Z[J] &= e^{iW[J]} = \int\calD\phi~ \exp\qty[i\int dx \qty(\calL[\phi]+J\phi)] \notag\\
     &= \exp\qty[i\qty(\Gamma[\varphi_v] + \int dx J(x)\varphi_v(x))].
\end{align}
Here we put the normalization condition $Z[0]=1$ to simplify the calculation.
This leads the effective action as 
\begin{align}
e^{i\Gamma[\varphi_v]} &= \int\calD\phi \exp\qty[i\int dx\qty(\calL[\phi] + J(x)(\phi(x)-\varphi(x)))] \notag\\
&= \int\calD\phi \exp\qty[i\int dx\qty(\calL[\phi] - \fdv{\Gamma[\varphi_v]}{\varphi_v}~(\phi-\varphi_v))].
\end{align}
As $\varphi_v$ is the vacuum expectation value of the field $\phi$, $\phi-\varphi_v =: \tilde\phi$ is the quantum fluctuation around the vacuum.
Rewriting the path integral for $\phi$ to $\tilde\phi$ we obtain
\begin{align}
e^{i\Gamma[\varphi_v]} = \int\calD\tilde\phi\exp\qty[i\int dx \qty(\calL[\tilde\phi+\varphi_v] - \fdv{\Gamma[\varphi_v]}{\varphi_v}\tilde\phi)].
\end{align}
To discuss the contribution of the second term in the space-time integral,
let us go back to the expansion of $\Gamma[\varphi_v]$ and the definition of the vertex function.
Effective action is the series of the vertex function $\Gamma^{(n)}$ with $n-$th number of vacuum expectation $\varphi_v$ on the external legs.
Thus $\fdv*{\Gamma[\varphi_v]}{\varphi_v}\tilde\phi$ is the tadpole diagram, which only has the single leg associating $\tilde\phi$.
This means the second term in the space-time integral generates all the contribution of single-particle {\it reducible} diagrams with the negative sign.
Therefore the effective action contains only the single-particle reducible diagram;
\begin{align}
e^{i\Gamma[\varphi_v]} = \int_{\rm 1PI}\calD\phi~\exp\qty[i\int dx \calL[\phi]]
\end{align}
where 1PI denotes the integral over the single-particle irreducible diagrams.

The discussion above can directly applied to the effective potential.
In the case of static vacuum expectation value $\varphi_v$ or $J(x)\to0$,
all the single-particle reducible diagrams had been killed by the gap equation.
Thus the normal path integral manifestly does not contain the single-particle reducible diagrams.
Thus we can get the following.
\begin{align}
e^{i\calV_4 V_{\rm eff}(\varphi_v)} = \int\calD\tilde\phi~\exp\qty[i\int dx \calL[\tilde\phi+\varphi_v]].
\end{align}

All together, after solving the gap equation
\begin{align}
	\left.\pdv{V_{\rm eff}}{\varphi_v}\right|_{\varphi_v=\varphi_v^*} = 0
\end{align}
we obtain the thermodynamic potential as
\begin{align}
	\Omega = V_{\rm eff}(\varphi_v^*).
\end{align}

\section{Thermodynamics and sound velocity}
In the case of the finite temperature and finite density QCD,
we obtain the thermodynamic potential as the function of the temperature $T$ and chemical potential $\mu$.
Once recall the grand canonical ensemble and the grand potential,
we can get the pressure $P$ as
\begin{align}
	P(\mu, T) = - \Omega(\mu, T)
\end{align}
and the entropy density $s$ and number density $n$ as
\begin{align}
	s(\mu, T) &= - \pdv{\Omega(\mu, T)}{T}, \\
	n(\mu, T) &= - \pdv{\Omega(\mu, T)}{\mu}.
\end{align}
Combining these equations we can get the energy density $\varepsilon$ as
\begin{align}
	\varepsilon(\mu, T) = T s(\mu, T) + \mu n(\mu, T) - P(\mu, T).
\end{align}

As mentioned in the Sec.$\,$\ref{sec:intro-sound-velocity}
the sound velocity $c_s$ is defined as
\begin{align}
	c_s^2 = \pdv{P}{\varepsilon},
\end{align}
but in the case of the finite temperature and finite density
there left the ambiguity to take the derivative on the $(\mu, T)$ plane.
Therefore we have to impose the additional condition to take the differentiation.

We first consider the simple case of $T=0$.
In this case both of pressure and energy density are the function of the chemical potential $\mu$ as
\begin{align}
	n(\mu) = \pdv{P(\mu)}{\mu}, ~~~ \varepsilon(\mu) = \mu n(\mu) - P(\mu)
\end{align}
and the sound velocity is well-defined as
\begin{align}
	c_s^2 &= \pdv{P(\mu)}{\varepsilon(\mu)} = \pdv{\mu}{\varepsilon(\mu)}\pdv{P(\mu)}{\mu}  \notag\\
	&= \frac{n(\mu)}{\mu \chi(\mu)}, ~~~ \chi(\mu) = \pdv{n(\mu)}{\mu}.
\end{align}
The same discussion can be applied to the case of $\mu=0$.

The other condition is for the first-derived quantity, entropy density and number density.
In this case we can consider the $(\mu, T)$ plane but the condition specifies a certain path on the plain
and we take the derivative along the path.
the calculation is performed using the Jacobian methods \cite{Carroll1965} as
\begin{align}
	c_n^2 &= \qty(\pdv{P}{\varepsilon})_n = \pdv{(P,n)}{(\varepsilon,n)} = \frac{s \chi_{\mu\mu} - n \chi_{\mu T}}{T\qty(\chi_{TT}\chi_{\mu\mu} - \chi_{\mu T}^2)} \\
	c_s^2 &= \qty(\pdv{P}{\varepsilon})_s = \pdv{(P,s)}{(\varepsilon,s)} = \frac{n \chi_{TT} - s \chi_{\mu T}}{\mu\qty(\chi_{TT}\chi_{\mu\mu} - \chi_{\mu T}^2)} ,
\end{align}
where $\chi_{xy} = \pdv*{P}{x}{y}$.
These two are usually used in the intermediate process of the hydrodynamical calculation.

The another condition is the {\it isentropic} condition in which $s_n$ is constant.
This condition corresponds to the ideal fluid and used in the calculation of the supernova explosion.
The sound velocity on this trajectory is given by the same fashion as
\begin{align}
	c_{s/n}^2 = \qty(\pdv{P}{\varepsilon})_{s/n} = \pdv{(P,s/n)}{(\varepsilon,s/n)} = \frac{c_n^2 Ts + c_s^2 \mu n}{Ts + \mu n}.
\end{align}

\chapter{Two-flavor Quark Meson model}
\label{chap:model}
In this chapter 
we explain the two-flavor Polyakov-loop quark meson (PQM) model.
This model is one of the effective models of isospin QCD,
where the baryon density $n_B$ is zero and the isospin density $n_I$ is non-zero.
The $u$- and $d$-quarks and spin-0 mesons $\vec\phi = (\sigma,\vec\pi)$ are included as the effective degrees of freedom
and the Polyakov-loop is introduced to describe the quark confinement.

\section{Quark meson model; zero temperature and finite density}
\label{sec:model-quarkmeson}

\subsection{Finite density Lagrangian}
\label{subsec:model-lagrangian}

The Lagrangian of the two-flavor quark-meson model is
\begin{align}
	\calL &= \frac{1}{2} \left(\partial_\mu \vec\phi\right)^2 
	- \frac{1}{2} m_0^2\vec\phi^2 -\frac{\lambda}{24}(\vec\phi^2)^2 + h\sigma + \psibar \left(\rmi \slashed{\partial} - g(\sigma + \rmi \gamma^5\vec\tau\cdot \vec\pi)\right)\psi ,
	\label{eq:model-lagrangian}
\end{align}
where $\psi$ is a quark field with $u$- and $d$-quark components
\begin{align}
	\psi = \left(\begin{matrix} u\\d\end{matrix}\right).
\end{align}
The $\vec\phi = (\sigma,\vec\pi)$ are meson fields 
which correspond to the isospin $\bf 1$ and $\bf 3$ representations.
The $\tau_i$'s are the Pauli matrices in flavor space.
The mesonic part has the $\rmO(4)$ symmetry and the explicit breaking term $h\sigma$ is included.
The quark part contains the scalar and pseudo-scalar coupling (Yukawa interaction) of the mesons and quarks.

To compute the Lagrangian with the isospin chemical potential $\mu_I$ or the thermodynamic potential $\Omega$ at the finite isospin density $n_I$
we will begin with the Hamiltonian formalism as partially mentioned in Sec.$\,$\ref{sec:intro-theoretical-finite-density}.
The Hamiltonian for the finite isospin density $n_I$ is given by
\begin{align}
	\calH_{\rm dense} = \calH - \mu_I n_I.
\end{align}

Here we introduced the isospin chemical potential $\mu_I$ as the Lagrange multiplier for the isospin density $n_I$.
But it is important to understand the meanings of $\mu_I$ in terms of the $u$- and $d$-quark chemical potential $\mu_{u/d}$.
The baryon density and isospin density are defined by
\begin{align}
	n_B = \frac{1}{3}\qty(n_u+n_d), ~~~ n_I = n_u - n_d,
\end{align}
which immediately leads the relation
\begin{align}
	\pdv{\Omega}{\mu_B} &= \pdv{\mu_u}{\mu_B} \pdv{\Omega}{\mu_u} + \pdv{\mu_d}{\mu_B} \pdv{\Omega}{\mu_d} = \frac{1}{3}\pdv{\Omega}{\mu_u} + \frac{1}{3}\pdv{\Omega}{\mu_d}\\
	\pdv{\Omega}{\mu_I} &= \pdv{\mu_u}{\mu_I} \pdv{\Omega}{\mu_u} + \pdv{\mu_d}{\mu_I} \pdv{\Omega}{\mu_d} = \pdv{\Omega}{\mu_u} - \pdv{\Omega}{\mu_d}.
\end{align}
Thus the quark chemical potentials can be written as 
\begin{align}
	\mu_u = \frac{1}{3}\mu_B + \mu_I, ~~~ \mu_d = \frac{1}{3}\mu_B - \mu_I
\end{align}
or equivalently
\begin{align}
	\mu_B = \frac{3}{2}(\mu_u + \mu_d), ~~~ \mu_I = \frac{1}{2}\qty(\mu_u-\mu_d).
\end{align}

The isospin density in terms of the fields can be introduced though the Noether's theorem.
The isospin transformation for the meson and quark fields are
\begin{align}
	\pi_a \mapsto \exp\qty(i\theta_i T_i)_{ab}\pi_b, ~~~ \psi \mapsto \exp\qty(i\theta_i \tau_i)\psi
\end{align}
and the corresponding Noether currents are
\begin{align}
	j^\mu_a = \epsilon_{abc}\pi_b\partial^\mu\pi_c + \delta^{\mu 0} \delta_{a3}\psibar \gamma^0 \tau_3 \psi
\end{align}
where $\epsilon_{abc}$ is the complete anti-symmetric tensor with $\epsilon_{123} = 1$.
Thus the isospin density is given by $a=3$ conserved charge as
\begin{align}
	n_I = j_{a=3}^0 = \rmi\pi_+ \partial^0\pi_- - \rmi\pi_-\partial^0\pi_+  + \psibar\gamma_0 \tau_3 \psi
\end{align}
where $\pi_\pm = (\pi_1 \pm \rmi\pi_2)/\sqrt{2}$.

Once the expression of the isospin density is obtained as a functional of $\Phi = (\vec\phi,\psi)$,
we can compute the partition function $Z$ as
\begin{align}
	Z = \int\calD \Pi_\Phi \calD \Phi \exp\left[ \rmi \int_x\left(\dot{\Phi}\cdot \Pi_\Phi - \calH + \mu_I n_I \right) \right]
\end{align}
where $(\Pi_\Phi)_i = \partial^0 \Phi_i$ is a conjugate field to $\Phi_i$.
After integrating out $\vec\Pi_{\Phi}$ field we obtain the Lagrangian with isospin density as
\begin{align}
	Z = \int\calD\Phi ~ \exp\qty(\rmi\int_x \calL_{\rm dense}) = e^{i\calV_4 \Omega}.
\end{align}
Here $\calV_4$ is the four-dimensional space-time volume and $\Omega$ is the thermodynamic potential
which is introduced in Chap.$\,$\ref{chap:thermo}.
Leaving the derivation of the Lagrangian at finite density to Appendix \ref{app:finite-density-lagrangian},
we obtain the tree-level Lagrangian of two-flavor QM model as
\begin{align}
	\calL_B [\Phi_B] &= \frac{1}{2}\qty[(\partial_\mu \sigma_B)^2+(\partial_\mu \pi_{B3} )^2] \notag\\
		&\quad + (\partial_\mu + 2 \rmi \mu_I\delta_\mu^0) \pi_B^+\qty(\partial^\mu - 2 \rmi \mu_I\delta^\mu_0)\pi_B^- \notag\\
		&\quad -\frac{1}{2}m_{0B}^2 \vec\phi_B^2  -{\lambda_B \over 24} ( \vec\phi_B^2 )^2  + h_B \sigma_B  \notag\\
		&\quad + \psibar_B \qty[ \rmi \slashed{\partial} + \mu_I \tau_3 \gamma^0 - g_B (\sigma_B + \rmi \gamma^5 \vec\tau \cdot \vec\pi_B) ]\psi_B ,
\end{align}
where the subscript $B$ denotes that the fields are not renormalized.

\subsection{Effective potential}
\label{subsec:model-quarkmeson-effective-potential}
To compute the thermodynamic potential $\Omega$ we take the mean-field approximation.
In this approach we separate the field to the condensate and the quantum fluctuation
and compute the effective potential as the function of the condensates.
In this model we assume that the $\sigma$ and $\pi_1$ field has the non-zero condensate as
\begin{align}
	\sigma = \braket{\sigma} + \tilde\sigma , ~~~ \pi_1 = \braket{\pi_1} + \tilde\pi_1,
\end{align}
where subscript tilde $\tilde{}$ denotes the quantum fluctuation.

With the auxiliary condensates
we can obtain the effective potential $V_{\rm eff}(\mu_I; M_q, \Delta)$ as the function of the condensates in addition to the isospin chemical potential.
The value of these condensates are determined by the minimum of the effective potential or the gap equation
\begin{align}
	\pdv{V_{\rm eff}}{M_q} = 0, ~~~ \pdv{V_{\rm eff}}{\Delta} = 0.
\end{align}

The role of these condensates can be seen by the Lagrangian in flavor space 
where the quark fields are involved in;
\begin{align}
	\psibar\qty[\rmi\slashed{\partial} + \mu_I\tau_3\gamma^0 - g\qty(\sigma+\rmi\gamma^5\vec\tau\cdot\vec\pi)]\psi 
	\longrightarrow 
	\psibar\qty[
	\begin{matrix}
		\rmi\slashed{\partial} + \mu_I\gamma^0 - M_q & \Delta\gamma^5 \\
		- \Delta\gamma^5 & \rmi\slashed{\partial} - \mu_I\gamma^0 - M_q
	\end{matrix}
	]\psi
	\label{eq:model-quark-lagrangian-flavormatrix}
\end{align}
where $M_q = g\braket{\sigma}$ and $\Delta = g\braket{\pi_1}$.
The $\braket{\sigma}$ is the chiral condensate which corresponds to the dynamical quark mass
which is made of quark-antiquark pairs with the same flavor and different chirality,
e.g., a left-handed quark and right-handed antiquark pairing.
The $\braket{\pi_1} = \braket{\ubar d} + \braket{\dbar u}$ is the quark mass-gap which possess the isospin-1 and mixes the $u$- and $d$-flavor.
The quark mass-gap comes from the $u$-quark and $d$-antiquark pairing near the Fermi surface.

The tree-level effective potential is derived by replacing the fields in the Lagrangian with the condensates as
\begin{align}
	V_0(\mu_I; M_q, \Delta) &=  \frac{m_{0}^2}{2g^2} M_q^2 +  \frac{m_{0}^2 - 4\mu_I^2}{2g^2} \Delta^2 
		 + \frac{\lambda}{24 g^4}\qty( M_q^2 + \Delta^2 )^2 - \frac{h}{g} M_q .
\end{align}
The gap equation in the tree-level have the isospin independent solution
\begin{align}
	M_q = g f_\pi, ~~~ \Delta = 0
\end{align}
at $\mu_I \leq m_\pi/2$.
This phase is called the vacuum phase.
In this phase all the physical quantities are independent of $\mu_I$
which is called the Silver Blaze property \cite{Cohen:2003kd}.
Now it is useful to notice for later use
that we can read off the relations between the parameters in the Lagrangian and physical parameters as
\begin{alignat}{2}
	m_0^2 &= - \frac{1}{2}\qty(m_\sigma^2 - 3 m_\pi^2), ~~~ & \lambda &= 3 \frac{m_\sigma^2 - m_\pi^2}{f_\pi^2}, \notag\\
	g^2 &= \frac{M_q^2}{f_\pi^2}, ~~~ & h &= m_\pi^2 f_\pi.
	\label{eq:model-parameters-relations}
\end{alignat}

In the tree-level effective potential the quark degrees of freedom is not taken into account.
We include the quark single particle energy (or Dirac sea contribution) as the leading order contribution of the quark degrees of freedom.
In the relativistic mean field theory (RMF),
the single particle energy is often neglected
assuming that the contribution is almost density independent.
But in this case the single particle energy depends on both chiral- and pion-condensates
thus its density dependence is not negligible.
Its energy can be read off from the zeros of the determinant of Dirac operator in \eqref{eq:model-quark-lagrangian-flavormatrix}
and its calculation is explained in Appendix \ref{app:quark-propagator}.
After the calculation we obtain the quark single particle energy as
\begin{align}
	E_u = E_{\dbar} = E(\mu_I), ~~~ E_d=E_{\ubar}=E(-\mu_I) ,
\end{align}
where
\begin{align}
	E(\mu_I) = \sqrt{\left(E_D-\mu_I\right)^2+\Delta^2}, ~~~ E_D = \sqrt{\vp^2 + M_q^2}.
\end{align}
Then the correction to the effective potential is given by
\begin{align}
	V_q = - N_c \int_\vp \qty(E_u+E_d+E_{\ubar}+E_{\dbar}).
	\label{eq:model-dirac-sea}
\end{align}
This momentum integral has the ultraviolet (UV) divergence
and the divergent part can be extracted by such as dimensional regularization $d=3 \to 3-2\epsilon$.
The momentum integral is amended as
\begin{align}
	\int_\vp = \qty(\frac{e^{\gamma_E \Lambda^2}}{4\pi})^\epsilon \int \frac{d^dp}{(2\pi)^d}
\end{align}
where  $\gamma_E = 0.577...$ is the Euler-Mascheroni constant and $\Lambda$ is the renormalizing scale introduced by the $\overline{\rm MS}$ scheme.
Applying the dimensional regularization and expanding the one-loop potential by $\mu_I$,
we can extract the divergent part from $V_q$ as
\begin{align}
	V_q^{\rm div} = \frac{4N_c}{(4\pi)^2} \qty( {e^{\gamma_E}\Lambda^2 \over M_q^2+\Delta^2} )^\epsilon \qty[ (M_q^2+\Delta^2)^2\Gamma(-2+\epsilon)-2\mu_I^2\Delta^2\Gamma(\epsilon)].
\end{align}
This divergent part is cancelled through the renormalized parameters and the counter terms
after the renormalization of the parameters and the fields.
Symbolically writing,
the renormalized effective potential $V_{\rm eff}$ can be written as
\begin{align}
	V_{\rm eff} = V_0 + \qty(V_q - V_q^{\rm div}) + \qty(V_q^{\rm div} + \delta V_q)
\end{align}
where $V_q - V_q^{\rm div}$.
The $\delta V_q$ is the counter terms for the effective potential
which kills the UV divergence from $V_q^{\rm div}$.

We have constructed the one-loop effective potential up to the leading order of the large $1/N_c$ expansion.
In this expansion we neglect the mesonic loop which has order $\calO(1)$ against the quark loop $\calO(N_c)$.
This means that our model emphasizes the role of the quark degrees of freedom
in the amplitude of the meson condensates and thus in the EOS.
The renormalization procedure is presented in the Appendix \ref{app:reno}
and the essential part is summarized in the following.

The first step is to rewrite the Lagrangian in terms of the renormalized parameters and fields.
The counter terms for the field is introduced as
\begin{align}
	\sigma_B = Z_\sigma^{1/2} \sigma, ~~~ \pi_B = Z_\pi^{1/2} \pi, ~~~ \psi_B = Z_\psi^{1/2} \psi.
\end{align}
The counter terms for the coupling constants $m, \lambda$ and $g$ are introduced as the substitution
but there remains the ambiguities.
We choose the two different substitution for the counter terms;
the one corresponds to the $\msbar$ scheme which manifestly preserves the original $\rmO(4)$ symmetry
and the other corresponds to the on-shell scheme which breaks the symmetry and the rescaling of the field is necessary to restore the symmetry.
In both cases,
the counter terms are determined by the renormalization conditions;
\begin{align}
	\rmi \Sigma_\phi(p^2 = m_\phi^2) = 0, ~~~ \left.\rmi\pdv{\Sigma_\phi}{p^2}\right|_{p^2=m_\phi^2} = \delta Z_\phi ~~~ (\phi = \sigma,~\pi).
\end{align}
The value of $h$ in the explicit breaking term $h\sigma$ is determined to make the chiral condensate $M_q$ takes the value $M_q^{\rm vac} = gf_\pi$ in the vacuum $\mu_I < m_\pi/2$;
\begin{align}
	\left.\pdv{V_{\rm 1-loop}}{M_q}\right|_{M_q = M_q^{\rm vac}} = 0.
\end{align}

Demanding these condition
we finally obtain the one-loop effective potential as
\begin{align}
	V_{\rm 1-loop} &= - \frac{1}{4}m_\sigma^2f_\pi^2 \qty[1 + \frac{4{M_q^{\rm vac}}^2N_c}{(4\pi)^2f_\pi^2}\qty{ - \frac{4{M_q^{\rm vac}}^2}{m_\sigma^2} F(m_\sigma^2) + \frac{4{M_q^{\rm vac}}^2}{m_\sigma^2} - (m_\sigma^2 - 4{M_q^{\rm vac}}^2)F'(m_\sigma^2)}]\frac{M_q^2+\Delta^2}{{M_q^{\rm vac}}^2} \notag\\
	&\quad + \frac{3}{4}m_\pi^2f_\pi^2 \qty[1 - \frac{4{M_q^{\rm vac}}^2N_c}{(4\pi)^2f_\pi^2}\bigg\{ - F(m_\pi^2) + F(m_\sigma^2) + (m_\sigma^2 - 4{M_q^{\rm vac}}^2)F'(m_\sigma^2)\bigg\}]\frac{M_q^2+\Delta^2}{{M_q^{\rm vac}}^2} \notag\\
	&\quad - 2\mu_I^2 f_\pi^2\qty[1 - \frac{4{M_q^{\rm vac}}^2N_c}{(4\pi)^2f_\pi^2}\qty(\ln\frac{M_q^2 + \Delta^2}{{M_q^{\rm vac}}^2} + F(m_\sigma^2) + (m_\sigma^2 - 4{M_q^{\rm vac}}^2)F'(m_\sigma^2))]\frac{\Delta^2}{{M_q^{\rm vac}}^2} \notag\\
	&\quad + \frac{1}{8}m_\sigma^2f_\pi^2 \left[1 - \frac{4{M_q^{\rm vac}}^2N_c}{(4\pi)^2f_\pi^2}\left\{\frac{4{M_q^{\rm vac}}^2}{m_\sigma^2}\qty(\ln\frac{M_q^2 + \Delta^2}{{M_q^{\rm vac}}^2} - \frac{3}{2}) \right.\right. \notag\\
	&\hspace{3cm} \left.\left. + \frac{4{M_q^{\rm vac}}^2}{m_\sigma^2}F(m_\sigma^2) + (m_\sigma^2 -4 {M_q^{\rm vac}}^2)F'(m_\sigma^2))\right\}\right] \frac{(M_q^2+\Delta^2)^2}{{M_q^{\rm vac}}^4} \notag\\
	&\quad - \frac{1}{8}m_\pi^2f_\pi^2\qty[1 - \frac{4{M_q^{\rm vac}}^2N_c}{(4\pi)^2f_\pi^2}\bigg\{- F(m_\pi^2) + F(m_\sigma^2) + (m_\sigma^2 - 4{M_q^{\rm vac}}^2)F'(m_\sigma^2)\bigg\}] \frac{(M_q^2+\Delta^2)^2}{{M_q^{\rm vac}}^4} \notag\\
	&\quad - m_\pi^2f_\pi^2\qty[1 - \frac{4{M_q^{\rm vac}}^2N_c}{(4\pi)^2f_\pi^2}\bigg\{F(m_\sigma^2) + (m_\sigma^2 - 4{M_q^{\rm vac}}^2)F'(m_\sigma^2)\bigg\}]\frac{M_q}{M_q^{\rm vac}} \notag\\
	&\quad -2N_c\int_p \left[\sqrt{\left(\sqrt{p^2+M_q^2}+\mu\right)^2+\Delta^2}+\sqrt{\left(\sqrt{p^2+M_q^2}-\mu\right)^2+\Delta^2}\right] \notag\\
	&\quad +4N_c\int_p\left[\sqrt{p^2+M_q^2+\Delta^2}+\frac{\mu^2\Delta^2}{2(p^2+M_q^2+\Delta^2)^{3/2}}\right].
\end{align}
The function $F(x)$ is defined as
\begin{align}
	F(p^2) &= 2-2r\arctan \left({1\over r}\right) \\
	p^2F'(p^2) &= {r^2+1 \over r}\arctan\left({1\over r}\right)-1
\end{align}
where $r=\sqrt{4{M_q^{\rm vac}}^2/p^2-1}$.

The appearance of the term $(M_q^2+\Delta^2)^2 \ln (M_q^2+\Delta^2)$ with negative coefficient means that the potential on the $M_q-\Delta$ plane is not bounded from below globally.
This indicates the existence of the higher order corrections $\ln^n$ for $n = 2,3,...$,
but does not immediately mean that the system $M_q^2+\Delta^2 \lesssim {M_q^{\rm vac}}^2$ is not physical.
In general,
in the case of the theory with several mass scales,
these logarithmic terms in effective potential can not be removed by choosing a single renormalization scale,
but can be removed by the renormalization group (RG) improvement \cite{Manohar:2020nzp}.

\section{Polyakov-loop quark meson model}
\label{sec:model-polyakov-quarkmeson}
In this section 
we consider the extension of the quark meson model at zero-temperature and finite density
to the non-zero temperature and finite density.
In the quantum field theory
the extension to the finite temperature is done by the Wick rotation of the time to the {\it imaginary time} $t = i\tau$
and the compactification of the time direction $\tau \sim \tau + 1/T = \tau + \beta$.
In this way the fields become periodic in the imaginary time direction as
\begin{align}
	\Phi(\tau+\beta,\vec x) = \xi \Phi(\tau,\vec x)
\end{align}
where $\xi = +1$ for bosons and $\xi = -1$ for Fermions.
This boundary condition discretizes the momentum of the fields in the imaginary time direction as
$\omega = 2\pi T n $ for bosons and $\omega = 2\pi T (n + 1/2)$ for Fermions where $n \in \mathbb{Z}$,
which is called the Matsubara frequency (Appendix.$\,$\ref{app:matsubara}).

Although we can derive the expression of the thermodynamic potential $\Omega_0$ at finite temperature using Matsubara sum,
the result does not directly describe the quark thermodynamics;
the lack of the quark confinement.
In this section we introduce the Polyakov loop to describe the quark confinement
and discuss the mean field treatment.

\subsection{Polyakov loop}
\label{subsec:model-polyakov-loop}
In the low energy region,
quarks must be color-neutral and bound together to form a color singlet state
and the thermal excitation of the quarks are suppressed.
However, 
as the energy increases, 
the vacuum structure changes and becomes filled with color-flux.
This allows quarks to become color-neutral with the color-flux in the vacuum rather than with other quarks
and allows the thermal excitation.
The deconfinement phase transition is characterized by the Polyakov loop \cite{Polyakov:1975rs}, 
which indicates a change in the vacuum structure.

The starting point of the confinement is the Wilson loop.
This is constructed by the integral of the gauge fields over the closed loop as
\begin{align}
	W[C] = \Tr\calP\exp\qty(i\oint_C dx^\mu A_\mu).
\end{align}
In order to adapt the discussion to the current case,
we take the special path for the Wilson loop,
which 3-vector position is fixed to $\vx$.
Then the Wilson loop is dependent on $\vx$ and has the form of
\begin{align}
	L(\vx) = \calP \exp\qty(ig\int_0^\beta d\tau A_0(\vx,\tau)).
\end{align}
This is called the Polyakov loop.

Leaving the formal discussion of the confinement and the Polyakov loop to the Appendix \ref{app:polyakov},
we will see the role of the Polyakov loop in the thermodynamics.
Assuming that the $A_0$ gauge field has the non-zero background,
the Lagrangian for the quark field and the gauge field is given by
\begin{align}
	\calL_{\rm QCD} = \psibar\qty[\rmi \gamma^0\qty(\pdv{t} - \rmi gA_0) + \rmi\vec\gamma\cdot\vec\nabla - m]\psi
\end{align}
neglecting the quantum fluctuation of the gauge field.
The only difference is that the extra term $+gA_0$ associates to the time derivative.
This term works as the imaginary chemical potential
and the thermodynamic potential $\Omega_0$ is modified as
\begin{align}
	\Omega_q = - \int_\vp \qty[\frac{\omega}{2} + T\tr_c \ln\qty(1 + e^{-\beta (\omega - \rmi gA_0)})].
\end{align}
Here $\tr_c$ denotes the trace for the color indices which $A_0$ field possesses.
Rewriting this expression in terms of the Polyakov loop $L$ we obtain
\begin{align}
	\Omega_q = - \int_\vp \qty[\frac{\omega}{2} + T\tr_c \ln\qty(1 + L e^{-\beta\omega})].
\end{align}
The Polyakov loop is associated with the quark thermal part of the potential,
concluding that the small value of the Polyakov loop suppresses the quark thermal excitation.

The discussion above is for the quark field, 
and the similar argument can be done for the antiquark field.
Only the difference is that we have to take Harmonic conjugate of the Polyakov loop as
\begin{align}
	\Omega_{\qbar} = - \int_\vp \qty[\frac{\omega}{2} + T\tr_c \ln\qty(1 + L^\dagger e^{-\beta\omega})].
\end{align}
As the off-diagonal part of the $A_0$ gauge field does not contribute to the thermodynamics,
$A_0$ can be expressed as the superposition of the $a=3,8$ component as
\begin{align}
	A_0 = A_0^3 T_3 + A_0^8 T_8
\end{align}
and the Polyakov loop becomes
\begin{align}
	L = \diag \qty(e^{\rmi q_1}, e^{\rmi q_2}, e^{\rmi q_3}), ~~~ 
	q_1 =   \frac{A_0^3}{2} + \frac{A_0^8}{2\sqrt{3}}, ~ 
	q_2 = - \frac{A_0^3}{2} + \frac{A_0^8}{2\sqrt{3}}, ~ 
	q_3 = - \frac{A_0^8}{\sqrt{3}}.
\end{align}
Computing the trace for the color indices we obtain
\begin{align}
	\Omega_{q\qbar} &= - T \int_\vp \qty[ \tr_c \ln\qty(1 + L e^{-\beta\omega}) + \tr_c \ln\qty(1 + L^\dagger e^{-\beta\omega}) ] \notag\\
	& = - T \int_\vp \left[ \ln\qty(1 + 3\Phi ~ e^{-\beta\omega} + 3\Phi^\dagger ~ e^{-2\beta\omega} + e^{-3\beta\omega})\right. \notag\\
	&\quad\quad + \left. \ln\qty(1 + 3\Phi^\dagger ~ e^{-\beta\omega} + 3\Phi ~ e^{-2\beta\omega} + e^{-3\beta\omega}) \right]
\end{align}
where $\Phi = \tr L/N_c$ and $0 \leq |\Phi| < 1$.
Here we can see that the thermal factor $e^{-\beta\omega}$ and $e^{-2\beta\omega}$ are suppressed by Polyakov loop
but the $e^{-3\beta\omega}$ is not.
This thermal factor corresponds to the baryonic thermal excitation which is color neutral.
As we can see from the expression of the thermal potential,
the Polyakov loop $\Phi$ is the indicator of the quark confinement.

\subsection{Polyakov loop potential}
\label{subsec:model-polyakov-loop-potential}
We introduce the Polyakov loop as the agent to suppress the quark thermal excitation,
but it is not easy to calculate the Polyakov loop as the dynamical object based on the $A_0$ gauge field.
To avoid this difficulty we have three alternative \cite{Fukushima:2017csk};
\begin{enumerate}[i)]
	\item parametrized Polyakov loop effective potential,
	\item inverted Wise potential,
	\item Polyakov loop matrix model.
\end{enumerate}
The common choice is the parametrized Polyakov loop effective potential
which treat the Polyakov loop as the variable of the effective potential as well as the condensates
and determine its value through the gap equation.

The manifestation of the Polyakov loop potential is follows;
\begin{enumerate}[i)]
	\item the value of the pressure and the Polyakov loop as a function of $T$ match to the LQCD results
	\item the Polyakov loop reproduces the large-$T$ behavior $\Phi \to 1$
	\item total pressure reproduces the Stefan-Boltzmann limit $p \to p_{\rm SB}$ at $T\to\infty$
\end{enumerate}
The common forms of the Polyakov loop potential are the polynomial type and the log type.
The first type is given by
\begin{align}
	\calU(T; \Phi,\Phibar)/T^4 = - \frac{b_2(T)}{2}\Phibar\Phi - \frac{b_3}{6}(\Phi^3 + \Phibar^3) + \frac{b_4}{4}(\Phibar\Phi)^2 
\end{align}
where $T$ dependent parameter is given by the power expansion $b_3(T) = a_0 + a_1t^{-1} + a_2 t^{-2} + a_3 t^{-3}$
for $t = T/T_c$ with the critical temperature $T_c$.
The common parameter set are given as \cite{Ratti:2005jh}
\begin{align}
	a_0 = 6.75, ~~~ a_1 = -1.95, ~~~ a_2 = 2.625, ~~~ a_3 = -7.44, ~~~ b_3 = 0.75, ~~~ b_4 = 7.5.
\end{align}
Another type is the log type potential which is given by
\begin{align}
	\calU(T; \Phi,\Phibar)/T^4 = - \frac{a(T)}{2}\Phibar\Phi + b(T)\ln\qty(1 - 6\Phibar\Phi + 4(\Phi^3 + \Phibar^3) - 3(\Phibar\Phi)^2)
	\label{eq:model-polyakov-loop-potential-log}
\end{align}
where $a(T)$ and $b(T)$ are given by the $t$ expansion as
\begin{align}
	a(T) = a_0 + a_1 t^{-1} + a_2 t^{-2}, ~~~ b(T) = b_3 t^{-3}.
\end{align}
This form of potential is based on the analytic expression of the gluon potential for the pure QCD using the strong coupling expansion.
The parametrization is simpler and better than the polynomial type
and the parameters are fixed as \cite{Roessner:2006xn},
\begin{align}
	a_0 = 3.51,~~~ a_1 = -2.47,~~~ a_2 = 15.2,~~~ b_3 = -1.75.
\end{align}
In both type
the critical temperature $T_c$ is defined by
\begin{align}
	T_c = T_\tau \exp\qty(- \frac{1}{\alpha_0 b}), ~~~ b = \frac{11N_c - 2N_f}{6\pi}.
\end{align}
The $T_\tau$ and $\alpha_0$ are determined from the $N_f$ dependence of the $T_c$ for the pure gluon \cite{Schaefer:2007pw} as
\begin{align}
	T_\tau = 1.770 \GeV, ~~~ \alpha_0 = 0.304.
\end{align}

As we can identify $\Phi = \Phibar$,
we have the effective potential of the Polyakov-loop quark meson model as
\begin{align}
	V_{\rm eff}(\mu_I, T; M_q, \Delta, \Phi) = V_{\rm 1-loop}(\mu_I;M_q,\Delta) + \Omega_{q\qbar}(\mu_I, T; M_q, \Delta, \Phi) + \calU(T;\Phi)
\end{align}
and the gap equations for each variables are given by
\begin{align}
	\left.\pdv{V_{\rm eff}}{M_q}\right|_{M_q^*,\Delta^*,\Phi^*} = 0, ~~~ 
	\left.\pdv{V_{\rm eff}}{\Delta}\right|_{M_q^*,\Delta^*,\Phi^*} = 0, ~~~ 
	\left.\pdv{V_{\rm eff}}{\Phi}\right|_{M_q^*,\Delta^*,\Phi^*} = 0.
\end{align}

\chapter{Zero temperature analysis}
\label{chap:zeroT}
In this chapter
we see the results of the numerical calculation of quark-meson model with vanishing temperature and finite density.
We have used the following vacuum parameters as the input:
\begin{gather}
	m_\pi^{\rm vac} = 140 \MeV, ~~~ m_\sigma^{\rm vac.} = 600 \MeV, ~~~ f_\pi^{\rm vac} = 90 \MeV, ~~ M_q^{\rm vac} = 300 \MeV.
\end{gather}
This parameter set corresponds to the following coupling constants for mesons
\begin{gather}
	g^{\rm vac} \simeq 3.333, ~~~ \lambda^{\rm vac} \simeq 126.1.
\end{gather}
The value of $\lambda^{\rm vac}$ is apparently too large to use the naive perturbation theory,
but this choice is the conventional one
and consistent with other analysis \cite{Meistrenko:2021aa, Brandt:2022aa}.

For comparison to the lattice data in Ref.\cite{Abbott:2023aa},
we also examine $m_\pi^{\rm vac}= 170 \MeV$ case
keeping other parameters $(m_\sigma, f_\pi, M_q)$ unchanged.

\section{Evolution of microscopic quantities}
\label{sec:zeroT-evolution}
\begin{figure}[thpb]
	\centering
	\includegraphics[width=0.7\linewidth]{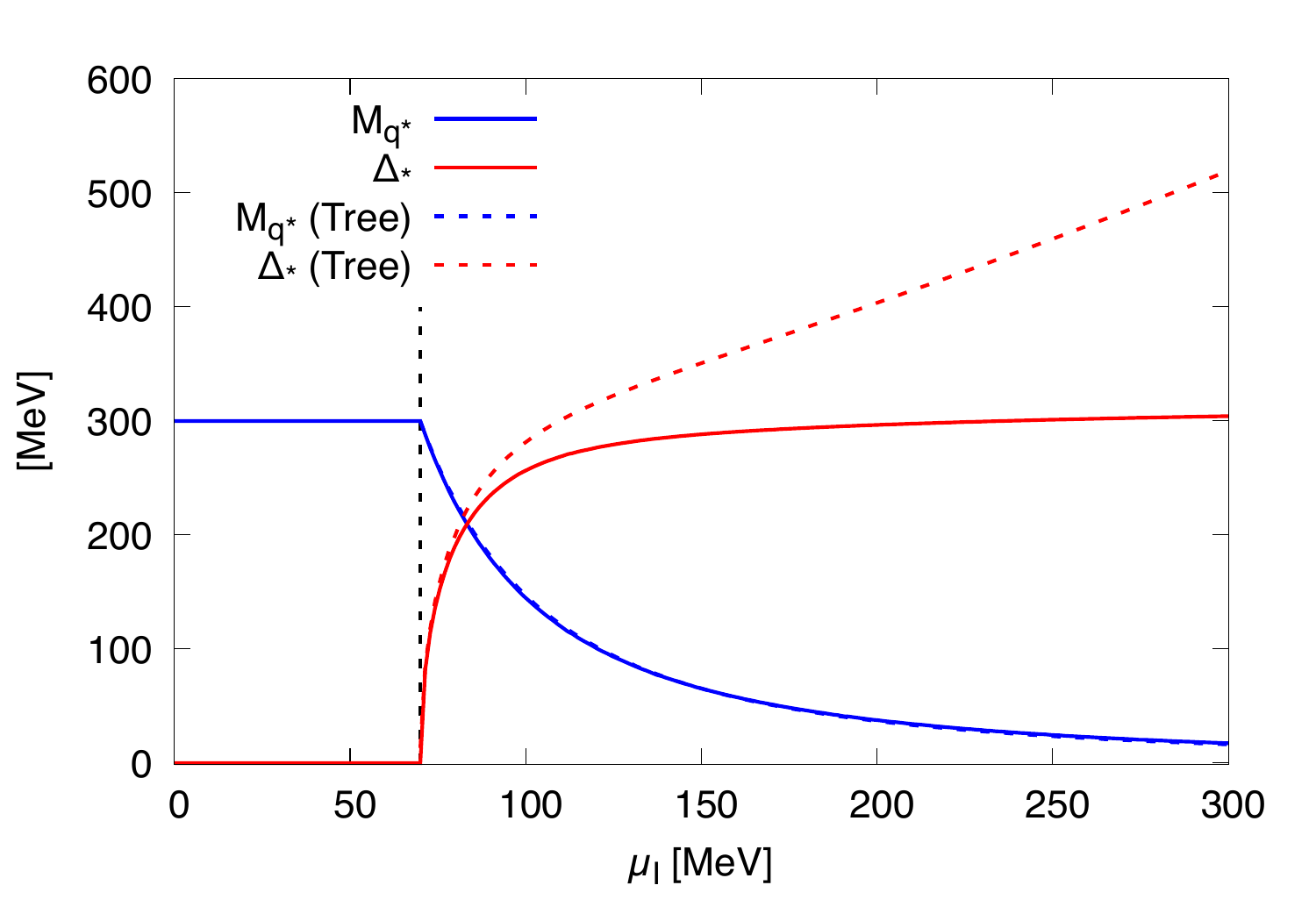}
	\caption{
		The constituent quark mass $M_q$ and the quark mass gap $\Delta$ as the function of $\mu_I/m_\pi$.
		The location of the second order phase transition $\mu_I = 70 \MeV$ is shown by the vertical dashed line.
	}
	\label{fig:zeroT-condensate}
\end{figure}

First,
we see the evolution of condensates as the function of $\mu_I$.
Fig.$\,$\ref{fig:zeroT-condensate} shows the constituent quark mass $M_q$ and the quark mass gap $\Delta$ associated with the pion condensate.
The results of tree-level are illustrated by the dashed lines and the one-loop results by the solid lines.
There are no illuminating differences in $M_q$ between tree-level and one-loop,
but the one-loop results of the quark mass gap $\Delta$ are significantly different from the tree-level ones;
the $\Delta$ increases linearly in $\mu_I$ at tree-level,
whereas it saturates $\Delta\sim 300 \MeV$ at one-loop.
This drastic change of $\mu_I$ dependence is beyond the range of the naive perturbation theory.
The reason should lie in the physical picture of mesons in the existence of quarks rather mere perturbative calculation.

At tree-level,
mesons are regarded as the fundamental degrees of freedom.
But adding the quark degrees of freedom,
they are no longer elementary and treated as composite particles made of quark and anti-quark.
Although the dynamical effects of quarks to meson fluctuations are neglected in our large $N_c$ approximation,
the quark degrees of freedom still have the critical roles on the condensed mesons.
Thus the evolution of the pion condensate is suppressed by the Pauli blocking of quarks.
Summed up,
the quark degrees of freedom couples to the condensed mesons
and change the quantitative behavior of the meson condensates,
by switching the fundamental degrees of freedom from mesons to quarks.

Now we closely examine the behavior of the quark mass gap $\Delta$
by considering the large chemical potential limit $\mu_I \to \infty$.
In this limit,
we assume chiral symmetry is asymptotically restored $M_q\to 0$.
Then the potential has the form of
\begin{align}
	V_0 \to -\frac{2\mu_I^2}{g^2}\Delta^2 + \frac{\lambda}{24g^4}\Delta^4.
\end{align}
After solving the gap equation,
this form immediately leads that $\Delta$ behaves as $\Delta\sim \mu_I$ in this limit.
In the high density limit only relevant scale is $\mu_I$
and the potential should satisfy the conformal scaling as $P\sim \mu_I^4$.
Therefore this scaling of $\Delta \sim \mu_I$ is inevitable by $\mu_I^2\Delta^2$ and $\Delta^4$ terms.

The linearity of $\Delta$ at the tree-level changes to the saturation at one-loop.
This is because the one-loop potential disfavors the large $\Delta$.
The one-loop potential in the same limit is given by
\begin{align}
	V_{\rm 1-loop} = V_0^{\msbar}  + \frac{2N_c}{(4\pi)^2}\qty[\qty(\Delta^4 - 4\mu_I^2\Delta^2)\ln\frac{\Lambda^2}{\Delta^2} + \frac{3}{2}\Delta^4] + V_q^{\rm fin}.
\end{align}
The $V_0^{\msbar}$ is the tree-level potential written in the renormalized parameters
and thus the structure of $\mu_I$ and $\Delta$ are unchanged;
the $\mu_I$ dependence is $V_0^{\msbar} \sim \mu_I^2 \Delta^2$.
The $V_q^{\rm fin}$ is the finite part of the quark single particle energy contribution.
We soon find that the coefficient of $\ln \Lambda^2/\Delta^2$ is negative
and that the one-loop potential disfavors the large $\Delta$.
In addition,
in the case of the tree-level the behavior $\Delta \sim \mu_I$ is necessary to satisfy the conformal scaling $P \sim \mu_I^4$,
however,
as the quark single particle energy $V_q^{\rm fin}$ directly provides the $\mu_I^4$ term
when kinetic energy of quark dominates the potential $E_{\rm kin} \sim \mu_I \gg \Delta$,
the behavior $\Delta \to \rm{const}$ is allowed.

\section{Equation of state and sound velocity}
\label{sec:zeroT-EOS}

Next we see the thermodynamic quantities.
Thermodynamic potential is given by the minimum of the thermodynamic potential $V_{\rm eff}$ as
\begin{align}
	P(\mu_I) = -V_{\rm eff}(\mu_I;{M_q}_*(\mu_I),\Delta_*(\mu_I))
\end{align}
with the solution of gap equation ${M_q}_*$ and $\Delta_*$.
The isospin density and energy density are given by
\begin{align}
	n_I = \pdv{P}{\mu_I}, ~~~ \epsilon = -P + \mu_I n_I.
\end{align}
We also see the sound velocity
\begin{align}
	c_s^2 = \pdv{P}{\epsilon} = \frac{n_I}{\mu_I \chi_I}, ~~~ \chi_I = \pdv[2]{P}{\mu_I}.
\end{align}

From now on
we compute the numerical results with LQCD data in Refs.\cite{Brandt:2022aa,Abbott:2023aa}.
The former is the results of $2+1$-flavor lattice QCD
which masses are at the physical point $m_\pi = 135 \MeV$.
The pion decay constant is $f_\pi \sim 92-96 \MeV$ (the  definition of $f_\pi$ differs from ours by factor $\sqrt{2}$ and the value above is converted to our definition).
The latter is the recent result of multi-pion correlation function
and the pion mass is $m_\pi \sim 170 \MeV$.
This calculation reaches to the high density region up to $\mu_I \sim 1 \GeV$ or $n_I \sim 400 n_0$
(again the definition of isospin chemical potential differs by factor $2$ and we corrected to ours).
Both of two lattice results are consistent to ChPT prediction at low density.

\begin{figure}[thpb]
	\centering
	\includegraphics[width=0.7\linewidth]{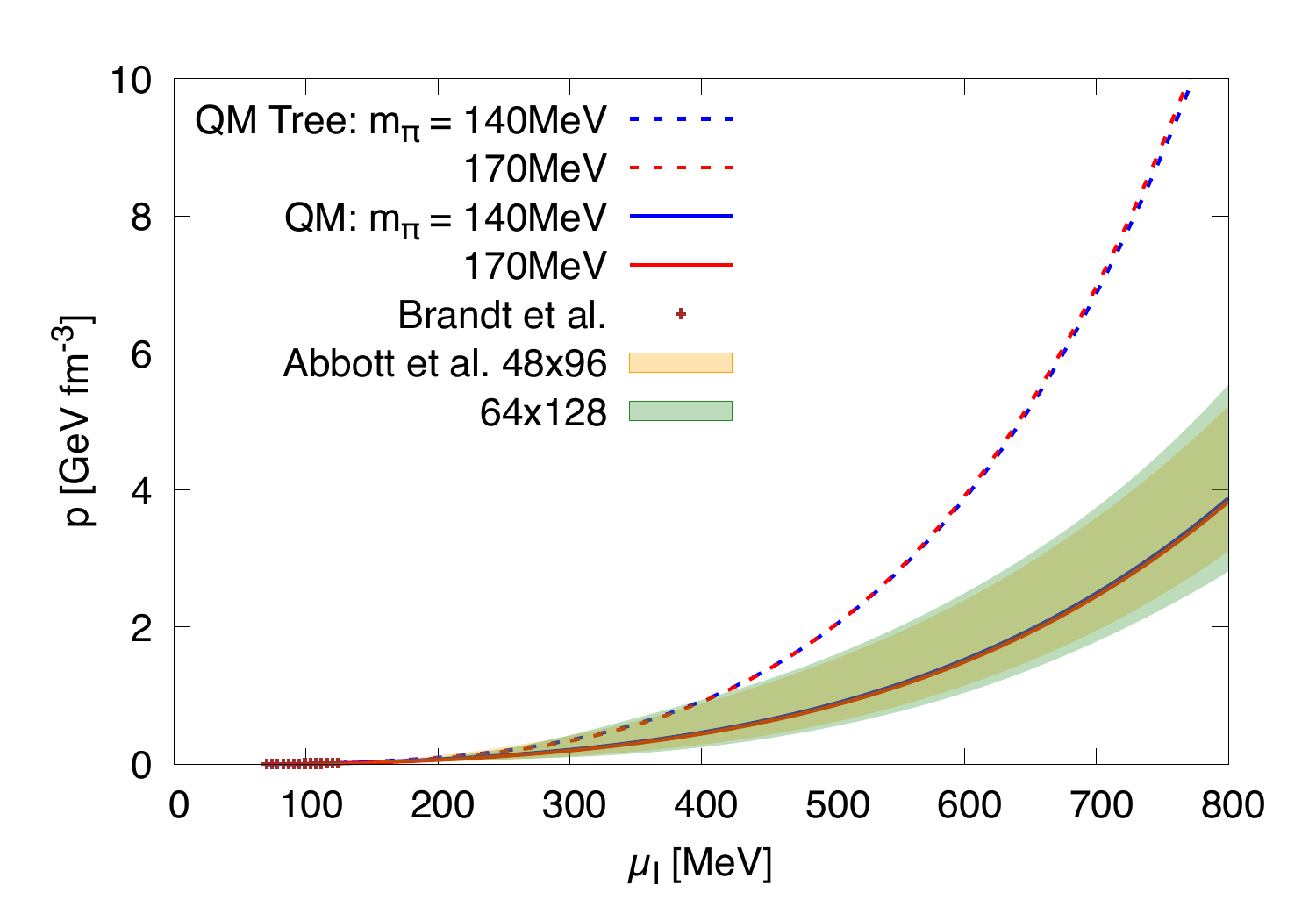}
	\includegraphics[width=0.7\linewidth]{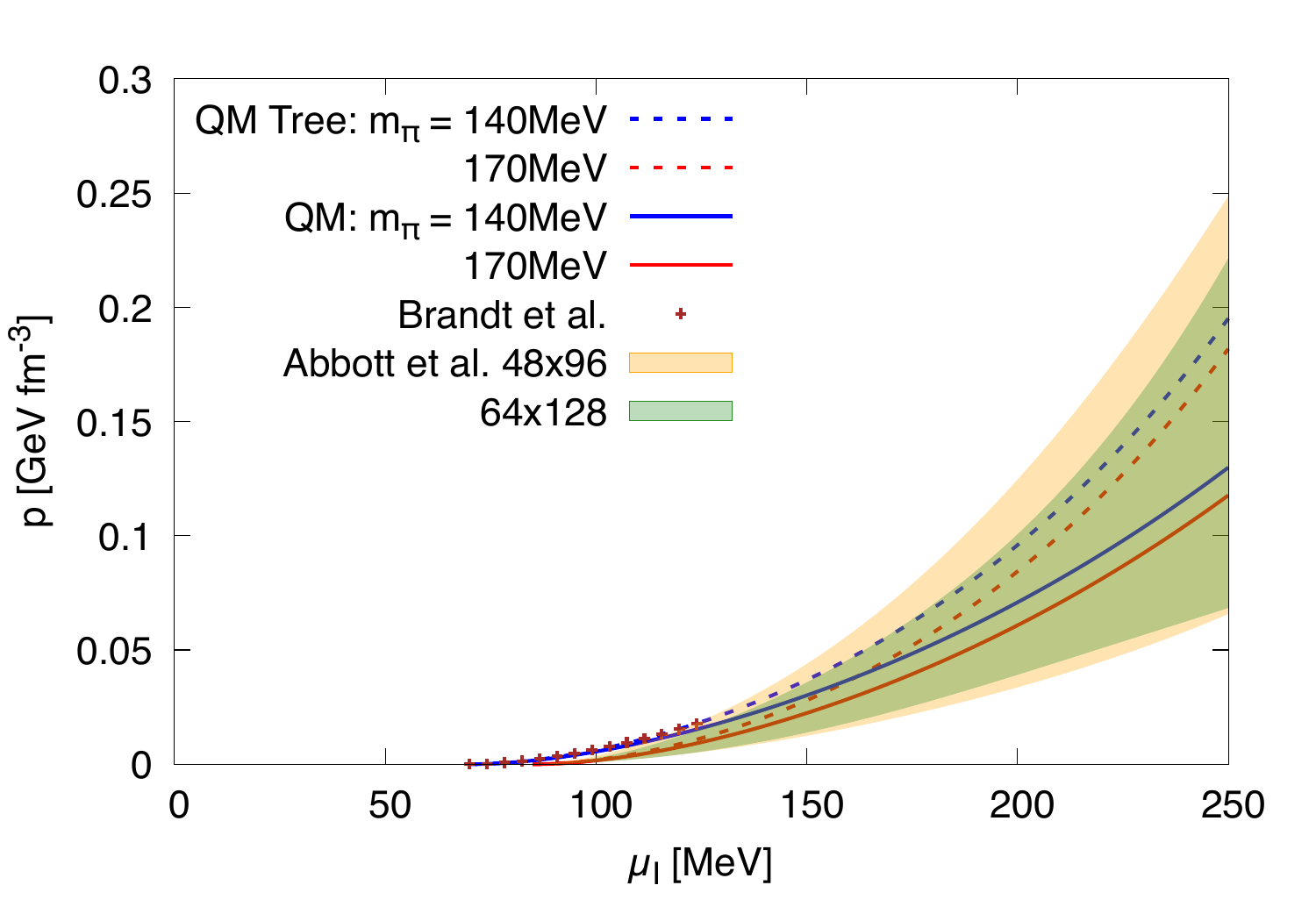}
	\caption{
		Pressure as a function of $\mu_I$.
		We also show the LQCD results in Refs.\cite{Brandt:2022aa,Abbott:2023aa}.
		The one-loop result of quark meson model with the pion mass $m_\pi = 170 \MeV$ matches to the LQCD results \cite{Abbott:2023aa} in wide range of chemical potential.
		The tree-level result increases rapidly 
		and has large discrepancy with one-loop and LQCD results.
		The upper is for the global behavior
		and the lower is for the low density behavior.
	}
	\label{fig:zeroT-pressure-mu}
\end{figure}

\begin{figure}[thpb]
	\centering
	\includegraphics[width=0.7\linewidth]{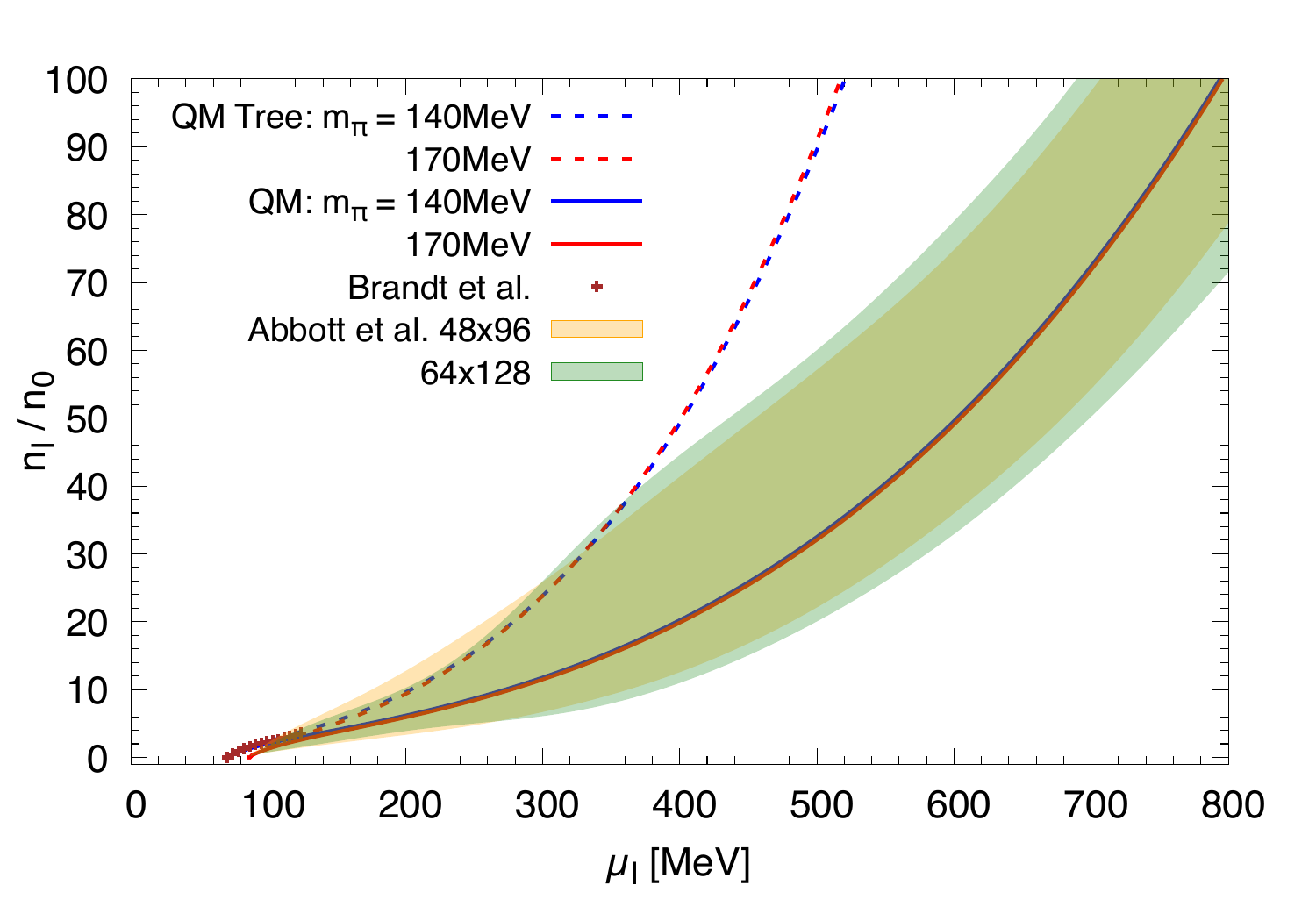}
	\includegraphics[width=0.7\linewidth]{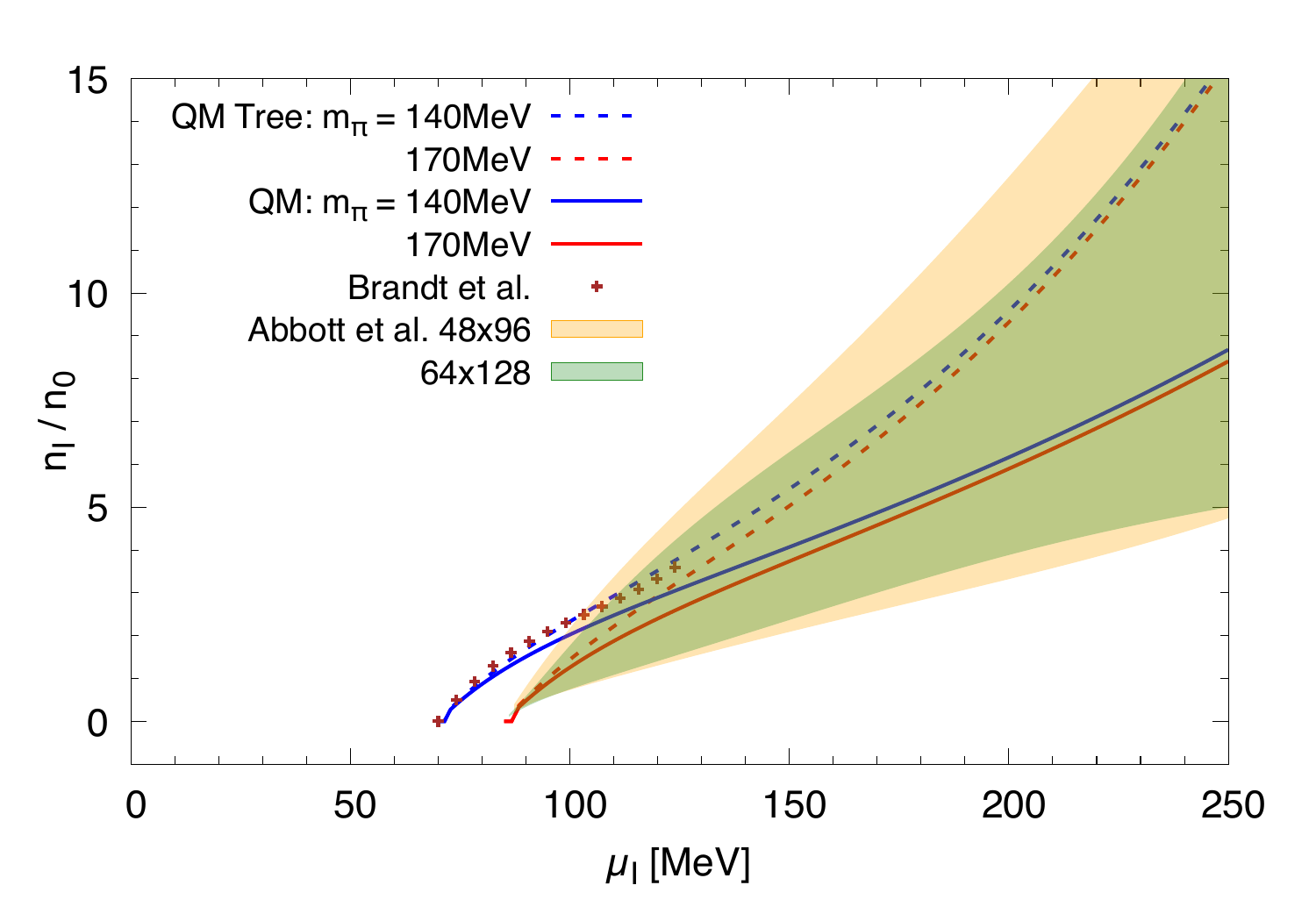}
	\caption{
		Isospin density as a function of $\mu_I$.
		We also show the LQCD results in Refs.\cite{Brandt:2022aa,Abbott:2023aa}.
		The one-loop result of quark meson model with the pion mass $m_\pi = 170 \MeV$ matches to the LQCD results \cite{Abbott:2023aa} in wide range of density.
		The tree-level result increases rapidly at high density region
		and has large discrepancy with one-loop and LQCD results.
		The upper is for the global behavior
		and the lower is for the low density behavior.
	}
	\label{fig:zeroT-isodensity}
\end{figure}

\begin{figure}[thpb]
	\centering
	\includegraphics[width=0.7\linewidth]{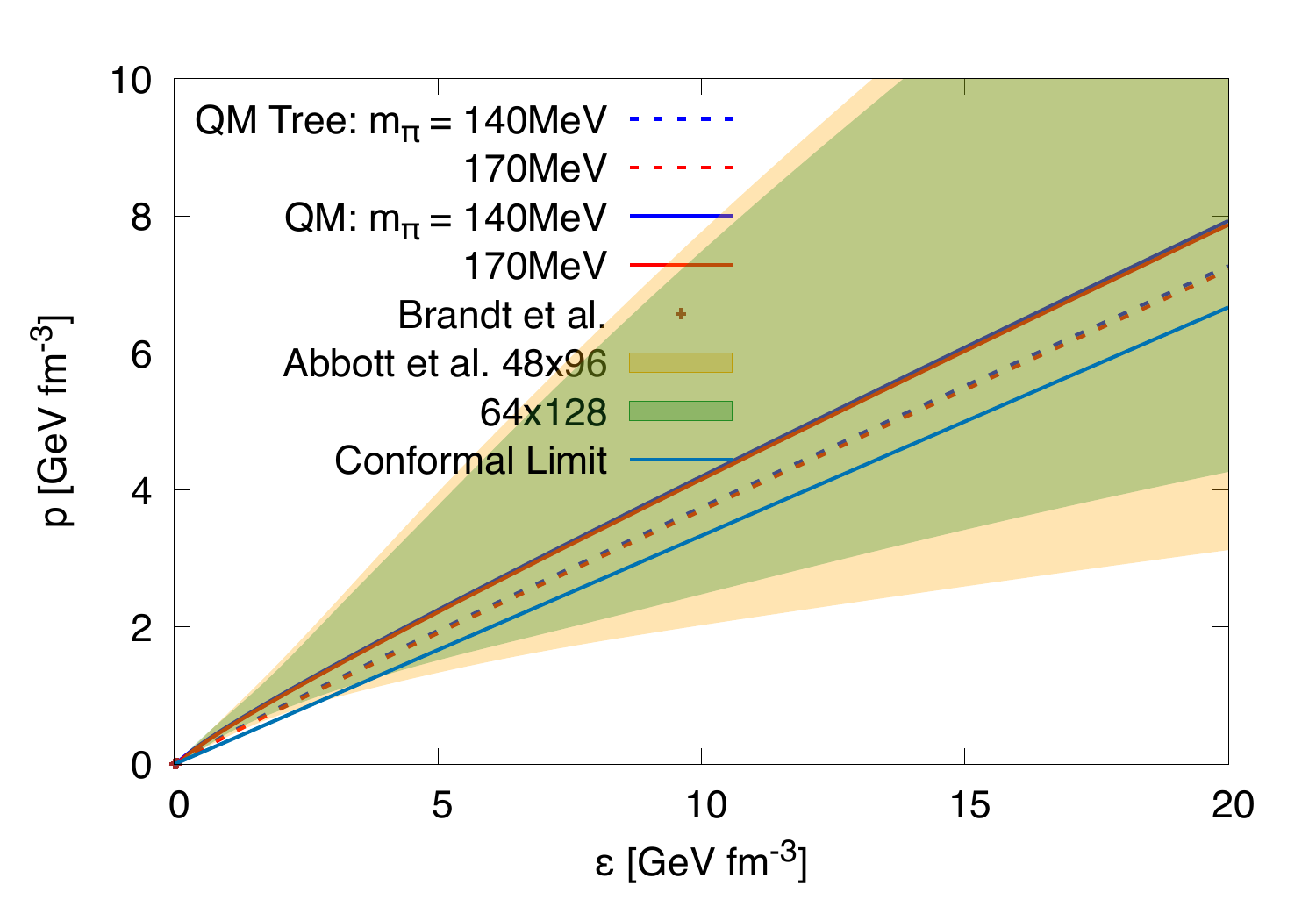}
	\includegraphics[width=0.7\linewidth]{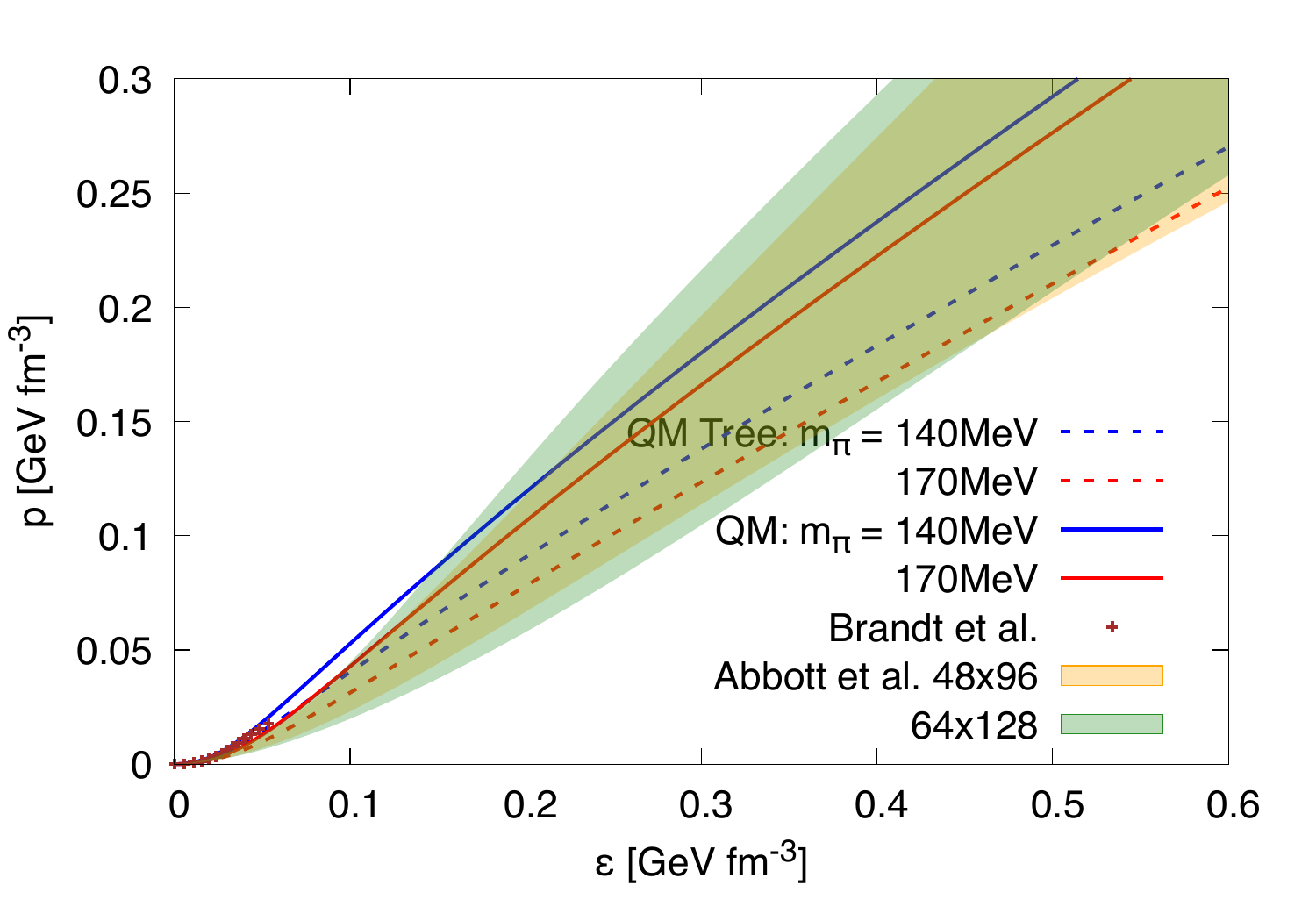}
	\caption{
		Pressure as a function of $\varepsilon$.
		We also show the LQCD results in Refs.\cite{Brandt:2022aa,Abbott:2023aa}.
		The upper is for the global behavior
		and the lower is for the low density behavior.
	}
	\label{fig:zeroT-pressure}
\end{figure}

Fig,$\,$\ref{fig:zeroT-pressure-mu} shows the pressure as a function of $\mu_I$.
Fig.$\,$\ref{fig:zeroT-isodensity} shows the isospin density as a function of $\mu_I$
and Fig.$\,$\ref{fig:zeroT-pressure} the pressure as a function of energy density $\epsilon$.
We also presents the results of tree-level and LQCD.
For comparison to LQCD results,
we also show the results of $m_\pi = 170 \MeV$ case.

In the former figure,
we can see the both tree-level and one-loop results have the consistency to LQCD results at low density.
But the tree-level results overestimate the density and have large discrepancy with LQCD and one-loop results at high density $n_I \sim 50 n_0$.
This too rapid growth of the density at tree-level is the consequence of the $\Delta$ behavior $\Delta \sim \mu_I$ at high density.
In contrast,
the one-loop result with the pion mass of $m_\pi = 170\MeV$ matches to the LQCD results \cite{Abbott:2023aa} in wide range of density.

In the latter figure of pressure as a function of energy density, or EOS,
the one-loop pressure are slightly larger than that of tree-level.
Both of them are consistent to the LQCD results in wide range of density
in contrast to the isospin density.

\begin{figure}[thpb]
	\centering
	\includegraphics[width=0.7\linewidth]{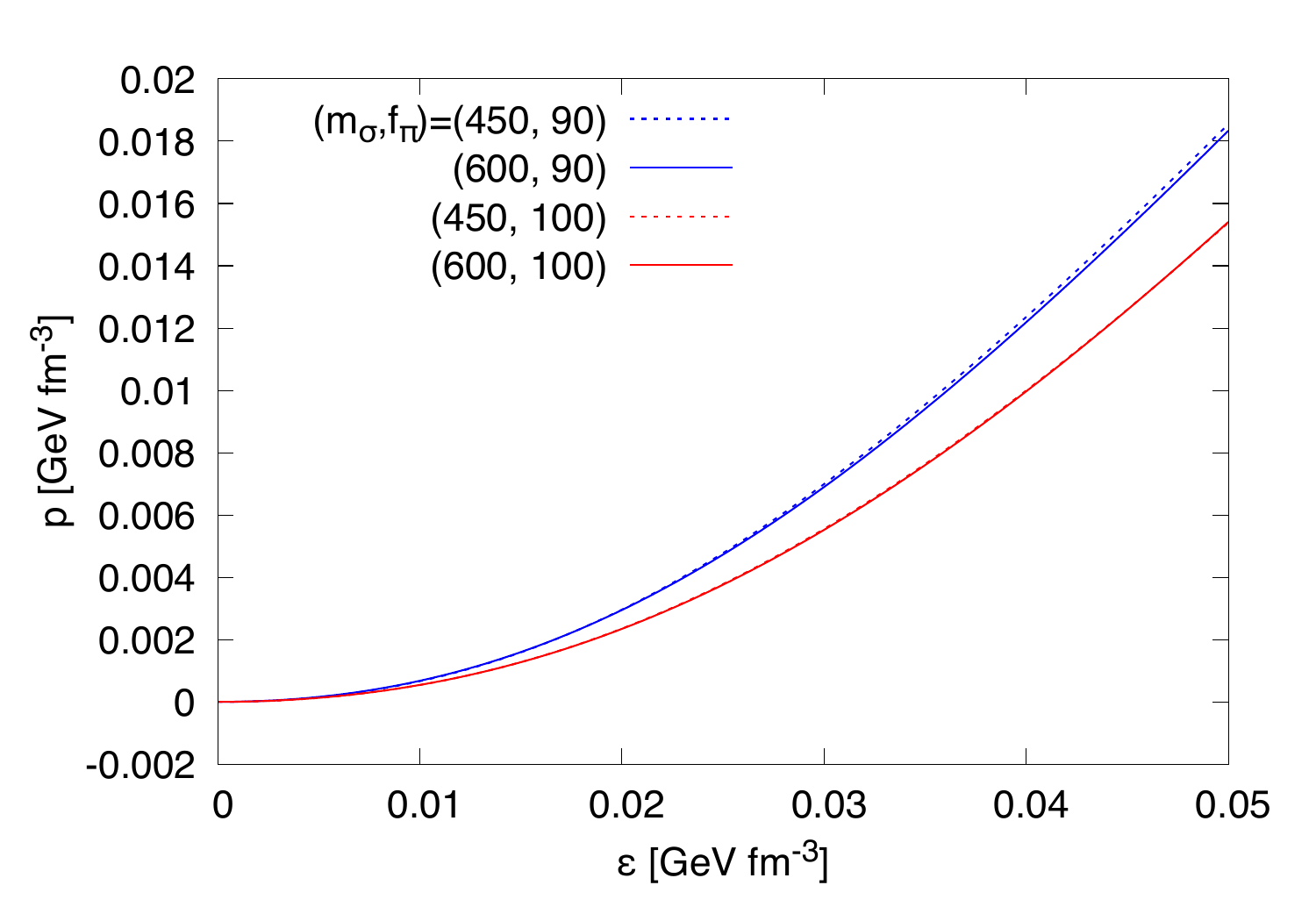}
	\caption{
		The pressure as a function of $\varepsilon$ with different model parameters $(m_\sigma,f_\pi)$.
		The results of $f_\pi = 90\MeV$ is shown by blue lines and those of $f_\pi = 100 \MeV$ by red lines.
		The results of $m_\sigma = 450 \MeV$ is shown by dashed lines and those of $m_\sigma = 600 \MeV$ by solid lines.
		With larger $m_\sigma$ and $f_\pi$ the pressure reduces.
	}
	\label{fig:zeroT-pressure-parameter}
\end{figure}

In \figref{fig:zeroT-pressure-parameter},
we illustrated the EOS at the low energy with different model parameters
$(m_\sigma,f_\pi)=(450,90)$, $(600,90)$, $(450,100)$ and $(600,100)$ MeV.
We can see the variation of the model parameters does not change the results qualitatively.
Especially the $m_\sigma$ does not change the results much.
But the larger $m_\sigma$ and $f_\pi$, or stronger chiral symmetry breaking in vacuum, reduce the pressure.
We get back this point in Sec.$\,$\ref{sec:zeroT-discussion}.

\begin{figure}[thpb]
	\centering
	\includegraphics[width=0.7\linewidth]{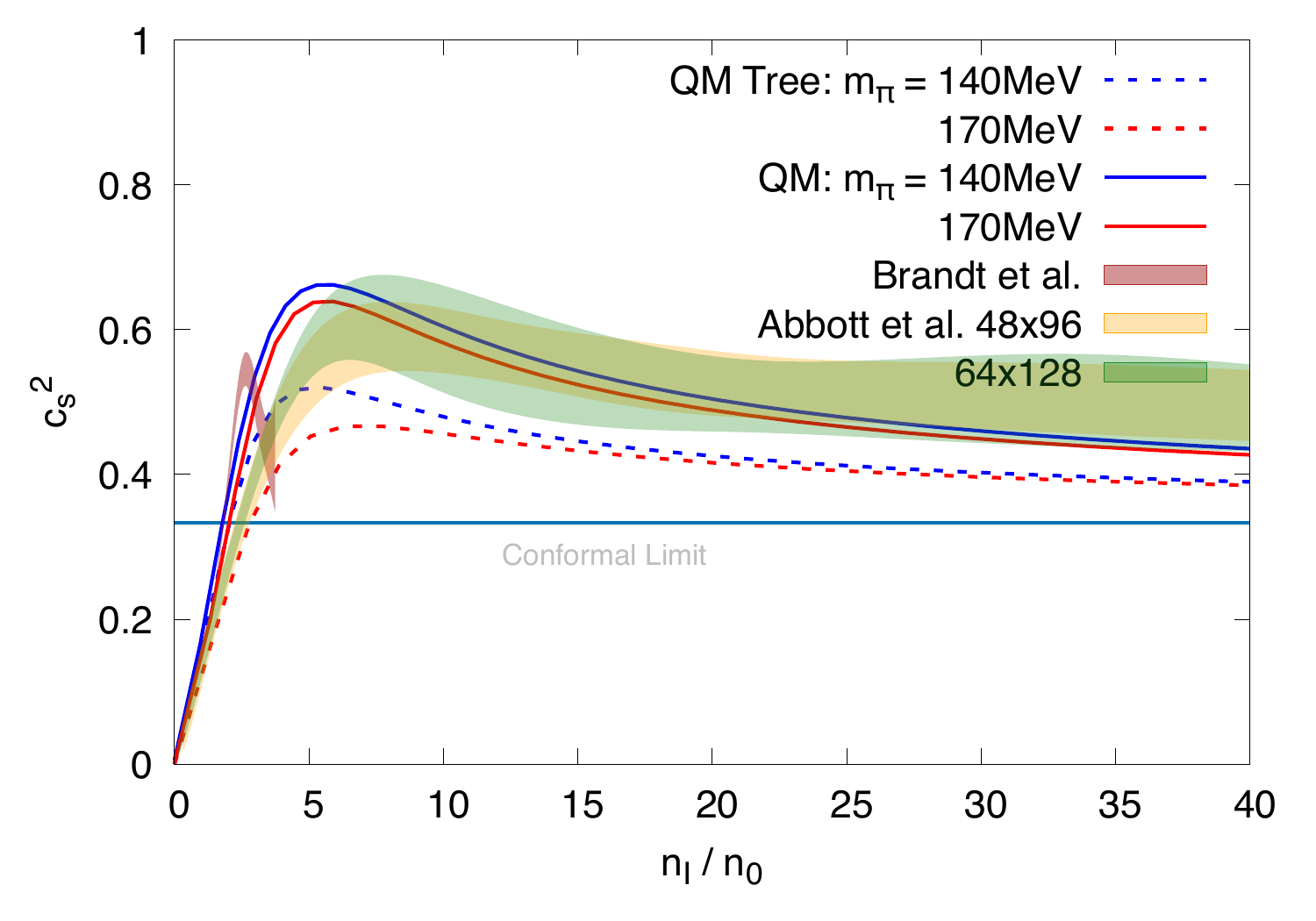}
	\caption{
		The squared sound velocity $c_s^2$ as a function of $n_I/n_0$.
		The qualitative behavior of $c_s^2$ is consistent to the both LQCD results,
		but the peak position and the height of the peak are slightly different with \cite{Brandt:2022aa}.
		Comparing to Ref. \cite{Abbott:2023aa} our results are consistent in wide range of density.
	}
	\label{fig:zeroT-cs2}
\end{figure}

For more detailed examination of the EOS,
we see the behavior of the squared sound velocity $c_s^2$ in Fig.$\,$\ref{fig:zeroT-cs2}.
The $c_s^2$ increases rapidly in low density region $n_I < 2 n_0$ and exceeds the conformal value $c_s^2 = 1/3$ around $c_s^2 \simeq 2 n_0$,
forms the peak at $n_I \simeq 5n_0$ and approaches to the conformal value from above at large density.
These qualitative features are consistent to the LQCD results in Refs.\cite{Brandt:2022aa,Abbott:2023aa}.
\begin{comment}
However the detailed results in the BCS region are slightly different.
The peak position of $c_s^2$ is $\mu_I/m_\pi \sim 1.3$ in our results which is consistent to the LQCD results in \cite{Abbott:2023aa},
while it is $\mu_I/m_\pi \sim 0.8$ in \cite{Brandt:2022aa}.
The height of the peak has the same tendency;
it is $c_s^2 \sim 0.5$ in our results and $c_s^2 \sim 0.3$ in \cite{Brandt:2022aa}
and the results of \cite{Abbott:2023aa} has agreement with our result.
\end{comment}
However,
the rapidity with which the $c_s^2$ decreases after making a peak is different between our results and \cite{Brandt:2022aa}.
In our results the $c_s^2$ decreases slowly and large $c_s^2 \simeq 0.5$ remains at high density region,
while in Ref.\cite{Brandt:2022aa} the $c_s^2$ decreases rapidly.
This rapid decrease to the conformal value indicates that the transition from hadronic to quark matter rapidly takes place.
The origin of this discrepancy in the sound velocity is not clear at present.

We should notice that the $c_s^2$ at tree-level also has the peak structure in the crossover region
and relax to the conformal value from above.
As we have seen in Sec.$\,$\ref{sec:intro-sound-velocity}
the term $a_2\mu_I^2$ in the pressure with positive $a_2$ in addition to $\mu_I^4$ term
concludes that the $c_s^2$ converges to $1/3$ from above.
Solving the gap equation at tree-level,
we can find that the tree-level pressure has the form
\begin{align}
	P_0(\mu_I) = \frac{m_\pi^4 f_\pi^2}{8\mu_I^2} + \frac{f_\pi^2 \qty(4\mu_I^2 - m_0^2)^2}{2(m_\sigma^2 - m_\pi^2)}.
\end{align}
This pressure has the $\mu_I^2$ term with positive coefficient,
thus the $c_s^2$ can exceed the conformal value $1/3$.
In addition the chiral restoration term $\sim \mu_I^{-2}$ suppresses $c_s^2$ at low density.
Combining these two effects,
the $c_s^2$ has the peak structure even at tree-level.
Even though the quantitative behavior is similar with one-loop level,
the peak structure at tree-level is physically fictitious.
This is because the peak structure of $c_s^2$ in middle density region is just the result of the competition between the dynamics of low and high density region.
As we have mentioned in Sec.$\,$\ref{sec:zeroT-evolution},
the underlying dynamics of high density region is not correctly described at tree-level.
Thus we conclude that the peak structure at tree-level is not physical.

The convergence to the conformal value from above is different tendency with pQCD
and we discuss this point in Sec.$\,$\ref{sec:zeroT-discussion}.

\section{Quark saturation}
\label{sec:zeroT-saturation}
As mentioned,
the low density physics is governed by meson degrees of freedom
and the internal structure of mesons is not appeared.
But at high density the distance of these mesons becomes short
and the quark degrees of freedom appears in its dynamics.
The physics of intermediate region is determined by the competition between these two degrees of freedom at low and high density.

To estimate the density where the quark degrees of freedom becomes important,
we introduce the overlap density of pion $n_I^{\rm overlap}$.
This is the density where the typical distance of pions becomes comparable to the pion size.
The experimental determination based on the $\pi e$ scattering and the $e^+e^- \to \pi^+\pi^-$ process \cite{Ananthanarayan:2017aa}
gives the estimate $\braket{r^2}_V = 0.434(5) \fm^2$ \cite{ParticleDataGroup:2018ovx}, or
\begin{align}
	r_\pi^V = \sqrt{\braket{r^2}_V} \sim 0.66 \fm,
\end{align}
which has been well reproduced by the lattice QCD calculation \cite{Koponen:2016aa, Wang:2021aa}.
Then the overlap density is given by
\begin{align}
	n_I^{\rm overlap} = \qty(\frac{4}{3}\pi r_\pi^3)^{-1} \sim 0.83 {\rm fm^{-3}} \sim 5.2 n_0.
\end{align}
Fig.$\,$\ref{fig:zeroT-isodensity} shows this density corresponds to 
\begin{align}
	\mu_I \sim 1.2 m_\pi \sim 168 \MeV.
\end{align}

We should notice that the quark wavefunction is percolated outside of $r_\pi^V$
and the internal quarks starts interacting with other quarks in other mesons at lower density.
This suggests that there should be more suitable measure which contains the information of quark wavefunction.
Following Refs.\cite{Kojo:2021ugu,Kojo:2019raj,McLerran:2018aa},
we interpret the peak structure as the signature of the {\it quark saturation}
or the onset of the quark matter.
This quark saturation is characterized by the occupation probability of the quark matter;
the quark states are occupied with the probability $\sim 1$ from low momentum 
and the newly added quarks by increasing the density must fill a state on top of the already occupied states.

\begin{figure}
	\centering
	\includegraphics[width=0.7\linewidth]{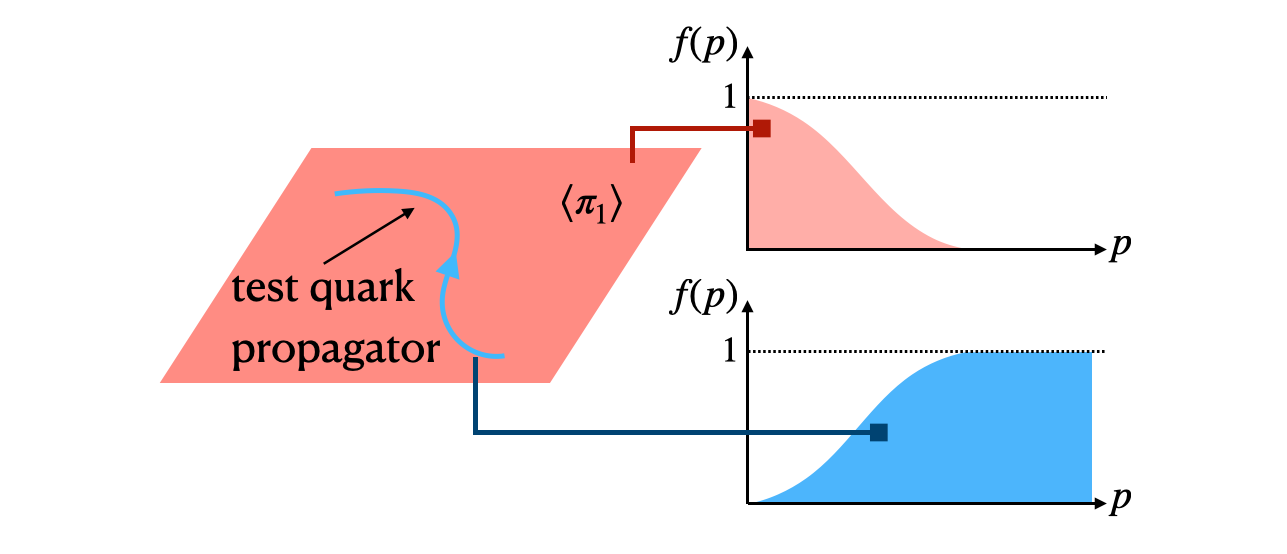}
	\caption{
		The test quark propagator in the pion condensed phase and the occupation probability $f(p)$.
		The $f(p)$ for the propagator reflect the occupation probability of the quark matter.
	}
	\label{fig:zeroT-occupation-cartoon}
\end{figure}

Practically,
the occupation probability of the quark matter is calculated through the test quark propagator,
as shown in \figref{fig:zeroT-occupation-cartoon}.
Because of the Pauli blocking,
the distribution of test quark propagator reflects the occupation probability of the quark matter.
With the pion condensate in the background,
the low momentum mode propagation of quark is suppressed
while that of the high momentum mode is not.
This means that the background condensate consists of the low-momentum quark 
and suppresses the low-momentum mode of the quark propagator through the Pauli blocking.

The quark occupation probability in the pion condensed phase can be computed in the standard Nambu-Gor\textquotesingle kov formalism.
The derivation is reviewed in Appendix \ref{app:quark-propagator}.
The occupation probability of $u, ~ d, ~ \bar u$ and $\bar d-$ quarks are given by
\begin{align}
	f(p) &= f_{u,\bar d}(p) = \frac{1}{2}\qty(1 + \frac{\mu_I - E_D(p)}{E(p)}) \\
	\bar f(p) &= f_{d,\bar u}(p) = \frac{1}{2}\qty(1 + \frac{\mu_I + E_D(p)}{E(p)}).
\end{align}
In the absence of mass gap $\Delta$,
$u-$ and $\bar d-$ quarks are occupied up to $p = \mu_I$
and the $d-$ and $\bar u-$ quarks are empty.
When the mass gap $\Delta$ is turned on,
the distribution of $u-$ and $\bar d-$ quarks becomes smoothly reducing distribution around $p\sim\mu_I$.

\begin{figure}[thpb]
	\centering
	\includegraphics[width=0.7\linewidth]{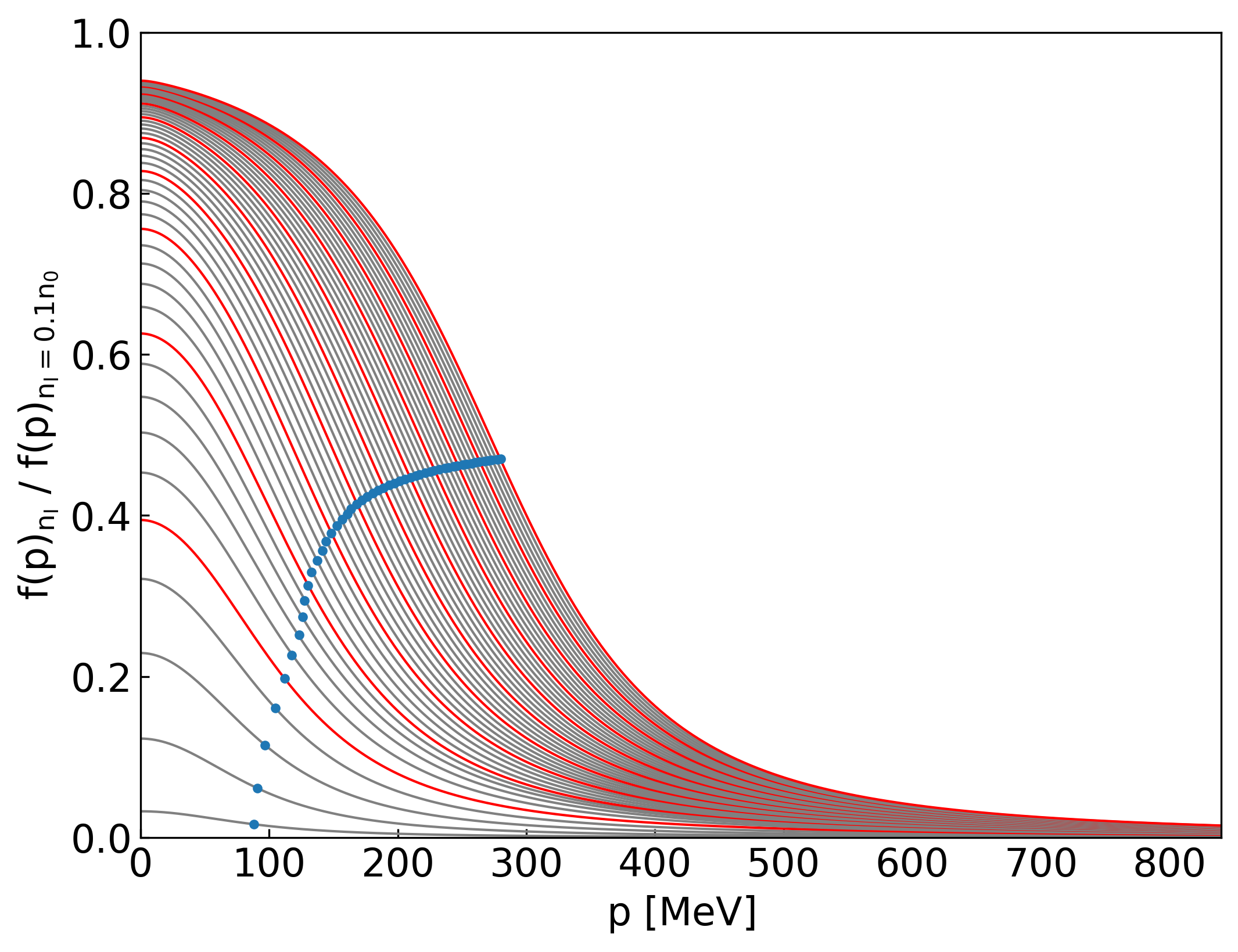}
	\caption{
		Quark occupation probability as a function of momentum $p$.
		The gray curves are the results of $n_I/n_0 = 0.2 - 10.0$ in $0.2$ increments.
		The red curves are the results of $1.0$ increments.
		The blue dots are the location of the surface of the distribution
		where $f(p)$ has the half maximum value.
	}
	\label{fig:zeroT-quark-occupation}
\end{figure}

The Fig.$\,$\ref{fig:zeroT-quark-occupation} shows the quark occupation probability as a function of momentum $p$
with isospin density from $n_I/n_0 = 0.2$ to $10$ in $0.2$ increments for gray curves and $1.0$ for red curves.
The blue dots are the location of the {\it surface} of the distribution
where $f(p)$ has the half maximum value; $f(p_{\rm surf}) = f(0)/2$.

\begin{figure}[thpb]
	\centering
	\includegraphics[width=0.7\linewidth]{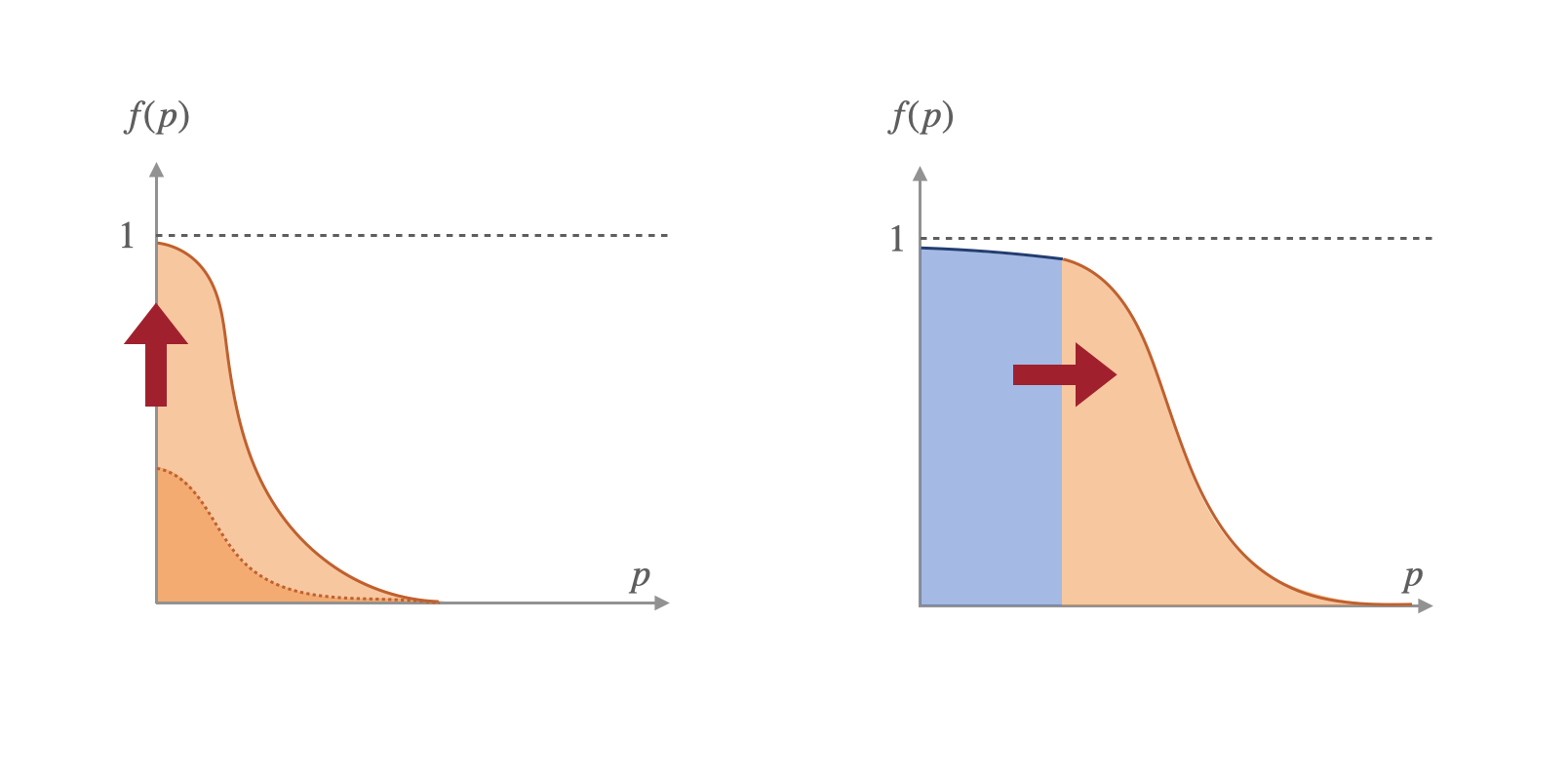}
	\caption{
		Schematic figures for the evolution of the occupation probability. 
		{\bf Left}: The vertical evolution;
		{\bf Right}: The horizontal evolution.
	}
	\label{fig:zeroT-occupation-evolution}
\end{figure}

To discuss the transition of the effective degrees of freedom in the context of the quark saturation,
it is useful to decompose the evolution of $f(p)$ into two modes.
The first mode is the {\it vertical} evolution (Fig.$\,$\ref{fig:zeroT-occupation-evolution}, {\bf Left})
in which the distribution $f(p)$ increases in monotonic way as $f(p) \sim n_I \varphi_\pi^{\rm vac}(p)$ with the quark distribution $\varphi_\pi^{\rm vac}(p)$ in a single pion.
This regime corresponds to the ideal pion gas regime at low density
where the pions are not so close to each other
and the quarks inside pions are largely unaffected by other quarks in other pions.
In this region $\epsilon/n_I$ is almost constant
and the pressure $P=n_I^2\partial(\epsilon/n_I)/\partial n_I$ is very small.
Here the quarks have the contributions to energy density but can not increase the pressure.
The second is the {\it horizontal} evolution (Fig.$\,$\ref{fig:zeroT-occupation-evolution}, {\bf Right})
in which the distribution $f(p)$ increases and the surface of $f(p)$ extends to the large momentum region.
This region corresponds to the ordinary quark matter region
where the quark interactions and Pauli blocking become important.
In this region the $\epsilon/n_I$ increases with $n_I$.
Thus the pressure can increase and the sound velocity becomes large.

\begin{figure}[thpb]
	\centering
	\includegraphics[width=0.7\linewidth]{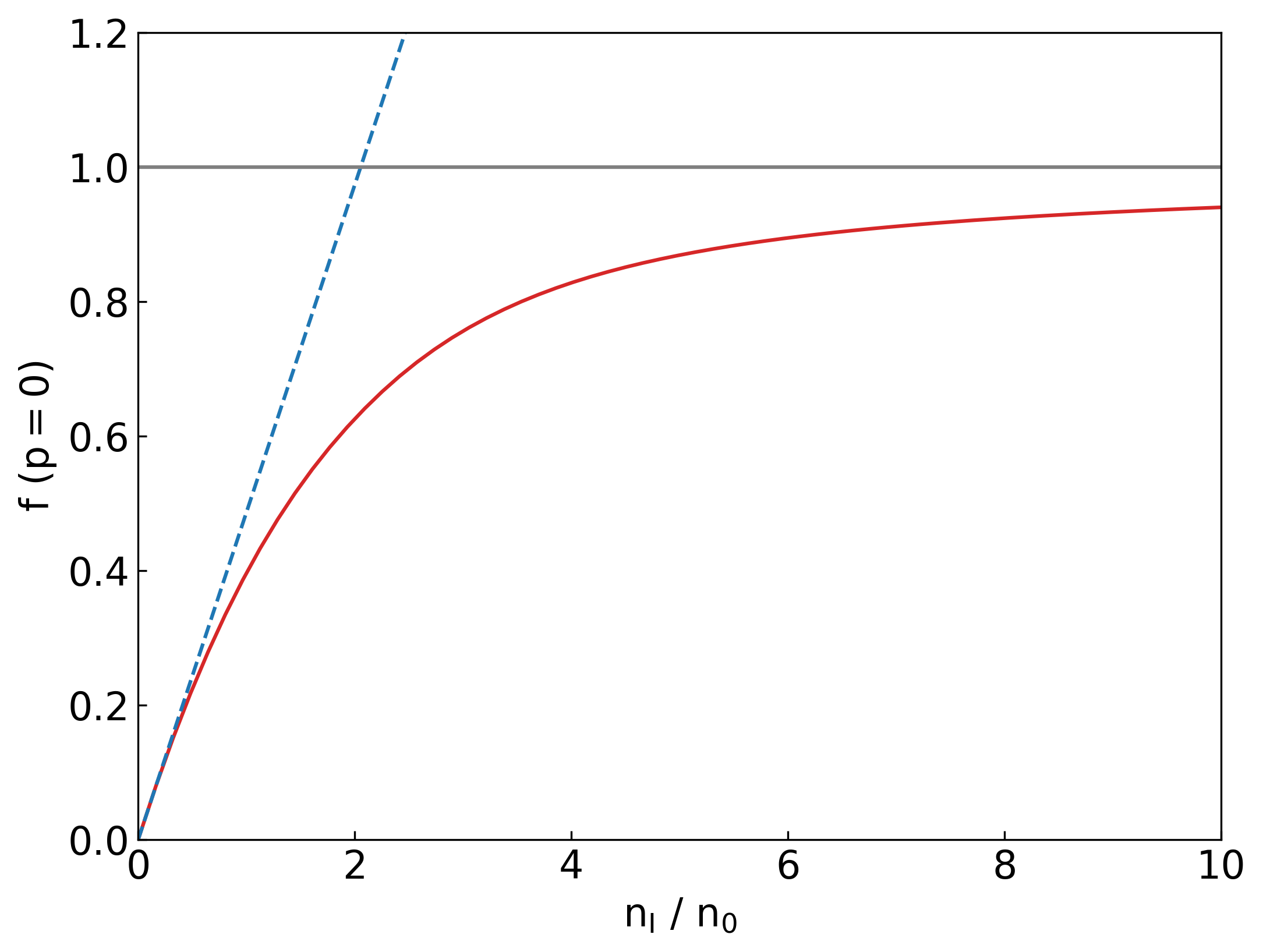}
	\caption{
		The evolution of the occupation probability for $p = 0$ (red line).
		If the momentum of increase near $n_I=0$ continues (blue dashed line), 
		the value of $f(p=0)$ exceed $1$ at $n_I \simeq 2.0 n_0$.
	}
	\label{fig:zeroT-occupation-zeromode}
\end{figure}

As we can see in Fig.$\,$\ref{fig:zeroT-quark-occupation},
the calculated distribution of $f(p)$ is the mixture of these two modes in our quark meson model.
In the low density region from $0$ to $\sim 3.0 n_0$,
the magnitude of $f(p)$ at $p=0$ grows rapidly and reaches the magnitude $\sim 0.8$,
which is considered as the result of the vertical evolution.
Even in this region we can see that 
the vertical evolution is gradually decelerating
and the horizontal evolution is slightly increasing.
If the momentum of increase near $n_I=0$ continues,
the value of $f(p=0)$ exceed $1$ at $n_I \simeq 2.0 n_0$,
as shown in Fig.$\,$\ref{fig:zeroT-occupation-zeromode}.
But the Pauli blocking does not allow $f(p)$ to exceed $1$,
the distribution must extend into the large momentum region.
Thus the horizontal evolution becomes dominant as increasing the density
while the vertical evolution is still remaining.
This smooth transition from the vertical to horizontal evolution suggests the smooth transition of the effective degrees of freedom \cite{Kojo:2021ugu}.

\begin{figure}[thpb]
	\centering
	\includegraphics[width=0.7\linewidth]{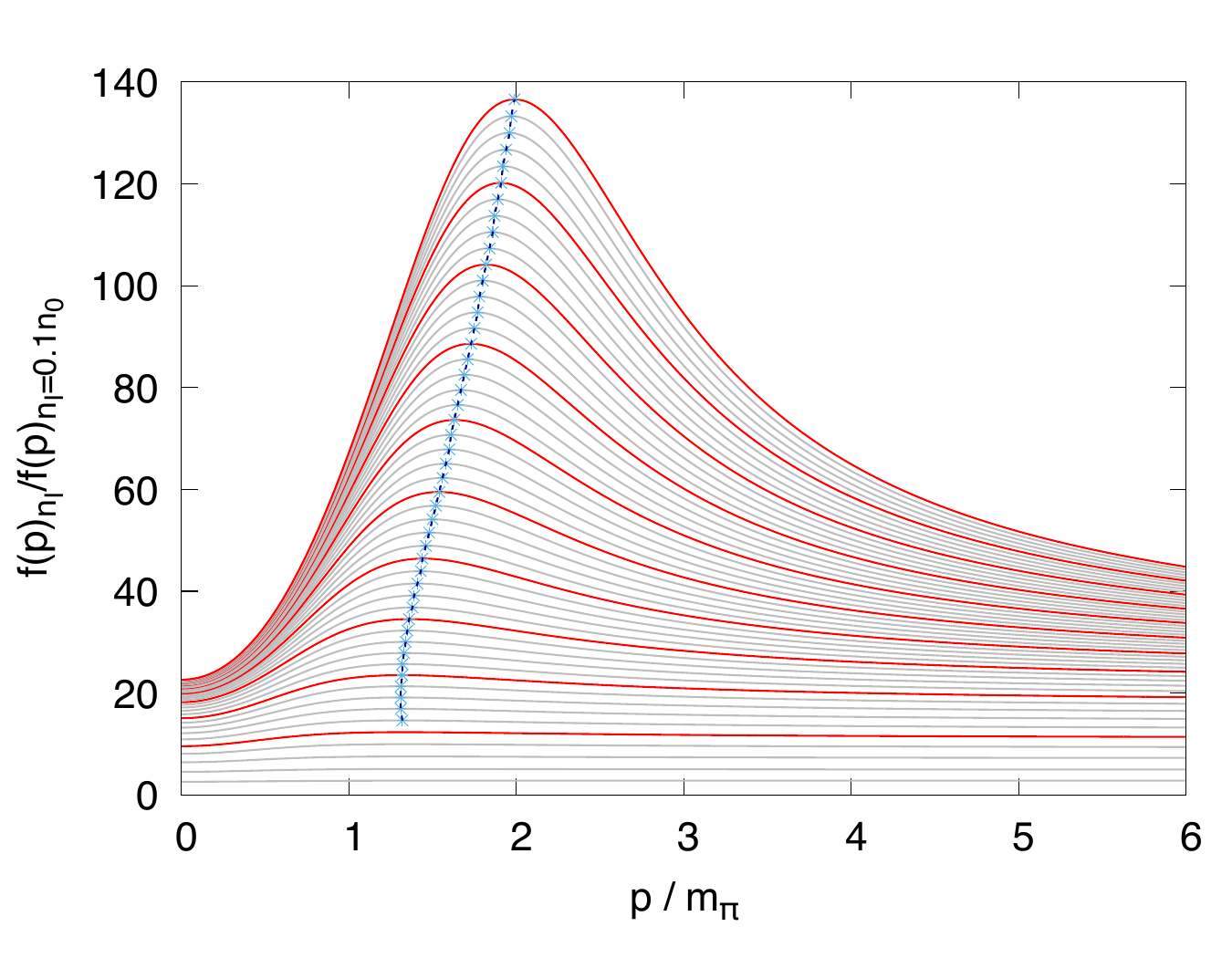}
	\caption{
		The ratio $f(p)_{n_I}/f(p)_{n_I=0.10n_0}$ as a function of momentum $p$.
		The gray and red curves are the same as Fig.$\,$\ref{fig:zeroT-quark-occupation}.
		The blue dots are the location of the peak of $f(p)_{n_I}/\varphi_\pi^{\rm vac}(p)$.
	}
	\label{fig:zeroT-surface-momentum}
\end{figure}

\begin{figure}[thpb]
	\centering
	\includegraphics[width=0.7\linewidth]{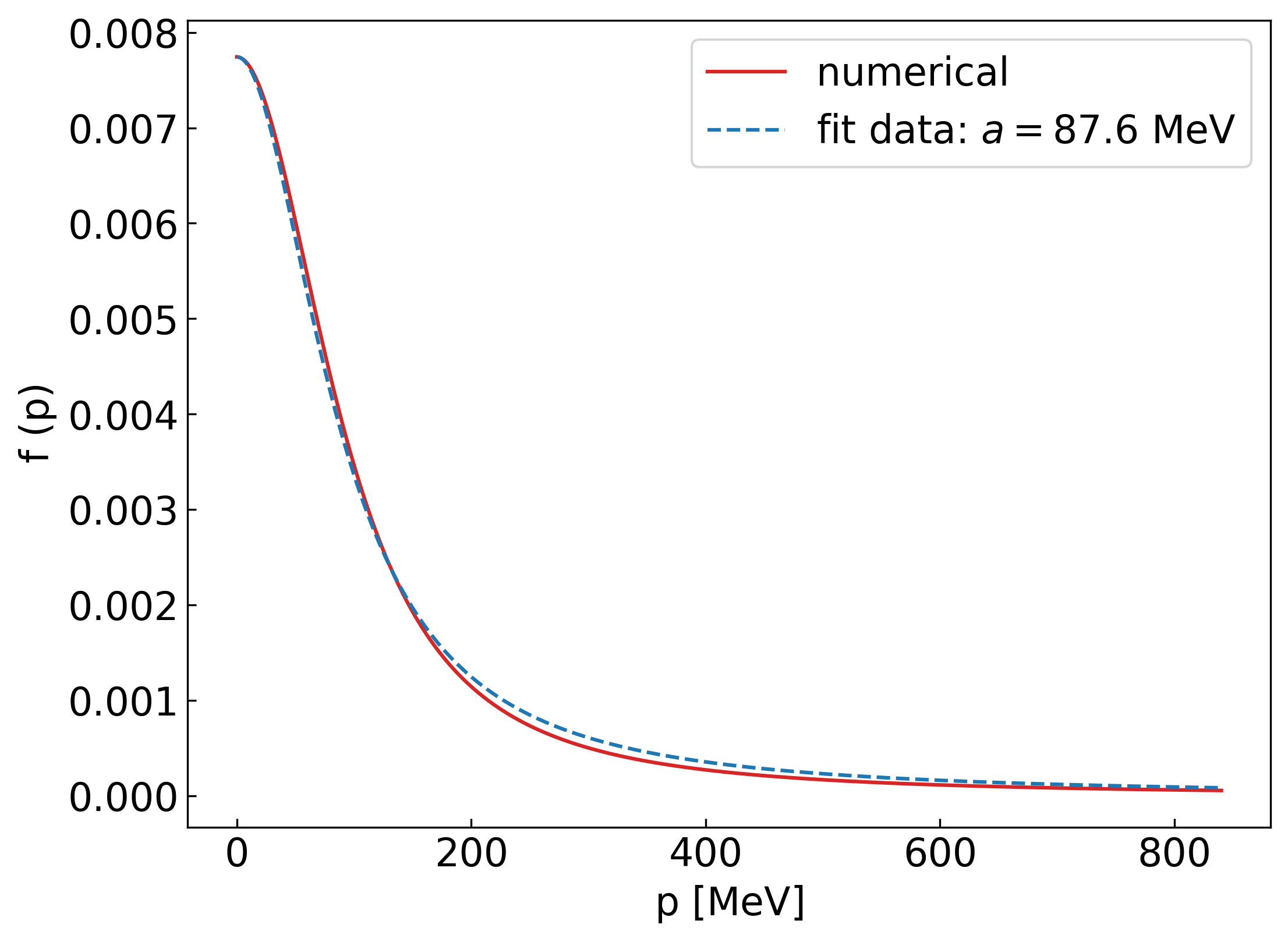}
	\caption{
		The occupation probability $f(p)$ at $n_I = 0.1 n_0$.
		The red line is the numerical result
		and the blue dashed line is the fit by the monopole Ansatz.
		This wavefunction can be regarded as the occupation probability of the vacuum $\varphi_\pi(p)$
		and is the reference distribution for the ratio $f(p)_{n_I}/\varphi_\pi(p)$ in Fig.$\,$\ref{fig:zeroT-surface-momentum}.
	}
	\label{fig:zeroT-occupation-vacuum}
\end{figure}

In order to observe more closely the transition from the BEC phase to the BCS phase,
we examine how much the horizontal evolution of $f(p)$ is realized in the intermediate region $n_I \sim 2.0 - 5.0 n_0$.
This is illustrated in Fig.$\,$\ref{fig:zeroT-surface-momentum}
by taking the ratio $f(p)_{n_I}/f(p)_{n_I=0.10n_0}$.
The vertical evolution keeps this ratio flat over all range of momentum.
In contrast the horizontal evolution increases this ratio at the large momentum region.
To characterize the evolution of the Fermi surface,
we trace $p_{\rm peak}$,
the position of the maximum of $f(p)/\varphi_\pi^{\rm vac}(p)$.
In practice,
we take $f(p)_{n_I=0.1n_0}$ instead of $\varphi_\pi^{\rm vac}$
expecting $n_I = 0.1n_0$ is low enough to be in the vertical evolution regime;
$f(p)_{n_I=0.1n_0} = 0.1n_0 \varphi_\pi^{\rm vac}(p)$.
The distribution of $f(p)_{n_I=0.1n_0}$ is shown in Fig.$\,$\ref{fig:zeroT-occupation-vacuum}
and described well by a simple monopole Ansatz or Breit-Wigner form
\begin{align}
	\varphi_\pi(p) \sim \frac{1}{1 + p^2/a^2}.
\end{align}
The value of $a$ is $\simeq 87.6$ MeV as determined by fit.
This form suggests the pion wavefunction in the coordinate space has the exponentially-decaying form.


Fig.$\,$\ref{fig:zeroT-surface-momentum} shows the position $p_{\rm peak}$ is almost constant in the low density region up to $n_I \sim 2.0 n_0$.
Beyond this density the $p_{\rm peak}$ starts to increase and the horizontal evolution becomes dominant.
The density region of horizontal evolution $n \gtrsim 2.0n_0$ is consistent with substantial increase of the sound velocity $c_s^2 \gtrsim 1/3$.
The onset of the horizontal evolution and the quark degrees of freedom in dynamics is relatively early comparing to the overlap density of pion $n_I^{\rm overlap}$.

We also examine the transition of the effective degrees of freedom by the half maximum value of $f(p)$
as illustrated in Fig.$\,$\ref{fig:zeroT-quark-occupation}
As increasing the density or chemical potential,
the width of the distribution becomes wider.
But in the vertical evolution the maximum value of $f(p)$ also increases.
Thus the value of $p_{\rm surf}$ does not increase much.
On the contrary,
in the vertical evolution where the low-momentum mode of the quarks are almost saturated,
the value of $f(p_{\rm surf})$ does not change
while the position of $p_{\rm surf}$ increases.
The transition point of these two evolution is around $n_I \simeq 2.0 n_0$,
which is the consistent result with the previous analysis.

This transition density is much smaller than the pion overlap density $n_I^{\rm overlap} \simeq 5.2n_0$.
This is because the pions does not have the solid boundary and the quark wavefunction is percolated outside of $r_\pi^V$.

\section{Discussion}
\label{sec:zeroT-discussion}
In this section we have short discussion on several issues which are not fully mentioned in the previous sections.
First we discuss the connection of the chiral symmetry restoration and softening of EOS at high density.
Next we compare our results to the pQCD calculation and discuss the possible origin of the discrepancy of the asymptotic behavior of sound velocity at high density region.
Finally we discuss the trace anomaly 
$\Delta_{\rm tr} = 1/3 - P/\varepsilon$ as the another measure of the conformality.

\subsection{Chiral symmetry restoration and softening}
\label{ssec:zeroT-chsb-softening}
In the Sec.$\,$\ref{sec:zeroT-EOS} we have seen that larger $f_\pi$ or $m_\sigma$ make the EOS softer at high density.
Here we discuss the connection of this softening to the chiral symmetry breaking in the vacuum and its restoration at high density region.
The stronger chiral symmetry breaking corresponds to larger $f_\pi$ and $m_\sigma$ in our model.

\begin{figure}[thpb]
	\centering
	\includegraphics[width=0.5\linewidth]{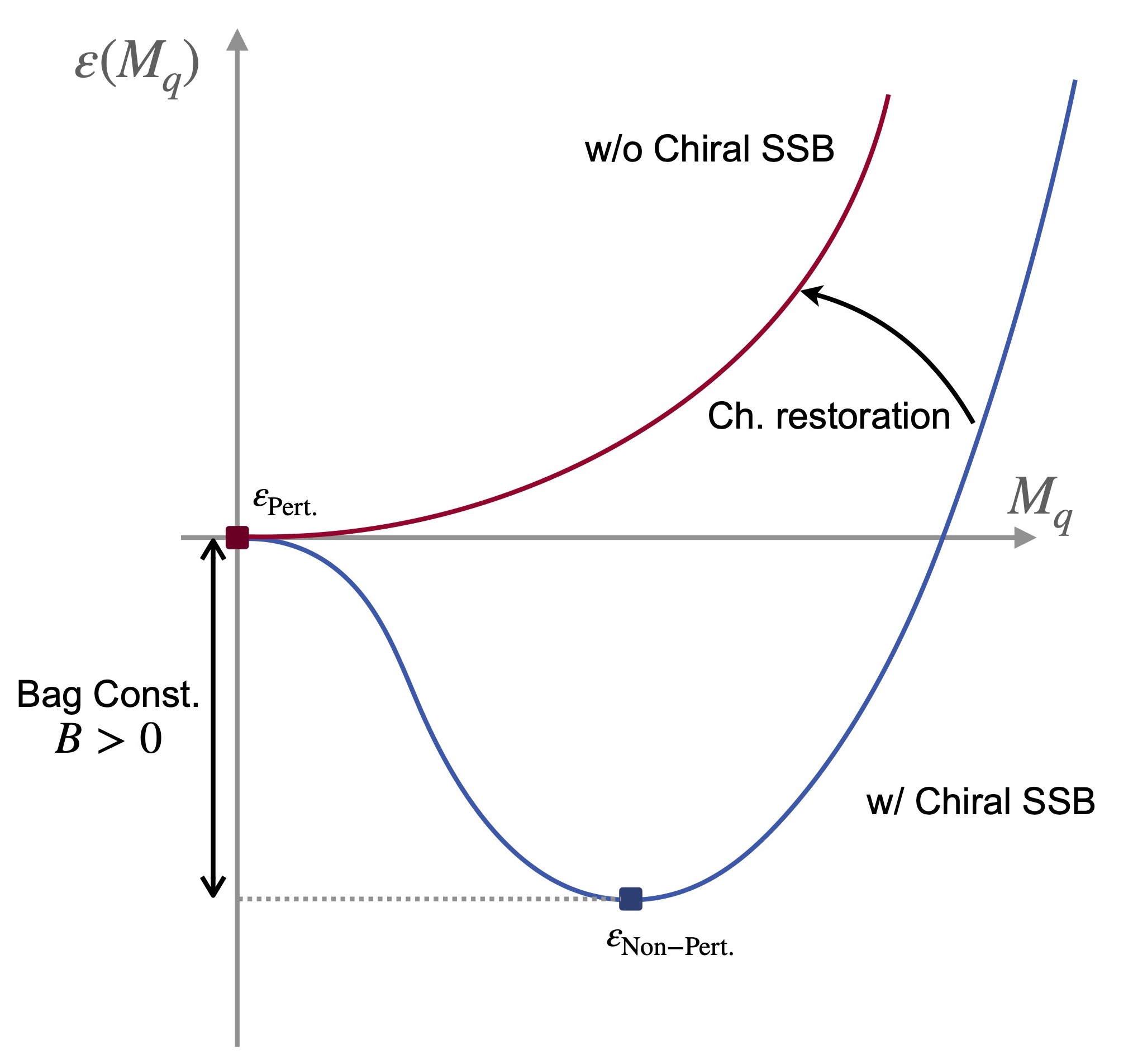}
	\caption{
		The energy density as a function of the chiral effective mass $M_q$.
		After chiral restoration the minimum energy is realized at $M_q = 0$,
		but with the broken chiral symmetry,
		the minimum is realized at $M_q \neq 0$ and the energy is smaller than that of $M_q = 0$.
		This gap in the zero-point energy density is the bag constant.
	}
	\label{fig:zeroT-bagconst1}
\end{figure}

\begin{figure}[thpb]
	\centering
	\includegraphics[width=0.5\linewidth]{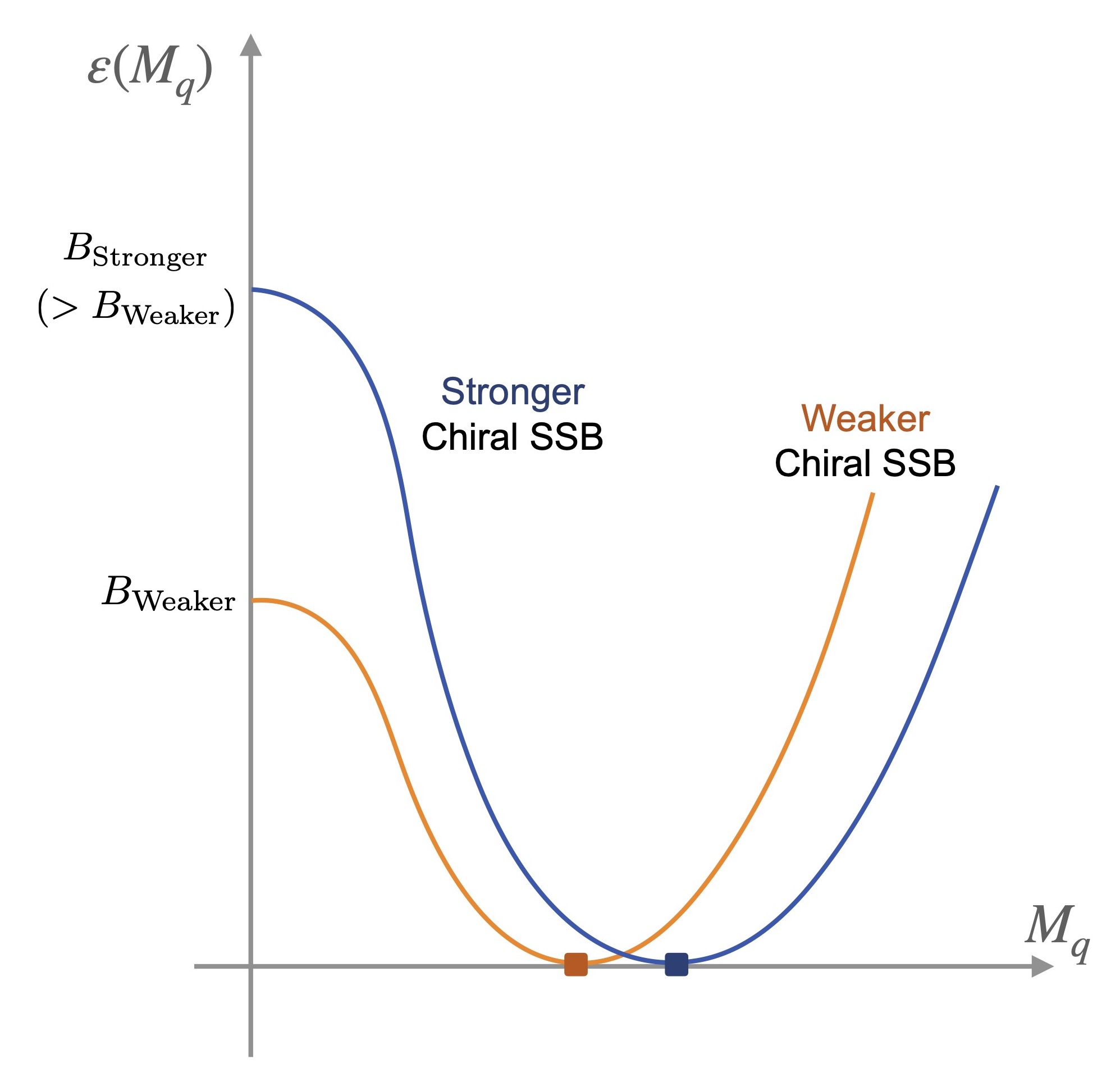}
	\caption{
		The energy density with different strength of the chiral symmetry breaking.
		The bag constant is larger for the stronger chiral symmetry breaking.
	}
	\label{fig:zeroT-bagconst2}
\end{figure}

The chiral symmetry breaking in the vacuum shifts the zero-chemical-potential value of thermodynamic potential as
\begin{align}
	B \equiv \Omega(M_q=0) - \Omega(M_q = M_q^{\rm vac}),
\end{align}
where the first term is the thermodynamic potential of the trivial vacuum
and the second term is that of the chiral symmetry broken vacuum.
The $B$ is the energy difference between these two vacuum which is called the bag constant.
As the bag constant is defined by the difference of the thermodynamic potential with symmetry-broken and unbroken vacuum,
its magnitude increases with stronger chiral symmetry breaking (\figref{fig:zeroT-bagconst2}).
\begin{comment}
In our model the bag constants $B(m_\sigma,f_\pi)$ in the $m_q^4$ unit are given as
\begin{align}
	B(600, 90) = 0.389, ~~~ B(600, 100) = 0.483 , \notag\\
	B(800, 90) = 0.394, ~~~ B(800, 100) = 0.489, 
\end{align}
\end{comment}
In our model the bag constants $B(m_\sigma,f_\pi)$ in the $m_\pi^4 = (140 \MeV)^4$ unit are given as
\begin{align}
	B(450, 90) = 0.874, ~~~ B(450, 100) = 1.035 , \notag\\
	B(600, 90) = 1.439, ~~~ B(600, 100) = 1.698, % in mpi^4 unit (mpi=140 MeV)
\end{align}
from which we can see stronger chiral symmetry breaking increase the bag constant $B$.

To see how the bag constant affects the stiffness of EOS,
it is useful to consider the EOS in the bag model with perturbative corrections.
Since EOS is normalized by $P=0$ at $\mu=0$ and $T=0$ for non-perturbative vacuum,
the perturbative evaluation of EOS should be normalized by the bag constant $B$ as
\begin{align}
	P_{\rm pert}^{\rm normalized} = P_{\rm pert} - B, ~~~ \varepsilon_{\rm pert}^{\rm normalized} = \varepsilon_{\rm pert} + B.
\end{align}
This shows that the bag constant $B$ associated with the chiral symmetry breaking in the vacuum
contributes to the EOS as the negative pressure and the positive energy density,
which makes the EOS softer.
The similar results on the sound velocity has been obtained the $2+1$ flavor NJL model \cite{He:2022kbc}.

\subsection{Comparison to pQCD with power corrections}
\label{ssec:zeroT-pQCD}
Now we get back the asymptotic behavior of the sound velocity at high density region.
In our calculation of QM model,
the $c_s^2$ converges to the conformal value $1/3$ from above as increasing density.
This behavior is qualitatively different from the pQCD prediction,
which is $c_s^2 \to 1/3$ from below.
The origin of this discrepancy is the existence of non-perturbative pressure $\sim \Lambda^2\mu_I^2$ in our model.
The QCD scale $\Lambda_{\rm QCD}$ has the non-analytic form for QCD coupling constant $\alpha_s$ as
\begin{align}
	\Lambda_{\rm QCD} = Q \exp\qty(-\frac{2\pi}{\beta_0 \alpha_s(Q^2)})
\end{align}
The $\mu_I^2$ term in the pressure need the energy scale 
but in the high density region the only relevant scale is $\Lambda_{\rm QCD}$.
Thus pQCD can not include the pressure $\sim \Lambda^2\mu_I^2$.
And as illustrated in Sec.$\,$\ref{sec:intro-sound-velocity},
this $\mu_I^2$ term makes the sound velocity $c_s^2$ converge to $1/3$ from above.

To examine in detail,
we calculate the pressure in the pQCD with power corrections
and how large power corrections are needed to change the asymptotic behavior of $c_s^2$.

The pQCD pressure up to $\rmO(\alpha_s^2)$ for single quark flavor $f$ is given as \cite{Graf:2016aa}
\begin{align}
	P_0^f  &= \frac{N_c}{ 12\pi^2 } \qty[|\mu_f| u_f  \qty( \mu_f^2 - \frac{ 5 }{2} M_q^2 ) + \frac{ 3 }{2} M_q^4 \ln \frac{ |\mu_f| + u_f }{ M_q } ] ,\\
	P_1^f  &= - \frac{ \alpha_s N_{\rm G} }{ 16 \pi^2 } 
	\begin{multlined}[t]
		\bigg[ 3 \qty( M_q^2 \ln \frac{ |\mu_f| + u_f }{ M_q }  - |\mu_f| u_f )^2 - 2 u_f^4 \\
		+ M_q^2 \qty( 6 \ln \frac{ \Lambda_{\rm reno} }{ M_q } + 4 ) \qty( |\mu_f| u_f - M_q^2 \ln  \frac{ |\mu_f| + u_f }{ M_q } ) \bigg]
	\end{multlined}
\end{align}
with $u_f = \sqrt{\mu_f^2 -M_f^2}$ and $N_{\rm G} = N_c^2 -1$.
The $M_q$ is the current quark mass for u/d quarks $M_q = 5.0 \MeV$.
The $\Lambda_{\rm reno}$ is the renormalization scale which commonly taken as $0.5\mu_f$ to $2\mu_f$.
In the isospin symmetric case the each flavor contributes equally to the pressure; $P^u = P^d$.

The running coupling constant $\alpha_s$ is given as \cite{Deur:2016aa}
\begin{align}
	\begin{split}
		\alpha_s (\Lambda_{\rm reno}) & = \frac{4 \pi}{\beta_0 L} \bigg[ 1 - 2 \frac{\beta_1}{\beta_0^2} \frac{\ln L}{L} %
		+ \frac{\beta_1^2}{\beta_0^4 L^2} \qty( \ln^2 L -\ln L - 1 + \frac{\beta_2\beta_0}{\beta_1^2} ) \\
		& \quad + \frac{\beta_1^3}{\beta_0^6 L^3} \qty( - \ln^3 L + \frac{5}{2}\ln^2 L + 2\ln L %
		- \frac{1}{2} - 3 \frac{\beta_2 \beta_0}{\beta_1^2} \ln L + \frac{\beta_3 \beta_0^2}{2\beta_1^3} ) \bigg]
	\end{split}
\end{align}
with $L = 2 \ln( \Lambda_{\rm reno}/\Lambda_{\msbar})$, $\beta_0 = 11-2N_f/3$, $\beta_1 = 102-19N_f/3$,
\begin{align}
	\beta_2 {} &= \frac{ 2857 }{ 2 } - \frac{ 5033 }{ 18 }N_f + \frac{ 325 }{ 54 }N_f^2, \\
	\begin{split}
		\beta_3 {} &=  
		\bigg( \frac{ 149753 }{ 6 } + 3564 \xi(3) \bigg) %
		- \bigg( \frac{ 1078361 }{ 162 } + \frac{ 6508 }{ 27 }\xi(3) \bigg) N_f \\
		& \quad + \bigg( \frac{ 50065 }{ 162 } + \frac{ 6472 }{ 81 }\xi(3) \bigg) N_f^2 %
		+ \frac{ 1093 }{ 729 } N_f^3
	\end{split}
\end{align}
and $\Lambda_{\msbar} \simeq 340  \MeV$.

Notice that this formula of $\alpha_s$ is valid only for $\Lambda_{\rm reno} > \Lambda_{\msbar}$.
We extrapolated $\alpha_s$ for smaller $\Lambda_{\rm reno}$ by using the Gaussian formula \cite{Deur:2014aa, Deur:2016aa}
\begin{align}
	\alpha_s^{\rm low} = \alpha_s^{\rm low} (0) e^{-Q^2/ 4\kappa^2}
\end{align}
with $\alpha_s^{\rm low}(0) \simeq 1.22$ and $\kappa \simeq 0.51$.
In the aim of calculating the sound velocity and thus the second derivative of the pressure,
we interpolate $\alpha_s$ by the smooth function $\alpha_s^{\rm mid}$ as
\begin{align}
	\alpha^{\rm low}_s(Q^2) = \theta(t_{\rm low}-Q^2) \alpha^{\rm mid}_s(Q^2) + \theta(t_{\rm high}-Q^2) \theta(Q^2-t_{\rm low}) + \alpha^{\rm high}_s(Q^2) \theta(Q^2 -t_{\rm high}) 
\end{align}
with $t_{\rm low}^{1/2} = 0.3 \GeV$ and $t_{\rm high}^{1/2} = 1.1 \GeV$.
The boundary condition requires that up to second derivative of $\alpha_s$ are smooth, which is
\begin{align}
	\left.\frac{\partial^n \alpha_s^{\rm low/high}}{(\partial Q^2)^n}\right|_{Q^2 = t_{\rm low/high}} = \left.\frac{\partial^n \alpha_s^{\rm mid}}{(\partial Q^2)^n}\right|_{Q^2 = t_{\rm low/high}}
\end{align}
for $n=0,1,2$.
And we choose $\alpha_s^{\rm mid}$ the polynomial function as 
\begin{align}
	\alpha_s^{\rm mid}(Q^2) = \sum_{m=0} c_m \mu_I^m.
\end{align}
Six coefficients $c_m$ are fully determined by six boundary conditions.

\begin{figure}[thpb]
	\centering
	\includegraphics[width=0.7\linewidth]{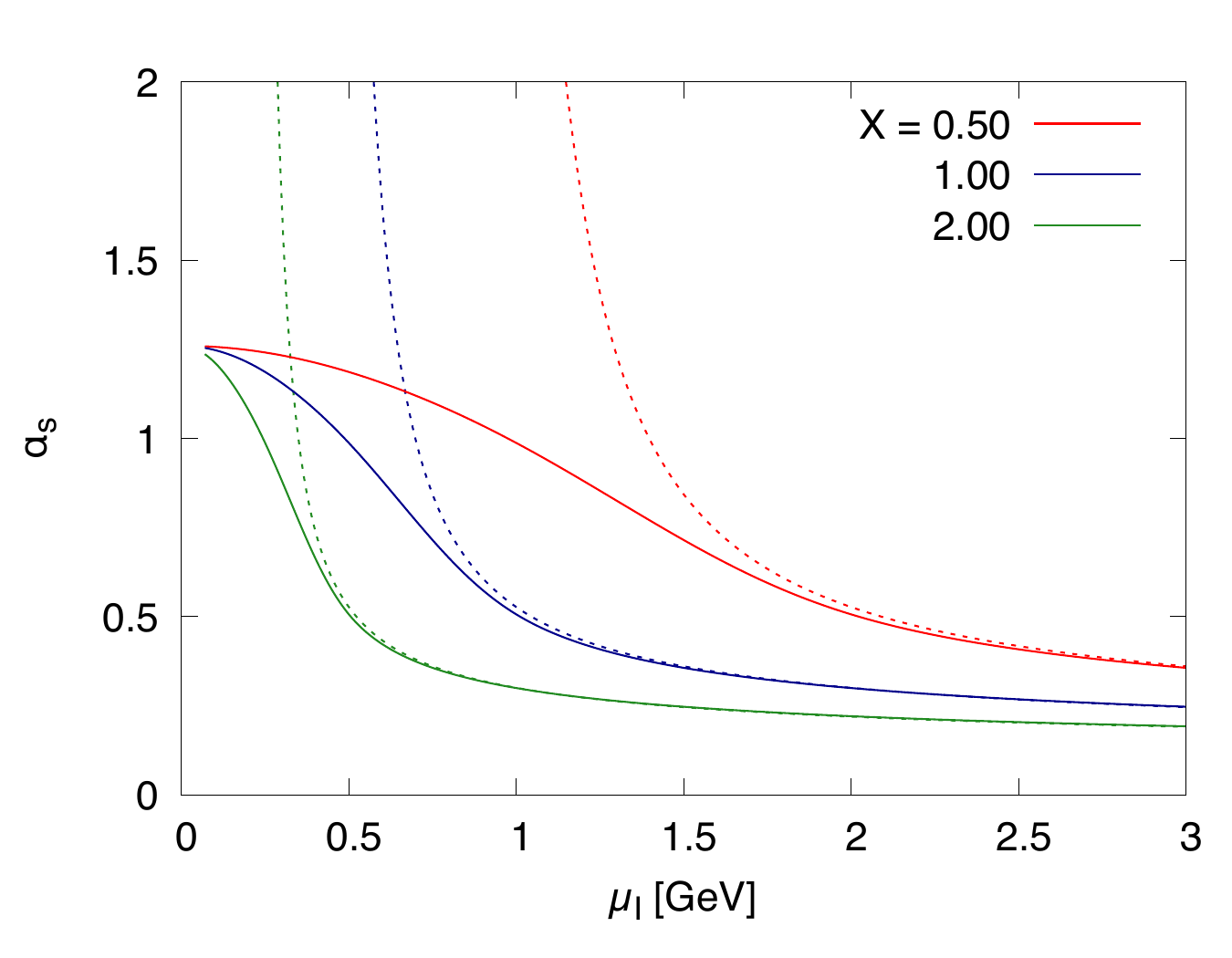}
	\caption{
		The running coupling constant $\alpha_s$ as a function of $\mu_I$.
		The solid curves are the results of interpolated expression.
		The dashed curve is the original pQCD formula.
		We took $X = 0.5, 1.0$ and $2.0$.
	}
	\label{fig:zeroT-alpha-s}
\end{figure}

\begin{figure}[thpb]
	\centering
	\includegraphics[width=0.7\linewidth]{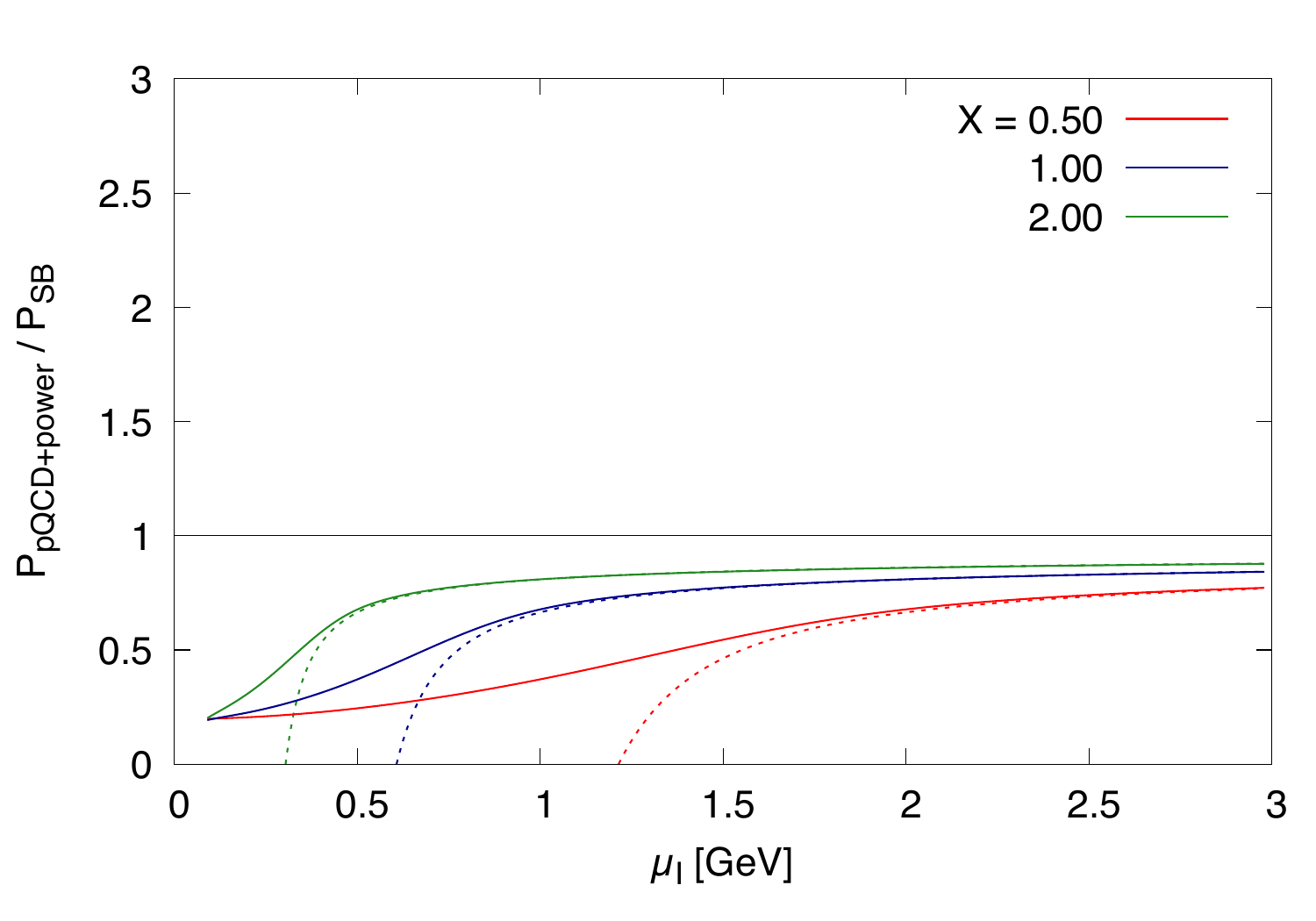}
	\caption{
		The perturbative pressure with the coupling constant $\alpha_s$ in Fig.$\,$\ref{fig:zeroT-alpha-s}.
		The pressure is normalized by the Stefan-Boltzmann limit $P_{\rm SB} = \pi^2 n_I^2/90$.
		The notation of the curves is same as Fig.$\,$\ref{fig:zeroT-alpha-s}.
	}
	\label{fig:zeroT-pert-pressure}
\end{figure}

Fig.$\,$\ref{fig:zeroT-alpha-s} illustrates the running coupling constant $\alpha_s$,
both the original pQCD formula and our interpolation.
We set $Q^2 = \Lambda_{\rm reno}^2$ and $\Lambda_{\rm reno} = X\mu_I$
and the scale factor $X$ are varied as $0.5, 1.0, 2.0$.
The pQCD pressure has the artificial rapid reduction in the low density region,
but freezing coupling keeps the pressure finite in the IR region
as shown in Fig.$\,$\ref{fig:zeroT-pert-pressure}.

Now we examine the impact of the power corrections to the sound velocity.
The power correction comes from the gap in the pion condensed phase.
The phase space factor $\sim 4\pi p_f^2 \Delta$ multiplying the gap $\Delta$ dividing $(2\pi)^3$ for each $u, \dbar$-quarks
gives the naive estimate of the power correction to the pressure
\begin{align}
	P_{\rm power} = C \frac{\mu_I^2 \Delta^2}{\pi^2}
\end{align}
with overall factor $C = \calO(1)$.
The similar examination has been done in Ref.\cite{Braun:2022aa},
which consider the diquark condensate $\Delta$ in 3-color QCD using the functional renormalization group.

\begin{figure}[thpb]
	\centering
	\includegraphics[width=0.7\linewidth]{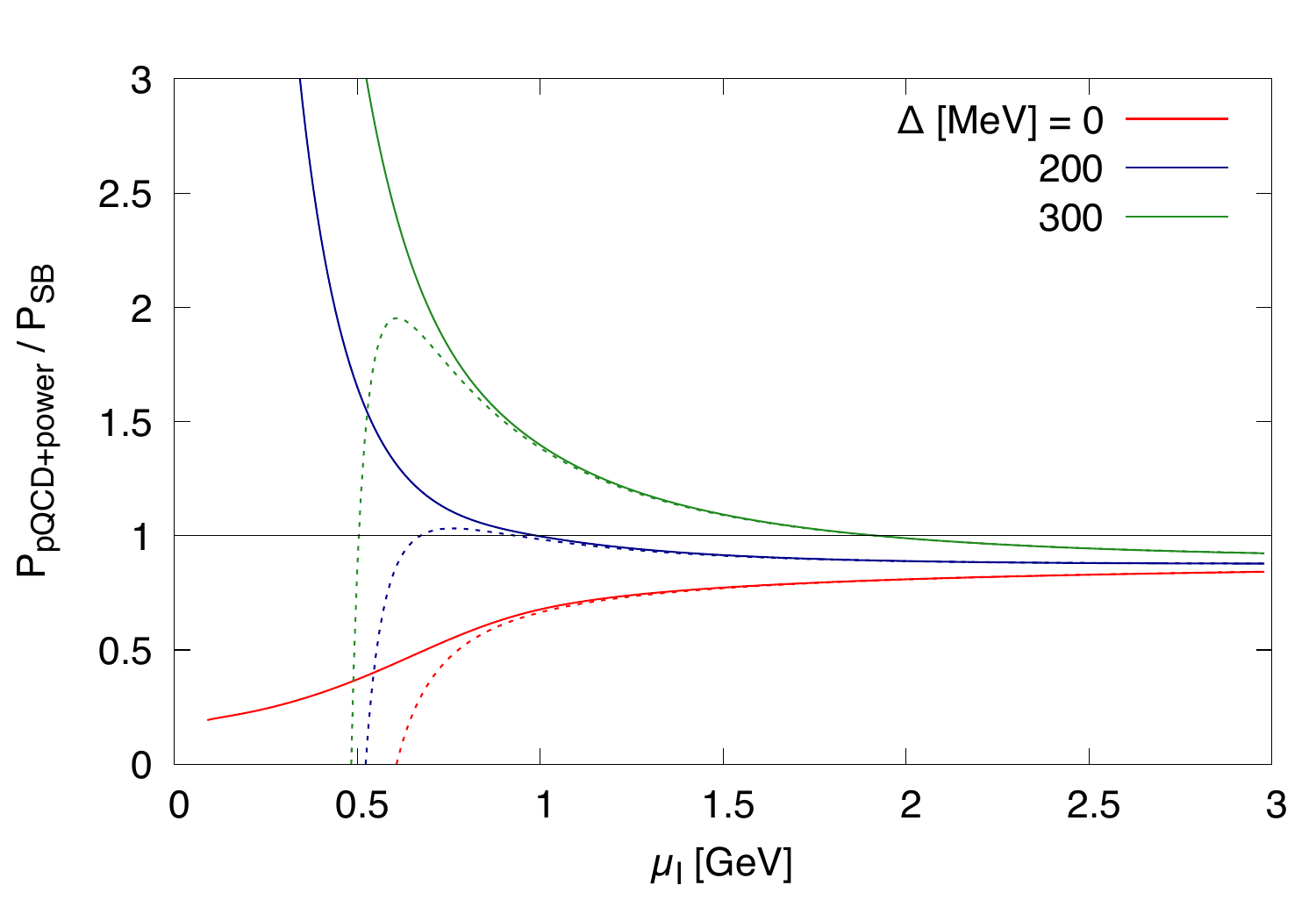}
	\caption{
		The perturbative pressure with ad-hoc power correction as a function of isospin density $n_0$.
		The scale factor $X$ is fixed to $1.0$.
		The solid and dashed line represents the freezing coupling and original pQCD formula, as in Fig.$\,$\ref{fig:zeroT-alpha-s}.
	}
	\label{fig:zeroT-pert-pressure-power}
\end{figure}

We can compute the gap $\Delta$ analytically in the weak-coupling and high density limit.
In the case of the chiral symmetric gap with $J^P = 0^+$
associating with the two-flavor color-superconductor (2SC) pairing,
the form of the gap is given as \cite{Son:aa,Pisarski:2000ab,Pisarski:2000aa}
\begin{align}
	\Delta \sim \mu_I g^{-5} \exp\qty(-\frac{3\pi^2}{\sqrt{2}g_s})
\end{align}
with running coupling constant $g_s = g_s(|\mu_I|)$.
In the case of the BCS gap at vanishing temperature
we have the following relation with gap $\Delta_{\rm BCS}$ and transition temperature $T_{\rm BCS}$
\begin{align}
	T_{\rm BCS} \simeq 0.57 \Delta_{\rm BCS}.
	\label{eq:zeroT-BCStemperature}
\end{align}
In our calculation the gap is about $\Delta\simeq 300 \MeV$ at high density
which corresponds to $T_{\rm BCS} \simeq 170 \MeV$.
This value has a good agreement with the LQCD result \cite{Brandt:2017oyy} $T_c^{\rm lattice} \simeq 161 \MeV$.
The pQCD pressure with ad-hoc power correction is shown in Fig.$\,$\ref{fig:zeroT-pert-pressure-power}.
We choose the size of gap $\Delta = 0, 200$ and $300 \MeV$.
With the original pQCD running coupling constant,
the pressure decreases to zero around $\mu_I \sim 0.3 \GeV$ as lowering the density.
This is the result of the IR divergence of the running coupling constant.
The freezing coupling constant keeps the pressure non-zero in the low density region.
The finite gap $\Delta$ makes the pressure seemingly diverge at low density,
while the pressure with vanishing gap $\Delta = 0$ is finite.

\begin{figure}[thpb]
	\centering
	\includegraphics[width=0.7\linewidth]{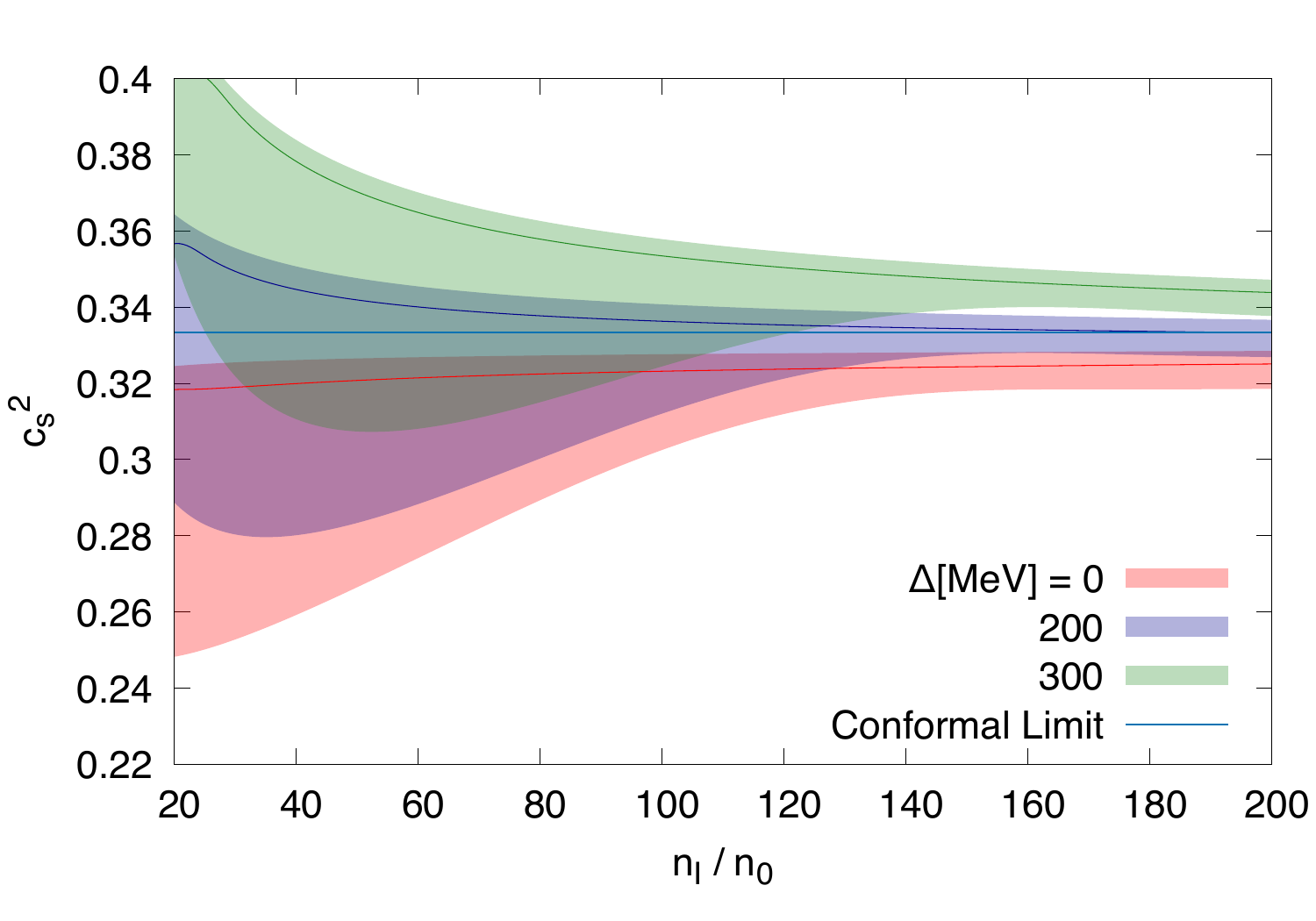}
	\caption{
		The sound velocity $c_s^2$ with ad-hoc power correction as a function of isospin density $n_0$.
		The band shows the variation of the scale factor $X = 0.5 - 2.0$.
		The solid line is the result of $X = 1.0$.	
	}
	\label{fig:zeroT-pert-cs2-power}
\end{figure}

The sound velocity $c_s^2$ is shown in Fig.$\,$\ref{fig:zeroT-pert-cs2-power}.
We varied the scale factor $X = 0.5 - 2.0$ which is shown in the band.
With the vanishing gap $\Delta = 0$,
we can see the sound velocity increases monotonically and converges to $1/3$ from below.
Adding the small gap $\Delta = 200 \MeV$ makes the sound velocity larger than $c_s^2 = 1/3$,
but the convergence from below is still alive at high density $n_I > 100 n_0$.
In contrast, with the large gap $\Delta = 300 \MeV$,
the sound velocity changes the trend and converges to $1/3$ from above.

As increasing the density
the power correction becomes relatively small comparing to the perturbative pressure.
This can be easily seen by taking ratio of these two terms as
\begin{align}
	\frac{P_{\rm power}}{P_{\rm pQCD}} \sim \qty(\frac{\Delta}{\mu_I})^2.
\end{align}
Around $\mu_I \simeq 1 \GeV$ or $n_I \simeq 80 n_0$,
the power correction has the small contribution in the total pressure
\begin{align}
	\frac{P_{\rm power}}{P_{\rm pQCD}} \sim \qty(\frac{300 \MeV}{1 \GeV})^2 = 0.09,
\end{align}
but change the qualitative behavior of the sound velocity.
This suggests that the power correction has the essential role even at the high density regime
and can be the main source of the discrepancy of the asymptotic behavior of the sound velocity.

\subsection{Trace anomaly}
\label{ssec:zeroT-trace-anomaly}
We have discussed the sound velocity $c_s^2$ to closely examine the EOS especially its stiffening.
Trace anomaly is another important quantity to characterize the EOS
and gets attention in the context of not only the NSs \cite{Marczenko:2022aa,Ma:2019aa,Fujimoto:2022aa}
but also in the mechanical properties of hadrons \cite{Sakai:2022aa,Polyakov:aa,Fujita:2022aa}.
The sound velocity is defined as $c_s^2 = \pdv*{P}{\varepsilon}$ and contains only the information of slope of EOS.
But in the context of the NSs,
the trace anomaly is the relation of $P$ and $\varepsilon$
thus contains the magnitude of EOS.

The trace anomaly is the indicator of the deviation from the conformal limit at high density.
It is defined as
\begin{align}
	\braket{T^\mu_{~\mu}} = \varepsilon - 3P,
\end{align}
Then the conformal limit concludes $\Delta_{\rm tr} = 0$.
It is useful to introduce the normalized trace anomaly divided by $3\varepsilon$ as
\begin{align}
	\Delta_{\rm tr} \equiv \frac{1}{3} - \frac{P}{\varepsilon}.
\end{align}

The trace anomaly $\Delta_{\rm tr}$ is conjectured to be positive in Ref.\cite{Fujimoto:2022aa},
but the positivity is broken with the non-perturbative effects as well as the sound velocity.
To examine the effect of the power correction to the trace anomaly,
we again use the parametric pressure as the function of $\mu_I$
\begin{align}
	P_\textrm{with powers} = a_0 \mu_I^4 + a_2 \mu_I^2.
	\label{eq:zeroT-parametric-pressure}
\end{align}
But this time we consider the running coefficients $a_0$ and $a_2$ as a function of $\mu_I$,
which comes from the $\mu_I$-dependence of  running coupling constant $\alpha_s(\mu_I)$.
Then the energy density $\varepsilon$ is derived as
\begin{align}
	\varepsilon = 3a_0\mu_I^4 + 2a_2\mu_I^2 + \pdv{a_0}{\ln \mu_I}\mu_I^4 + \pdv{a_2}{\ln \mu_I}\mu_I^2
\end{align}
and the trace anomaly is given as
\begin{align}
	\braket{T^\mu_{~\mu}}_\textrm{with powers} &= -2a_2\mu_I^2 + \pdv{a_0}{\ln \mu_I}\mu_I^4 + \pdv{a_2}{\ln \mu_I}\mu_I^2.
\end{align}
The original pQCD pressure favors the positive trace anomaly,
but as is the case for the sound velocity
the attractive power correction $a_2 > 0$ can make the trace anomaly negative.

\begin{figure}
	\centering
	\includegraphics[width=0.7\linewidth]{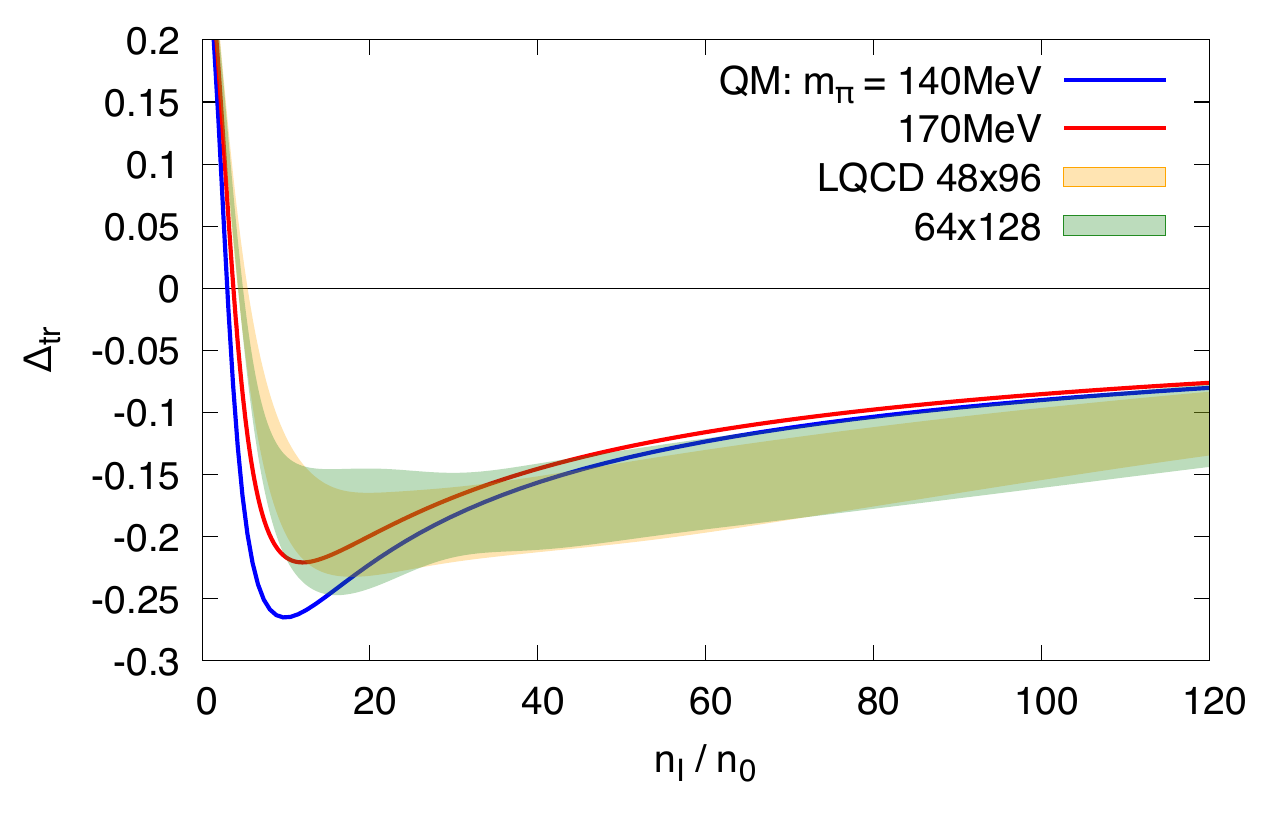}
	\caption{
		The trace anomaly $\Delta_{\rm tr}$ as a function of isospin density $n_I / n_0$.
		The qualitative behavior of the trace anomaly is consistent with the LQCD results \cite{Abbott:2023aa}.
	}
	\label{fig:zeroT-traceanomaly-qm-lqcd}
\end{figure}

Our numerical result of the trace anomaly as a function of $n_I/n_0$ is shown in Fig.$\,$\ref{fig:zeroT-traceanomaly-qm-lqcd}
with the LQCD result \cite{Abbott:2023aa}.
The trace anomaly starts with the positive value at $n_I \simeq 0$
and decreases to under zero around $n_I \simeq 3 n_0$.
At the high density around $n_I \simeq 10 n_0$,
$\Delta_{\rm tr}$ starts to increase and converges to zero from below.

\begin{figure}[thpb]
	\centering
	\includegraphics[width=0.7\linewidth]{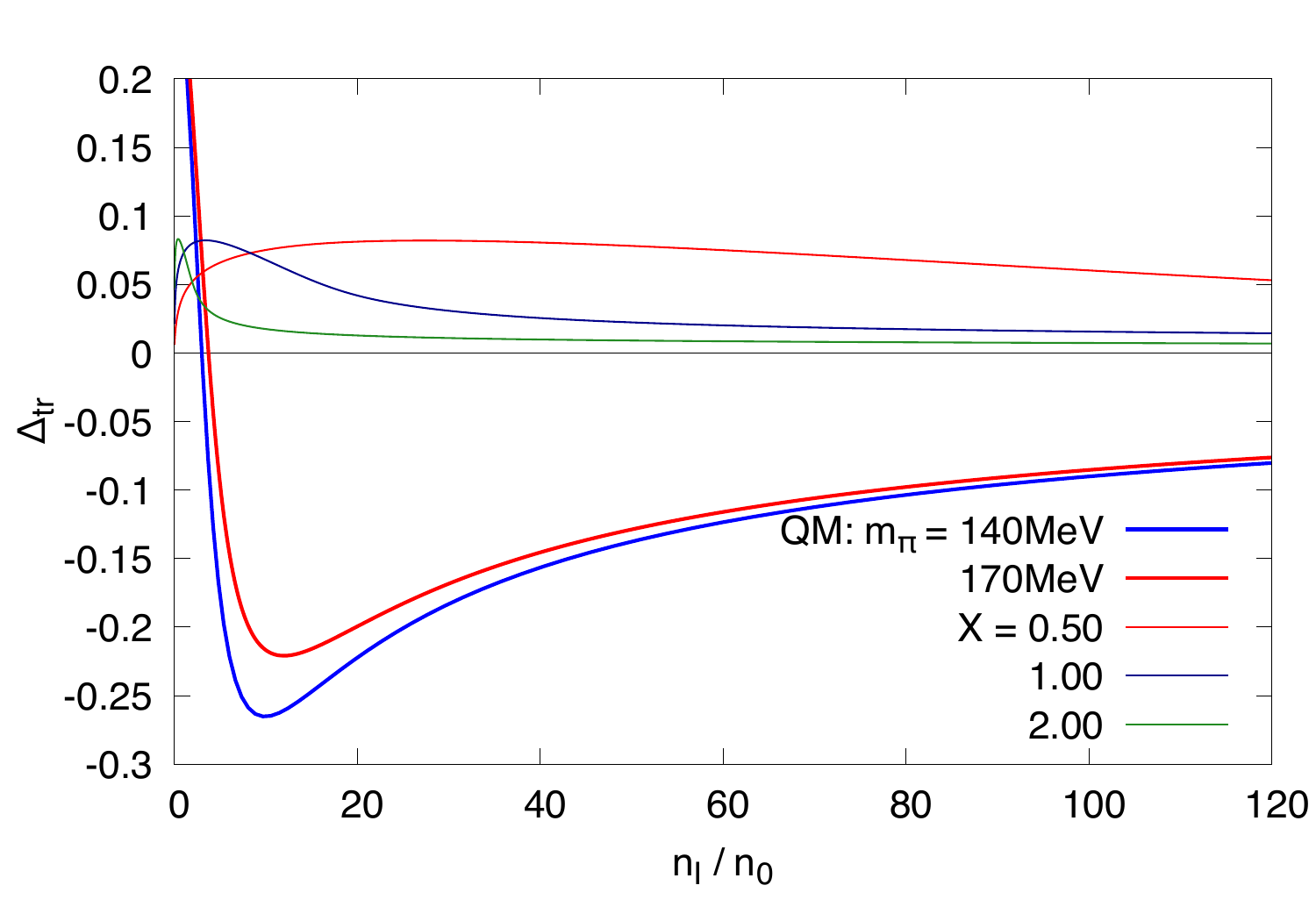}
	\caption{
		The trace anomaly $\Delta_{\rm tr}$ without gap $\Delta$ as a function of isospin density $n_I$.
		The notation of the curves is same as Fig.$\,$\ref{fig:zeroT-alpha-s}.
	}
	\label{fig:zeroT-pqcd-trace-anomaly}
\end{figure}

\begin{figure}[thpb]
	\centering
	\includegraphics[width=0.7\linewidth]{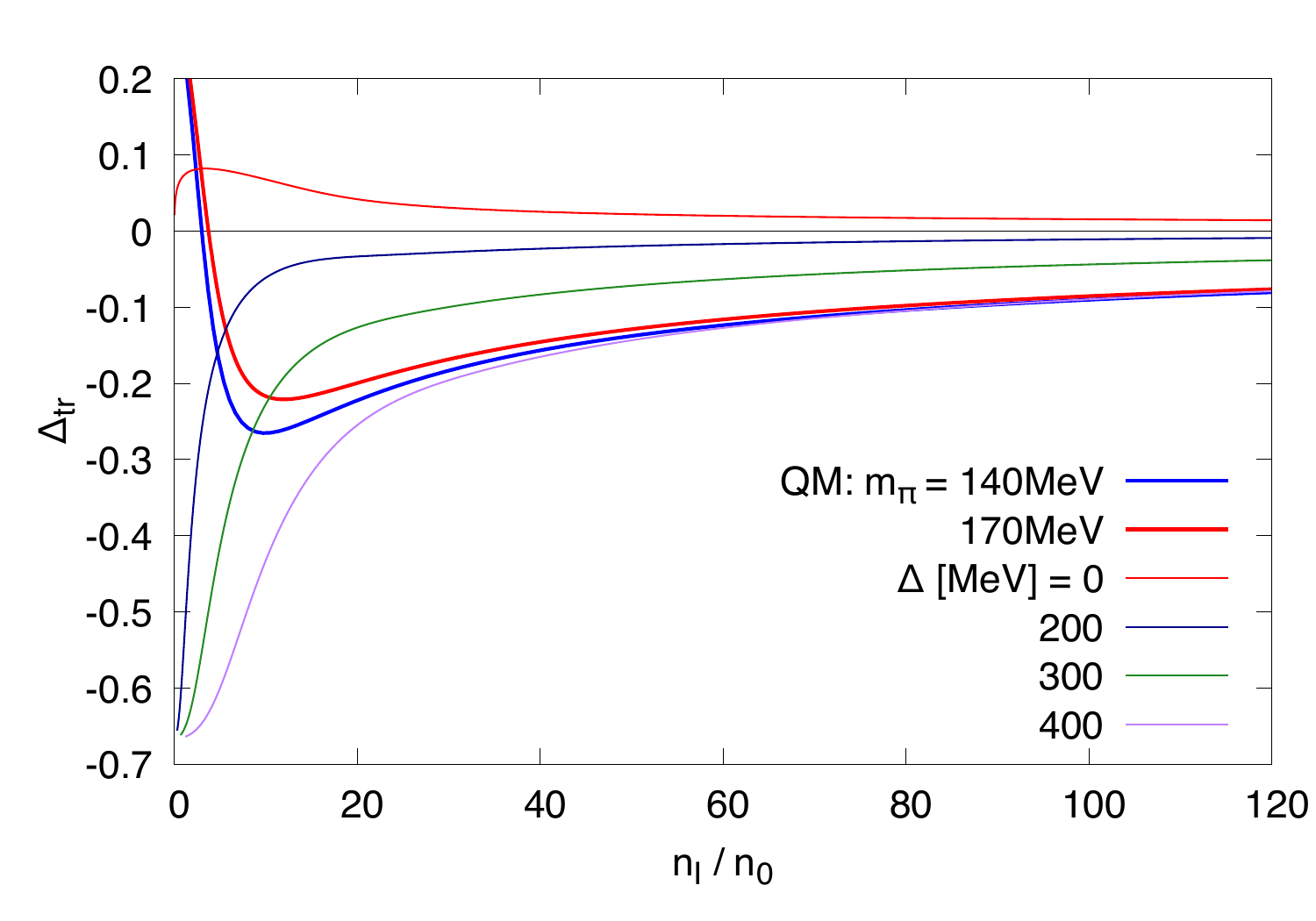}
	\caption{
		The trace anomaly with the power correction $\Delta = 0, 200, 300$ and $400  \MeV$.
		The notation of the curves is same as Fig.$\,$\ref{fig:zeroT-alpha-s}.
		Nonzero gap favors the negative trace anomaly.
	}
	\label{fig:zeroT-power-trace-anomaly}
\end{figure}

The trace anomaly calculated by the pQCD adding the power correction is shown in Fig.$\,$\ref{fig:zeroT-pqcd-trace-anomaly} and Fig.$\,$\ref{fig:zeroT-power-trace-anomaly}.
along with our model calculations.
Fig.$\,$\ref{fig:zeroT-pqcd-trace-anomaly} shows the trace anomaly without the power correction with different renormalization scale $X$,
which has the positive value in the whole range of $\mu_I$, independent on $X$.
But adding the power correction with $\Delta \sim 200 \MeV$ or larger makes the trace anomaly negative 
as shown in Fig.$\,$\ref{fig:zeroT-power-trace-anomaly}.
These results suggests that the sign of the trace anomaly is sensitive to the power correction
and can be useful indicator of the non-perturbative effects.

\chapter{Non-zero temperature analysis}
\label{chap:finiteT}
In this chapter
we examine the impact of the thermal excitation of quarks on the thermodynamics on the $\mu_I-T$ plane.
In addition we consider the add-hoc thermal meson contribution to the pressure
to discuss the impact of the thermal mesons to the EOS
and the validity and limit of our approximation which neglect the thermal mesons.

The choice of parameters are the same as Chap.$\,$\ref{chap:zeroT},
and we use the Polyakov loop 
to describe the quark confinement and its potential is given by \eqref{eq:model-polyakov-loop-potential-log}.

\section{Phase diagram}
\label{sec:finiteT-phasediagram}
\begin{figure}[thpb]
	\centering
	\begin{minipage}{0.45\hsize}
		\centering	
		\includegraphics[width=0.95\linewidth]{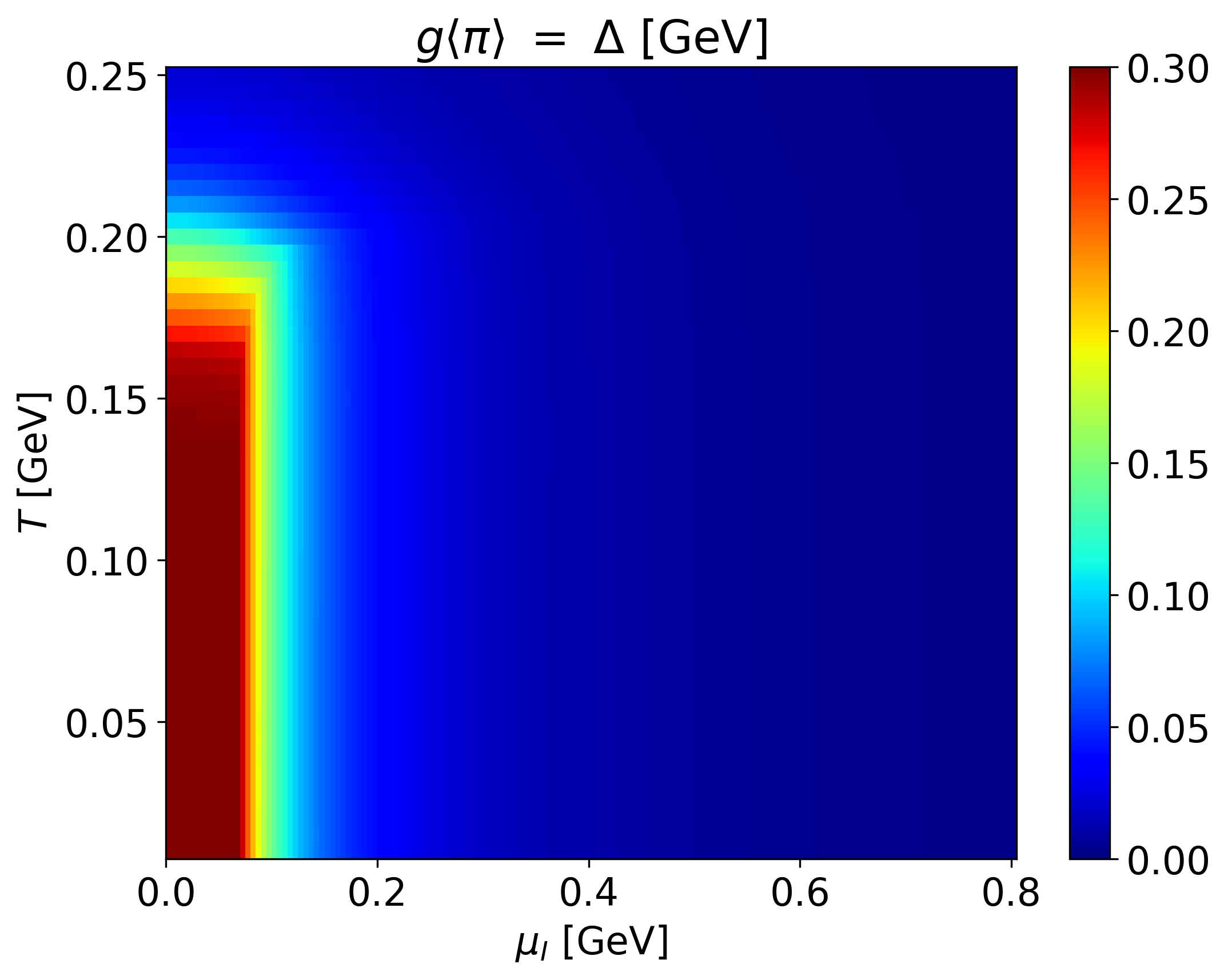}
		\label{fig:finiteT-phasediagram-chiral}
	\end{minipage}
	\begin{minipage}{0.45\hsize}
		\centering	
		\includegraphics[width=0.95\linewidth]{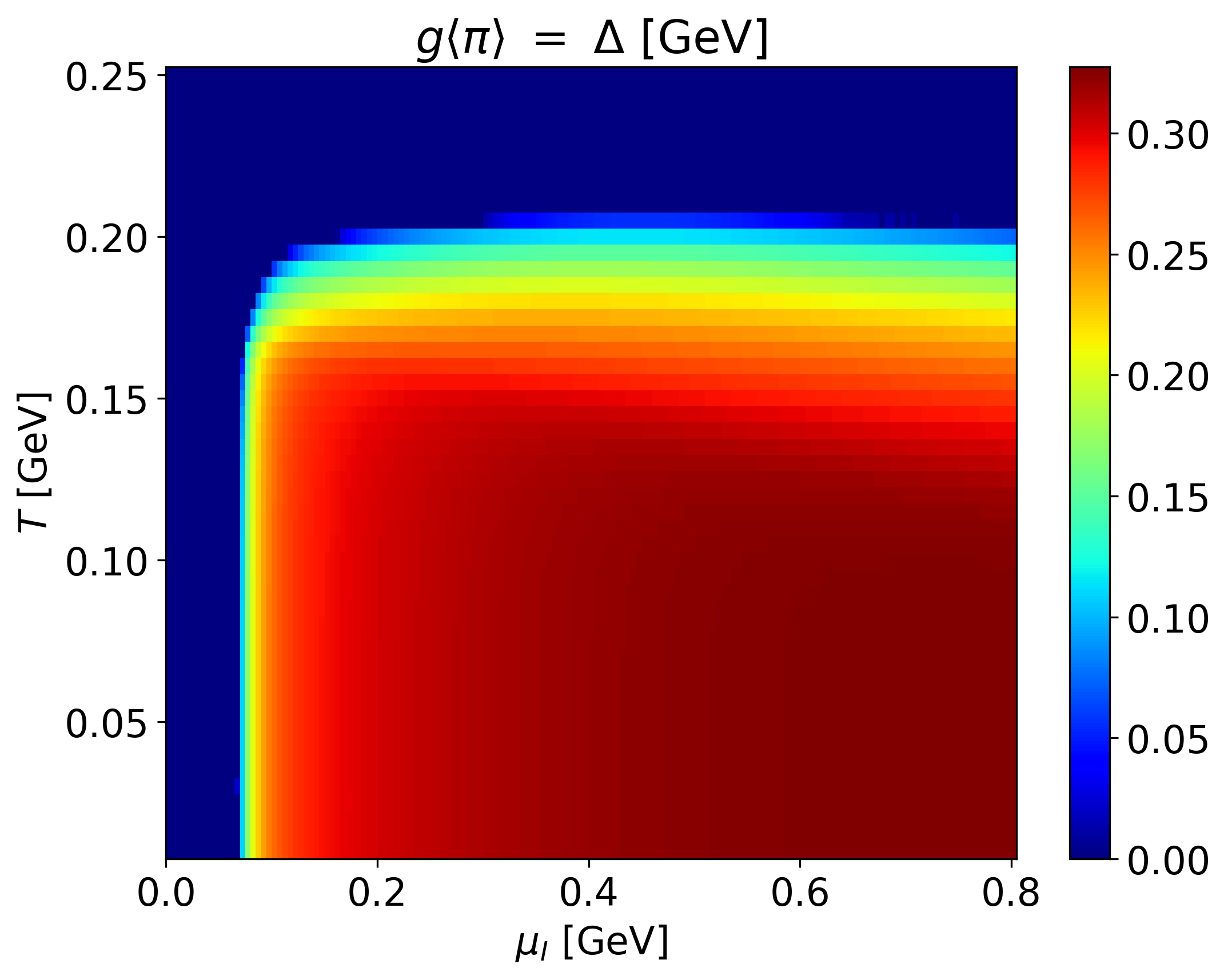}
		\label{fig:finiteT-phasediagram-pion}
	\end{minipage}
	\\
	\begin{minipage}{0.45\hsize}
		\centering	
		\includegraphics[width=0.95\linewidth]{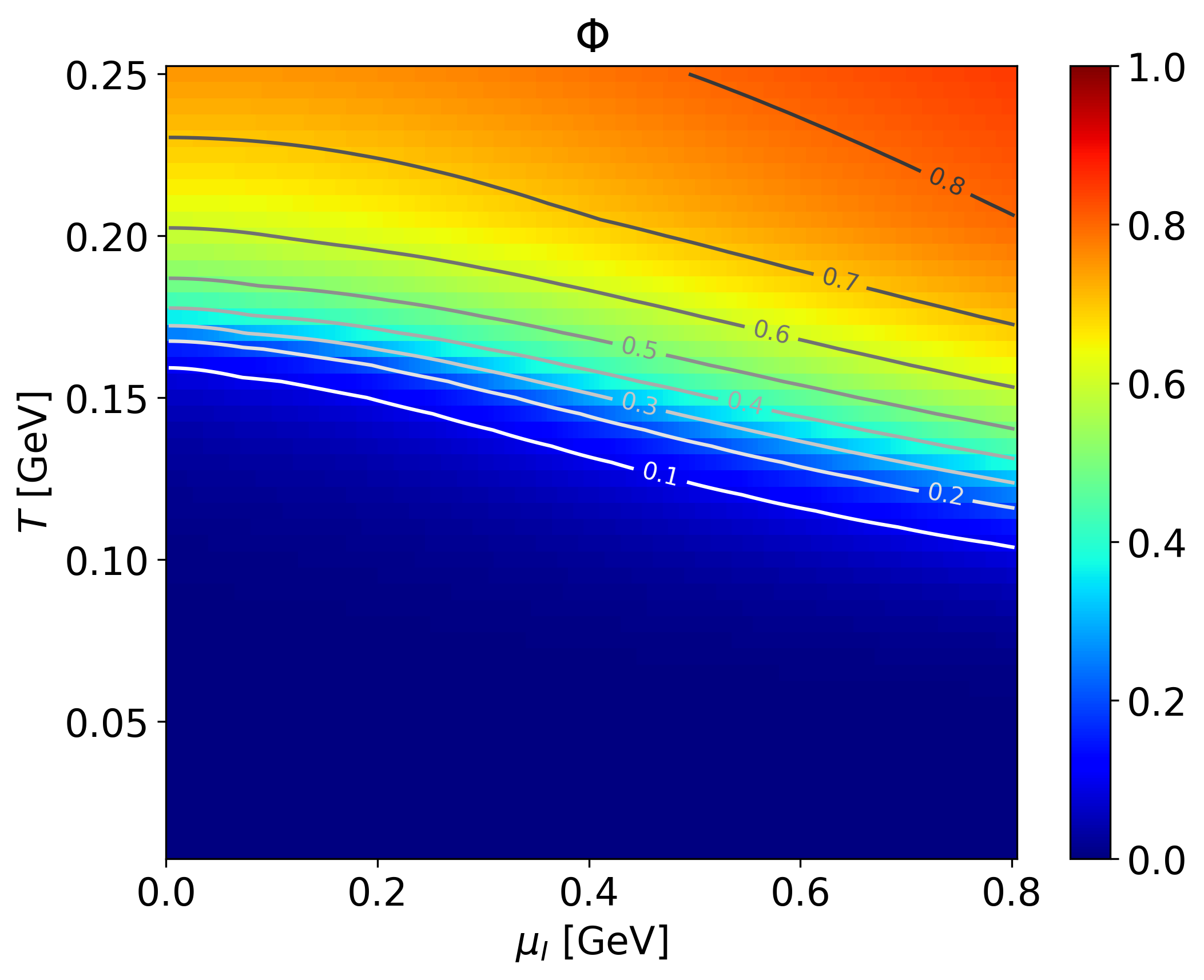}
		\label{fig:finiteT-phasediagram-polyakov}
	\end{minipage}
	\caption{
		The phase diagram.
		a) Scaled chiral condensate $M_q = g\braket{\sigma}$;
		b) Scaled pion condensate $\Delta = g\braket{\pi_1}$;
		c) Polyakov loop $\Phi$.
	}
	\label{fig:finiteT-phasediagram}
\end{figure}

First we examine the phase structure,
chiral and pion condensates and Polyakov loop on the $\mu_I-T$ plane,
as shown in Fig.$\,$\ref{fig:finiteT-phasediagram}.
In the small temperature region $T < 100 \MeV$,
the Polyakov loop has the small value $\Phi \sim 0.1$
and the phase structure is similar to the zero temperature case.
For small $\mu_I < m_\pi/2$
the chiral condensate has the non-zero value $M_q = g\braket{\sigma} \simeq 300 \MeV$
and the pion condensate is zero.
As $\mu_I$ increases
the chiral symmetry is restored $M_q \to 0$ and the pion condensate appears and get saturated $\Delta \simeq 300\MeV$.
Increasing the temperature with constant $\mu_I$
the value of Polyakov loop increases toward $\Phi = 1$.
Chiral and pion condensates start decreasing from $T \sim 150 \MeV$,
and chiral condensate disappears around $T_\chi \sim 200 \MeV$ smoothly
while pion condensate at $T_\pi \simeq 200 \MeV$ with the second order phase transition.

The value of mass gap at $T = 0$ is related to the temperature at which $\Delta$ vanishes
as shown in \eqref{eq:zeroT-BCStemperature}.
The resulting temperature without Polyakov loop is $T_\pi \simeq 170 \MeV$.
In the presence of Polyakov loop the melting temperature is higher by $30 \MeV$.

% figure: chiral, pion condensate + polyakov loop

In contrast to the chiral condensate,
the temperature at which pion condensate starts melting has the weak $\mu_I$-dependence.
As increasing $\mu_I$ the pion condensate starts melting at slightly lower temperature.
This dependency corresponds to the contour of the Polyakov loop.
The pion condensate is formed from the $u$- and $\dbar$-quarks with binding energy $\sim \Delta$.
By adding the thermal energy to the system the quark-antiquark pairing will break to thermally excited quark and antiquark,
but the Polyakov loop suppresses the final state of excited quarks and antiquarks.
Thus the smaller Polyakov loop suppresses
the melting of the pion condensate.
This mechanism is also the origin of the increase of $T_\pi$.

\begin{figure}[thpb]
	\centering	
	\includegraphics[width=0.7\linewidth]{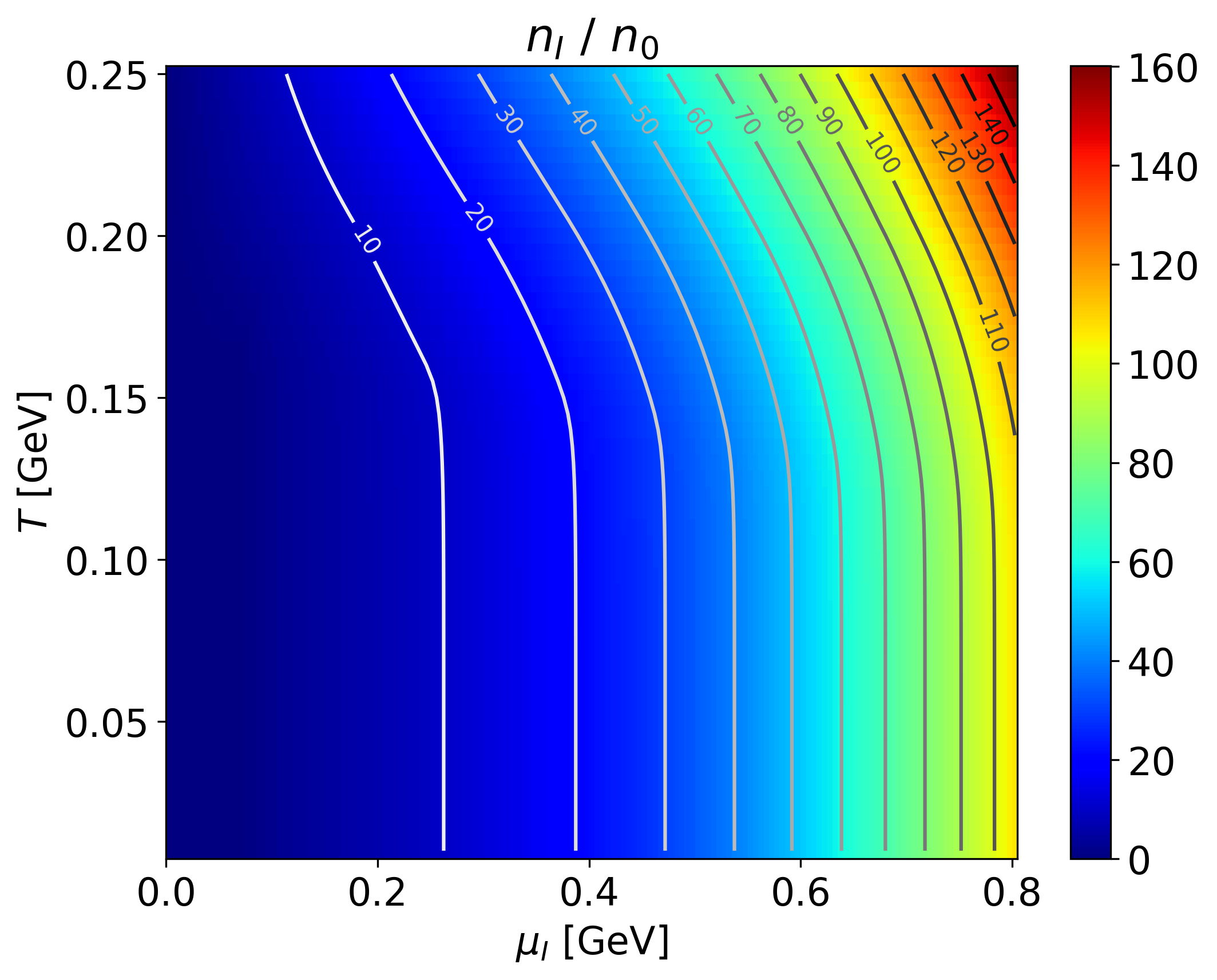}
	\caption{
		The isospin density $n_I$ normalized by $n_0$.
	}
	\label{fig:finiteT-density}
\end{figure}

\begin{figure}[thpb]
	\centering	
	\includegraphics[width=0.7\linewidth]{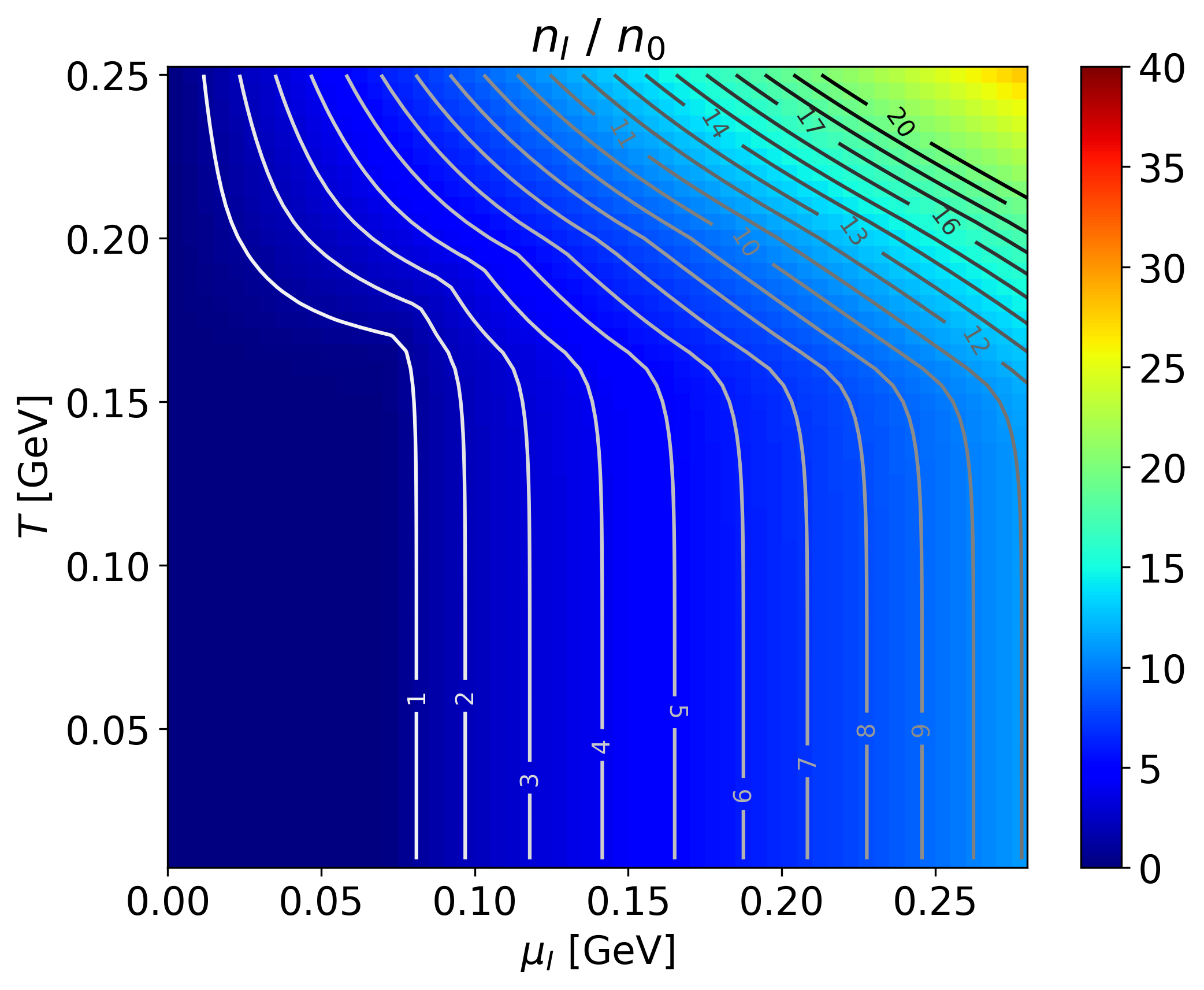}
	\caption{
		The isospin density $n_I$ normalized by $n_0$, zoomed for $\mu_I < 2m_\pi$ region.
	}
	\label{fig:finiteT-density-tight}
\end{figure}

\figref{fig:finiteT-density} shows the isospin density normalized by $n_0$ on the $\mu_I-T$ plain.
\figref{fig:finiteT-density-tight} is the zoomed figure for $\mu_I < 2m_\pi$ region.
Fixing $\mu_I$,
the isospin density has the small temperature dependence in the condensed phase $T \lesssim 150 \MeV$,
and after melting the mass gap the density increases by the thermal excitation of quarks.

% figure: density (µ < 2m_pi so far) 

\begin{figure}[thpb]
	\centering	
	\includegraphics[width=0.7\linewidth]{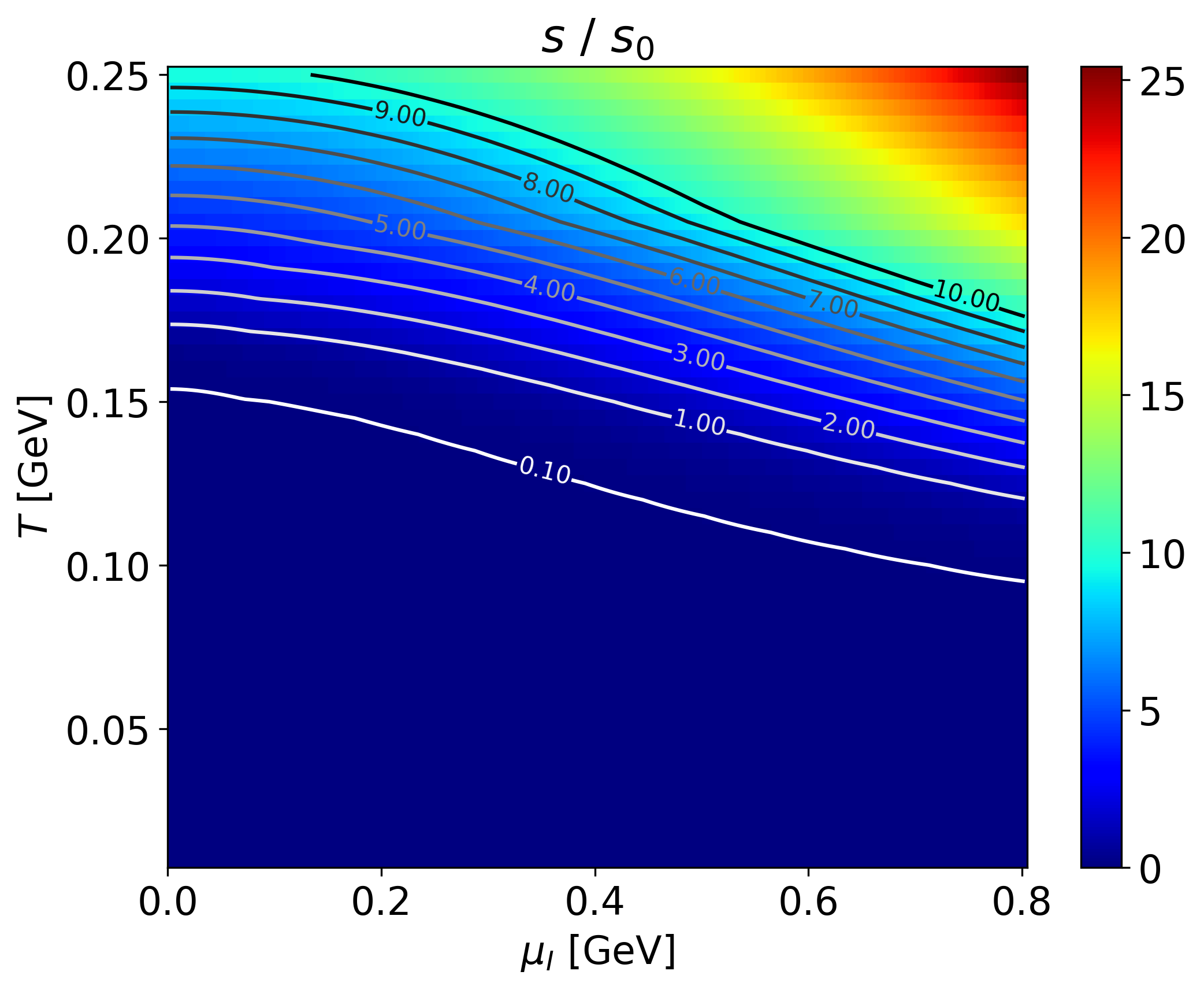}
	\caption{
		The entropy density $s$ normalized by $s_0 = 2 \fm^{-3} \simeq 0.0153 \GeV^3$.
		The $s_0$ is the typical value of entropy density for the hadron to overlap.
	}
	\label{fig:finiteT-entropy}
\end{figure}

\figref{fig:finiteT-entropy} shows the entropy density normalized by $s_0 = 2 \fm^{-3} \simeq 0.0153 \GeV^3$.
Hadrons contain two or three quarks and antiquarks in the volume $\sim 1\fm^3$.
Thus $2-3 \fm^{-3}$ is the typical value of entropy density for the hadron to overlap.

% figure: entropy (s0 scaling)

In case of massless and gap-less quark
at high density
the thermodynamic potential is given by the Stefan-Boltzmann limit.
The entropy density is given by
\begin{align}
	s_{\rm SB} = \frac{N_cN_f}{6}\qty(\mu_I^2 T + \frac{7\pi^2 T^3}{15})
\end{align}
where SB denotes the Stefan-Boltzmann limit.
Within this assumption the entropy density reaches $s_0$ at $T \simeq 15 \MeV$ when $\mu_I \simeq 1\GeV$,
which result in the much lower temperature than the finite mass gap case.
In case of non-zero mass gap in the confined phase at the low temperature,
the second term is neglected and the first term\ is modified as
\begin{align}
	s \sim \Phi \mu_I^2 T e^{-\Delta/T}.
\end{align}
This means that both the Polyakov loop $\Phi$ and the mass gap $\Delta$ suppress the entropy density.

\section{On isentropic trajectories}
\label{sec:finiteT-isentropic}
In this section we discuss the EOS on isentropic condition $s/n_I=\text{const}$.
The process on this trajectory is an idealized process without dissipation which is both adiabatic and reversible.
The supernova and neutron star mergers are the typical examples of the isentropic process.

\subsection{Isentropic trajectories}
\label{ssec:finiteT-isentropic-trajectories}

\begin{figure}[thpb]
	\centering	
	\includegraphics[width=0.95\linewidth]{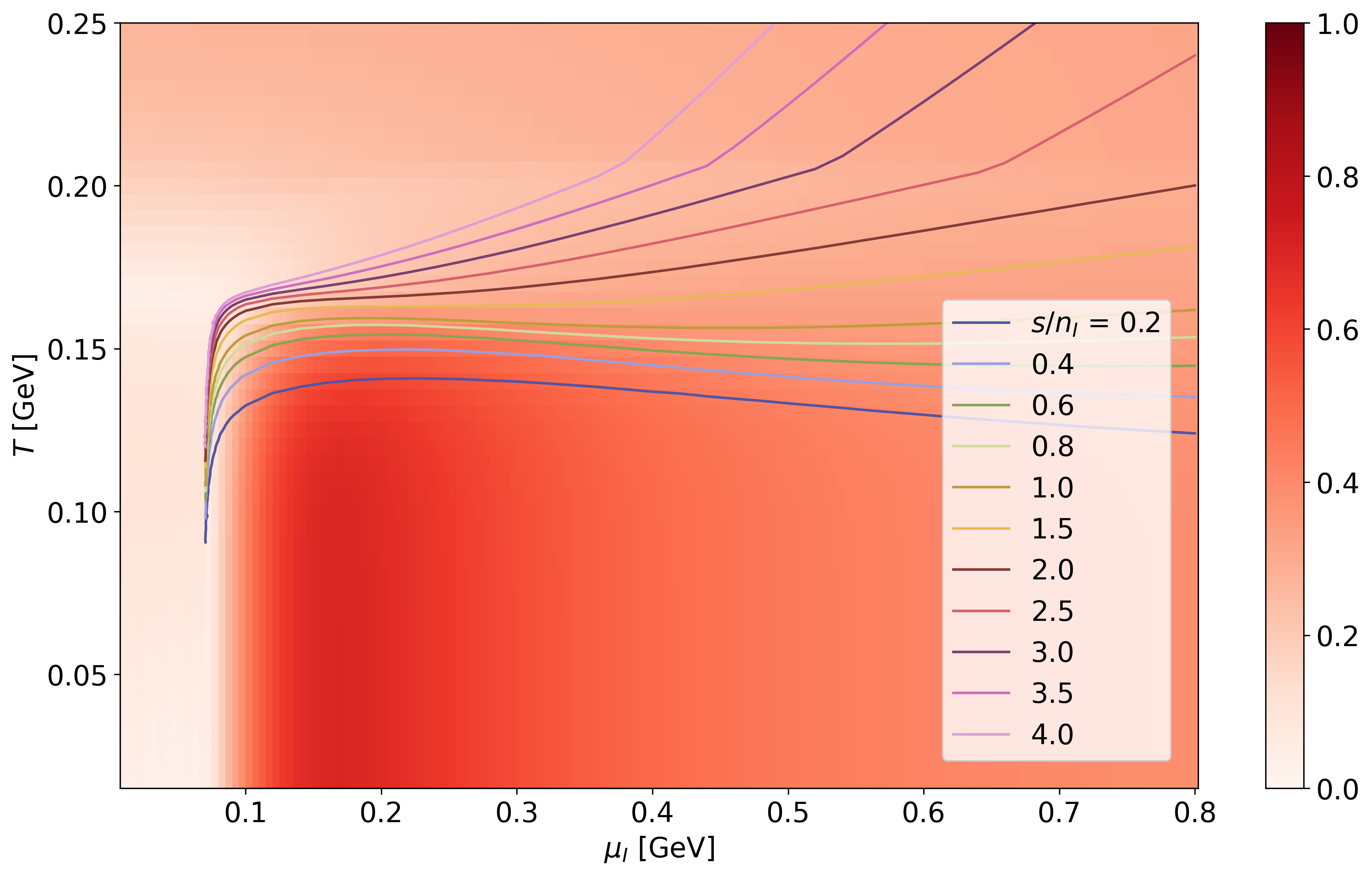}
	\caption{
		The isentropic trajectories with $s/n_I=0.2 - 4.0$ and the isentropic sound velocity $c_{s/n_I}^2$ on $(\mu_I-T)$ plane.
		The bending of the trajectories around $T\simeq 200\MeV$ for $s/n_I = 2.5-4.0$ is due to the second order phase transition.
	}
	\label{fig:finiteT-isentropic}
\end{figure}

\figref{fig:finiteT-isentropic} illustrates the isentropic sound velocity $c_{s/n_I}^2$ and the isentropic trajectories for different value of $s/n_I$.
The sound velocity is shown by the color map
and the isentropic trajectories are shown by the lines.
In the low density around $\mu_I \simeq m_\pi/2$
the trajectories are almost parallel to the $T$-axis,
which means the growth of the entropy density is highly suppressed by the gap and the Polyakov loop.
Therefore
the isentropic trajectories require the high temperature in the pion condensed phase.

% figure: sound velocity map

The trajectories with $s/n_I = 0.2 - 1.0$ run though the condensed phase and have the small $\mu_I$ dependence.
To examine this behavior we should get back to the parametric expression of the entropy density with the mass gap $\Delta$
\begin{align}
	s \sim \mu_I^2 T e^{-\Delta/T}.
\end{align}
With large density and fixed temperature,
the isospin density behaves as $n_I \sim \mu_I^3$.
This indicates the temperature on the isentropic trajectory is given as $(s/n_I)_{\rm fix} \sim \textrm{(phase space)} \times e^{-\Delta/T}$,
therefore
\begin{align}
	\left.T(\mu_I)\right|_{s/n_I} \simeq \frac{\Delta(\mu_I,T)}{\ln\qty[\,\textrm{(phase space)} \times \left.(s/n_I)\right|_\textrm{fixed}]\,}
\end{align}
The the phase space factor contributes to the temperature through the logarithm,
and its variation is modest.
This indicates that the temperature on the isentropic trajectory is almost governed by the behavior of the mass gap $\Delta$,
which has small $\mu_I$ dependence.

Once the trajectory enters the normal phase $\Delta \sim 0$ ($s/n_I = 1.5 - 4.0$)
the thermodynamic quantities are given by the SB limit.
As the entropy and the isospin density are given by $s \sim \mu_I T^2, ~ n_I \sim \mu_I^2 T$,
both $T$ and $\mu_I$ increase with the same rate on the trajectory.
When entering the normal phase from the condensed phase
trajectories meet the second order phase transition,
thus the line has the kink around $T \sim 200 \MeV$.

\subsection{Sound velocity}
\label{ssec:finiteT-soundvelocity}

\begin{figure}[thpb]
	\centering	
	\includegraphics[width=0.7\linewidth]{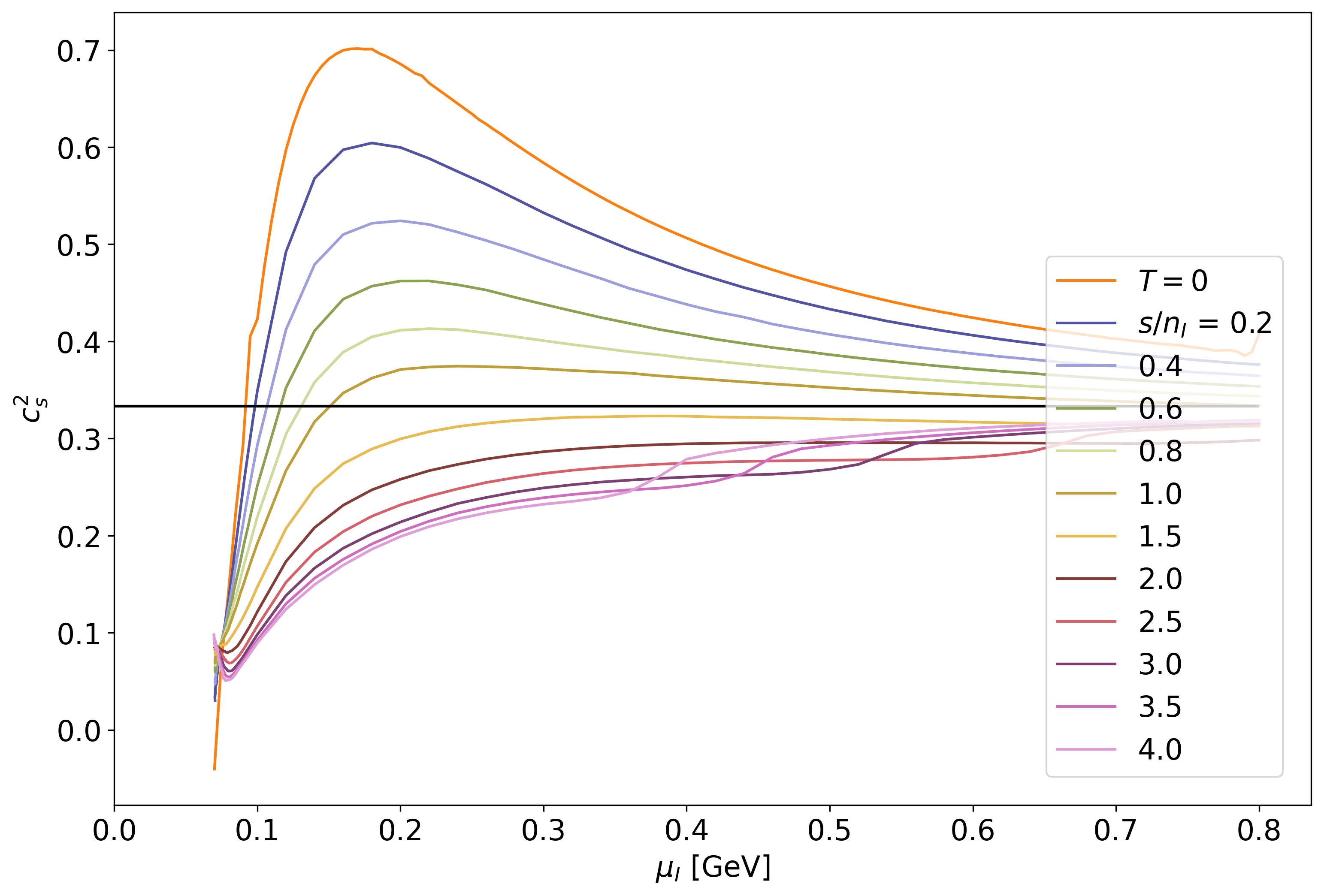}
	\caption{
		The isentropic sound velocity $c_{s/n_I}^2$ along with the isentropic trajectories.
		The $T=0$ case is also shown for the comparison.
	}
	\label{fig:finiteT-cs2-snap}
\end{figure}

To examine the thermal correction to EOS,
we see the sound velocity on the isentropic trajectories in \figref{fig:finiteT-cs2-snap}.
The result of $T=0$ is also shown for the comparison.
As increasing the $s/n_I$, which corresponds to the increasing $T$,
results in the decreasing of sound velocity and softer EOS.
Especially on the $s/n_I \geq 1.5$ trajectories,
the value of $c_{s/n_I}^2$ does not exceed $1/3$ and the peak structure get smeared.
In contrast for $s/n_I \geq 2.5$,
the sound velocity has the sudden increase near the second order phase transition;
The disappearance of the mass gap favors the larger sound velocity and stiffer EOS.

At the zero temperature,
the non-zero mass gap $\Delta$ favors the sound velocity $c_{s/n_I}^2 > 1/3$
and the melting-out of the mass gap decrease $c_s^2$ to $1/3$
as discussed in Chap.$\,$\ref{chap:zeroT}.
In addition,
the thermal quarks with finite mass gap behaves as the non-relativistic gas
and the sound velocity favors $c_{s/n_I}^2 < 1/3$.
Thus for $s/n_I > 1.5$,
the peak structure of $c_{s/n_I}^2$ disappears and $c_s^2$ approaches to $1/3$ from below.
When the mass gap is melted out,
the thermal quark begins to behave as the relativistic gas
and the sound velocity favors $c_{s/n_I}^2 = 1/3$.
When it happens to the trajectory where the sound velocity is less than $1/3$,
the sound velocity increases to $1/3$ which is shown in $s/n_I > 2.5$ trajectories.

\section{Trace anomaly}
\label{sec:finiteT-traceanomaly}

\begin{figure}[thpb]
	\centering	
	\includegraphics[width=0.7\linewidth]{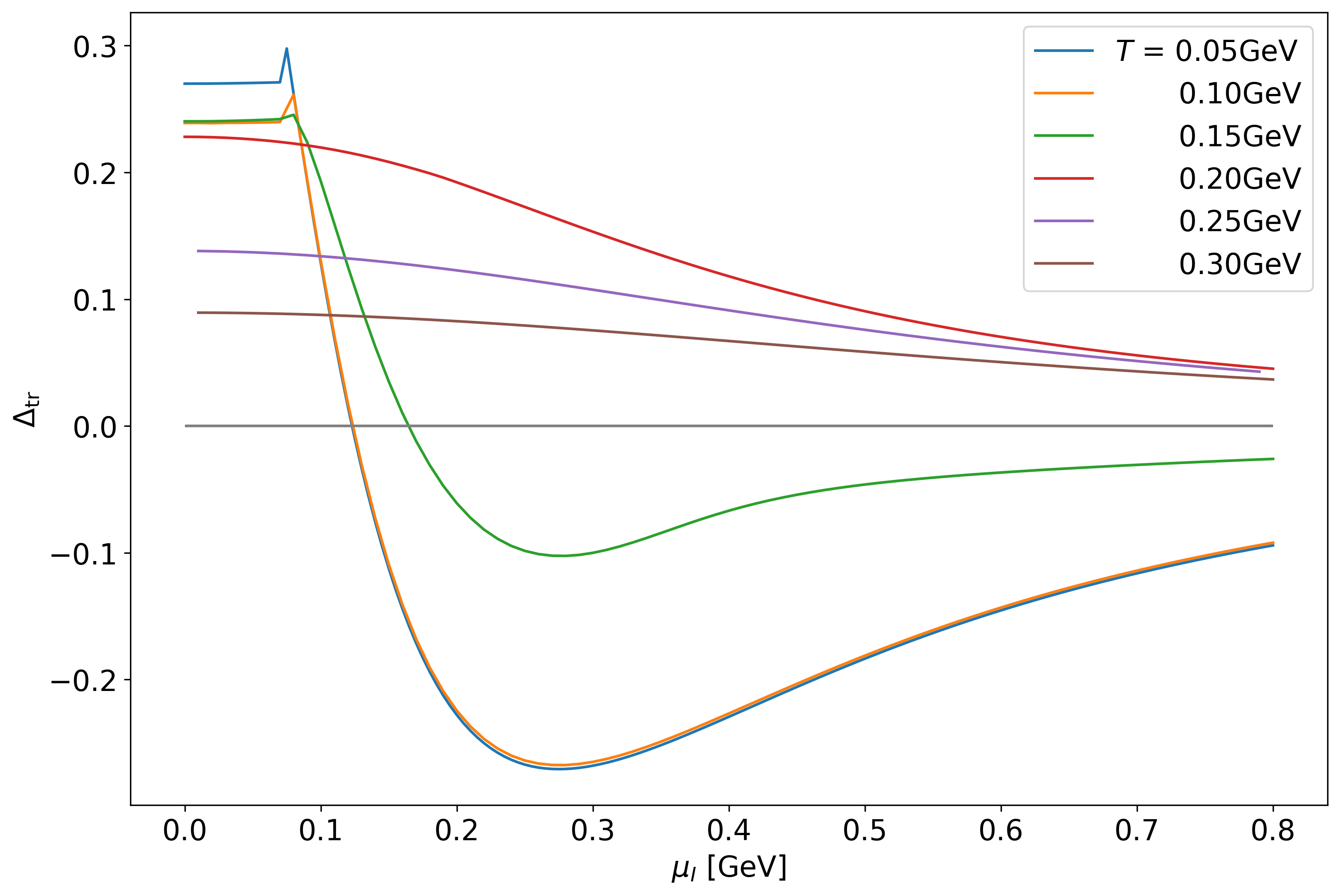}
	\caption{
		The trace anomaly $\Delta_{\rm tr} = 1/3 - P/\varepsilon$ as a function of $\mu_I$ for fixed temperature.
	}
	\label{fig:finiteT-traceanomaly}
\end{figure}

As discussed in Sec.$\,$\ref{ssec:zeroT-trace-anomaly},
the trace anomaly $\Delta_{\rm tr} = 1/3 - P/\varepsilon$ is the deviation from the conformal limit
and its asymptotic behavior is the measure of non-perturbative effects.
The trace anomaly for fixed temperature as a function of $\mu_I$ is shown in \figref{fig:finiteT-traceanomaly}.

We first consider the low density $\mu_I \leq m_\pi/2$.
In this region the system is dilute from low to high temperature
and $3\varepsilon > P$ is satisfied,
thus the trace anomaly is positive $\Delta_{\rm tr} > 0$.
Especially for low temperature $T < T_\chi \simeq 170 \MeV$,
both pressure and energy density does not have the strong $\mu_I$ dependence
and the trace anomaly is almost constant.

The situation changes as the system enters the condensed phase.
For $T < 150\MeV$ the pion condensate is formed at $\mu_I = m_\pi/2$.
At small $\mu_I$ the BEC phase is realized and $\varepsilon \simeq m_\pi n_I \gg P$,
thus the trace anomaly is still positive.
As increasing $\mu_I$, however,
the system soon crossovers to the BCS phase and the trace anomaly decreases rapidly to falls below $\Delta_{\rm tr} = 0$.
As increasing the temperature,
the trace anomaly keeps positive for larger $\mu_I$ because of the small $\Delta$.
But the quark excitation gains the pressure $P \simeq T^4$ and the trace anomaly approaches to the conformal value $\Delta_{\rm tr} = 0$ for all the $\mu_I$ region.

\begin{figure}[thpb]
	\centering	
	\includegraphics[width=0.7\linewidth]{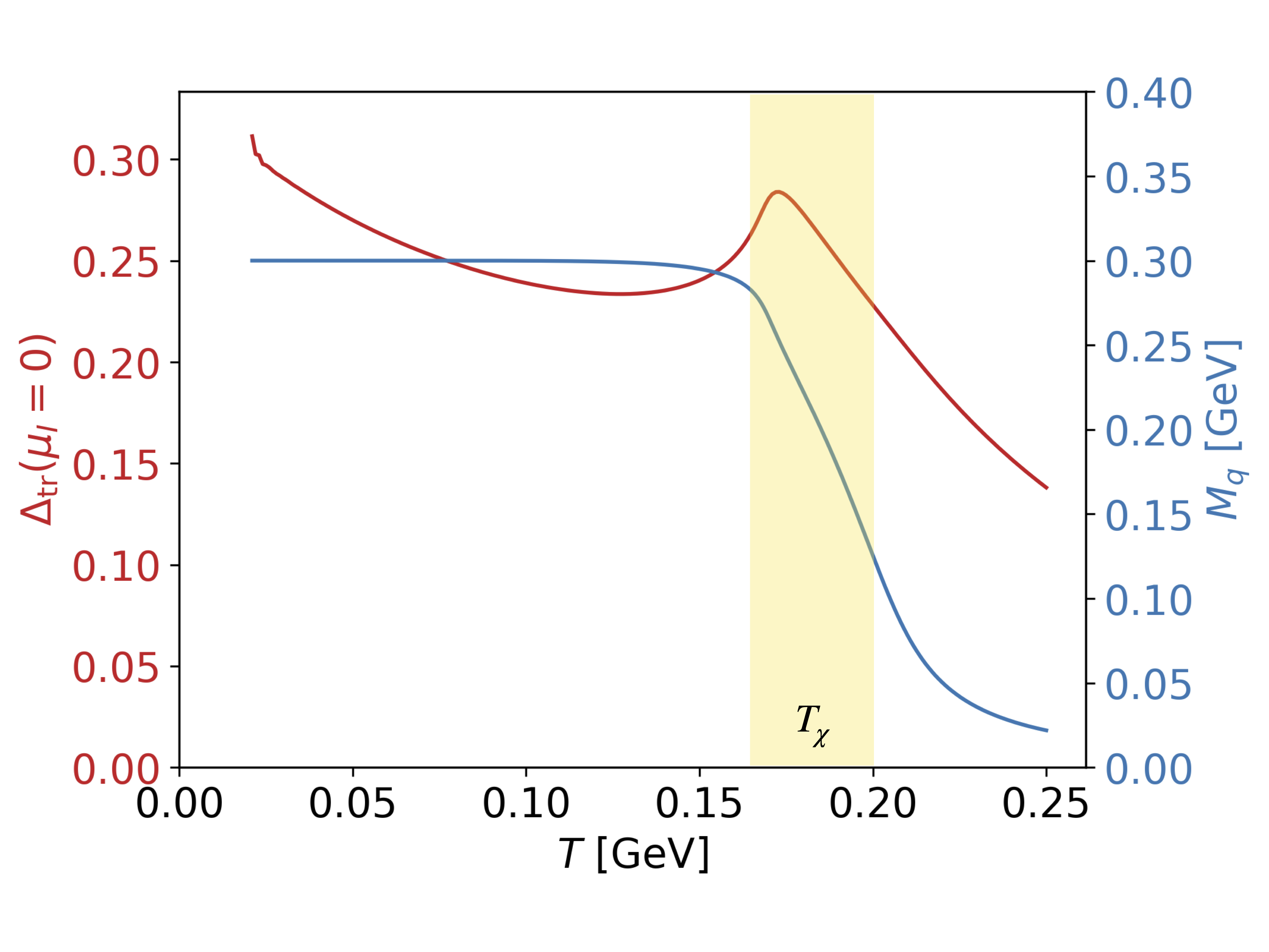}
	\caption{
		The trace anomaly $\Delta_{\rm tr}$ at $\mu_I = 0$ as a function of temperature.
		The thermal quark pressure favors $\Delta_{\rm tr} = 0$
		while the chiral restoration around $T \sim T_\chi$ softens the EOS and increases $\Delta_{\rm tr}$.
		By the competition of these two, the trace anomaly has the non-monotonic behavior.
	}
	\label{fig:finiteT-traceanomaly_mu0}
\end{figure}

In addition
we can see that the value of $\Delta_{\rm tr}$ at $\mu_I = 0$ is not much changed with temperature at $T \sim 100 - 200 \MeV$.
This is the consequence of the competition of the chiral restoration and the increasing of thermal quark pressure.
As mentioned 
the chiral restoration leads the softer EOS and larger $\Delta_{\rm tr}$
while the relativistic quark pressure favors the conformal value $\Delta_{\rm tr} = 0$.
The competition of these two effects leads the plateau and the peak of $\Delta_{\rm tr}(\mu_I = 0)$
as shown in \figref{fig:finiteT-traceanomaly_mu0}.

\section{Discussions}
\label{sec:finiteT-discussion}

\subsection{Thermal mesons on thermodynamics}
\label{ssec:finiteT-thermalmesons}
In our calculation of the quark meson model we neglected the possibly important thermal meson contributions.
The LQCD calculation with the vanishing density and finite temperature \cite{HotQCD:2014kol} shows that
the thermal mesons have the significant contribution to the EOS.
In this section we examine whether and how the thermal mesons contribute to the EOS in the condensed phase.

Within the straightforward mean field approximation,
the pion condensate breaks the isospin symmetry and requires the consistent treatment of the meson mixing and the multi-channel interactions of quarks and mesons.
This treatment is formulated as generalized Beth-Uhlenbeck approach \cite{Schmidt:1990oyr,Blaschke:2016fdh,Blaschke:2013zaa}.

Instead of this consistent treatment,
we evaluate the thermal meson contribution in an add-hoc manner;
we first solve the gap equation in the absence of the thermal mesons
and evaluate the thermal meson contribution using that solution afterwords.
If the add-hoc contribution is small the approximation is valid.
Otherwise the consistent treatment is required.

The pressure of the thermal mesons is given by
\begin{align}
	\Omega_\rmM = T \int_\vp \ln\qty(1 - e^{-\beta E_\rmM(\vp)})
\end{align}
where $\rmM$ denotes the meson eigenstates and $E_\rmM$ is its eigenenergy energy including $\mu_I$.
The $E_\rmM$ can be read off from the pole of the meson propagator
or the solution of Klein-Gordon equation
\begin{align}
	\det G(p_0,\vp) = 0, ~~~ \calL_{\rm EFT} = \frac{1}{2}\phi_i G_{ij} \phi_j + \calO(\phi^3)
\end{align}
for the quantum fluctuation of mesons $\phi_j$.

In the normal phase the mesons are not mixed and their energy are given as
\begin{align}
	E_\sigma &= \sqrt{m_\sigma^2 + \vp^2}\\
	E_{\pi_0} &= \sqrt{m_\pi^2 + \vp^2}\\
	E_{\pi_+} &= \sqrt{m_\pi^2 + \vp^2} + 2\mu_I\\
	E_{\pi_-} &= \sqrt{m_\pi^2 + \vp^2} - 2\mu_I.
\end{align}
In the condensed phase,
the mesons $\vec\phi = (\sigma,\vec\pi)$ are mixed because of the finite density and the sigma and pion condensation.
The bi-linear term of meson Lagrangian is given as
\begin{align}
	\calL_{\sigma\sigma} &= \frac{1}{2}\qty(\partial_\mu\sigma)^2 - \frac{1}{2}\qty(m_0^2 + \frac{\lambda}{6g^2}\qty(3M_q^2 + \Delta^2)) \sigma^2 \\
	\calL_{\pi_1\pi_2} &= \frac{1}{2}\qty(\partial_\mu\pi_1)^2 + \frac{1}{2}\qty(\partial_\mu\pi_2)^2 - 2\mu_I\qty(\partial_0\pi_1\pi_2 - \pi_1 \partial_0\pi_2) \notag\\
	&\quad - \frac{1}{2}\qty(m_0^2 - 4\mu_I^2 + \frac{\lambda}{6g^2}\qty(M_q^2 + 3\Delta^2)) (\pi_1)^2 \notag\\
	&\quad - \frac{1}{2}\qty(m_0^2 - 4\mu_I^2 + \frac{\lambda}{6g^2}\qty(M_q^2 +  \Delta^2)) (\pi_2)^2 \\
	\calL_{\pi_3\pi_3} &= \frac{1}{2}\qty(\partial_\mu\pi_3)^2 - \frac{1}{2}\qty(m_0^2 + \frac{\lambda}{6g^2}\qty(M_q^2 + \Delta^2)) (\pi_3)^2 \\
	\calL_{\sigma\pi_1} &= - \frac{\lambda}{3g^2}M_q\Delta \sigma\pi_1,
\end{align}
which shows that the $\sigma \leftrightarrow \pi_1$ is mixing due to the both the chiral and pion condensate
and $\pi_1 \leftrightarrow \pi_2$ is mixing due to the isospin chemical potential.
The $\pi_3$ is the neutral pion and does not mix with other mesons.
After the chiral symmetry restored $M_q \to 0$
the $\sigma$ and $\pi_1$ are decoupled.
In this limit only $\pi_1$ and $\pi_2$ are mixed
and the off-diagonal component of the Klein-Gordon operator is given as
\begin{align}
	G_{\pi_1\pi_2} &= \qty(
	\begin{matrix}
		p_0^2 - \vp^2 - 8\mu_I^2 + 2m_0^2 & 4i\mu_I p_0 & \\
		-4i\mu_I p_0 & p_0^2 - \vp^2
	\end{matrix}
	)
\end{align}
after using the analytic solution of $\Delta$;
\begin{align}
	\Delta = \sqrt{\frac{6g^2}{\lambda}\qty(4\mu_I^2 - m_0^2)}.
\end{align}
Solving the equation $\det G_{\pi_1\pi_2}=0$,
we obtain the eigenenergy for mixed state of $\pi_1$ and $\pi_2$ as
\begin{align}
	E_{\tilde\pi_+} &= v_\pi|\vp| \\
	E_{\tilde\pi_-} &= \sqrt{\bar m^2 + \qty(1 + \frac{16\mu_I^2}{\bar m^2})\vp^2}. 
\end{align}
with 
\begin{align}
	\bar m^2 = 24\mu_I^2 - 2m_0^2, ~~~ v_\pi = \sqrt{\frac{4\mu_I^2 - m_0^2}{12\mu_I^2 - m_0^2}}
\end{align}
and the expansion for $16\mu_I^2\vp^2 / \bar m^4 \ll 1$.

Here the mass-less mode $\tilde \pi_+$ corresponds to the Nambu-Goldstone mode for the $I_3$ symmetry breaking.
$v_\pi$ corresponds to the {\it velocity}
and its value converges to $v_\pi^2 \to 1/3$ in the high density limit $\mu_I \to \infty$.
The remaining non-mixed modes have the energy given as
\begin{align}
	E_\sigma &= \sqrt{m_\sigma^2 + \vp^2}\\
	E_{\pi_0} &= \sqrt{4\mu_I^2 + \vp^2}.
\end{align}
As the mesons can decay into the quark and antiquark with gap $\Delta$,
the following constraint is imposed for the momentum integral.
\begin{align}
	E_\rmM (\vp) \leq 2\Delta.
	\label{eq:finiteT-meson-integral-constraint}
\end{align}

In the following we examined the thermal meson contribution in the two region;
{\bf a)} the normal phase at $\mu_I = 10 \MeV$ where $M_q > 0$ and $\Delta = 0$,
{\bf b)} the condensed phase at $\mu_I = 240 \MeV$ where $M_q \simeq 0$ and $\Delta > 0$.

\begin{figure}[thpb]
	\centering
	\begin{minipage}{0.45\hsize}
		\centering	
		\includegraphics[width=0.95\linewidth]{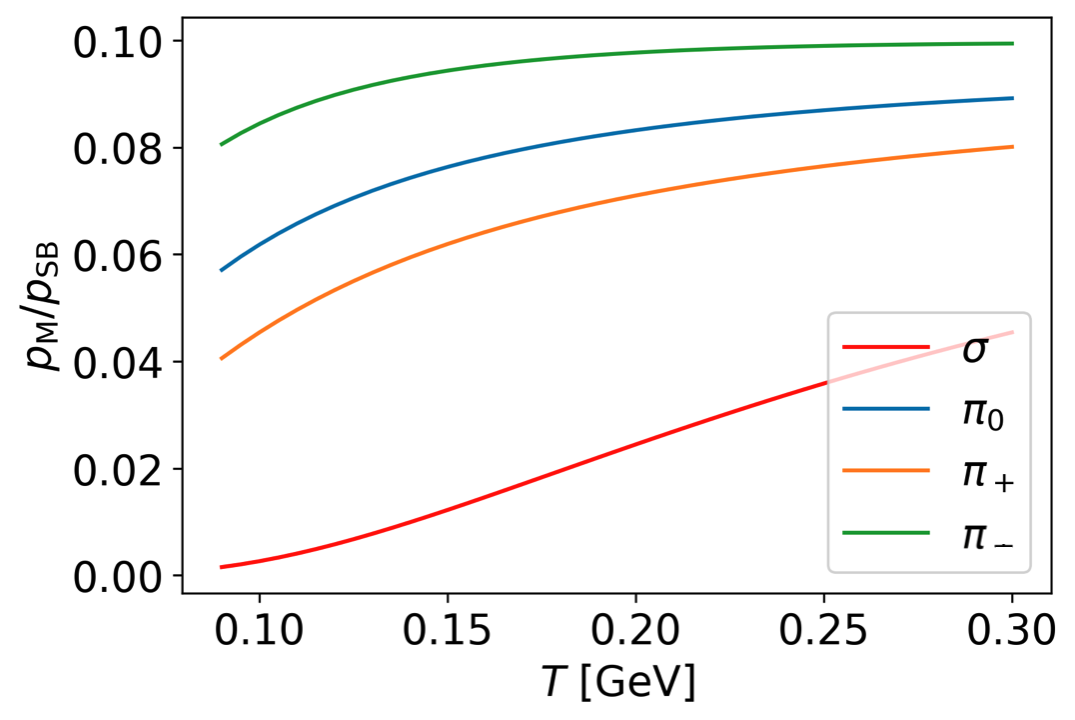}
	\end{minipage}
	\begin{minipage}{0.45\hsize}
		\centering	
		\includegraphics[width=0.95\linewidth]{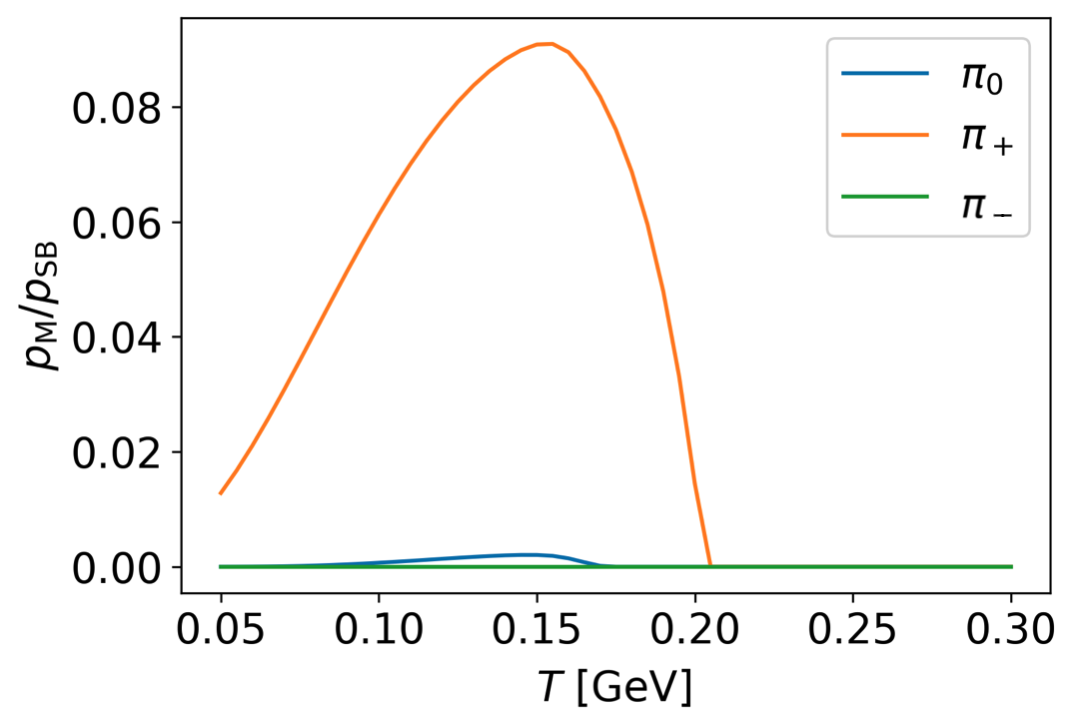}
	\end{minipage}
	\caption{
	The pressure for the mesons normalized by the Stefan-Boltzmann pressure
	at {\bf a)} $\mu_I = 10 \MeV$; {\bf b)} $\mu_I = 240\MeV$.
	}
	\label{fig:finiteT-thermalmesons-pressure}
\end{figure}

\figref{fig:finiteT-thermalmesons-pressure} shows the pressure of the each meson modes.
In the both region
the order of the pressure matches to the order of the mass.
In the normal phase
the sigma meson pressure has non-zero contribution 
but much smaller than the pion pressure.
In the condensed phase
it has zero contribution because of the constraint in \eqref{eq:finiteT-meson-integral-constraint},
or in other words,
the sigma meson is unstable and soon decay to the quark and antiquark.

% figure: pressure of each modes

\begin{figure}[thpb]
	\centering
	\begin{minipage}{0.45\hsize}
		\centering	
		\includegraphics[width=0.95\linewidth]{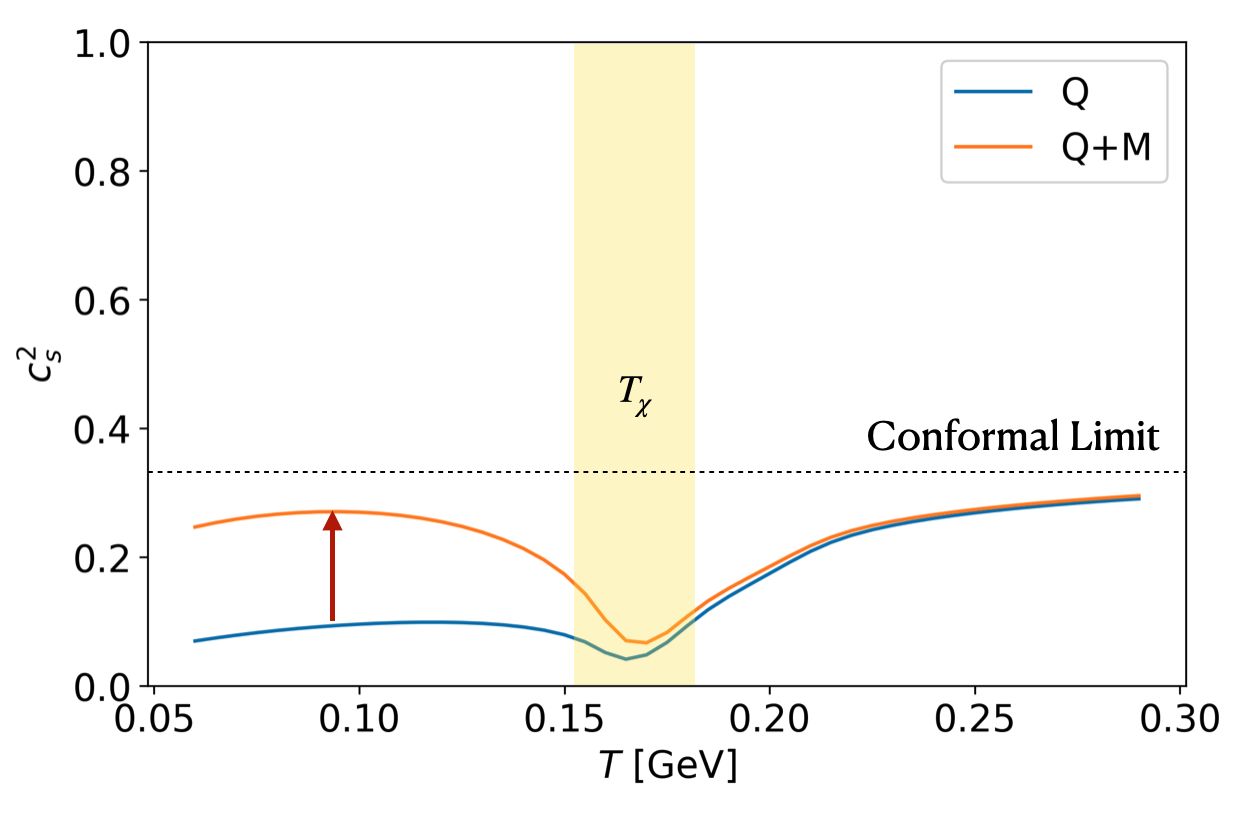}
	\end{minipage}
	\begin{minipage}{0.45\hsize}
		\centering	
		\includegraphics[width=0.95\linewidth]{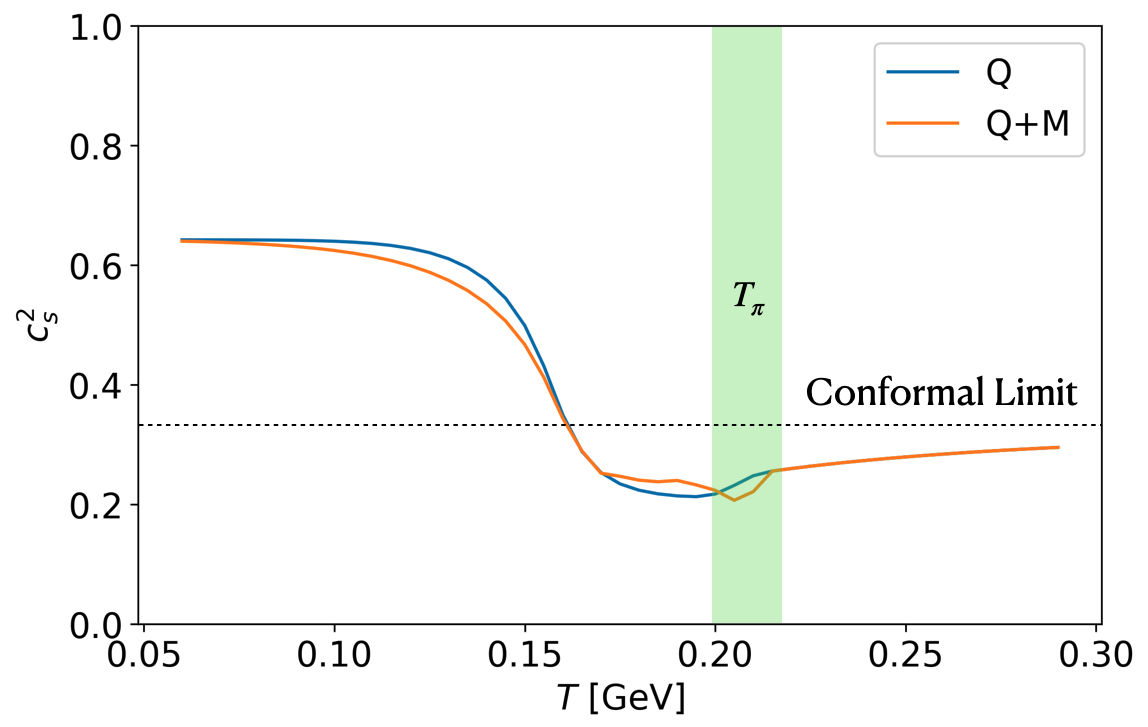}
	\end{minipage}
	\caption{
		The isentropic sound velocity with the thermal mesons at {\bf a)} $\mu_I = 10 \MeV$ and {\bf b)} $\mu_I = 240\MeV$.
		The label ``Q+M'' and ``Q'' denotes the calculation with and without the thermal mesons, respectively.
	}
	\label{fig:finiteT-thermalmesons-cs2}
\end{figure}

We examine the impact of the thermal mesons on the isentropic sound velocity,
illustrated in \figref{fig:finiteT-thermalmesons-cs2}.
In the normal phase with vanishing density a),
the thermal mesons have the significant contribution to the thermodynamics
resulting the $c_{s/n_I}^2$ two times bigger.
This enhancement leads the dip structure of $c_{s/n_I}^2$ around $T \sim T_\chi$
which is also seen in the lattice QCD calculation \cite{HotQCD:2014kol}.
In the condensed phase b),
on the other hand,
the thermal mesons have negligible contribution to $c_{s/n_I}^2$.

% figure: sound velocity (fixed µ)

These calculation indicates that the thermal mesons have the significant contribution to the EOS in the normal phase.
In the condensed phase, however,
the thermal mesons does not have a large contribution to EOS
and our current treatment to ignore the thermal meson seems valid.

\chapter{Summary and conclusions}
\label{chap:summary}
% introduction of summary
This thesis examines the EOS of isospin QCD using a two-flavor quark meson model
which includes both the mesons and quark degrees of freedom
comparing to the LQCD results. 
We examined the impacts of quark degrees of freedom on the EOS,
which are introduced as the substructure of the pion condensate.
We also examined the thermal quark and meson contributions to the EOS.

% pion condensate, tree vs 1-loop
The model describes the BEC-BCS crossover of pion condensate and is renormalizable, 
making it applicable from low to high density region. 
At the tree-level
the pion condensate behaves as a bosonic object and increases linearly with $\mu_I$,
however,
at the one-loop,
the pion condensate is treated as the quark-antiquark pairing
and its growth is suppressed by the Pauli blocking.
This microscopic analysis reveals the limitation of the dense matter description with only the meson degrees of freedom.

% zero temperature
% sound velocity and its peak, BEC->BCS
% non-perturbative effects on sound velocity and trace anomaly
Next we studied the EOS at zero temperature.
The sound velocity $c_s^2$ exceeds the conformal value $1/3$ around $n_I \simeq 2 n_0$,
which is because of the quark mass gap $\Delta \simeq 300 \MeV$.
The maximum value of $c_s^2$ is reached at around $n_I \simeq 5n_0$, with $c_s^2 \simeq 0.7$
and relaxes to the conformal value $1/3$ from above,
which is consistent to the LQCD.
This asymptotic behavior differs from the pQCD results,
however,
we show that this discrepancy is due to the existence of the non-perturbative contribution to the pressure $\sim \Delta^2\mu_I^2$ in QM model.
This result indicates that the power correction is not negligible
even at the high density $\sim 40 n_0$ where pQCD is considered to be valid.
This power correction also changes the sign of the trace anomaly $\Delta_{\rm tr}$.
The trace anomaly is positive in the pQCD,
it can be negative with the non-perturbative contribution.
In other words,
$c_s^2 > 1/3$ and $\Delta_{\rm tr} < 0$ at the high density indicate the existence of the non-perturbative physics.

We also discussed the quark saturation of the matter.
The evolution of the occupation probability of the quark in the matter is decomposed into the vertical evolution and horizontal evolution
which characterize the BEC phase and BCS phase respectively.
The occupation probability is calculated in our model,
and it is found that the transition from BEC to BCS occurs around $n_I \simeq 2 n_0$
and that the transition is smooth.
The transition density is smaller than the pion overlap density $n_I \simeq 5.2 n_0$.
This simple analysis suggests that the quark wavefunction percolates outside the charged radius of pion even at the low density $n \lesssim n_0$.

% finite temperature
% phase diagram, especially melting of pion condensate
% thermal effects on sound velocity and trace anomaly
% thermal meson contribution
Next we studied the EOS on the $T-\mu_I$ plane with the Polyakov loop,
especially on the isentropic trajectory where $s/n_I$ is fixed.
The thermal effects of quarks with the Polyakov loop background reduce the pion condensate and the sound velocity,
and the peak structure of $c_s^2$ is washed out for large $T$.
The melting-out of the pion condensate also makes the trace anomaly positive.
We also examined the thermal meson contribution
which is suggested by LQCD to be important in the normal phase where density is almost zero.
The effects of the thermal mesons are confirmed to be large in the normal phase,
and the sound velocity is greatly enhanced by the thermal mesons below the chiral transition temperature $T_\chi$.
In contrast,
the thermal meson contribution to the EOS is found to be small in the pion condensed phase.

% prospects
% 1. towards QGP with mesonic excitation and gluon dynamics
In this thesis,
we include the thermal quarks and mesons in addition to the condensed mesons made of quarks.
The gluon contribution is included through the Polyakov loop potential,
however,
the dynamics of thermal gluons must be taken into account for the QGP phase.
The thermal mesons are not treated in a self-consistent way.
Although we made a rough estimate of the thermal meson contribution
and found that it does not have the large impact on the EOS in the pion condensed phase,
its application is restricted to the high density and low temperature.
The consistent treatment of the thermal mesons and quarks is necessary
for the analysis of the hadronic excitation in dense matter.

% 2. gluonic contribution in finite density
Another question is the finite density effect to the gluon dynamics and the Polyakov loop potential.
The Polyakov loop potential is constructed from the gluon propagators
which which may be modified by the finite density effects.
This can change the phase structure, 
particularly the melting of the pion condensate at high density.
Since the non-perturbative contribution from the pion condensate has a important role on $c_s^2$ to exceed the conformal value,
it is important to study when and how such effects disappear.
Although the perturbative treatment of the finite density effects on the gluon propagator in the analogy of the Debye screening is conducted,
it is important to examine the impact of non-perturbative effects at high density.

% 3. RG imporvement

These analyses will provide important insights into the universal concept of QCD-like theories at finite density
which can be applied to QCD at finite baryon density.

\clearpage
\backmatter
\setcounter{counterforappendices}{\value{page}}
\mainmatter
\setcounter{page}{\value{counterforappendices}}

\appendix

\chapter{Tolman - Oppenheimer - Volkoff equation}
\label{app:TOV}
In this section,
we derive the Tolman - Oppenheimer - Volkoff (TOV) equation \cite{Oppenheimer:1939ne},
which describes the non-rotating and spherically symmetric massive star.
The detailed discussion is shown in Ref.\cite{Weber:1999qn}.

\section{Statics of the massive star}
In the assumption of the spherical symmetry and no rotation,
all the stars including Neutron star can be understood 
by the balance of pressure and gravity.
In the classical mechanics,
this balance can be written as 
\begin{align}
	\frac{dP}{dr} = -\frac{Gm(r)\rho(r)}{r^2}
\end{align}
with the pressure $P(r)$and density $\rho(r)$ at radius $r$ and mass $m(r)$ up to radius $r$.

But in the case of Neutron star
the relativistic effect is not negligible.
To estimate this effect,
let us consider the ratio of the light velocity and gravitational effect $GM/R$ on the surface of Neutron star.
Assuming that the mass and radius are $M \sim 2.0 M_\odot, ~ R \sim 10 {\rm km}$,
this ratio will be
\begin{align}
	\frac{GM/R}{c^2} \simeq 0.295
\end{align}
which is not small.
Thus we should take the relativistic effect into account.

The relativistic version of the above equation is called Tolman - Oppenheimer - Volkoff (TOV) equation;
\begin{align}
	\dv{P}{r} = -\frac{Gm(r)\rho(r)}{r^2}\qty(1+\frac{P(r)}{\rho(r)c^2})\qty(1+\frac{4\pi r^3 P(r)}{m(r)c^2})\qty(1-\frac{2Gm(r)}{rc^2})^{-1}
\end{align}
In this section,
we will firstly review the Einstein equation on Schwarzschild metric.
Then we will derive the TOV equation.

\section{Einstein equation on Schwarzschild metric}
To begin with,
we will roughly review the Einstein equation
and consider the spherically symmetric case, Schwarzschild solution.

In the given spacetime with metric $g_{\mu\nu}$,
Einstein equation is the relativistic (Lorentz-covariant) differential equation
with the form of
\begin{align}
	G_{\mu\nu} = R_{\mu\nu} - \frac{1}{2}g_{\mu\nu}R = \frac{8\pi G}{c^4}T_{\mu\nu}
\end{align}
with Ricci tensor $R_{\mu\nu}$, Ricci scalar $R=g^{\mu\nu} R_{\mu\nu}$ and energy-momentum tensor $T_{\mu\nu}$.
Ricci tensor is given in terms of Christoffel symbol $\Gamma^\alpha_{\mu\nu}$ as
\begin{align}
	R_{\mu\nu} &\coloneqq \Gamma_{\mu\nu,\alpha}^\alpha - \Gamma_{\mu\alpha,\nu}^\alpha + \Gamma_{\beta\alpha}^\alpha \Gamma_{\mu\nu}^\beta - \Gamma_{\beta\mu}^\alpha \Gamma_{\alpha\nu}^\beta \\
	\Gamma_{\alpha\beta}^\mu &\coloneqq  \frac{1}{2}g^{\mu\nu}\qty(g_{\nu\alpha,\beta}+g_{\nu\beta,\alpha}-g_{\alpha\beta,\nu}).
\end{align}
Here we used the notation of differentiation
\begin{align}
	\partial_\alpha X_{\mu\nu} = X_{\mu\nu,\alpha}.
\end{align}

The left hand side of the Einstein equation is fully determined
once the stricture of spacetime, indeed metric $g^{\mu\nu}$, is determined.

Now we will consider the Schwarzschild solution
which metric is
\begin{align}
	g_{\mu\nu} = \mqty(\dmat{-e^{2\Psi(r)}, e^{2\Phi(r)}, r^2, r^2\sin\theta^2})
\end{align}
and its inverse is
\begin{align}
	g^{\mu\nu} = \mqty(\dmat{-e^{-2\Psi(r)}, e^{-2\Phi(r)}, \frac{1}{r^2}, \frac{1}{r^2\sin\theta^2} }).
\end{align}
This metric respects the rotational symmetry around the axis with angle $\psi$.
In addition we assumed the static system thus the metric is time-independent.

From now on
we will calculate the Christoffel symbols.
Non-vanishing terms are the following.
\begin{align}
	\Gamma_{tr}^t &= \frac{1}{2}g^{t\mu}\qty(g_{\mu t,r}+g_{\mu r,t}-g_{tr,\mu}) = \frac{1}{2}g^{tt}g_{tt,r} = \Psi'\\
	\Gamma_{tt}^r &= \frac{1}{2}g^{r\mu}\qty(g_{\mu t,t}+g_{\mu t,t}-g_{tt,\mu}) = -\frac{1}{2}g^{rr}g_{tt,r} = \Psi' e^{2(\Psi-\Phi)}\\
	\Gamma_{rr}^r &= \frac{1}{2}g^{r\mu}\qty(g_{\mu r,r}+g_{\mu r,r}-g_{rr,\mu}) = -\frac{1}{2}g^{rr}g_{rr,r} = \Phi'\\
	\Gamma_{r\theta}^\theta &= \frac{1}{2}g^{\theta\mu}\qty(g_{\mu r,\theta}+g_{\mu \theta, r}-g_{r\theta,\mu}) = \frac{1}{2}g^{\theta\theta}g_{\theta\theta,r} = \frac{1}{r}\\
	\Gamma_{r\psi}^\psi &= \frac{1}{2}g^{\psi\mu}\qty(g_{\mu r,\psi}+g_{\mu \psi, r}-g_{r\psi,\mu}) = \frac{1}{2}g^{\psi\psi}g_{\psi\psi,r} = \frac{1}{r}\\
	\Gamma_{\theta\theta}^r &= \frac{1}{2}g^{\mu r}\qty(g_{\mu \theta,\theta}+g_{\mu\theta,\theta}-g_{\theta\theta,\mu}) = -\frac{1}{2}g^{rr}g_{\theta\theta,r} = -re^{-2\Phi}\\
	\Gamma_{\theta\psi}^\psi &= \frac{1}{2}g^{\mu \psi}\qty(g_{\mu \theta,\psi}+g_{\mu\psi,\theta}-g_{\psi\theta,\mu}) = \frac{1}{2}g^{\psi\psi}g_{\psi\psi,\theta} = \frac{\cos\theta}{\sin\theta}\\
	\Gamma_{\psi\psi}^r &= \frac{1}{2}g^{\mu r}\qty(g_{\mu\psi,\psi}+g_{\mu\psi,\psi}-g_{\psi\psi,\mu}) = -\frac{1}{2}g^{rr}g_{\psi\psi,r} = - r\sin\theta^2 e^{-2\Phi} \\
	\Gamma_{\psi\psi}^\theta &= \frac{1}{2}g^{\mu \theta}\qty(g_{\mu\psi,\psi}+g_{\mu\psi,\psi}-g_{\psi\psi,\mu}) = -\frac{1}{2}g^{\theta\theta}g_{\psi\psi,\theta} = - \sin\theta\cos\theta
\end{align}
Here we should notice that the Christoffel symbol is symmetric as
\begin{align}
	\Gamma_{\alpha\beta}^\mu = \Gamma_{\beta\alpha}^\mu.
\end{align}

Next we calculate the Ricci tensor.
To get the Ricci scalar we only need the diagonal elements of Ricci tensor $R_{tt},R_{rr},R_{\theta\theta}$ and $R_{\psi\psi}$.
\begin{align}
	R_{tt} &= \Gamma_{tt,\mu}^\mu - \Gamma_{t\mu,t}^\mu + \Gamma_{\nu\mu}^\mu \Gamma_{tt}^\nu - \Gamma_{\nu t}^\mu \Gamma_{\mu t}^\nu \notag\\
	&= \Gamma_{tt,r}^r + \Gamma_{r\mu}^\mu \Gamma_{tt}^r - 2\Gamma_{tt}^r\Gamma_{rt}^t \notag\\
	&= \qty(\Psi'' + 2(\Psi'-\Phi')\Psi' + \qty(\Psi' + \Phi' + \frac{2}{r})\Psi' - 2{\Psi'}^2)e^{2(\Psi-\Phi)} \notag\\
	&= \qty(\Psi'' + {\Psi'}^2 - \Psi'\Phi' + \frac{2}{r}\Psi')e^{2(\Psi-\Phi)}.\\
	%%%%%%%%%%%%
	R_{rr} &= \Gamma_{rr,\mu}^\mu - \Gamma_{r\mu,r}^\mu + \Gamma_{\nu\mu}^\mu \Gamma_{rr}^\nu - \Gamma_{\nu r}^\mu \Gamma_{\mu r}^\nu \notag\\ 
	&= \Gamma_{rr,r}^r - \Gamma_{r\mu,r}^\mu + \Gamma_{r\mu}^\mu \Gamma_{rr}^r - \qty( {\Gamma_{tr}^t}^2 + {\Gamma_{rr}^r}^2 + {\Gamma_{\theta r}^\theta}^2 + {\Gamma_{\phi r}^\phi}^2)\notag\\
	&= \Phi'' - \qty(\Psi'' + \Phi'' - \frac{2}{r^2}) + \qty(\Psi'+\Phi' + \frac{2}{r})\Phi' - \qty({\Psi'}^2 + {\Phi'}^2 + \frac{2}{r^2}) \notag \\
	&= -\Psi'' - {\Psi'}^2 + \Psi'\Phi' + \frac{2}{r}\Phi'. \\
	%%%%%%%%%%%%
	R_{\theta\theta} &= \Gamma_{\theta\theta,\mu}^\mu - \Gamma_{\theta\mu,\theta}^\mu + \Gamma_{\nu\mu}^\mu \Gamma_{\theta\theta}^\nu - \Gamma_{\nu \theta}^\mu \Gamma_{\mu \theta}^\nu \notag\\ 
	&= \Gamma_{\theta\theta,r}^r - \Gamma_{\theta\psi,\theta}^\psi + \Gamma_{r\mu}^\mu \Gamma_{\theta\theta}^r - 2\Gamma_{r\theta}^\theta\Gamma_{\theta\theta}^r - {\Gamma_{\psi\theta}^\psi}^2 \notag\\
	&= (-1 + 2r\Phi')e^{-2\Phi} + \frac{1}{\sin^2\theta} - \qty(\Psi' + \Phi' + \frac{2}{r})re^{-2\Phi} + 2e^{-2\Phi} - \frac{\cos^2\theta}{\sin^2\theta} \notag\\
	&= \qty(-1 -r\Psi' + r\Phi' + e^{2\Phi})e^{-2\Phi} . \\
	%%%%%%%%%%%%
	R_{\psi\psi} &= \Gamma_{\psi\psi,\mu}^\mu - \Gamma_{\psi\mu,\psi}^\mu + \Gamma_{\nu\mu}^\mu \Gamma_{\psi\psi}^\nu - \Gamma_{\nu \psi}^\mu \Gamma_{\mu \psi}^\nu \notag\\ 
	&= \Gamma_{\psi\psi,r}^r + \Gamma_{\psi\psi,\theta}^\theta + \Gamma_{r\mu}^\mu \Gamma_{\psi\psi}^r + \Gamma_{\theta\psi}^\psi\Gamma_{\psi\psi}^\theta - 2\Gamma_{r\psi}^\psi\Gamma_{\psi\psi}^r - 2\Gamma_{\theta\psi}^\psi\Gamma_{\psi\psi}^\theta \notag\\
	&= (-1 + 2r\Phi')\sin^2\theta e^{-2\Phi} - \cos^2\theta + \sin^2\theta - \qty(\Psi'+\Phi'+\frac{2}{r})r\sin^2\theta e^{-2\Phi} + 2\sin^2\theta e^{-2\Phi} + 2\cos^2\theta \notag\\
	&= \sin^2\theta \qty(-1 -r\Psi' + r\Phi' + e^{2\Phi})e^{-2\Phi} \notag \\
	&= \sin^2\theta R_{\theta\theta}.
\end{align}
Thus the Ricci scalar is given by
\begin{align}
	R &= -e^{-2\Psi}R_{tt} + e^{-2\Phi}R_{rr} + \frac{1}{r^2} R_{\theta\theta} + \frac{1}{r^2\sin^2\theta} R_{\psi\psi} \\
	&= \begin{multlined}[t]
		- \qty(\Psi'' + {\Psi'}^2 - \Psi'\Phi' + \frac{2}{r}\Psi')e^{-2\Phi} 
		+ \qty(-\Psi'' - {\Psi'}^2 + \Psi'\Phi' + \frac{2}{r}\Phi')e^{-2\Phi}\\
		+ \frac{2}{r^2}\qty(-1 -r\Psi' + r\Phi' + e^{2\Phi})e^{-2\Phi}
	\end{multlined}\\
	&= \qty(-2\Psi'' - 2{\Psi'}^2 + 2 \Psi'\Phi' + \frac{4}{r}(-\Psi'+\Phi') - \frac{2}{r^2} + \frac{2}{r^2}e^{2\Phi} )e^{-2\Phi}
\end{align}

\section{Tolman - Oppenheimer - Volkoff equation}
Tolman - Oppenheimer - Volkoff equation (TOV equation) is the equation of the massive star,
especially the equation of the hydrostatic equilibrium with rotational symmetry.

To derive the TOV equation,
we will continue our discussion of the Schwarzschild solution
considering the energy-momentum tensor $T_{\mu\nu}$ of the perfect fluid.
This is given by
\begin{align}
	T^{\mu\nu} = (\epsilon + P)u^\mu u^\nu + P g^{\mu\nu}
	\label{eq:em_tensor_def}
\end{align}
with energy density $\epsilon$, pressure $P$ and four-vector velocity
\begin{align}
	u^{\mu} = \dv{x^\mu}{\tau} = \gamma(1,\vec v).
\end{align}
In the static system they are reduced to
\begin{align}
	u^\mu = \qty(\dv{t}{\tau}, \vec 0) = \qty(e^{-\Psi}, \vec 0),
\end{align}
thus the energy-momentum tensor has non-vanishing components only in diagonal part as
\begin{align}
	T^{\mu\nu} = \rm{diag}\qty(e^{-2\Psi}\epsilon, e^{-2\Phi}P, \frac{1}{r^2}P, \frac{1}{r^2\sin^2\theta}P),
\end{align}
or more conveniently
\begin{align}
	T^\mu_{~\nu} = \rm{diag}\qty(-\epsilon, P, P, P).
\end{align}
This means we should focus on the diagonal part of the Einstein equation
to get the non-trivial equation.

we will consider the space-time structure, 
the left hand side of the Einstein equation.
After the messy calculation we get the diagonal elements as 
\begin{align}
	G_{tt} &= \qty(\frac{2}{r}\Phi' - \frac{1}{r^2} + \frac{1}{r^2}e^{2\Phi})e^{2(\Psi-\Phi)} \\
	%%%%%%%%%%%%
	G_{rr} & = \frac{2}{r}\Psi' + \frac{1}{r^2} - \frac{1}{r^2}e^{2\Phi} \\
	%%%%%%%%%%%%
	G_{\theta\theta} &= r^2\qty(\frac{1}{r}\qty(\Psi'-\Phi') - \Psi'\Phi' + \Psi'' + {\Psi'}^2)e^{-2\Phi} \\
	%%%%%%%%%%%%
	G_{\psi\psi} &= \sin^2\theta G_{\theta\theta}.
\end{align}
From all the above we can write down the diagonal components of the Einstein equation as follows.
\begin{align}
	G^t_{~t} &= -\qty(\frac{2}{r}\Phi' - \frac{1}{r^2} + \frac{1}{r^2}e^{2\Phi})e^{-2\Phi} = -\frac{8\pi G}{c^4}\epsilon 
	\label{eq:Gtt}\\
	%%%%%%%%%%%%
	G^r_{~r} &= \qty(\frac{2}{r}\Psi'+\frac{1}{r^2}-\frac{1}{r^2}e^{2\Phi})e^{-2\Phi} = \frac{8\pi G}{c^4}P 
	\label{eq:Grr}\\
	%%%%%%%%%%%%
	G^\theta_{~\theta} &= \qty(\frac{1}{r}\qty(\Psi'-\Phi') - \Psi'\Phi' + \Psi'' + {\Psi'}^2) = \frac{8\pi G}{c^4}P .
	\label{eq:Gthth}
\end{align}
Notice that the ``$\psi-\psi$'' component is not independent from the ``$\theta-\theta$'' one.

Next we will consider the conservation of energy-momentum tensor.
In the flat space-time,
the conservation raw is given by
\begin{align}
	T^\mu_{~\nu;\mu} = 0.
\end{align}
On the curved space-time, however,
we need to introduce the covariant differentiation for the tensor $X^{\mu\nu}$.
Its definition is
\begin{align}
	\nabla_\mu X^\mu_{~\nu} &= X^\mu_{~\nu;\mu} = \partial_\mu X^\mu_{~\nu} + \Gamma_{\alpha \mu}^\mu X^\alpha_{~\nu} - \Gamma_{\nu\mu}^\alpha X^\mu_{~\alpha}.
\end{align}
We can apply this formula to the energy-momentum tensor,
however,
we should use the definition of $T^{\mu\nu}$ \eqref{eq:em_tensor_def}
and use the definition of covariant derivative for vector;
\begin{align}
	\nabla_\mu a_\nu = a_{\nu;\mu} = \partial_\mu a_\nu - \Gamma_{\mu\nu}^\alpha a_\alpha.
\end{align}
Then we obtain
\begin{align}
	T^\mu_{\nu;\mu} = \qty(\epsilon+P)_{;\mu}u^\mu u_\nu + \qty(\epsilon+P)\qty(u^\mu_{~;\mu}u_\nu + u^\mu u_{\nu;\mu}) + P_{;\mu}\delta^\mu_{~\nu} = 0.
\end{align}
Now we have four equations which might be independent,
however,
there is only one non-trivial equation with static configuration.

To see this,
we will consider the case of $\nu=0$ and $\nu=j = 1,2,3$.
For the case of $\nu=0$,
we obtain
\begin{align}
	0 &= \partial_0 (\epsilon + P) u^0u_0 + (\epsilon+P)\qty(\partial u_0 u^0 + u_0(\partial_0 u^0 + \Gamma_{\mu\alpha}^0 g^{\mu\alpha})) + \partial_0 P \notag \\
	&= -\dv{\epsilon}{t}.
\end{align}
This is the conservation of energy which is obvious for the configuration.

For the case of $\nu=j$, we obtain
\begin{align}
	0 &= (\epsilon+P)\qty(\partial_0 u_j  - \Gamma_{0j}^0u_0)u^0 + \partial_j P\notag\\
	&= \qty[(\epsilon+P) \Psi' + \dv{P}{r}] \delta_{jr}.
\end{align}
The information of the space-time stricture is obtained in $\psi$
which can be determined by the Einstein equation.
For future calculation
we will define the mass function as integral of the energy density as
\begin{align}
	m(r) = \int_0^r 4\pi r^2 dr \epsilon(r)/c^2 
\end{align}
or equivalently
\begin{align}
	\dv{m}{r} = 4\pi r^2 \epsilon(r)/c^2
\end{align}
with the boundary condition $m(0)=0$.
Then the equation \eqref{eq:Gthth} is reduced to
\begin{align}
	\frac{2G}{c^2}\dv{m}{r} &= \qty(2r\Phi' - 1 + e^{2\Phi})e^{-2\Phi} \notag\\
	&= - \dv{r}\qty(re^{-2\Phi} - r)
\end{align}
and we obtain
\begin{align}
	e^{-2\Phi} = 1 - \frac{2Gm}{c^2r}.
\end{align}
Now we can write down \eqref{eq:Grr} without $\Phi$.
After the simple calculation we get
\begin{align}
	\Psi' = \frac{G}{c^4 r^2}\qty(4\pi r^3 P + mc^2)\qty(1-\frac{2Gm}{c^2r})^{-1}.
\end{align}
Combining the all equations,
we obtain the TOV equation as
\begin{align}
	\dv{P}{r} = -\frac{G}{c^4 r^2}(\epsilon + P)\qty(4\pi r^3 P + mc^2)\qty(1-\frac{2Gm}{c^2r})^{-1}.
\end{align}
Introducing mass density $\rho(r)$ instead of energy density $\epsilon(r)$ as
\begin{align}
	\rho(r) c^2 = \epsilon(r)
\end{align}
we concludes
\begin{align}
	\dv{P}{r} = -\frac{Gm(r)\rho(r)}{r^2}\qty(1+\frac{P(r)}{\rho(r)c^2})\qty(1+\frac{4\pi r^3 P(r)}{m(r)c^2})\qty(1-\frac{2Gm(r)}{rc^2})^{-1}.
\end{align}

\chapter{Derivation of finite density Lagrangian}
\label{app:finite-density-lagrangian}
We begin with calculating the Hamiltonian corresponds to the Lagrangian in \eqref{eq:model-lagrangian}.
The conjugate fields to $\vec\phi$ and $\psi,\psibar$ are given by
\begin{align}
	\Pi_{\vec\pi} &= \fdv{\calL}{\partial_0 \vec\pi} = \partial^0 \vec\pi, \\ 
	\Pi_{\sigma} &= \fdv{\calL}{\partial_0 \sigma} = \partial^0 \sigma, \\
	\Pi_{\psi} &= \fdv{\calL}{\partial_0 \psi} = \rmi\psibar \gamma^0, \\
	\Pi_{\psibar} &= \fdv{\calL}{\partial_0 \psibar} = 0.
\end{align}
The Hamiltonian is obtained by the Legendre transformation of the fields as 
\begin{align}
	\calH &= \Pi_{\Psi}\partial_0 \Psi - \calL \notag\\
	&= \frac{1}{2}\qty(\partial_0\vec\phi)^2 + \frac{1}{2}\qty(\nabla_j\vec\phi)^2 + \frac{1}{2}m_0^2 \vec\phi^2 + \frac{\lambda}{24}\qty(\vec\phi^2)^2 - h\sigma  \notag\\
	&\quad\quad + \psibar\qty(-\rmi \gamma^j\nabla_j + g(\sigma + \rmi \gamma^5\vec\tau\cdot\vec\pi))\psi.
\end{align}
The next step is to calculate the Hamiltonian at the finite isospin density $\calH_{\rm dense} = \calH - \mu n_I$
and construct the finite density Lagrangian thought the partition function.
As we have seen in Sec. \ref{subsec:model-lagrangian},
the isospin density in terms of the field variable can be written as
\begin{align}
	n_I &= \pi_1\partial^0\pi_2 - \pi_2\partial^0\pi_1  + \psibar\gamma_0\tau3\psi \notag\\
	&= \pi_1 \Pi_{\pi_2} - \pi_2 \Pi_{\pi_1} - \rmi \Pi_{\psi} \tau3 \psi.
\end{align}
Then we obtain the Hamiltonian at finite density as
\begin{align}
	\calH_{\rm dense} &= \calH - 2\mu_I n_I \notag\\
	&= \frac{1}{2}\Pi_{\vec\phi}^2 - 2\mu_I(\pi_1 \Pi_{\pi_2} - \pi_2 \Pi_{\pi_1}) + \calV \notag\\
	&\quad\quad - \rmi \Pi_\psi \gamma^0\qty(-\rmi \gamma^j\nabla_j + g(\sigma + \rmi \gamma^5\vec\tau\cdot\vec\pi))\psi 
	+ \mu_I \rmi \Pi_\psi \tau_3 \psi. \\
	\calV &= \frac{1}{2}\qty(\nabla_j\vec\phi)^2 + \frac{1}{2}m_0^2 \vec\phi^2 + \frac{\lambda}{24}\qty(\vec\phi^2)^2 - h\sigma
\end{align}
Through this Hamiltonian the partition function is defined as
\begin{align}
	Z = \int\calD\Pi_{\phi} \calD\Psi ~ \exp\qty[ \rmi\int_\vx \qty(\partial^0\Psi \cdot\Pi_\Psi - \calH_{\rm dense}) ].
\end{align}
Here we should note that the conjugate filed $\Pi_\psi$ is nothing but the $\psibar$ field
and we do not have the explicit integral over $\Pi_\psi$.

For the mesonic field $\vec\phi$
we can perform the integral over $\Pi_{\vec\phi}$ by the Gaussian integral.
Completing the square for the $\Pi_{\vec\phi}$ as
\begin{align}
	&- \frac{1}{2}\qty(\Pi_\sigma - \partial^0\sigma)^2 - \frac{1}{2}\qty(\Pi_{\pi_3} - \partial^0\pi_3)^2 \notag\\
	&\quad - \frac{1}{2}\qty(\Pi_{\pi_1} - \partial^0\pi_1 + 2\mu_I\pi_2)^2 - \frac{1}{2}\qty(\Pi_{\pi_2} - \partial^0\pi_2 - 2\mu_I\pi_1)^2 \notag\\
	&\quad + \frac{1}{2}\qty(\partial^0\sigma)^2  + \frac{1}{2}\qty(\partial^0\pi_1 - 2\mu_I\pi_2)^2 + \frac{1}{2}\qty(\partial^0\pi_2 + 2\mu_I\pi_1)^2 + \frac{1}{2}\qty(\partial^0\pi_3)^2
\end{align}
As the squared terms for $\Pi_\Psi$ give only the overall factor for the partition function after the integral,
the final expression of the partition function is given by
\begin{align}
	Z \propto \int\calD\Phi ~ \exp\qty(\rmi \int_\vx \calL_{\rm dense})
\end{align}
where
\begin{align}
	\calL_{\rm dense} &= \frac{1}{2}\qty[\qty(\partial_\mu\sigma)^2 + \qty(\partial_\mu\pi_3)^2] \notag\\
	&\quad + \frac{1}{2}\qty(\partial^0\pi_1 - 2\mu_I\pi_2)^2 + \frac{1}{2}\qty(\partial^0\pi_2 + 2\mu_I\pi_1)^2 - \calV \notag\\
	&\quad + \psibar\qty[\rmi\slashed{\partial} + \mu_I\tau_3\gamma^0 - g(\sigma + \rmi\gamma^5\vec\tau\cdot\vec\pi)]\psi \\
	&= \frac{1}{2}\qty[(\partial_\mu \sigma)^2+(\partial_\mu \pi_{3} )^2] \notag\\
	&\quad + (\partial_\mu + 2 \rmi \mu_I\delta_\mu^0) \pi^+\qty(\partial^\mu - 2 \rmi \mu_I\delta^\mu_0)\pi^- \notag\\
	&\quad -\frac{1}{2}m_{0}^2 \vec\phi^2  -{\lambda \over 24} ( \vec\phi^2 )^2  + h \sigma  \notag\\
	&\quad + \psibar \qty[ \rmi \slashed{\partial} + \mu_I \tau_3 \gamma^0 - g (\sigma + \rmi \gamma^5 \vec\tau \cdot \vec\pi) ]\psi.
\end{align}

\chapter{Quark propagators}
\label{app:quark-propagator}
We calculate the mean field quark propagator in the presence of the chiral and pion condensates.
From the propagator one can read off the excitation energy from the pole of the propagator and the occupation probability from the residue of the propagator.

It is convenient to introduce the projection operators for particle and antiparticles,
\begin{align}
	\Lambda_{p,a} = {1\over 2} \pm {\gamma_{j}p^j+M_q \over 2E_D}\gamma_0 \,,
\end{align}
which satisfy
\begin{align}
	\Lambda_p + \Lambda_a = 1\,,~ \Lambda_{p,a}\Lambda_{p,a} = \Lambda_{p,a}\,, ~\Lambda_{p,a}\Lambda_{a,p} = 0 \,,
\end{align}
as they should.
The propagator of quarks can be written as 
\begin{align}
	{ i \over \slashed{p}+\mu_f\gamma_0-M_q} &= S_p(p)\gamma_0\Lambda_p+S_a(p)\gamma_0\Lambda_a \,, \notag\\
	S_{p,a}(p) &= { i \over p_0+\mu_f \mp E_D} \,, 
\end{align}
where $S_{p,a}$ is the propagator for a particle and an antiparticle, respectively.

The inverse of the propagator can also be separated by $\Lambda_{p,a}$ as 
\begin{align}
	\slashed{p}+\mu_f\gamma_0-M_q &= (p_0+\mu_f-E_D)\Lambda_p\gamma_0\notag\\
		& + (p_0+\mu_f+E_D)\Lambda_a\gamma_0 \,.
\end{align}
The inverse of the Dirac operator is the quark propagator $S(p)$, and we can write
\begin{align}
	S(p)^{-1} \equiv - i \left(\begin{matrix}\slashed{p} + \mu_u \gamma_{0}-M_q & -i\gamma_5\Delta \\ -i\gamma_5\Delta & \slashed{p} + \mu_d \gamma_{0}-M_q\end{matrix}\right)
\end{align}
and consider its inverse.
To simplify the discussion, we introduce the single-particle propagator
\begin{align}
	(G_{u,d}^0)^{-1} = -i \left(\slashed{p} + \mu_{u,d} \gamma_0 -M_q \right) \,,
\end{align}
and write off-diagonal term $\Xi = \gamma_5\Delta$.
Then our propagator must satisfy
\begin{align}
	S(p)^{-1}S(p) = \left(\begin{matrix}(G_u^0)^{-1} & \Xi\\ \Xi & (G_d^0)^{-1}\end{matrix}\right)S(p) = {\bf 1}.
\end{align}
Introducing $(G_{u,d})^{-1} \equiv (G_{u,d}^0)^{-1}-\Xi G_{d,u}^0 \Xi$, the propagator $S(p)$ can be written as follows.
\begin{align}
	S(p) = \left(\begin{matrix}G_u & -G_u^0 \Xi G_d\\ -G_d^0 \Xi G_u& G_d\end{matrix}\right).
\end{align}
What we are interested in is the diagonal part of $S(p)$ which corresponds to $\langle u\overline{u}\rangle$ and $\langle d\overline{d}\rangle$.
Let us see the detail of $G_{u,d}$.
Its definition is 
\begin{align}
	(G_{u,d})^{-1} &\equiv (G_{u,d}^0)^{-1}-\Xi G_{d,u}^0 \Xi \notag\\
		&= - i \left(\slashed{p}+\mu_{u,d}\gamma_0 - M_q \right) -\gamma_5\Delta { i \over\slashed{p} + \mu_{d,u} \gamma_0 - M_q }\gamma_5\Delta \,. \notag\\
\end{align}
To calculate the inverse we rewrite this formula using the projection operators.
Performing some calculations we can find
\begin{align}
	\gamma_5\gamma_0 \Lambda_{p,a}\gamma_5 = -\Lambda_{a,p}\gamma_0.
\end{align}
From the above, we obtain
\begin{align}
\hspace{-1.0cm}
(G_{u,d})^{-1}
	&= - i \bigg[
		{\, p_0^2 -(E_D-\mu_{u,d})^2-\Delta^2 \over p_0-\mu_{u,d}+E_D}\Lambda_p\gamma_0 %
		+\, {\, p_0^2-(E_D+\mu_{u,d})^2-\Delta^2\over p_0-\mu_{u,d}-E_D}\Lambda_a\gamma_0 
	\bigg].
\end{align}
Now we could separate the diagonal elements $(G_{u,d})^{-1}$ by projection operators,
and each part is not mixed by the inverse operation.

Introducing the excitation energy
\begin{align}
	\xi_{p,a}^f(\vec p) = \sqrt{(E_D\mp\mu_f)^2+\Delta^2},
\end{align}
we obtain the propagator
\begin{align}
G_f &= i \bigg[{|u_{p}^f(\vec p)|^2\over \, p_0-\xi_{p}^f(\vec p)}+{|v_{p}^f(\vec p)|^2\over \, p_0+\xi_{p}^f(\vec p)} \bigg]\gamma_0 \Lambda_p + i \bigg[{|u_{a}^f(\vec p)|^2\over \, p_0-\xi_{a}^f(\vec p)}+{|v_{a}^f(\vec p)|^2\over \, p_0+\xi_{a}^f(\vec p)} \bigg]\gamma_0 \Lambda_a \,.
\end{align}
Here we have used $\mu_{u,d} = -\mu_{d,u}$.
The residues are
\begin{align}
	|u_{p,a}^f(\vec p)|^2 &= {1\over 2}\left(1+{\pm E_D-\mu_f\over \xi_{p,a}^f}\right)\\
	|v_{p,a}^f(\vec p)|^2 &= {1\over 2}\left(1-{\pm E_D-\mu_f\over \xi_{p,a}^f}\right).
\end{align}
They correspond to the occupation probability and satisfy $|u_p|^2+|v_p|^2=|u_a|^2+|v_a|^2=1$ as expected.
In addition,
these probabilities become the step function when $\Delta$ vanish.
\begin{align}
	|u_{p,a}^f(\vec p)|^2 &= {1\over 2}\left(1 + \frac{\pm E_D-\mu_f}{|E_D \mp \mu_f|}\right) = \theta (\pm E_D - \mu_f)\\
	|v_{p,a}^f(\vec p)|^2 &= {1\over 2}\left(1 - \frac{\pm E_D-\mu_f}{|E_D \mp \mu_f|}\right) = \theta (\mp E_D + \mu_f).
\end{align}

\chapter{Renormalization}
\label{app:reno}

In this chapter
we perform the renormalization procedure of the effective potential.
The one-loop correction evolves the divergence
which can be cancelled by the finite number of counter terms.

The Lagrangian with isospin chemical potential is the following.
\begin{align}
	\calL_B[\vec\phi_B, \psi_B, \bar\psi_B] &= %
		\begin{multlined}[t]
			\frac{1}{2}\qty(\partial_\mu \sigma_B)^2 + \frac{1}{2}\qty(\partial_\mu {\pi_3}_B)^2 + \qty(\partial_\mu + 2i\mu_I\delta^0_{~\mu}) \pi_B^+ \qty(\partial^\mu - 2i\mu_I \delta^\mu_{~0}) \pi_B^- \\
			- \frac{1}{2}{m_0}_B^2 \vec\phi_B^2 - \frac{\lambda_B}{4!}(\vec\phi_B)^4 + h_B \sigma_B\\
			+ \bar\psi_B \qty(\rmi\slashed{\partial} - \gamma_0\tau_3\mu_I - g_B\qty(\sigma_B + i\gamma_5 \vec\tau \cdot \vec\pi_B)) \psi_B.
		\end{multlined}
\end{align}
Here we attached the symbol ``B'' representing the bare fields and parameters.

We take into account the Dirac sea energy presented in \eqref{eq:model-dirac-sea} as the leading order of the quark contribution.
\begin{align}
	V_q = - N_c \int_\vp \qty(E_u+E_d+E_{\ubar}+E_{\dbar}),
\end{align}
which raises the UV divergence.
The divergence part $V_q^{\rm div}$ can be extracted as the first and second term of the expansion series of $\mu_I$ as
\begin{align}
	V_q^{\rm div} = - 4N_c \int_\vp \qty(\sqrt{\vp^2 + M_q^2} + \frac{1}{2}\frac{\Delta^2\mu_I^2}{(\vp^2 + M_q^2)^{3/2}}).
\end{align}
To handle these divergence we use the dimensional regularization $d=3 \to 3 - 2\epsilon$,
\begin{align}
	\int_{\vp} = \qty(\frac{e^{\gamma_E}\Lambda^2}{4\pi})^{\epsilon}\frac{d^dp}{(2\pi)^d}
\end{align}
where $\Lambda$ is the renormalization scale introduced by $\msbar$ scheme
and $\gamma_E = 0.577...$ is the Euler-Mascheroni constant.
Each momentum integral have the explicit $\Lambda$ dependence,
but the final result of the effective potential is manifestly independent of $\Lambda$.
After calculation of $d$-dimensional integral we obtain
\begin{align}
	V_q^{\rm div} &= \frac{4N_c}{(4\pi)^2} \qty( {e^{\gamma_E}\Lambda^2 \over M_q^2+\Delta^2} )^\epsilon \qty[ (M_q^2+\Delta^2)^2\Gamma(-2+\epsilon)-2\mu_I^2\Delta^2\Gamma(\epsilon)] \notag\\
	&= - \frac{2N_c}{(4\pi)^2}\qty[\qty((M_q^2+\Delta^2)^2 - 4\mu_I^2\Delta^2)\qty(\frac{1}{\epsilon} + \ln\frac{\Lambda^2}{M_q^2+\Delta^2}) + \frac{3}{2}\qty(M_q^2+\Delta^2)^2 + \calO(\epsilon)].
\end{align}
where we used
\begin{align}
	\Gamma(\epsilon) &= \frac{1}{\epsilon} - \gamma_E + \calO(\epsilon), \\
	\Gamma(-2+\epsilon) &= \frac{1}{2}\qty[\frac{1}{\epsilon} - \gamma_E + \frac{3}{2} + \calO(\epsilon)].
\end{align}
This divergence should be cancelled by the finite number of counter terms.
Denoting the counter potential as $\delta V_q$,
the renormalized one-loop effective potential $V_{\rm 1-loop}$ is symbolically obtained as
\begin{align}
	V_{\rm 1-loop} &= V_0 + V_q + \delta V_q \notag\\
	&= V_0 + V_q^{R} + V_q^{\rm div} + \delta_\msbar V_q + \delta_{\rm fin} V_q
\end{align}
with subtracted Dirac sea energy $V_q^{R} = V_q - V_q^{\rm div}$.
The symbol $\delta_\msbar$ denotes the counter terms only contain the divergent $1/\epsilon$ term
which are determined by the $\overline{\rm MS}$ scheme,
and $\delta_{\rm fin}$ denotes the remaining finite part;
\begin{align}
	\delta = \delta_\msbar + \delta_{\rm fin}.
\end{align}

There are ambiguities in the choice of the counter terms.
One choice is presented in the Refs.\cite{Adhikari:2016eef,Adhikari:2017ydi,Adhikari:2018aa},
where the counter terms for the coupling constants does not couple to the wavefunction renormalization.
This choice of the counter terms manifestly preserves the original $\rmO(4)$ symmetry in the Lagrangian.
The counter terms for the coupling constants are
\begin{align}
	m_{0B}^2 &= Z_{m^2}m_0^2 = m_0^2 + \delta m_0^2, ~~~ \lambda_B = Z_\lambda \lambda = \lambda + \delta\lambda, \notag\\
	g_B^2 &= Z_{g^2} g^2 = g^2 + \delta g^2,
	\label{eq:reno-cts-andersen-01}
\end{align}
and for the wave functions are
\begin{align}
	\phi_B^2 = Z_\phi \phi^2 = (1 + \delta Z_\phi)\phi^2, ~~~ \psibar_B\psi_B = Z_\psi \psibar\psi = (1 + \delta Z_\psi) \psibar\psi.
	\label{eq:reno-cts-andersen-02}
\end{align}

Another choice couples the wavefunction renormalization and the counter terms,
presented as followings.
\begin{align}
	\sigma_B^2 = Z_\sigma\sigma^2 = (1 + \delta Z_\sigma)\sigma^2 &, ~~~ {m_0}_B^2 Z_\sigma = m_0^2 + \delta m_\sigma^2 \notag \\
	{\pi_i}_B^2 = Z_\pi \pi_i^2 = (1 + \delta Z_\pi)\pi_i^2 &, ~~~ {m_0}_B^2 Z_\pi = m_0^2 + \delta m_\pi^2 \notag \\
	\psibar_B\psi_B = Z_\psi \psibar\psi = (1 + \delta Z_\psi) \psibar\psi &, ~~~ h_B Z_\sigma^{1/2} = h + \delta h.
	\label{eq:reno-cts-kojo}
\end{align}
This choice breaks the $\rmO(4)$ symmetry
as the wavefunction of mesons are evaluated at the different mass.
In this case,
we need to rescale the wavefunction
to restore the symmetry of the final expression of the effective potential.

In the following,
we demonstrate the renormalization for these two choices of the counter terms
and obtain the same expression of the effective potential
of quark meson model with the Dirac sea energy.
For both choices,
all the counter terms are fixed in the vacuum with the renormalization condition
for the pole and the residue of the meson propagator
\begin{align}
	\rmi \Sigma_\phi(p^2 = m_\phi^2) = 0, ~~~ \left.\rmi\pdv{\Sigma_\phi}{p^2}\right|_{p^2=m_\phi^2} = \delta Z_\phi ~~~ (\phi = \sigma,~\pi).
	\label{eq:reno-renocondition-propagator}
\end{align}
The value of $h$ in the explicit breaking term $h\sigma$ is determined to make the chiral condensate $M_q$ takes the value $M_q^{\rm vac} = gf_\pi$ in the vacuum $\mu_I < m_\pi/2$;
\begin{align}
	\left.\pdv{V_{\rm 1-loop}}{M_q}\right|_{M_q = M_q^{\rm vac}} = 0.
	\label{eq:reno-renocondition-breaking}
\end{align}
The expression of counter terms determined in the vacuum cancel the divergence even in the medium.

\section{$\rmO(4)$ symmetry preserved procedure}
In the following
we consider the counter terms in \eqsref{eq:reno-cts-andersen-01}{eq:reno-cts-andersen-02}
and apply the renormalization conditions.
The Lagrangian is rewritten by the renormalized fields and parameters, 
decomposed into the renormalized part and the counter terms as
\begin{align}
	\calL_{\rm B}[\Phi_B] &= \calL[\vec \phi, \psi, \bar\psi] + \calL_{\rm c.t.}
\end{align}
where $\calL$ is the renormalized Lagrangian
which is the same as $\calL_{\rm B}$ omitted the subscript ``B''.
The counter Lagrangian is given by
\begin{align}
	\calL_{\rm c.t.} &= \frac{\delta Z_\sigma}{2}\qty(\partial_\mu \sigma)^2 + \frac{\delta Z_\pi}{2}\qty(\partial_\mu \pi_2)^2 + \delta Z_\pi \qty(\partial_\mu + 2i\mu_I\delta^0_\mu)\pi^+\qty(\partial^\mu - 2i\mu_I\delta^\mu_0)\pi^- \notag \\
	&\quad - \frac{1}{2}\qty(\delta m_0^2 \vec\phi^2 + m_0^2\delta Z_\sigma \sigma^2 + m_0^2 \delta Z_\pi \vec\pi^2) \notag\\
	&\quad - \frac{1}{24}\qty(\delta \lambda \vec\phi^4 + 2\lambda\delta Z_\sigma \sigma^4 + 2\lambda\qty(\delta Z_\sigma + \delta Z_\pi)\sigma^2\vec\pi^2 + 2\lambda \delta Z_\pi\vec\pi^4) \notag\\
	&\quad - \delta Z_\psi \psibar\qty(i\slashed{\partial} - \gamma_0\tau_3\mu_I - g(\sigma + i\gamma_5\vec\tau \cdot \vec\pi))\psi \notag\\
	&\quad - \psibar \delta g\qty(\sigma + i\gamma_5\vec\tau\cdot \vec\pi)\psi \notag\\
	&\quad - \psibar g \qty(\frac{1}{2}\delta Z_\sigma \sigma + \frac{1}{2}\delta Z_\pi \vec\tau\cdot\vec\pi)\psi.
\end{align}

As these parameters are related to the physical parameters in vacuum as \eqref{eq:model-parameters-relations},
the counter terms can be written as
\begin{align}
	\delta m_0^2 &= - \frac{1}{2}\qty(\delta m_\sigma^2 - 3\delta m_\pi^2), ~~~ \delta\lambda = 3\frac{\delta m_\sigma^2 - \delta m_\pi^2}{f_\pi^2} - \lambda\frac{\delta f_\pi^2}{f_\pi^2}, \notag\\
	\delta g^2 &= \frac{\delta M_q^2}{f_\pi^2} - g^2 \frac{\delta f_\pi^2}{f_\pi^2}.
\end{align}
The mass counter terms $\delta m_\pi^2,~ \delta m_\sigma^2$ and wavefunction counter terms $\delta Z_\sigma,~ \delta Z_\pi$ are determined by the renormalization condition for the pole and the residue of the meson propagators.
The remaining counter terms $\delta g^2, \delta M_q^2$ and $\delta f_\pi^2$ are related to the other counter terms in the large $N_c$ limit.
The vacuum mass counter term $\delta M_q^2$ and quark wavefunction counter term $\delta Z_\psi$ vanish in this limit.
As the $f_\pi$ is defined by $f_\pi = \braket{\sigma}$ at the vacuum,
the $f_\pi$ counter term is related to $\delta Z_\sigma$ as $\delta f_\pi^2 = f_\pi^2\delta Z_\sigma$.
All together we obtain $\delta g^2 = - g^2 \delta f_\pi^2 / f_\pi^2 = - g^2 \delta Z_\sigma$.

% renormalization conditino
In the following we calculate the meson self-energy
and apply the renormalization conditions at vacuum.
As mentioned in Sec.$\,$\ref{sec:thermo-effective-potential},
we have to compute only the 1PI diagrams for the meson self-energy for renormalization.
This means that the tadpole diagrams does not contribute to the self-energy.
In addition
we evaluate the renormalization condition with {\it bare} field $\phi_B$ instead of renormalized ones
to preserve the $\rmO(4)$ symmetry.

In the end of the renormalization procedure,
we assume that the renormalized effective potential concludes the tree-level relation for the quark mass in the vacuum $M_q = gf_\pi$.
Within this assumption
we will attach the symbol ``vac'' to all the quark mass.

\begin{figure}[thpb]
	\centering
	\includegraphics[width=0.7\textwidth]{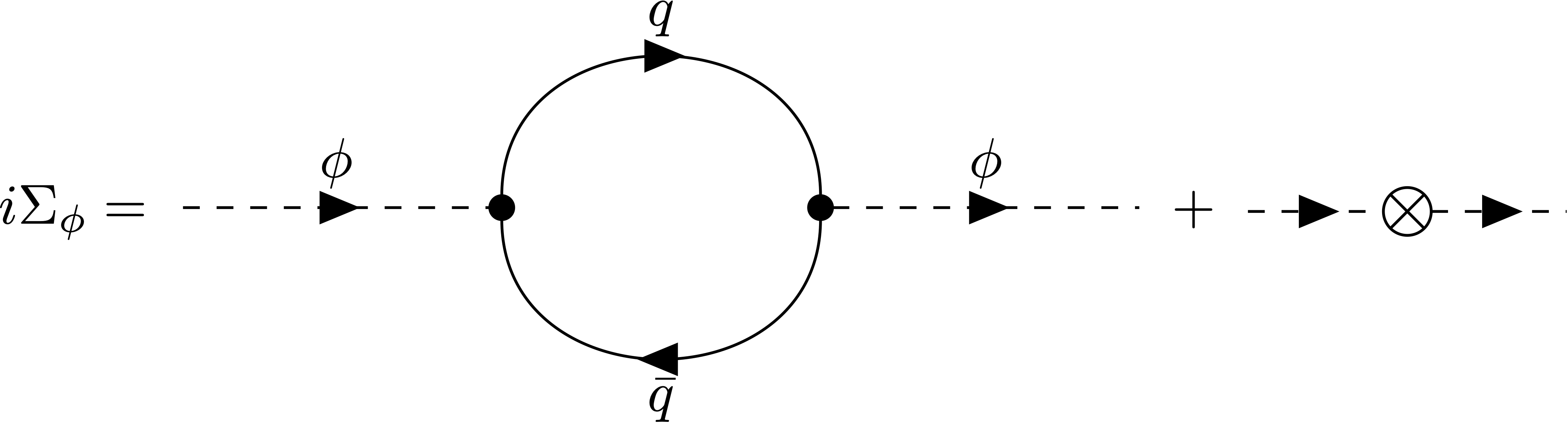}
	\caption{
		One-loop 1PI diagrams for the self-energy of the mesons.
		The vertex of the meson-quark interaction is $-ig$ for $\phi = \sigma$ and $g\gamma_5\tau_i$ for $\phi = \pi_i$.
		The second term is the counter term for the meson mass in the presence of the chiral condensate $\braket{\sigma}$.
	}
	\label{fig:remo-1pi}
\end{figure}

The 1PI one-loop diagrams are shown in \figref{fig:remo-1pi}
which are written as
\begin{align}
	\rmi\Sigma_{\sigma}(p^2) &= \rmi\bar\Sigma_{\sigma}(p^2) + \delta m_0^2 + \frac{1}{2}\qty(\delta\lambda + \lambda\delta Z_\sigma)\qty(\frac{M_q}{g})^2\\
	\rmi\Sigma_{\pi}(p^2) &= \rmi\bar\Sigma_{\pi}(p^2) + \delta m_0^2 + \frac{1}{6}\qty(\delta\lambda + \lambda\delta Z_\sigma)\qty(\frac{M_q}{g})^2 \\
	\rmi\bar\Sigma_{\sigma}(p^2) &= - 8\rmi g^2N_c\qty[A - \frac{1}{2}(p^2 - 4{M_q^{\rm vac}}^2)B(p^2)] \\
	\rmi\bar\Sigma_{\sigma}(p^2) &= - 8\rmi g^2 N_c\qty[A - \frac{1}{2}p^2 B(p^2)]
\end{align}
where the quarks for the loops have the tree-level mass $M_q^{\rm vac} = gf_\pi$.

Here $A$ and $B(p^2)$ are given by
\begin{align}
	A &= \int_{\vp} \frac{1}{p^2 - {M_q^{\rm vac}}^2} \notag\\
	&= \frac{\rmi {M_q^{\rm vac}}^2}{(4\pi)^2}\qty[\frac{1}{\epsilon} + 1 + \ln \frac{\Lambda^2}{{M_q^{\rm vac}}^2} + \calO(\epsilon)] \\
	B(p^2) &= \int_{\vk} \frac{1}{(k^2 - {M_q^{\rm vac}}^2)}\frac{1}{((k+p)^2 - {M_q^{\rm vac}}^2)} \notag\\
	&= \frac{\rmi}{(4\pi)^2}\qty[\frac{1}{\epsilon} + \int_0^1 dx \ln\frac{\Lambda^2}{x(1-x)p^2 - {M_q^{\rm vac}}^2} + \calO(\epsilon)] \notag\\
	&= \frac{\rmi}{(4\pi)^2}\qty[\frac{1}{\epsilon} + \ln\frac{\Lambda^2}{{M_q^{\rm vac}}^2} + F(p^2) + \calO(\epsilon)] \\
	B'(p^2) &= \frac{\rmi}{(4\pi)^2}F'(p^2)
\end{align}
using the function $F(p^2)$ as 
\begin{align}
	F(p^2) &= - \int_0^1 dx \ln\qty[\frac{p^2}{{M_q^{\rm vac}}^2}x(x-1) - 1] = 2 - 2r\arctan\qty(\frac{1}{r}) \\
	p^2 F'(p^2) &= \frac{r^2+1}{r}\atan\qty(\frac{1}{r}) - 1
\end{align}
where $r = \sqrt{4{M_q^{\rm vac}}^2/p^2 - 1}$.

The renormalization conditions for the meson propagator in \eqref{eq:reno-renocondition-propagator} are equivalent to
\begin{align}
	\delta m_0^2 &= \frac{\rmi\Sigma_\sigma - 3i \Sigma_\pi}{2}
	\label{eq:app-reno-mass}\\
	\delta \lambda &= - \frac{3}{f_\pi^2}\qty(\rmi\Sigma_\sigma - \rmi\Sigma_\pi) - \lambda \delta Z_\sigma,
	\label{eq:app-reno-lambda}
\end{align}
and
\begin{align}
	\rmi\left.\pdv{\Sigma_\sigma}{p^2}\right|_{p^2 = m_\sigma^2} - \delta Z_\sigma = 0 & : 4\rmi g^2N_c \qty(B(m_\sigma^2) + (m_\sigma^2 - 4{M_q^{\rm vac}}^2)B'(m_\sigma^2)) - \delta Z_\sigma = 0 
	\label{eq:app-reno-residue-sigma}\\
	\rmi\left.\pdv{\Sigma_\pi}{p^2}\right|_{p^2 = m_\pi^2} - \delta Z_\pi = 0 & : 4\rmi g^2N_c \qty(B(m_\pi^2) + m_\pi^2 B'(m_\pi^2)) - \delta Z_\pi = 0
	\label{eq:app-reno-residue-pi}.
\end{align}
Combining the results \eqref{eq:app-reno-mass}{eq:app-reno-residue-pi}
and the detailed expression of functions $A$ and $B$
we get
\begin{align}
	\delta m_0^2 &= 8\rmi g^2N_c\qty[A + \frac{1}{4}\qty(m_\sigma^2 - 4M_q^{\rm vac})B(m_\sigma^2) - \frac{3}{4}m_\pi^2 B(m_\pi^2)] \notag\\
	&= - \frac{1}{2}m_\sigma^2 ~ \frac{4{M_q^{\rm vac}}^2N_c}{(4\pi)^2f_\pi^2}\qty(\frac{1}{\epsilon} + \ln\frac{\Lambda^2}{{M_q^{\rm vac}}^2} + \frac{4{M_q^{\rm vac}}^2}{m_\sigma^2} + \qty(1 - \frac{4{M_q^{\rm vac}}^2}{m_\sigma^2})F(m_\sigma^2)) \notag\\
	&\quad + \frac{3}{2}m_\pi^2 ~ \frac{4{M_q^{\rm vac}}^2N_c}{(4\pi)^2f_\pi^2}\qty(\frac{1}{\epsilon} + \ln\frac{\Lambda^2}{{M_q^{\rm vac}}^2} + F(m_\pi^2)) \\
	\delta \lambda &= -\frac{12\rmi g^2N_c}{f_\pi^2}\qty[(m_\sigma^2 - {M_q^{\rm vac}}^2)B(m_\sigma^2) - m_\pi^2B(m_\pi^2)] - 4\rmi g^2N_c\lambda\qty[B(m_\pi^2) + m_\pi^2B'(m_\pi^2)] \notag\\
	&= \frac{3}{f_\pi^2}m_\sigma^2 ~ \frac{4{M_q^{\rm vac}}^2N_c}{(4\pi)^2f_\pi^2}\left\{\frac{1}{\epsilon} - \frac{4{M_q^{\rm vac}}^2}{m_\sigma^2}\ln \frac{\Lambda^2}{{M_q^{\rm vac}}^2} \right. \notag\\
	&\hspace{3cm} \left. + \qty(1 - \frac{4{M_q^{\rm vac}}^2}{m_\sigma^2})F(m_\sigma^2) + F(m_\sigma^2) + \qty(m_\sigma^2 - 4{M_q^{\rm vac}}^2)F'(m_\sigma^2)\right\} \notag\\
	&\quad - \frac{3}{f_\pi^2}m_\pi^2 ~ \frac{4{M_q^{\rm vac}}^2N_c}{(4\pi)^2f_\pi^2}\qty(\frac{1}{\epsilon} + 2\ln\frac{\Lambda^2}{{M_q^{\rm vac}}^2} + F(m_\pi^2) + F(m_\sigma^2) + \qty(m_\sigma^2 - 4 {M_q^{\rm vac}}^2)F'(m_\sigma^2)) \\
	\delta g^2 &= -4\rmi g^4N_c\qty[B(m_\sigma^2) + (m_\sigma^2 - {M_q^{\rm vac}}^2)B'(m_\sigma^2)] \notag\\
	&= \frac{{M_q^{\rm vac}}^2}{f_\pi^2} ~ \frac{4{M_q^{\rm vac}}^2N_c}{(4\pi)^2f_\pi^2}\qty(\frac{1}{\epsilon} + \ln \frac{\Lambda^2}{{M_q^{\rm vac}}^2} + F(m_\sigma^2) + (m_\sigma^2 - 4{M_q^{\rm vac}}^2)F'(m_\sigma^2))
\end{align}

The effective potential with the renormalized parameters can be written as
\begin{align}
	V_{1-loop} &= \frac{m_0^2}{2g^2}(M_q^2 + \Delta^2) + \frac{1}{2g^2}(-4\mu_I^2\Delta^2) + \frac{\lambda}{24g^4}(M_q^2 + \Delta^2)^2 + V_q \notag\\
	&= \frac{m_{0B}^2}{2g_B^2}\qty(1 + \frac{\delta m_0^2}{m_{0B}^2} - \frac{\delta g^2}{g_B^2}) \
	+ \frac{1}{2g_B^2}\qty(1 - \frac{\delta g^2}{g_B^2})(-4\mu_I^2\Delta^2) \
	+ \frac{\lambda_B}{24g_B^4}\qty(1 + \frac{\delta \lambda}{\lambda_B} - 2 \frac{\delta g^2}{g_B^2})(M_q^2 + \Delta^2)^2 \notag\\
	&\quad + V_q^R + V_q^{\rm div}. 
\end{align}
The divergence terms $1/\epsilon$ in counter terms should be cancelled with the divergence in $V_q^{\rm div
}$
and we obtain the effective potential at a vacuum as
\begin{align}
	V_{\rm 1-loop} &= - \frac{1}{4}m_\sigma^2f_\pi^2 \qty[1 + \frac{4{M_q^{\rm vac}}^2N_c}{(4\pi)^2f_\pi^2}\qty{\qty( - \frac{4{M_q^{\rm vac}}^2}{m_\sigma^2})F(m_\sigma^2) + \frac{4{M_q^{\rm vac}}^2}{m_\sigma^2} - (m_\sigma^2 - 4{M_q^{\rm vac}}^2)F'(m_\sigma^2)}]\frac{M_q^2}{{M_q^{\rm vac}}^2} \notag\\
	&\quad + \frac{3}{4}m_\pi^2f_\pi^2 \qty[1 - \frac{4{M_q^{\rm vac}}^2N_c}{(4\pi)^2f_\pi^2}\bigg\{- F(m_\pi^2) + F(m_\sigma^2) + (m_\sigma^2 - 4{M_q^{\rm vac}}^2)F'(m_\sigma^2)\bigg\}]\frac{M_q^2}{{M_q^{\rm vac}}^2} \notag\\
	&\quad + \frac{1}{8}m_\sigma^2f_\pi^2\left[1 - \frac{4{M_q^{\rm vac}}^2N_c}{(4\pi)^2f_\pi^2}\left\{\frac{4{M_q^{\rm vac}}^2}{m_\sigma^2}\qty(\ln\frac{M_q^2}{{M_q^{\rm vac}}^2} - \frac{3}{2}) \right.\right. \notag\\
	&\hspace{3cm} \left.\left. + \frac{4{M_q^{\rm vac}}^2}{m_\sigma^2}F(m_\sigma^2) + (m_\sigma^2 -4 {M_q^{\rm vac}}^2)F'(m_\sigma^2)\right\}\right] \frac{M_q^4}{{M_q^{\rm vac}}^4} \notag\\
	&\quad - \frac{1}{8}m_\pi^2f_\pi^2\qty[1 - \frac{4{M_q^{\rm vac}}^2N_c}{(4\pi)^2f_\pi^2}\bigg\{- F(m_\pi^2) + F(m_\sigma^2) + (m_\sigma^2 - 4{M_q^{\rm vac}}^2)F'(m_\sigma^2)\bigg\}] \frac{M_q^4}{{M_q^{\rm vac}}^4}.
\end{align}
Adding the explicit breaking term $-\frac{h}{g}M_q$ in this potential
and solving the gap equation in \eqref{eq:reno-renocondition-breaking},
we obtain
\begin{align}
	\frac{h}{g} &= m_\pi^2f_\pi^2\qty[1 - \frac{4{M_q^{\rm vac}}^2N_c}{(4\pi)^2f_\pi^2}\qty(-F(m_\pi^2) + F(m_\sigma^2) + (m_\sigma^2 - 4{M_q^{\rm vac}}^2)F'(m_\sigma^2))]\frac{1}{M_q^{\rm vac}}.
\end{align}

As all the counter terms are fixed in a vacuum,
the finial task is to apply these results to the effective potential in the medium.
Using the counter terms
we obtain the following.
\begin{align}
	V_{\rm 1-loop} &= - \frac{1}{4}m_\sigma^2f_\pi^2 \qty[1 + \frac{4{M_q^{\rm vac}}^2N_c}{(4\pi)^2f_\pi^2}\qty{\qty( - \frac{4{M_q^{\rm vac}}^2}{m_\sigma^2})F(m_\sigma^2) + \frac{4{M_q^{\rm vac}}^2}{m_\sigma^2} - (m_\sigma^2 - 4{M_q^{\rm vac}}^2)F'(m_\sigma^2)}]\frac{M_q^2+\Delta^2}{{M_q^{\rm vac}}^2} \notag\\
	&\quad + \frac{3}{4}m_\pi^2f_\pi^2 \qty[1 - \frac{4{M_q^{\rm vac}}^2N_c}{(4\pi)^2f_\pi^2}\bigg\{ - F(m_\pi^2) + F(m_\sigma^2) + (m_\sigma^2 - 4{M_q^{\rm vac}}^2)F'(m_\sigma^2)\bigg\}]\frac{M_q^2+\Delta^2}{{M_q^{\rm vac}}^2} \notag\\
	&\quad - 2\mu_I^2 f_\pi^2\qty[1 - \frac{4{M_q^{\rm vac}}^2N_c}{(4\pi)^2f_\pi^2}\qty(\ln\frac{M_q^2 + \Delta^2}{{M_q^{\rm vac}}^2} + F(m_\sigma^2) + (m_\sigma^2 - 4{M_q^{\rm vac}}^2)F'(m_\sigma^2))]\frac{\Delta^2}{{M_q^{\rm vac}}^2} \notag\\
	&\quad + \frac{1}{8}m_\sigma^2f_\pi^2 \left[1 - \frac{4{M_q^{\rm vac}}^2N_c}{(4\pi)^2f_\pi^2}\left\{\frac{4{M_q^{\rm vac}}^2}{m_\sigma^2}\qty(\ln\frac{M_q^2 + \Delta^2}{{M_q^{\rm vac}}^2} - \frac{3}{2}) \right.\right. \notag\\
	&\hspace{3cm} \left.\left. + \frac{4{M_q^{\rm vac}}^2}{m_\sigma^2}F(m_\sigma^2) + (m_\sigma^2 -4 {M_q^{\rm vac}}^2)F'(m_\sigma^2))\right\}\right] \frac{(M_q^2+\Delta^2)^2}{{M_q^{\rm vac}}^4} \notag\\
	&\quad - \frac{1}{8}m_\pi^2f_\pi^2\qty[1 - \frac{4{M_q^{\rm vac}}^2N_c}{(4\pi)^2f_\pi^2}\bigg\{- F(m_\pi^2) + F(m_\sigma^2) + (m_\sigma^2 - 4{M_q^{\rm vac}}^2)F'(m_\sigma^2)\bigg\}] \frac{(M_q^2+\Delta^2)^2}{{M_q^{\rm vac}}^4} \notag\\
	&\quad - m_\pi^2f_\pi^2\qty[1 - \frac{4{M_q^{\rm vac}}^2N_c}{(4\pi)^2f_\pi^2}\bigg\{-F(m_\pi^2) + F(m_\sigma^2) + (m_\sigma^2 - 4{M_q^{\rm vac}}^2)F'(m_\sigma^2)\bigg\}]\frac{M_q}{M_q^{\rm vac}} \notag\\
	&\quad -2N_c\int_p \left[\sqrt{\left(\sqrt{p^2+M_q^2}+\mu\right)^2+\Delta^2}+\sqrt{\left(\sqrt{p^2+M_q^2}-\mu\right)^2+\Delta^2}\right] \notag\\
	&\quad +4N_c\int_p\left[\sqrt{p^2+M_q^2+\Delta^2}+\frac{\mu^2\Delta^2}{2(p^2+M_q^2+\Delta^2)^{3/2}}\right] \,.
	\label{eq:reno-1loop-final}
\end{align}

\section{$\rmO(4)$ symmetry broken procedure}
Even in the case of the counter terms in \eqref{eq:reno-cts-kojo},
the renormalization procedure is the same as previous one.
The Lagrangian is decomposed into the two part 
and the counter Lagrangian is given by
\begin{align}
	\calL_{\rm c.t.} &= \begin{multlined}[t]
		\frac{\delta Z_\sigma}{2} (\partial_\mu\sigma)^2+  \frac{\delta Z_\pi}{2} (\partial_\mu\pi_3)^2 
		+ \delta Z_\pi  (\partial_\mu+2 \rmi \mu_I\delta_\mu^0)\pi^+\left(\partial^\mu-2 \rmi \mu_I\delta^\mu_0\right)\pi^- \\
		- \frac{1}{2} \delta m_{0\sigma}^2 \sigma^2 - \frac{1}{2} \delta m_{0\pi}^2 \vec\pi^2 + \delta h\sigma  %
		 - \frac{1}{24} \big[ \delta \lambda_{4\sigma} (\sigma^2 )^2 + \delta \lambda_{4\pi} ( \vec \pi^2 )^2 %
			+ 2 \delta \lambda_{2\sigma \pi} \sigma^2  \vec \pi^2 \big] \\
		+ \overline{\psi} \delta Z_\psi \big( \rmi \slashed{\partial} + \mu_I \tau_3 \gamma^0 \big) \psi %
		- \overline{\psi}\big(  \delta g_\sigma \sigma + \rmi  \delta g_\pi \gamma^5\vec\tau\cdot\vec\pi \big) \psi 
	\end{multlined}
\end{align}

The difference is the expression of the self-energy;
\begin{align}
	\rmi\Sigma_{\sigma}(p^2) &= \rmi\bar\Sigma_{\sigma}(p^2) + \delta m_{0\sigma}
	^2  + \frac{\delta\lambda_{4\sigma}}{2}\qty(\frac{M_q}{g})^2 - \delta Z_\sigma m_\sigma^2\\
	\rmi\Sigma_{\pi}(p^2) &= \rmi\bar\Sigma_{\pi}(p^2) + \delta m_{0\pi}^2 + \frac{\delta\lambda_{2\sigma\pi}}{6}\qty(\frac{M_q}{g})^2 - \delta Z_\pi m_\pi^2.
\end{align}
The expression of $\rmi\bar\Sigma_\sigma$ and $\rmi\bar\Sigma_\pi$ are determined from \figref{fig:remo-1pi} which are the same.

Using $M_q^{\rm vac} = gf_\pi$ and the relation of counter terms in the beginning of this chapter,
the mass counter terms and $\lambda$ counter terms are determined as follows.
\begin{align}
	\delta m_\sigma^2 &= \frac{\rmi \bar\Sigma_\sigma - 3\rmi \bar\Sigma_\pi}{2} - \delta Z_\sigma \frac{m_\sigma^2 -3m_\pi^2}{2} \notag\\
	&= - \frac{2g^2N_c}{(4\pi)^2}\qty[4{M_q^{\rm vac}}^2 + (3m_\pi^2 - 4{M_q^{\rm vac}}^2)F(m_\sigma^2) - 3m_\pi^2 F(m_\pi^2) - (m_\sigma^2 - 4{M_q^{\rm vac}}^2)(m_\sigma^2 - 3m_\pi^2)F'(m_\sigma^2)] \\
	\delta m_\pi^2 &= \frac{\rmi \bar\Sigma_\sigma - 3\rmi \bar\Sigma_\pi}{2} - \delta Z_\pi \frac{m_\sigma^2 -3m_\pi^2}{2} \notag\\
	&= - \frac{2g^2N_c}{(4\pi)^2}\qty[4{M_q^{\rm vac}}^2 + (m_\sigma^2 - 4{M_q^{\rm vac}}^2)F(m_\sigma^2) - m_\sigma^2 F(m_\pi^2) - (m_\sigma^2 - 3m_\pi^2)m_\pi^2F'(m_\pi^2)] \\
	\delta \lambda_{4\sigma} &= -\frac{3}{f_\pi^2}\qty[\rmi \bar\Sigma_\sigma - \rmi \bar\Sigma_\pi - (m_\sigma^2 - m_\pi^2)\delta Z_\sigma] \notag\\
	&= - \frac{24g^4 N_cN_f}{(4\pi)^2} \qty[\frac{1}{\epsilon} + \ln\frac{\Lambda^2}{{M_q^{\rm vac}}^2}]\notag\\
	& \quad + \frac{12g^2N_c}{f_\pi^2 (4\pi)^2}\qty[(m_\pi^2 - 4{M_q^{\rm vac}}^2)F(m_\sigma^2) - (m_\sigma^2 - 4{M_q^{\rm vac}}^2)(m_\sigma^2 - m_\pi^2)F'(m_\sigma^2) - m_\pi^2F(m_\pi^2) ] \\
	\delta \lambda_{2\sigma\pi} &= -\frac{3}{f_\pi^2}\qty[\rmi \bar\Sigma_\sigma - \rmi \bar\Sigma_\pi - (m_\sigma^2 - m_\pi^2)\delta Z_\pi] \notag\\
	&= - \frac{24g^4 N_cN_f}{(4\pi)^2} \qty[\frac{1}{\epsilon} + \ln\frac{\Lambda^2}{{M_q^{\rm vac}}^2}]\notag\\
	&\quad + \frac{12g^2N_c}{f_\pi^2(4\pi)^2}\qty[(m_\sigma^2 - 4{M_q^{\rm vac}}^2)F(m_\sigma^2) - m_\sigma^2 F(m_\pi^2) - (m_\sigma^2 - m_\pi^2)m_\pi^2 F'(m_\pi^2)] \\
	\delta \lambda_{4\pi} &= - \frac{3}{f_\pi^2}\qty[\rmi \bar\Sigma_\sigma - \rmi \bar\Sigma_\pi + (m_\sigma^2 - m_\pi^2)\qty(\delta Z_\sigma - 2\delta Z_\pi)] \notag\\
	&= - \frac{24g^4 N_cN_f}{(4\pi)^2} \qty[\frac{1}{\epsilon} + \ln\frac{\Lambda^2}{{M_q^{\rm vac}}^2}]\notag\\
	&\quad + \frac{12g^2N_c}{f_\pi^2(4\pi)^2}\left[(2m_\sigma^2 - m_\pi^2 - 4{M_q^{\rm vac}}^2)F(m_\sigma^2) + (m_\sigma^2 - 4{M_q^{\rm vac}}^2)(m_\sigma^2 - m_\pi^2)F'(m_\sigma^2) \right.\notag\\
	&\quad\quad \left. - (2m_\sigma^2 - m_\pi^2)F(m_\pi^2) - 2(m_\sigma^2 - m_\pi^2)m_\pi^2 F'(m_\pi^2)\right].
\end{align}

The finite part of the counter terms in the effective potential is written as
\begin{align}
	\delta_{\rm fin} V_q &= \frac{\delta_{\rm fin} Z_\pi}{2g^2}(-4\mu_I^2 \Delta^2) + \frac{1}{2g^2}\qty(\delta_{\rm fin} m_\sigma^2 M_q^2 + \delta_{\rm fin} m_\pi^2 \Delta^2) \notag\\
	& \quad + \frac{1}{24g^4}\qty(\delta_{\rm fin} \lambda_{4\sigma}M_q^4 + \delta_{\rm fin} \lambda_{4\pi}\Delta^2 + 2\delta_{\rm fin}\lambda_{2\sigma\pi}M_q^2\Delta^2) - \frac{\delta_{\rm fin} h}{g}M_q.	
\end{align}
In this expression the effective potential does not have original $\rmO(4)$ symmetry even when $h = 0$.
To restore the symmetry
we rescale the field condensates as
\begin{align}
	M_q \to (1 + \delta X)M_q, ~~~ \Delta \to (1 + \delta Y)\Delta
\end{align}
with order $\calO(\hbar)$ factor $\delta X$ and $\delta Y$.
The conditions for restoring the symmetry are
\begin{gather}
	\delta_{\rm fin} m_\sigma^2 + m_0^2 \delta X = \delta_{\rm fin} m_\pi^2 + m_0^2 \delta Y, \\
	\delta_{\rm fin} \lambda_{4\sigma} + 2\lambda\delta X = \delta_{\rm fin} \lambda + 2\lambda \delta Y = \delta_{\rm fin} \lambda + \lambda \delta X + \lambda \delta Y.
\end{gather}
This condition is valid as long as the following relations for the mass and $\lambda$ counter terms are satisfied;
\begin{align}
	\frac{\delta_{\rm fin}\lambda_{4\pi} - \delta_{\rm fin}\lambda_{4\sigma}}{\lambda} = -2\frac{\delta_{\rm fin} m_\sigma^2 - \delta_{\rm fin} m_\pi^2}{m_0^2} \\
	\frac{\delta_{\rm fin}\lambda_{2\sigma\pi} - \delta_{\rm fin}\lambda_{4\sigma}}{\lambda} = -\frac{\delta_{\rm fin} m_\sigma^2 - \delta_{\rm fin} m_\pi^2}{m_0^2}.
\end{align}
These conditions are sustained
and we can restore the original symmetry of the effective potential.

Choosing the scale factor $\delta X,~ \delta Y$ to reproduce the previous section,
we obtain
\begin{align}
	\delta X &= 0, \\
	\delta Y &= \frac{4g^2N_c}{(4\pi)^2}\qty[- F(m_\sigma^2) - (m_\sigma^2 - 4{M_q^{\rm vac}}^2)F'(m_\sigma^2) + F(m_\pi^2) + m_\pi^2F'(m_\pi^2)].
\end{align}
The first condition means that the $\sigma$ field is not rescaled.
Applying these scaling for the field condensates
and solve the gap equation at vacuum to determine the factor $h$ for explicit breaking term $ -\frac{h}{g}M_q$ in the effective potential,
we get the same expression of the effective potential in \eqref{eq:reno-1loop-final}.

\chapter{Boundary condition and Matsubara formalism}
\label{app:matsubara}

In the Chap.$\,$\ref{chap:model},
we have seen the imaginary time formulation of the quantum mechanics.
The finite temperature is introduced by the Euclideanization and the $\calS^1$ compactization of the imaginary time.
This compactization associates the boundary condition of the fields and discritization of the frequency $\omega$.

In the following we discuss the boundary condition of the fields and the Matsubara formalism.

\section{Bosonic boundary condition}
Consider the bosonic field $\phi(x)$.
We can move from $(\vx, \tau)$ space to $(\vp, \omega_n)$ space by the Fourier transformation as
\begin{align}
	\phi(\vx, \tau) = \sum_{n}\int_{\vp}\tilde\phi(\vp,\omega_n)e^{i(\vp\cdot\vx + \omega_n\tau)}.
\end{align}
This is true even for the fermionic field $\psi(x)$,
but in the case of bosons the frequency $\omega_n$ is discretized as $\omega_n = 2n\pi T$ for integer $n$.
This can be verified by examining the thermal Green's function defined by two-point function as
\begin{align}
	G(\vx_1,\vx_2;\tau_1,\tau_2) &= \braket{T_\tau\phi(\vx_1,\tau_1)\phi(\vx_2,\tau_2)}_{\beta} \\
	&= \frac{1}{Z}\Tr \qty{e^{-\beta H} T_\tau\qty[\phi(\vx_1,\tau_1)\phi(\vx_2,\tau_2)]}.
\end{align}

In the canonical formalism the filed is quantized by the equal-time commutation relation
\begin{align}
	\qty[\phi(t, \vx), \phi(t, \vy)] = i\delta^3(\vx-\vy).
\end{align}
and the imaginary-time-ordered correlation function is defined by 
\begin{align}
	T_\tau \qty[\phi(\vx_1,\tau_1)\phi(\vx_2,\tau_2)] = \phi(\tau_1)\phi(\tau_2)\theta(\tau_1-\tau_2) + \phi(\tau_2)\phi(\tau_1)\theta(\tau_2-\tau_1)
\end{align}
with the Heaviside step function $\theta(x)$.

Therefore we can find the Green's function is
\begin{align}
	G_B(\vx_1,\vx_2; \tau, 0) &= \frac{1}{Z}\Tr\qty[e^{-\beta H}\phi(\vx_1, \tau)\phi(\vx_2, 0)] \notag\\
	&= \frac{1}{Z} \Tr\qty[e^{-\beta H}e^{\beta H}\phi(\vx_2,0)e^{-\beta H}\phi(\vx_1, \tau)] \notag\\
	&= \frac{1}{Z} \Tr\qty[e^{-\beta H}\phi(\vx_2,\beta)\phi(\vx_1, \tau)] \notag\\
	&= \frac{1}{Z} \Tr\qty{e^{-\beta H} T_\tau\qty[\phi(\vx_1,\tau)\phi(\vx_2,\beta)]}\notag\\
	&= G_B(\vx_1,\vx_2;\beta,\tau).
\end{align}
This result shows that the field $\phi(x)$ is periodic $\phi(\vx,0) = \phi(\vx,\beta)$ and hence the frequency is discretized as $\omega_n = 2n\pi T$.

Next we compute the partition function and free energy for bosonic field with finite temperature.
For the free theory, the Euclidean action integral is
\begin{align}
	S_E= - \frac{1}{2}\int_0^\beta d\tau \int_{\vx,\vy} \phi(x) \qty(-\partial_\tau^2 - \partial_\vy^2 + m^2) \phi(y).
\end{align}
The partition function is defined by
\begin{align}
	Z &= \int\calD \calD\phi ~ e^{-S_E[\phi]} \notag\\
	&= \int\calD \tilde\phi(\vp,\omega_n) \exp\qty(-\frac{1}{2}\sum_n\int_\vp\tilde\phi (\omega_n^2+\omega^2)\tilde\phi)
\end{align}
where $\omega = \sqrt{\vp^2+m^2}$.
This Gaussian integral can be performed as
\begin{align}
	Z &= \Det\qty[\omega_n^2 + \omega^2]^{-1/2} \notag\\
	&= \sum_n \int_\vp \qty(\omega_n^2 + \omega^2)^{-1/2}.
\end{align}
The summation over discrete $n$ is called the Matsubara summation.

\begin{figure}[thpb]
	\centering
	\begin{minipage}{0.45\hsize}
		\centering	
		\includegraphics[width=0.95\linewidth]{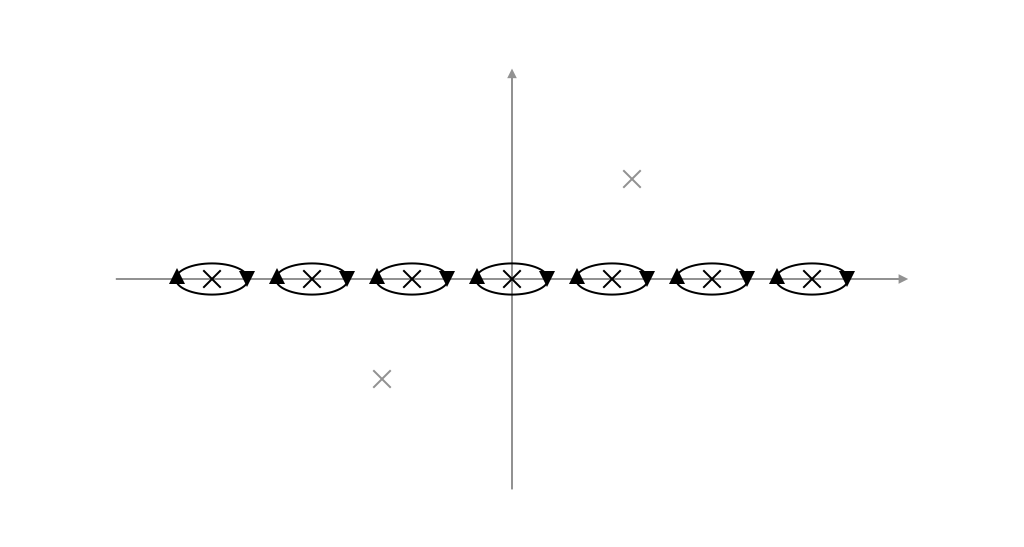}
		\subcaption{
			The original path of the integral.
			The summation for $n$ is obtained by the residue of the function $F(z)$.
		}
		\label{fig:matsubara-path1}
	\end{minipage}
	\begin{minipage}{0.45\hsize}
		\centering	
		\includegraphics[width=0.95\linewidth]{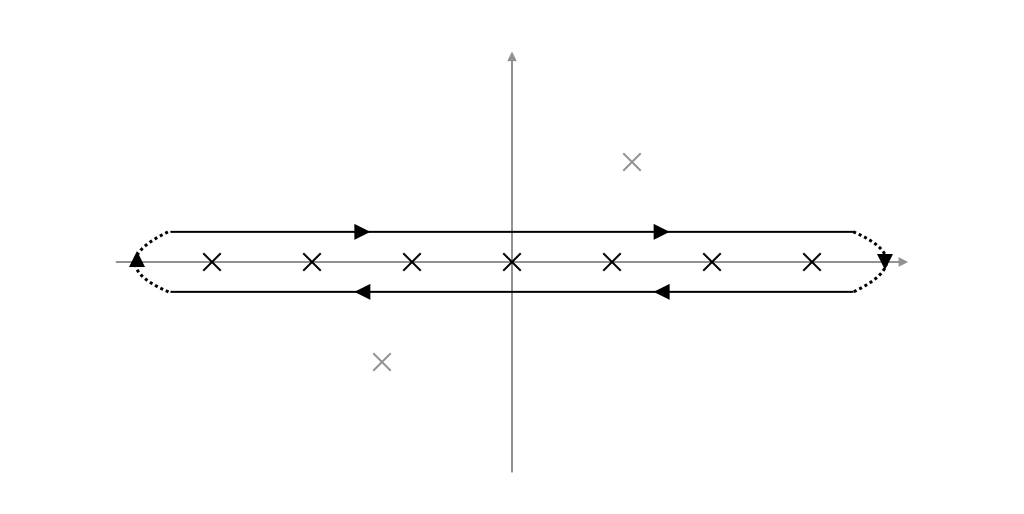}
		\subcaption{
			Deformed path of the integral.
			Each path for every $n$ is combined and formed a single closed path.
		}
		\label{fig:matsubara-path2}
	\end{minipage}
	\\
	\begin{minipage}{0.45\hsize}
		\centering	
		\includegraphics[width=0.95\linewidth]{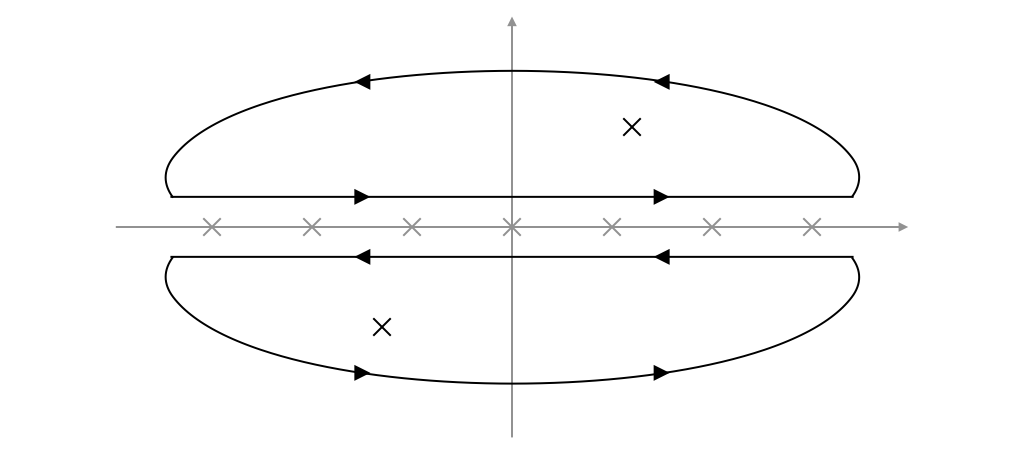}
		\subcaption{
			The final path of the integral.
			The imaginary residue of $F(z)$ is picked up.
		}
		\label{fig:matsubara-path3}
	\end{minipage}
	\caption{}
\end{figure}

Now we compute the free energy.
Using the relation
\begin{align}
	\ln \det A = \Tr \ln A,
\end{align}
we obtain
\begin{align}
	\ln Z = -\frac{1}{2}\sum_n \sum_\vp \ln \qty(\omega_n^2 + \omega^2).
\end{align}
Its calculation can be performed analytically 
but we calculate its differentiation by $\omega$ instead.
As $\omega_n = 2\pi nT$ for integer $n$, we obtain
\begin{align}
	\pdv{\omega} \ln Z = -\sum_n\sum_\vp \frac{\omega}{4\pi^2T^2 n^2+\omega^2}.
\end{align}
This summation for $n$ equals to the summation of residue of the function
\begin{align}
	F(z) = f(z)\frac{\pi\cos\pi z}{\sin\pi z},~~~ f(z)=\frac{\omega}{4\pi^2T^2 z^2 + \omega^2}.
\end{align}
This statement can be easily checked by naive calculation of the residue at $z=n$.
\begin{align}
	\underset{z=n}{\Res} F(z) dz &= \lim_{z\to n}\frac{\pi f(z)\cos\pi z}{\dv{z}\sin\pi z}\notag \\
	&=\lim_{z\to n} \frac{\pi f(z)\cos\pi z}{\pi\cos\pi z} = f(n).
\end{align}
Therefore we obtain $\sum_n$ as the summation of the integral over the small circle around $z = n$ as shown in \figref{fig:matsubara-path1}.
This path can be modified as long as the closed path of integral does not contain another pole in the function $f(z)$.
Thus we can deform the path as \figref{fig:matsubara-path2} which is equivalent to the integral in the \figref{fig:matsubara-path3}.

The final path in \figref{fig:matsubara-path3} picks up the imaginary residue of $F(z)$, which is
\begin{align}
	z = \pm \frac{2\pi iT}{\omega} = \pm\frac{i\omega_0}{\omega} = z_\pm.
\end{align}
Hence we obtain
\begin{align}
	\sum_n f(z) &= \int_C \frac{dz}{2\pi i} f(z) \frac{\pi\cos\pi z}{\sin\pi z} \notag\\
	&= -\underset{z=z_+}{\Res} f(z)\frac{\pi\cos\pi z}{\sin\pi z} %
	- \underset{z=z_-}{\Res} f(z)\frac{\pi\cos\pi z}{\sin\pi z} \notag\\
	&= 2\times \frac{-\pi}{2i\omega_0}\frac{\cos(\pi i\omega/\omega_0)}{\sin(\pi i\omega/\omega_0)} \notag\\
	&= \frac{\pi}{\omega_0}\frac{\cosh(\pi\omega/\omega_0)}{\sinh(\pi\omega/\omega_0)}.
\end{align}
This was obtained by differentiating $\ln Z$ by $\omega$,
we can integrate it back to obtain
\begin{align}
	\ln Z &= -\int^\omega d\omega' \frac{\pi}{\omega_0}\frac{\cosh(\pi\omega'/\omega_0)}{\sinh(\pi\omega'/\omega_0)} \notag\\
	&= - \int_\vp \ln \sinh(\pi \omega/\omega_0) + C \notag\\
	&= - \int_\vp \qty[\frac{1}{2}\beta\omega + \ln \qty(1+e^{-\beta\omega})] + C .
\end{align}
Here the factor $\ln 2$ is absorbed into the constant $C$.
The thermodynamic potential is obtained by dividing $\ln Z$ by $\beta$
\begin{align}
	\Omega = -\frac{1}{\beta} \ln Z = \int_\vp\qty[\frac{1}{2}\omega + T\ln\qty(1 - e^{-\beta\omega})] + C.
\end{align}
The first term corresponds to the zero-point energy,
and the second term is the bosonic thermal correction
with the Bose-Einstein distribution function 
\begin{align}
	n_B(\omega) = \frac{1}{e^{\beta\omega}-1}.
\end{align}

\section{Fermionic boundary condition}
\label{ssub:fermionic_boundary_condition}
Next we consider the fermionic field $\psi(x)$.
The only difference with bosonic case is the anti-periodic boundary condition
which comes from the anti-commutation relation.
\begin{align}
	\qty{\psi(t,\vx),\psi(t,\vy)} = i\delta^3(\vx-\vy).
\end{align}
This affects to the imaginary-time-ordered correlation function as
\begin{align}
	&T_\tau \qty[\psi(\vx_1,\tau_1)\psi(\vx_2,\tau_2)] = \psi(\tau_1)\psi(\tau_2)\theta(\tau_1-\tau_2) - \psi(\tau_2)\psi(\tau_1)\theta(\tau_2-\tau_1) \\
	&G_F(\vx_1,\vx_2; \tau, 0) = -G_F(\vx_1,\vx_2;\beta,\tau)
\end{align}
which indicates the anti-periodic boundary condition $\psi(\vx,0) = -\psi(\vx,\beta)$
and discritization of the frequency $\omega_n = (2n+1)\pi T$.

The Lagrangian of the free theory for Fermion is
\begin{align}
	\calL = \psibar (i\slashed{\partial}-m)\psi
\end{align}
or written in explicitly
\begin{align}
	\calL = \psibar \qty(i\gamma^0\pdv{t} + i \bm{\gamma}\cdot\bm{\nabla} - m)\psi.
\end{align}

The partition function is defined via Hamiltonian,
thus we calculate the conjugate field of $\psi(x)$. 
That is
\begin{align}
	\Pi  = \pdv{\calL}{\dot\psi} = i\psibar \gamma^0
\end{align}
and Hamiltonian is
\begin{align}
	\calH = \Pi \dot\psi - \calL = \psibar \qty(-i\bm{\gamma}\cdot\bm{\nabla} + m)\psi.
\end{align}
We can determine $\psibar = \psi^\dagger\gamma^0$ once field $\psi$ is determined,
however,
we need to treat $\Pi=i\psibar \gamma^0$ and $\psi$ as independent fields.
Noticing this fact we can get the partition function as the path integral formalism of $\psibar$ and $\psi$,
\begin{align}
	Z = \int\calD\psibar\calD\psi ~ \exp\qty[-\int_0^\beta d\tau\int_{\vx,\vy} \psibar(x)\qty(\gamma^0\pdv{\tau} - i\bm{\gamma}\cdot\bm{\nabla}_\vy + m)\psi(y)]
\end{align}

Tracing our the case of bosonic field,
we first introduce the Fourier transformation of the field $\psi$
\begin{align}
	\psi(\vx,\tau) = \sum_n\int_\vp \tilde\psi(\vp,\omega_n)e^{i(\vp\cdot\vx+\omega_n\tau)}
\end{align}
and rewrite the partition function in terms of the path integral for $\tilde\psibar,~\tilde\psi$ as
\begin{align}
	Z = \int\calD\tilde\psibar(\vp,\omega_n)\calD\tilde\psi(\vp,\omega_n) ~ \exp\qty[-\sum_n\int_\vp \tilde\psibar\qty(i\gamma^0\omega_n + \bm{\gamma}\cdot\vp + m)\tilde\psi].
\end{align}
As the fermionic field $\psi$ is Grassmann variable
this path integral concludes
\begin{align}
	Z = \Det\qty[-i\gamma^0\omega_n - \bm{\gamma}\cdot\vp - m].
\end{align}
As we have done in the bosonic case,
the logarithm of partition function is obtained by
\begin{align}
	\ln Z &= \Tr\ln \qty[-i\gamma^0\omega_n - \bm{\gamma}\cdot\vp - m] \notag\\
	&= \frac{1}{2}\Tr\qty{\ln\qty[-i\gamma^0\omega_n - \bm{\gamma}\cdot\vp - m] + \ln[-i\gamma^0\omega_n - \bm{\gamma}\cdot\vp - m]^\dagger} \notag\\
	&= \frac{1}{2} \sum_n \int_\vp \ln \qty(\omega_n^2 + \omega^2).
\end{align}
For the second equality we used the fact that the value is real.
The computation of the summation over $n$ is the same as the bosonic case,
and at first, we obtain the $\omega$ differentiation of $\ln Z$ as
\begin{align}
	\pdv{\omega}\ln Z &= \sum_n \int_\vp \frac{\omega}{\omega_n^2 + \omega^2}
\end{align}
and Matsubara summation of $n$ is obtained by the residue of the function as
\begin{align}
	\sum_n \frac{\omega}{\omega_n^2 + \omega^2} &= \sum_n \underset{z=n}{\Res} F(z)dz, ~~~ F(z) = \frac{\omega}{\omega_0^2(n+1/2)^2 + \omega^2}\frac{\pi\cos\pi z}{\sin\pi z}
\end{align}
for $\omega_0 = 2\pi T$.
Using the same argument for the residue and contour as the bosonic case,
we obtain
\begin{align}
	\sum_n \frac{\omega}{\omega_n^2 + \omega^2} &= - \underset{z=z_+}{\Res}F(z)dz - \underset{z=z_-}{\Res}F(z)dz \notag\\
	&= -\frac{2}{2i\omega_0}\frac{\pi\cos(\pi(i\omega/\omega_0 - 1/2))}{\sin(\pi(i\omega/\omega_0 - 1/2))} \notag\\
	&= -\frac{\pi}{\omega_0} \frac{\sinh \frac{1}{2}\beta\omega}{\cosh \frac{1}{2}\beta\omega}
\end{align}
with $z = z_\pm$ is the imaginary pole of $F(z)$,
\begin{align}
	z_\pm = \pm \frac{i\omega}{\omega_0} - \frac{1}{2}.
\end{align}
Finally we can get back to the partition function by integrating the result by $\omega$
\begin{align}
	\ln Z = - \int_\vp \qty[\frac{1}{2}\beta\omega + \ln\qty(1+e^{-\beta\omega})] + C,
\end{align}
and the free energy is obtained by
\begin{align}
	\Omega  = -\frac{1}{\beta}\ln Z = \int_\vp \qty[\frac{1}{2}\omega + T\ln\qty(1+e^{-\beta\omega})].
\end{align}
We can see that the second term indicates the Fermi-Dirac distribution function
\begin{align}
	n_F(\omega) = \frac{1}{e^{\beta\omega}+1}.
\end{align}

\chapter{Polyakov loop}
\label{app:polyakov}
In Chap.$\,$\ref{chap:model},
we introduced the Polyakov loop to describe the quark confinement.
In this section,
we give the more detailed explanation from the theoretical aspects. 

\section{Massive quark energy}
We mentioned that the Polyakov loop $\Phi$ is introduced as the imaginary chemical potential in the quark thermal potential
and the extra factor to the thermal factor of quarks. 
This is one of the rolls of the Polyakov loop as the agent to suppress the thermal quark excitation.
In addition,
we can show that the Polyakov loop is the exact order parameter of the confinement 
in the case that the quark mass if infinite. 

We consider the color-averaged free energy of the quark
which indicate the quark confinement and deconfinement.

To begin with, 
we introduce the creation and annihilation operators of the colored quarks
\begin{align}
	\psi_a^\dagger(\vx,\tau), ~~~ \psi_a(\vx,\tau)
\end{align}
and assume that they are static and immovable.
These operators creates/annihilates the quark with color index $a$ at position $x=(\vx,\tau)$.
The quantization is determined via equal-time anti-commutation relations
\begin{align}
	\qty{\psi_a(\vx,\tau),\psi_a^\dagger(\vy,\tau)} = \delta_{ab}\delta^{(3)}(\vx-\vy).
\end{align}
As they are time dependent operators in Heisenberg picture,
their time evolution is governed by the imaginary-time Dirac equation.
\begin{align}
	\qty(-i\pdv{\tau} + gA_0(\vx,\tau))\psi(\vx,\tau) = 0.
\end{align}
By integrating this equation we can get integral equation
\begin{align}
	\psi(\vx,\tau) = \calP\exp\qty(ig\int_0^\tau d\tau'A_0(\tau',\vx))\psi(0,\vx).
\end{align}

Now we consider the color-averaged free energy or quark potential.
\begin{align}
	e^{-\beta F_q(\vx)} = \frac{1}{N_c} \sum_{|s\rangle} \expv<s| e^{-\beta H} |s>
\end{align}
where $|s\rangle$ indicates the state with the heavy single quark.
Introducing the quark fields,
we can get
\begin{align}
	e^{-\beta F_q(\vx)} = \frac{1}{N_c} \sum_{|s'\rangle} \sum_a \expv<s'| \psi_a(0,\vx) e^{-\beta H} \psi_a^\dagger(0,\vx) |s'>
\end{align}
where the sum is over all the stetes $|s'\rangle$ with no heavy quarks.
Here we have set $\tau=0$ for convenience but choice of imaginary-time has nothing to do with this discussion.
The free energy of test quark can be calculated as follows.
\begin{align}
	e^{-\beta F_q(\vx)} &= \frac{1}{N_c} \sum_{|s'\rangle} \sum_a \expv<s'| e^{-\beta H} \psi_a(\beta,\vx)\psi_a^\dagger(0,\vx) |s'> \notag\\
	&= \frac{1}{N_c} \sum_{|s'\rangle} \sum_{a,b} \expv<s'| e^{-\beta H} \calP \exp\qty(ig\int_0^\beta d\tau A_0(\vx,\tau))_{ab}\psi_b(0,\vx)\psi_a^\dagger(0,\vx) |s'> \notag\\
	&= \frac{1}{N_c} \braket{\Tr\calP\exp\qty(ig\int_0^\beta d\tau A_0(\vx,\tau)) }_\beta \notag\\
	& =: \frac{1}{N_c}\braket{\Tr L(\vx)}.
\end{align}
Thus the expectation value of the Polyakov loop is related to the quark potential $F_q(\vx)$.
In case of $\braket{\Tr L(\vx)} = 0$
the free energy of the single quark must be infinite.
This means that the single quark is prohibited and must be confined.
On the other hand,
if $\braket{\Tr L(\vx)}$ has non-zero value,
the free energy of the single quark must be finite and the single quark can exist
which means the deconfinement.

The same calculation can be done for single test quark and anti-quark
and we can get
\begin{align}
	e^{-\beta F_{q\bar q}(\vx_1,\vx_2)} &= \frac{1}{N_c^2}\sum_{|s'\rangle} \sum_{a,a'} \expv<\psi_a(\vx_1,0),\psi_{a'}(\vx_2,0)|e^{-\beta H}|\psi_a(\vx_1,0),\psi_{a'}(\vx_2,0)> \notag\\
	&= \frac{1}{N_c^2}\sum_{|s'\rangle} \sum_{a,a'}\expv<s'|\psi_a(\vx_1,0)\psibar_{a'}(\vx_2,0)e^{-\beta H}\psi^\dagger_a(\vx_1,0)\psibar^\dagger_{a'}(\vx_2,0)|s'> \notag\\
	&= \frac{1}{N_c^2} \braket{L(\vx_1)L^\dagger(\vx_2)}.
\end{align}
This can be generalized to the assembly of $N_q$ quarks and $N_{\bar q}$ anti-quarks in a simple way as
\begin{align}
	\exp\qty(-\beta F_{N_q N_{\bar q}}) = \frac{1}{N_c^{N_q + N_{\bar q}}} \braket{L(\vx_1)\dots L(\vx_{N_q}) L^\dagger(\vx_1')\dots L^\dagger(\vx_{N_{\bar q}}')}
\end{align}

In the current discussion with the pinned quark
the Polyakov loop is the exact order parameter of the confinement,
however,
it is not the case for the finite mass quark.
Nevertheless,
the Polyakov loop is still the good measure for the confinement.

\section{Center symmetry}
To see how the value of Polyakov loop is determined from the aspect of the symmetry,
we should consider the gauge transformation on the space-time manifold $\calS^1 \times \mathbb{R}^3$.
The gauge transformation of the gauge field for the unitary operator $U(x)$ is
\begin{align}
	\calG: A_\mu(x) \mapsto A_\mu'(x) = U(x)\qty(A_\mu(x) + \frac{i}{g}\partial_\mu)U^\dagger(x).
\end{align}
We have to notice that the $A_\mu'(x)$ after transformation does not satisfy the boundary condition any longer,
or that the theory is no longer on the original manifold $\calS^1 \times \mathbb{R}^3$.

If we insist on the original condition or the manifold,
we should restrict the gauge field to the specific one
which satisfies
\begin{align}
	\calG \supset \calG_{\rm C.S.} : A_\mu(x) \mapsto A_\mu'(x) ~~~ \rm{s.t.} ~~~ A_\mu'(\beta,\vx) = A_\mu'(0,\vx).
\end{align}
This restriction for the gauge fields is called center symmetry, actually does not require the periodicity on the unitary matrix $U(x)$.

The condition of the periodicity for the transformed gauge fields $A_\mu'$ can be expressed in terms of the original $A_\mu$ as 
\begin{align}
	U(0,\vx)A_\mu(0,\vx)U^\dagger(0,\vx) = U(\beta,\vx)A_\mu(\beta,\vx)U^\dagger(\beta,\vx).
\end{align}
Because $A_\mu(\beta,\vx)=A_\mu(0,\vx)$ the $A_\mu(0,\vx)$ in the left hand side can be extracted as
\begin{align}
	A_\mu(0,\vx) &= U^\dagger(0,\vx)U(\beta,\vx)A_\mu(0,\vx)U^\dagger(\beta,\vx)U(0,\vx)\notag\\
	&= V A_\mu(0,\vx) V^\dagger
\end{align}
where $V=U^\dagger(0,\vx)U(\beta,\vx)$ is an unitary operator.
This operator transform $A_\mu(0,\vx)$ to itself,
thus $V$ must be proportional to identity.
Using $\det V = 1$ we can get
\begin{align}
	V = z_k = e^{2\pi ik/N_c}, ~~~ k = 0,1,...,N_c-1.
\end{align}
This means $U(x)$ must satisfy at least the twisted boundary condition
\begin{align}
	U(\beta,\vx) = z_k U(0,\vx), ~~~ z_k = e^{2\pi ik/N_c}
\end{align}
for the integer $k=0,1,...,N_c-1$.
$z_k$s are belong to the center group $\mathbb{Z}_{N_c}$ of the SU($N_c$) gauge group,
that is why this gauge transformation is called the {\it center symmetry}.

Let us see the effect of this symmetry on the Polyakov loop.
Because of the definition of it,
we can soon conclude that
\begin{align}
	\calG_{\rm C.S.} : L(x) \mapsto L'(x) = z_k L(x).
	\label{eq:center_symmetry_l3}
\end{align}
From this transformation rule,
the traced Polyakov loop expectation value
\begin{align}
	\Phi = \frac{1}{N_c}\braket{\Tr L}  
\end{align}
must be zero as long as the center symmetry is preserved.
As mentioned in the previous section,
this concludes that the energy of the single quark diverges;
\begin{align}
	\Phi = 0 ~~~ \Rightarrow ~~~ F_q \to \infty.
\end{align}
Therefore this phase corresponds to the confined phase.

In conclusion of the above discussion,
the center symmetric phase is the confined phase
and the Polyakov loop works as the order parameter of the confinement.
\begin{align}
	\begin{cases}
		\Phi = 0 & \text{confined phase}\\
		\Phi \neq 0 & \text{deconfined phase}\\
		|\Phi| \to 1 & \text{high temperature limit}
	\end{cases}
\end{align}

\printbibliography[
heading=bibintoc,
title={Bibliography}
]
%\bibliography{ref}

\end{document}

\chapter{Renormalization}
\label{app:renormalization}

In this chapter
we perform the renormalization procedure of the effective potential.
The one-loop correction evolves the divergence
which can be cancelled by the infinite number of counter terms.

The Lagrangian with isospin chemical potential is the following.
\begin{align}
	\calL_B[\vec\phi_B, \psi_B, \bar\psi_B] &= %
		\begin{multlined}[t]
			\frac{1}{2}\qty(\partial_\mu \sigma_B)^2 + \frac{1}{2}\qty(\partial_\mu {\pi_3}_B)^2 + \qty(\partial_\mu + 2i\mu_I\delta^0_{~\mu}) \pi_B^+ \qty(\partial^\mu - 2i\mu_I \delta^\mu_{~0}) \pi_B^- \\
			- \frac{1}{2}{m_0}_B^2 \vec\phi_B^2 - \frac{\lambda_B}{4!}(\vec\phi_B)^4 + h_B \sigma_B\\
			+ \bar\psi_B \qty(\rmi\slashed{\partial} - \gamma_0\tau_3\mu_I - g_B\qty(\sigma_B + i\gamma_5 \vec\tau \cdot \vec\pi_B)) \psi_B.
		\end{multlined}
\end{align}
Here we attached the symbol ``B'' representing the bare fields and parameters.
The renormalized fields and parameters are given by
\begin{align}
	\sigma_B = Z_\sigma^{1/2}\sigma &, ~~~ {m_0}_B^2 Z_\sigma = m_0^2 + \delta m_\sigma^2 \\
	{\pi_i}_B = Z_\pi^{1/2} {\pi_i} &, ~~~ {m_0}_B^2 Z_\pi = m_0^2 + \delta m_\pi^2 \\
	\psi_B = Z_\psi \psi &, ~~~ h_B Z_\sigma^{1/2} = h + \delta h
\end{align}
and 
\begin{align}
	\lambda_B Z_\sigma^2 = \lambda + \delta\lambda_{4\sigma} &, ~~~ \lambda_B Z_\pi^2 = \lambda + \delta\lambda_{4\pi} \\
	\lambda_B Z_\pi Z_\sigma = \lambda + \delta\lambda_{2\sigma\pi} &, \\
	g_B Z_\sigma^{1/2} Z_\psi = g + \delta g_\sigma &, ~~~ g_B Z_\pi^{1/2} Z_\psi = g + \delta g_\pi.
\end{align}
Because $\sigma$ and $\vec\pi$ are originally included symmetric way,
the counter terms are partly dependent each other.
We write them in parallel for clarifying the meanings.

Here we have the eleven counter terms and the four relations between them.
Therefore we need to prepare seven independent conditions.

Now the Lagrangian is written by the renormalized fields and parameters,
decomposed into the renormalized part and the counter terms as
\begin{align}
	\calL_{\rm B}[\Psi_B] &= \calL[\vec \phi, \psi, \bar\psi] + \calL_{\rm c.t.}
\end{align}
where $\calL$ is the renormalized Lagrangian
whose stracture is the same as $\calL_{\rm B}$ omitted the subscript ``B''.
Counter Lagrangian is given by
\begin{align}
	\calL_{\rm c.t.} [\Psi] &= \begin{multlined}[t]
		\frac{\delta Z_\sigma}{2} (\partial_\mu\sigma)^2+  \frac{\delta Z_\pi}{2} (\partial_\mu\pi_3)^2 
		+ \delta Z_\pi  (\partial_\mu+2 \rmi \mu_I\delta_\mu^0)\pi^+\left(\partial^\mu-2 \rmi \mu_I\delta^\mu_0\right)\pi^- \\
		- \frac{1}{2} \delta m_{0\sigma}^2 \sigma^2 - \frac{1}{2} \delta m_{0\pi}^2 \vec\pi^2 + \delta h\sigma  %
		 - \frac{1}{24} \big[ \delta \lambda_{4\sigma} (\sigma^2 )^2 + \delta \lambda_{4\pi} ( \vec \pi^2 )^2 %
			+ 2 \delta \lambda_{2\sigma \pi} \sigma^2  \vec \pi^2 \big] \\
		+ \overline{\psi} \delta Z_\psi \big( \rmi \slashed{\partial} + \mu_I \tau_3 \gamma^0 \big) \psi %
		- \overline{\psi}\big(  \delta g_\sigma \sigma + \rmi  \delta g_\pi \gamma^5\vec\tau\cdot\vec\pi \big) \psi 
	\end{multlined}
\end{align}

These counter terms are determined to cancel the divergence of the loop diagrams in the effective potential
under on-mass shell condition.
We calculate the one-loop correction to the effective potential
with large $N_c$ approximation, 
which we neglect the loop diagram of the mesons.

In case of the on-mass shell renormalization,
the wavefunction renormalization for the mesons $\delta Z_\sigma$ and $\delta Z_\pi$ will take the different value
because they are evaluated at the different mass.
This asymmetry breaks the original $\rmO(4)$ symmetry of the effective potential.
In the end of the procedure 
we rescale the fields to restore the symmetry.

In the following
we determine the counter terms in a vacuum; $\mu_I = 0$ and $M_q \neq 0,~ \Delta = 0$.
After that we use the relations of counter terms to determine the remaining counter terms and the renormalized effective potential in a medium. 

\section{Renormalization in a vacuum}
\label{sec:reno-vacuum}
We begin with the vacuum renormalization at $\mu_I = 0$.
Here only $\sigma$ has the non-zero expectation value $\braket{\sigma}$ and can be written as $\sigma = \tilde \sigma + \braket{\sigma}$.
The meson masses are given from the quadratic terms of the Lagrangian for $\tilde\sigma$ and $\vec \pi$ as
\begin{align}
	m_\sigma^2 = m_0^2 + \frac{\lambda}{2}\braket{\sigma}, ~~~ m_\pi^2 = m_0^2 + \frac{\lambda}{6}\braket{\sigma}.
\end{align}
This $\braket{\sigma}$ is acquired as the mass for quark part
\begin{align}
	\bar\psi(\rmi\slashed{\partial} - M_q + \mu_I\tau_3\gamma^0)\psi , ~~~ M_q = g\braket{\sigma}.
\end{align}
Below we rewrite $\braket{\sigma}$ in terms of $M_q$ and omit the symbol $\tilde \ast$ from $\tilde \sigma$ for simplicity.

With this shift of $\sigma$ field,
Lagrangian can be separated into three parts as
\begin{align}
	\calL = \calL_0 + \calL_2 + \calL_{\rm int}.
\end{align}
The first term is the classical potential part which corresponds to the zeroth order of the effective potential.
\begin{align}
	- \calL_0 = V_0 = \frac{m_0^2}{2g^2}M_q^2 + \frac{\lambda}{24g^4}M_q^4 - \frac{h}{g}M_q.
\end{align}
The quantum corrections are calculated from the kinetic terms
\begin{align}
	\calL_2 = \frac{1}{2}\qty[(\partial_\mu\sigma)^2 + (\partial_\mu\vec\pi)^2] - \frac{1}{2}m_\sigma^2\sigma^2 - \frac{1}{2}m_\pi^2 \vec\pi^2 + \bar\psi \qty(\rmi\slashed{\partial} - M_q)\psi
\end{align}
and the interaction terms
\begin{align}
	\calL_{\rm int} = -\frac{\lambda}{24}(\vec\phi^2)^2 - \frac{\lambda M_q}{6g}\sigma \vec\phi^2 - g\bar\psi\qty(\sigma + \rmi\gamma^5 \vec\tau\cdot\vec\pi)\psi.
	\label{eq:interactions}
\end{align}
The counter terms for effective potential is obtained as
\begin{align}
	\delta V = \frac{\delta m_{0\sigma}^2}{2g^2}M_q^2 + \frac{\delta \lambda_{4\sigma}}{24g^4}M_q^4 - \frac{\delta h}{g}M_q.
\end{align}

In our model the one-loop contribution to effective potential is the quark zero point energy.
This has the form of 
\begin{align}
	V_q = - 2N_fN_c \int_{\vp} E_D(\vp), ~~~ E_D(\vp) = \sqrt{\vp^2 + M_q^2}
\end{align}
which diverges in the ultraviolet region.
To handle these divergence we use the dimensional regularization $d=3 \to 3 - 2\epsilon$,
\begin{align}
	\int_{\vp} = \qty(\frac{e^{\gamma_E}\Lambda^2}{4\pi})^{\epsilon}\frac{d^dp}{(2\pi)^d}
\end{align}
where $\Lambda$ is the renormalization scale introduced by $\msbar$ scheme
and $\gamma_E = 0.577...$ is the Euler constant.
Each momentum integral have the explicit $\Lambda$ dependence,
but the final result of the effective potential is manifestly independent of $\Lambda$.
After calculation of $d$-dimensional integral we obtain
\begin{align}
	V_q &= \frac{2N_fN_c}{(4\pi)^2}\qty(\frac{e^{\gamma_E}\Lambda^2}{M_q^2})^\epsilon\Gamma(-2+\epsilon)M_q^4 \notag\\
	&= \frac{N_fN_c}{(4\pi)^2}\qty[\frac{1}{\epsilon} + \frac{3}{2} + \ln\frac{\Lambda^2}{M_q^2}]M_q^4 + \calO(\epsilon)
\end{align}
where we used
\begin{align}
	\Gamma(-2+\epsilon) = \frac{1}{2}\qty[\frac{1}{\epsilon} - \gamma_E + \frac{3}{2} + \calO(\epsilon)].
\end{align}
This expression shows that the effective potential only has the quadratic divergence for $M_q$, not $M_q^1, M_q^2$.
We can determine the divergent part of the counter terms by comparing the divergent part of $V_q$ with $\delta V$
and obtain
\begin{align}
	\delta_\msbar\lambda_{4\sigma} = -\frac{24g^4 N_fN_c}{(4\pi)^2}\frac{1}{\epsilon} , ~~~ \delta_\msbar m_{0\sigma}^2 = 0, ~~~ \delta_\msbar h = 0.
\end{align}
The symbol $\delta_\msbar$ denotes the counter terms are determined by the $\overline{\rm MS}$ scheme
and we write the remaining finite part as $\delta_{\rm fin}$;
\begin{align}
	\delta = \delta_\msbar + \delta_{\rm fin}.
\end{align}
Here we should notice that the $\delta_\msbar$ counter terms can also be determined
by the renormalization conditions discussed below
with the consistent treatment of the one-loop correction.

After determining the infinite part of the counter terms,
we can write the renormalized one-loop effective potential as
\begin{align}
	V_{\rm 1-loop} = V_0 + \frac{N_fN_c}{(4\pi)^2}\qty[\frac{3}{2} + \ln\frac{\Lambda^2}{M_q^2}]M_q^4 + \delta_{\rm fin} V_q.
\end{align}

Rest of the counter terms are determined by the on-mass shell renormalization.
We perform one-loop calculation neglecting the meson loop contribution under the large $N_c$ approximation.
First we consider the quark propagator.
Quark self-energy has the scalar and vector part as $\Sigma_\psi = \slashed{p}\Sigma_V(p^2) + \Sigma_S(p^2)$,
and they satisfy the renormalization condition as follows.
\begin{align}
	\Sigma_V(p^2 = M_q^2) + \delta Z_\psi &= 0 \\
	\Sigma_S(p^2 = M_q^2) + \delta g_\sigma \frac{M_q}{g} &= 0.
\end{align}
One loop contributions of quark self energy come from only meson loop
and they are suppressed by $1/N_c$.
Therefore we conclude $\delta Z_\psi, \delta g_\sigma \to 0$.

This means the nine counter terms and four relations are remained.
To determine the remaining counter terms 
we apply the following renormalization conditions to the meson self-energy $\Sigma_\phi$ as
\begin{align}
	\Sigma_\phi(p^2 = m_\phi^2) = 0, ~~~ \rmi\left.\pdv{\Sigma_\phi}{p^2}\right|_{p^2 = m_\phi^2} - \delta Z_\phi= 0.
\end{align}
The first condition is the on-mass shell condition and the second one is the wave function renormalization.
In addition 
we apply the renormalization condition to the effective potential
\begin{align}
	\left.\pdv{V_{\rm 1-loop}}{M_q}\right|_{M_q = M_q^{\rm vac} = gf_\pi} = 0
	\label{eq:reno-condition-gap}
\end{align}
which requires the gap equation to reproduce the tree-level relation for $M_q$ in the vacuum.
With these five renormalization condition
we can determine the remaining counter terms and obtain the renormalized effective potential.

In the following we calculate the meson self-energy
and apply the renormalization conditions.
As mentioned in Sec.$\,$\ref{sec:thermo-effective-potential},
we have to compute only the 1PI diagrams for the meson self-energy for renormalization.

In the end of the renormalization procedure,
we require that the renormalized effective potential concludes the tree-level relation for the quark mass in the vacuum $M_q^{\rm vac} = gf_\pi$.
With this treatment
we will attach the symbol ``vac'' to all the quark mass.

\begin{figure}[thpb]
	\centering
	\includegraphics[width=0.7\textwidth]{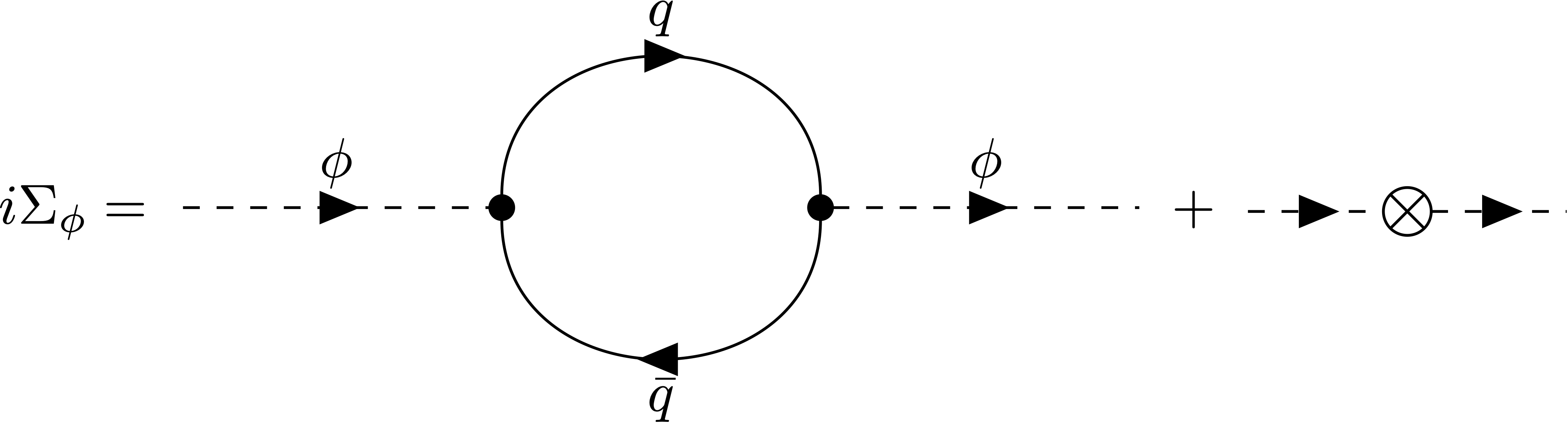}
	\caption{
		One-loop 1PI diagrams for the self-energy of the mesons.
		The vertex of the meson-quark interaction is $-ig$ for $\phi = \sigma$ and $g\gamma_5\tau_i$ for $\phi = \pi_i$.
		The second term is the counter term for the meson mass in the presence of the chiral condensate $\braket{\sigma}$.
	}
	\label{fig:remo-1pi}
\end{figure}

The 1PI one-loop diagrams are shown in \figref{fig:remo-1pi}
which are written as
\begin{align}
	\rmi\Sigma_{\sigma}(p^2) &= \rmi\bar\Sigma_{\sigma}(p^2) + \qty[\delta m_{0\sigma}
	^2  + \frac{\delta\lambda_{4\sigma}}{2}\qty(\frac{M_q}{g})^2] - \delta Z_\sigma m_\sigma^2\\
	\rmi\Sigma_{\pi}(p^2) &= \rmi\bar\Sigma_{\pi}(p^2) + \qty[\delta m_{0\pi}^2 + \frac{\delta\lambda_{2\sigma\pi}}{6}\qty(\frac{M_q}{g})^2] - \delta Z_\pi m_\pi^2, \\
	\rmi\bar\Sigma_{\sigma}(p^2) &= - 8\rmi g^2N_c\qty[A - \frac{1}{2}(p^2 - 4{M_q^{\rm vac}}^2)B(p^2)] \\
	\rmi\bar\Sigma_{\pi}(p^2) &= - 8\rmi g^2 N_c\qty[A - \frac{1}{2}p^2 B(p^2)].
\end{align}

Here $A$ and $B(p^2)$ are given by
\begin{align}
	A &= \int_{\vp} \frac{1}{p^2 - {M_q^{\rm vac}}^2} \notag\\
	&= \frac{\rmi {M_q^{\rm vac}}^2}{(4\pi)^2}\qty[\frac{1}{\epsilon} + 1 + \ln \frac{\Lambda^2}{{M_q^{\rm vac}}^2} + \calO(\epsilon)] \\
	B(p^2) &= \int_{\vk} \frac{1}{(k^2 - {M_q^{\rm vac}}^2)}\frac{1}{((k+p)^2 - {M_q^{\rm vac}}^2)} \notag\\
	&= \frac{\rmi}{(4\pi)^2}\qty[\frac{1}{\epsilon} + \int_0^1 dx \ln\frac{\Lambda^2}{x(1-x)p^2 - {M_q^{\rm vac}}^2} + \calO(\epsilon)] \notag\\
	&= \frac{\rmi}{(4\pi)^2}\qty[\frac{1}{\epsilon} + \ln\frac{\Lambda^2}{{M_q^{\rm vac}}^2} + F(p^2) + \calO(\epsilon)] \\
	B'(p^2) &= \frac{\rmi}{(4\pi)^2}F'(p^2)
\end{align}
using the function $F(p^2)$ as 
\begin{align}
	F(p^2) &= - \int_0^1 dx \ln\qty[\frac{p^2}{{M_q^{\rm vac}}^2}x(x-1) - 1] = 2 - 2r\arctan\qty(\frac{1}{r}) \\
	p^2 F'(p^2) &= \frac{r^2+1}{r}\atan\qty(\frac{1}{r}) - 1
\end{align}
where $r = \sqrt{4{M_q^{\rm vac}}^2/p^2 - 1}$.

The renormalization conditions for the meson propagator are given as
\begin{align}
	\rmi\Sigma_\sigma (p^2 = m_\sigma^2) = 0, ~~~ \rmi\Sigma_\pi (p^2 = m_\pi^2) = 0 
	\label{eq:reno-condition-pole}
\end{align}
and
\begin{align}
	\rmi\left.\pdv{\Sigma_\sigma}{p^2}\right|_{p^2 = m_\sigma^2} - \delta Z_\sigma = 0 & : 4\rmi g^2N_c \qty(B(m_\sigma^2) + (m_\sigma^2 - 4{M_q^{\rm vac}}^2)B'(m_\sigma^2)) - \delta Z_\sigma = 0 
	\label{eq:app-reno-residue-sigma}\\
	\rmi\left.\pdv{\Sigma_\pi}{p^2}\right|_{p^2 = m_\pi^2} - \delta Z_\pi = 0 & : 4\rmi g^2N_c \qty(B(m_\pi^2) + m_\pi^2 B'(m_\pi^2)) - \delta Z_\pi = 0
	\label{eq:app-reno-residue-pi}.
\end{align}
These condition determine all the remaining counter terms.

Using $M_q^{\rm vac} = gf_\pi$ and the relation of counter terms in the beginning of this chapter,
the mass counter terms and $\lambda$ counter terms are determined as follows.
\begin{align}
	\delta m_\sigma^2 &= \frac{\rmi \bar\Sigma_\sigma - 3\rmi \bar\Sigma_\pi}{2} - \delta Z_\sigma \frac{m_\sigma^2 -3m_\pi^2}{2} \notag\\
	&= - \frac{2g^2N_c}{(4\pi)^2}\qty[4{M_q^{\rm vac}}^2 + (3m_\pi^2 - 4{M_q^{\rm vac}}^2)F(m_\sigma^2) - 3m_\pi^2 F(m_\pi^2) - (m_\sigma^2 - 4{M_q^{\rm vac}}^2)(m_\sigma^2 - 3m_\pi^2)F'(m_\sigma^2)] \\
	\delta m_\pi^2 &= \frac{\rmi \bar\Sigma_\sigma - 3\rmi \bar\Sigma_\pi}{2} - \delta Z_\pi \frac{m_\sigma^2 -3m_\pi^2}{2} \notag\\
	&= - \frac{2g^2N_c}{(4\pi)^2}\qty[4{M_q^{\rm vac}}^2 + (m_\sigma^2 - 4{M_q^{\rm vac}}^2)F(m_\sigma^2) - m_\sigma^2 F(m_\pi^2) - (m_\sigma^2 - 3m_\pi^2)m_\pi^2F'(m_\pi^2)] \\
	\delta \lambda_{4\sigma} &= -\frac{3}{f_\pi^2}\qty[\rmi \bar\Sigma_\sigma - \rmi \bar\Sigma_\pi - (m_\sigma^2 - m_\pi^2)\delta Z_\sigma] \notag\\
	&= - \frac{24g^4 N_cN_f}{(4\pi)^2} \qty[\frac{1}{\epsilon} + \ln\frac{\Lambda^2}{{M_q^{\rm vac}}^2}]\notag\\
	& \quad + \frac{12g^2N_c}{f_\pi^2 (4\pi)^2}\qty[(m_\pi^2 - 4{M_q^{\rm vac}}^2)F(m_\sigma^2) - (m_\sigma^2 - 4{M_q^{\rm vac}}^2)(m_\sigma^2 - m_\pi^2)F'(m_\sigma^2) - m_\pi^2F(m_\pi^2) ] \\
	\delta \lambda_{2\sigma\pi} &= -\frac{3}{f_\pi^2}\qty[\rmi \bar\Sigma_\sigma - \rmi \bar\Sigma_\pi - (m_\sigma^2 - m_\pi^2)\delta Z_\pi] \notag\\
	&= - \frac{24g^4 N_cN_f}{(4\pi)^2} \qty[\frac{1}{\epsilon} + \ln\frac{\Lambda^2}{{M_q^{\rm vac}}^2}]\notag\\
	&\quad + \frac{12g^2N_c}{f_\pi^2(4\pi)^2}\qty[(m_\sigma^2 - 4{M_q^{\rm vac}}^2)F(m_\sigma^2) - m_\sigma^2 F(m_\pi^2) - (m_\sigma^2 - m_\pi^2)m_\pi^2 F'(m_\pi^2)] \\
	\delta \lambda_{4\pi} &= - \frac{3}{f_\pi^2}\qty[\rmi \bar\Sigma_\sigma - \rmi \bar\Sigma_\pi + (m_\sigma^2 - m_\pi^2)\qty(\delta Z_\sigma - 2\delta Z_\pi)] \notag\\
	&= - \frac{24g^4 N_cN_f}{(4\pi)^2} \qty[\frac{1}{\epsilon} + \ln\frac{\Lambda^2}{{M_q^{\rm vac}}^2}]\notag\\
	&\quad + \frac{12g^2N_c}{f_\pi^2(4\pi)^2}\left[(2m_\sigma^2 - m_\pi^2 - 4{M_q^{\rm vac}}^2)F(m_\sigma^2) + (m_\sigma^2 - 4{M_q^{\rm vac}}^2)(m_\sigma^2 - m_\pi^2)F'(m_\sigma^2) \right.\notag\\
	&\quad\quad \left.+ (-2m_\sigma^2 + m_\pi^2)F(m_\pi^2) - 2(m_\sigma^2 - m_\pi^2)m_\pi^2 F'(m_\pi^2)\right].
\end{align}

The remaining factor $\delta h$ is determined to reproduce the vacuum relation $M_q = gf_\pi$
in the final step of the renormalization.

\section{Renormalization in medium}
\label{sec:reno-medium}
In this section we extend the renormalization procedure to the medium.
As all the counter terms are fixed in a vacuum,
the finial task is to apply these results to the effective potential in the medium.

The one-loop effective potential before the renormalization is given by
\begin{align}
	V_q = - N_c N_f \int_{\vp} \qty[\sqrt{\qty(E_D - \mu_I)^2 + \Delta^2} + \sqrt{\qty(E_D + \mu_I)^2 + \Delta^2}]
\end{align}
and its divergent part is extracted as
\begin{align}
	V_q^{\rm div} = - 2N_c  N_f \int_\vp \qty[\sqrt{E_D^2 + \Delta^2} + \frac{\mu_I^2 \Delta^2}{2(E_D^2 + \Delta^2)^{3/2}}].
\end{align}
These terms can be evaluated by the dimensional regularization as
\begin{align}
	V_q^{\rm div} &= - \frac{N_c N_f}{(4\pi)^2}\qty[\qty((M_q^2+\Delta^2)^2 - 4\mu_I^2\Delta^2)\qty(\frac{1}{\epsilon} + \ln\frac{\Lambda^2}{M_q^2+\Delta^2}) + \frac{3}{2}\qty(M_q^2+\Delta^2)^2 + \calO(\epsilon)].
\end{align}
Finally,
the effective potential in the medium is given by
\begin{align}
	V_{\rm 1-loop} = V_0 + V_q^R + \delta_{\rm fin} V_q
\end{align}
where $V_q^R$ and $\delta_{\rm fin} V_q$ are
\begin{align}
	V_q^R &= \qty(V_q - V_q^{\rm div}) + \qty(V_q^{\rm div} + \delta_\msbar V_q)\\
	\delta_{\rm fin} V_q &= \frac{\delta_{\rm fin} Z_\pi}{2g^2}(-4\mu_I^2 \Delta^2) + \frac{1}{2g^2}\qty(\delta_{\rm fin} m_\sigma^2 M_q^2 + \delta_{\rm fin} m_\pi^2 \Delta^2) \notag\\
	& \quad + \frac{1}{24g^4}\qty(\delta_{\rm fin} \lambda_{4\sigma}M_q^4 + \delta_{\rm fin} \lambda_{4\pi}\Delta^2 + 2\delta_{\rm fin}\lambda_{2\sigma\pi}M_q^2\Delta^2) - \frac{\delta_{\rm fin} h}{g}M_q.
\end{align}

In this expression the effective potential does not have original $\rmO(4)$ symmetry even when $h = 0$.
To restore the symmetry
we rescale the field condensates as
\begin{align}
	M_q \to (1 + \delta X)M_q, ~~~ \Delta \to (1 + \delta Y)\Delta
\end{align}
with order $\calO(\hbar)$ factor $\delta X$ and $\delta Y$.
The conditions for restoring the symmetry are
\begin{gather}
	\delta_{\rm fin} m_\sigma^2 + m_0^2 \delta X = \delta_{\rm fin} m_\pi^2 + m_0^2 \delta Y, \\
	\delta_{\rm fin} \lambda_{4\sigma} + 2\lambda\delta X = \delta_{\rm fin} \lambda + 2\lambda \delta Y = \delta_{\rm fin} \lambda + \lambda \delta X + \lambda \delta Y.
\end{gather}
This condition is valid as long as the following relations for the mass and $\lambda$ counter terms are satisfied;
\begin{align}
	\frac{\delta_{\rm fin}\lambda_{4\pi} - \delta_{\rm fin}\lambda_{4\sigma}}{\lambda} = -2\frac{\delta_{\rm fin} m_\sigma^2 - \delta_{\rm fin} m_\pi^2}{m_0^2} \\
	\frac{\delta_{\rm fin}\lambda_{2\sigma\pi} - \delta_{\rm fin}\lambda_{4\sigma}}{\lambda} = -\frac{\delta_{\rm fin} m_\sigma^2 - \delta_{\rm fin} m_\pi^2}{m_0^2}.
\end{align}
These conditions are sustained
and we can restore the original symmetry of the effective potential.

Choosing the scale factor $\delta X,~ \delta Y$ to reproduce the previous results \cite{Adhikari:2018aa},
we obtain
\begin{align}
	\delta X &= 0, \\
	\delta Y &= \frac{4g^2N_c}{(4\pi)^2}\qty[- F(m_\sigma^2) - (m_\sigma^2 - 4{M_q^{\rm vac}}^2)F'(m_\sigma^2) + F(m_\pi^2) + m_\pi^2F'(m_\pi^2)].
\end{align}
The first condition means that the $\sigma$ field is not rescaled.
Applying these scaling for the field condensates
and solve the gap equation at vacuum to determine the factor $h$ for explicit breaking term $ -\frac{h}{g}M_q$ in the effective potential,
we get the final expression of the effective potential as following.
\begin{align}
	V_{\rm 1-loop} &= - \frac{1}{4}m_\sigma^2f_\pi^2 \qty[1 + \frac{4{M_q^{\rm vac}}^2N_c}{(4\pi)^2f_\pi^2}\qty{ - \frac{4{M_q^{\rm vac}}^2}{m_\sigma^2} F(m_\sigma^2) + \frac{4{M_q^{\rm vac}}^2}{m_\sigma^2} - (m_\sigma^2 - 4{M_q^{\rm vac}}^2)F'(m_\sigma^2)}]\frac{M_q^2+\Delta^2}{{M_q^{\rm vac}}^2} \notag\\
	&\quad + \frac{3}{4}m_\pi^2f_\pi^2 \qty[1 - \frac{4{M_q^{\rm vac}}^2N_c}{(4\pi)^2f_\pi^2}\bigg\{ - F(m_\pi^2) + F(m_\sigma^2) + (m_\sigma^2 - 4{M_q^{\rm vac}}^2)F'(m_\sigma^2)\bigg\}]\frac{M_q^2+\Delta^2}{{M_q^{\rm vac}}^2} \notag\\
	&\quad - 2\mu_I^2 f_\pi^2\qty[1 - \frac{4{M_q^{\rm vac}}^2N_c}{(4\pi)^2f_\pi^2}\qty(\ln\frac{M_q^2 + \Delta^2}{{M_q^{\rm vac}}^2} + F(m_\sigma^2) + (m_\sigma^2 - 4{M_q^{\rm vac}}^2)F'(m_\sigma^2))]\frac{\Delta^2}{{M_q^{\rm vac}}^2} \notag\\
	&\quad + \frac{1}{8}m_\sigma^2f_\pi^2 \left[1 - \frac{4{M_q^{\rm vac}}^2N_c}{(4\pi)^2f_\pi^2}\left\{\frac{4{M_q^{\rm vac}}^2}{m_\sigma^2}\qty(\ln\frac{M_q^2}{{M_q^{\rm vac}}^2} - \frac{3}{2}) \right.\right. \notag\\
	&\hspace{3cm} \left.\left. + \frac{4{M_q^{\rm vac}}^2}{m_\sigma^2}F(m_\sigma^2) + (m_\sigma^2 -4 {M_q^{\rm vac}}^2)F'(m_\sigma^2))\right\}\right] \frac{(M_q^2+\Delta^2)^2}{{M_q^{\rm vac}}^4} \notag\\
	&\quad - \frac{1}{8}m_\pi^2f_\pi^2\qty[1 - \frac{4{M_q^{\rm vac}}^2N_c}{(4\pi)^2f_\pi^2}\bigg\{- F(m_\pi^2) + F(m_\sigma^2) + (m_\sigma^2 - 4{M_q^{\rm vac}}^2)F'(m_\sigma^2)\bigg\}] \frac{(M_q^2+\Delta^2)^2}{{M_q^{\rm vac}}^4} \notag\\
	&\quad - m_\pi^2f_\pi^2\qty[1 - \frac{4{M_q^{\rm vac}}^2N_c}{(4\pi)^2f_\pi^2}\bigg\{F(m_\sigma^2) + (m_\sigma^2 - 4{M_q^{\rm vac}}^2)F'(m_\sigma^2)\bigg\}]\frac{M_q}{M_q^{\rm vac}} \notag\\
	&\quad -2N_c\int_p \left[\sqrt{\left(\sqrt{p^2+M_q^2}+\mu\right)^2+\Delta^2}+\sqrt{\left(\sqrt{p^2+M_q^2}-\mu\right)^2+\Delta^2}\right] \notag\\
	&\quad +4N_c\int_p\left[\sqrt{p^2+M_q^2+\Delta^2}+\frac{\mu^2\Delta^2}{2(p^2+M_q^2+\Delta^2)^{3/2}}\right] \,.
\end{align}